\documentclass[preprint,12pt]{elsarticle}
\usepackage[a4paper, total={7in, 9in}]{geometry}

\usepackage{lineno}
\usepackage[    T1]{fontenc}
\usepackage[utf8]{inputenc}
\usepackage[english]{babel}
\usepackage{amssymb,amsmath,amsthm,mathtools,bbm,bm,amsfonts}
\usepackage[]{graphicx}
\usepackage[]{subfigure}
\usepackage{tensor}
\usepackage[usenames,dvipsnames]{xcolor}
\usepackage{cancel}
\usepackage{setspace}
\usepackage{fancyhdr}
\usepackage[bookmarks,linktocpage, colorlinks=true, plainpages = false, citecolor = blue,  linkcolor=blue, urlcolor = blue, filecolor = blue]{hyperref} 
\usepackage{graphicx}
\usepackage[normalem]{ulem}
\usepackage{natbib}

 \usepackage{array}
\usepackage{longtable}

\newcommand*{\bra}[1]{\langle #1\rvert}
\newcommand*{\ket}[1]{\lvert #1 \rangle}
\newcommand*{\braket}[2]{\langle #1 \lvert #2 \rangle}
\newcommand*{\ketbra}[2]{\lvert #1 \rangle\!\langle #2 \rvert}

\def\tr           {\mathop{\rm tr}}
\newcommand{\LParen}{ \bm{(} }
\newcommand{\RParen}{ \bm{)} }
\newcommand{\tar}{{\rm T}}  
\def\id{\mathbbm{1}}   
\newcommand{\refer}{{\rm R}}
\newcommand{\be}{\begin{equation}}
\newcommand{\ee}{\end{equation}}
\newcommand{\bea}{\begin{eqnarray}}
\newcommand{\eea}{\end{eqnarray}}
\newcommand{\Tr}{\text{Tr}}
\newcommand{\cev}[1]{\reflectbox{\ensuremath{\vec{\reflectbox{\ensuremath{#1}}}}}}
\usepackage{enumitem,kantlipsum}

\modulolinenumbers[5]

\biboptions{numbers,sort&compress}

\newcommand{\eg}{e.g.,\,}
\newcommand{\ie}{i.e.,\,}

\begin{document}

\begin{frontmatter}

\title{Quantum complexity in gravity,  quantum field theory, \\ 
and quantum information science}

\author[PG,BS]{Stefano Baiguera}
\ead{stefano.baiguera@pg.infn.it}

\author[UPEN,SantaFe,VUB,Oxford]{Vijay Balasubramanian}
\ead{vijay@physics.upenn.edu}

\author[StockholmU,YITP,Warsaw]{Pawel Caputa}\ead{pawel.caputa@fysik.su.se}

\author[BGU]{Shira Chapman}\ead{schapman@bgu.ac.il}

\author[Cambridge,Saarland]{Jonas Haferkamp}\ead{jhaferkamp42@gmail.com}

\author[MHaddress]{Michal P. Heller}
\ead{michal.p.heller@ugent.be}

\author[UoM1,UoM2]{Nicole Yunger Halpern}\ead{nicoleyh@umd.edu}

\address[PG]{INFN Sezione di Perugia, Via A. Pascoli, 06123 Perugia, Italy}
\address[BS]{Dipartimento di Matematica e Fisica, UCSC, Via della Garzetta 48, 25133 Brescia, Italy} 
\address[UPEN]{David Rittenhouse Laboratory, University of Pennsylvania,
209 S. 33rd Street, Philadelphia, PA 19104, USA}
\address[SantaFe]{Santa Fe Institute,
1399 Hyde Park Road, Santa Fe, NM 87501, USA}
\address[VUB]{Theoretische Natuurkunde, Vrije Universiteit Brussel,
Pleinlaan 2, B-1050 Brussels, Belgium}
\address[Oxford]{Rudolf Peierls Centre for Theoretical Physics, University of Oxford,Oxford OX1 3PU, UK}
\address[StockholmU]{The Oscar Klein Centre and Department of Physics, Stockholm University, AlbaNova, 106 91 Stockholm, Sweden}
\address[YITP]{Yukawa Institute for Theoretical Physics, Kyoto University, Kitashirakawa Oiwakecho, Sakyo-ku, Kyoto 606-8502, Japan}
\address[Warsaw]{Faculty of Physics, University of Warsaw, Pasteura 5, 02-093 Warsaw, Poland}
\address[BGU]{Department of Physics, Ben-Gurion University of the Negev, Beer-Sheva 84105, Israel}
\address[Cambridge]{School of Engineering and Applied Sciences, Harvard University, Cambridge, MA 02318, USA}
\address[Saarland]{Department of Mathematics, Saarland University, Saarbruecken, Germany}
\address[MHaddress]{Department of Physics and Astronomy,
 Ghent University, 9000 Gent, Belgium}
\address[UoM1]{Joint Center for Quantum Information and Computer Science, NIST/University of Maryland, College Park, Maryland 20742, USA}
\address[UoM2]{Institute for Physical Science and Technology, University of Maryland, College Park, MD 20742, USA}

\begin{abstract}
Quantum complexity quantifies the difficulty of preparing a state or implementing a unitary transformation with limited resources.  Applications range from quantum computation to condensed matter physics and quantum gravity. We seek to bridge the approaches of these fields, which define  and study complexity using different frameworks and tools. We describe several definitions of complexity, along with their key properties. In quantum information theory, we focus on complexity growth in random quantum circuits. In quantum many-body systems and quantum field theory (QFT), we discuss a geometric definition of complexity in terms of geodesics on the unitary group.  In dynamical systems, we explore a definition of complexity in terms of state or operator spreading, as well as concepts from tensor-networks.  We also outline applications to simple quantum systems, quantum many-body models, and QFTs including conformal field theories (CFTs). Finally, we explain the proposed relationship between complexity and gravitational observables within the holographic anti-de Sitter (AdS)/CFT correspondence.
\end{abstract}

\end{frontmatter}

\begin{flushright}
Preprint number: YITP-25-39
\end{flushright}

\tableofcontents

\sloppy 

\section*{}

\section{Preface}

The goal of this review is to facilitate the construction of new bridges  between different fields of physics that study dynamical phenomena through the lens of quantum complexity.  These fields define and use different notions of complexity, which require distinct methods and tools for analysis.  Many of these  developments are individually reviewed in \cite{Susskind:2018pmk,Chapman:2021jbh,Nandy:2024htc}.
We hope to offer a complementary, integrative perspective at the intersection of  gravity, quantum information theory, quantum field theory (QFT), and quantum many-body systems.  This is a fast moving field and our review captures the state of the art in early 2025.

Historically, the study of quantum complexity originated in the field of quantum computation, generalizing ideas from classical computation.  The basic question was, ``How difficult is it to solve instances of a particular problem class?''   Here, difficulty was measured in terms of a scarce resource like the minimal amount of time,  space, or  number of circuit gates necessary to implement an  algorithm to solve problems exactly, approximately, or even just with high probability. Defined in this way, the concept of complexity functioned as a mathematical tool guiding the design of optimal problem-solving algorithms or circuits, or in the language of quantum mechanics, unitary transformations.  Given the arbitrariness inherent in any computational model, the output of complexity theory was usually given in the form of bounds on the minimal amount of resource required to solve general instances of the problem class, as a function of some measure of the problem size.  Many related questions have been asked in quantum information theory.  For example, how many local operations are required to build a given entangled state of $N$ spins starting with a product state?  The answer will depend quantitatively on the ``amount'' and ``kind'' of entanglement in the target state, both quantified by information theoretic measures.  Local operations are privileged in such analyses because physical processes act locally, and nonlocal coordination or structures must be constructed from local ones.  Many of these notions are reviewed in \cite{Aaronson:2016vto}.

Over the last two decades, these sorts of questions have also turned out to be relevant for quantum gravity.  This is because, in the ``holographic'' duality between $d+1$-dimensional quantum gravity in Anti-de Sitter (AdS) spacetimes and $d$-dimensional Conformal Field Theories (CFTs), entanglement patterns in the CFT state are geometrically encoded in the dual gravity theory \cite{Ryu:2006bv}.\footnote{While the AdS/CFT is the most well-developed duality between a gravitational and a non-gravitational
theory, other proposals have also been discussed in the literature, including different proposals for de Sitter holography (which we refer to in section \ref{ssec:generalization_holo_complexity}) see \eg the review \cite{Galante:2023uyf}, celestial holography, see \eg the review \cite{Raclariu:2021zjz} and Matrix model of M theory, see \eg the reviews \cite{Brahma:2022ikl,Lin:2025iir} among others.} Matrix product and tensor network constructions of quantum states provide cartoon models for visualizing  the holographic encoding of quantum information, with coarser structure built up deeper in the network and finer structure at the periphery. Questions about the structure, dynamical construction, and time evolution of entangled quantum states can then also be geometrized.  We will discuss aspects of the rich literature on this subject as appropriate.

Within this holographic quantum gravity setting, important impetus for thinking about ``complexity'' as a physical quantity came from the observation that some aspects of geometry on the gravity side of the AdS/CFT duality could not possibly  be related to quantum information or entanglement-related quantities in the CFT \cite{Susskind:2014moa}.  Specifically, it turns out that the volumes of spatial sections of  black hole spacetimes keep increasing through to times exponential in the black hole entropy, far after the  timescale in the dual theory where all entanglement related quantities have  equilibrated.  Now, if we think of time evolution in a physical theory as a quantum computation enacted by a local Hamiltonian, then the complexity of that computation, understood for example as the size of the minimal local circuit that could perform it, can increase for exponential time.  This suggested the idea that some notion of quantum complexity may be geometrized as the volume of spatial section in the AdS gravities dual to CFTs  -- see Chapter 6 of the review \cite{Aaronson:2016vto} -- and seeded many investigations of alternative notions of complexity in physical systems.  The resulting ideas have been applied to an enormous range of problems ranging from thermalization in bounded systems, to quantum information in many-body physics, to quantum chaos and integrability, and to the quantization of gravity.   Some of these ideas are also shedding light on foundational problems such as the origin of the entropy of black hole horizons, and whether information is lost during black hole evaporation. Some solutions to these problems suggest that black holes appear to have a horizon and to destroy information simply because the fine structure required to decode the information is inaccessible to simple probes (older approaches are discussed in \cite{Balasubramanian:2011dm} and new developments are reviewed in \cite{Almheiri:2020cfm}).   The new tools that are now available to define and measure quantum complexity may enable us to study such ideas quantitatively.

\begin{figure}[ht]
    \centering
    \includegraphics[scale=0.9]{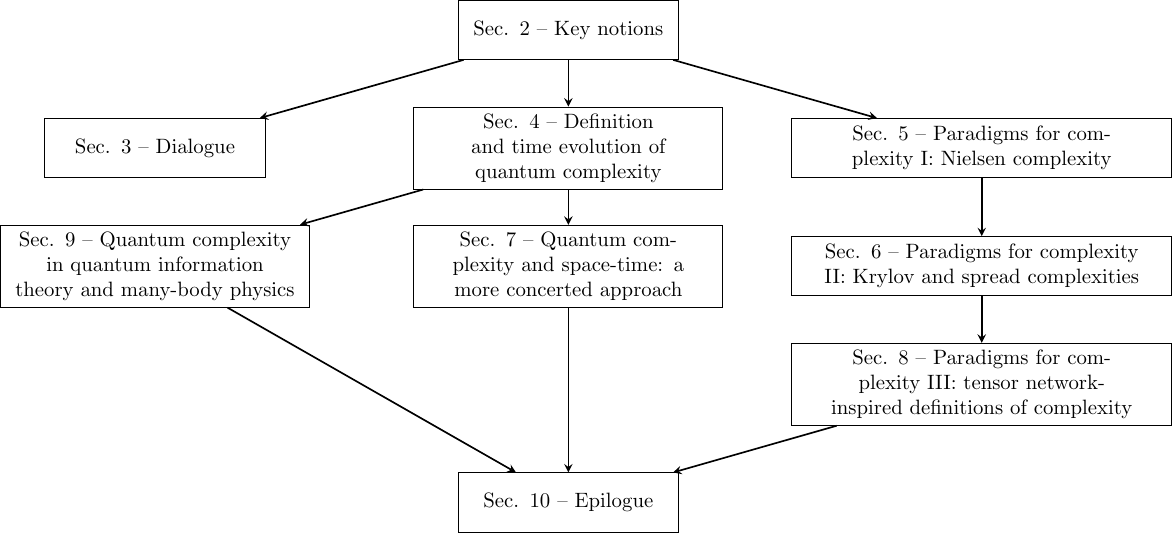}
    \caption{Structure of the review. }
    \label{fig:preface}
    \end{figure}

Figure~\ref{fig:preface} depicts the review's structure; the arrows indicate possible paths through the material.  All paths begin in  Sec.~\ref{sec:key_notions}, a glossary for key concepts. A reader may study this section more or less thoroughly, depending on their background, and return to the section later as necessary. Let us now outline the paths consecutively.

The leftmost arrow in Fig.~\ref{fig:preface} points to Sec.~\ref{sec:dialogue}, a  dialogue inspired by Galilei's
famous discourse on cosmology. The dialogue is modeled on discussions at a conference: not all terms are sharply defined, and different speakers present different perspectives. Later sections provide greater rigor and  detail. The dialogue terminates the first, briefest path through the review. 

The second path, marked by the second-leftmost sequence of arrows in Fig.~\ref{fig:preface}, focuses on quantum information theoretic features of  complexity. This path defines circuit complexity in Sec.~\ref{sec:what_is_quantum_complexity}. Section~\ref{sec:qi} presents applications of complexity to quantum information theory and many-body physics.

The third path, marked by the second-rightmost sequence of arrows in Fig.~\ref{fig:preface}, considers time evolution of complexity. Section~\ref{sec:what_is_quantum_complexity} uses counting arguments to analyze the effects of Hamiltonian dynamics and of random circuits transforming a state.  Some of this material has a bearing on proposed geometric duals of quantum complexity in the correspondence between quantum theories of gravity in asymptotically AdS spacetimes and Conformal Field Theories, as discussed in Sec.~\ref{sec:holography}.

The right-hand side of Fig.~\ref{fig:preface} illustrates the fourth path through the review. It focuses on quantum complexity definitions that are suited to continuous-time evolution.   Section~\ref{ssec:Nielsen_complexity} introduces the idea that complexity of time evolution can be measured in terms of lengths of geodesic paths on the unitary group from the identity to the time evolution operator.   Sec.~\ref{sec:Krylov} introduces approaches that measure complexity of dynamics in terms of the spread of operators or states over time. Sec.~\ref{sec:complexity_QFT} introduces approaches inspired by tensor networks.  

Some of the main results and open problems are summarized in another dialogue in Sec.~\ref{sec:conclusions}.  \ref{app_Acronyms} contains a guide to acronyms  used in this review.   

Many ideas about complexity in physical systems that are being studied today have antecedents in the classic works of Landauer, Bennet, Zurek and other pioneers of the application of information and computation theory to physics.  Interested readers can learn more in the collection \cite{zurek2018complexity} where questions like ``What is complexity?'',  ``What causes its increase?'', and ``Is there a limit to its increase?'' are discussed along with topics of current interest like the onset of chaos and the formation of complex states from simple initial conditions.

\clearpage
\section{Key notions}
\label{sec:key_notions}

Below we survey  vocabularies from high-energy physics and quantum gravity through many-body physics and quantum information science to assist readers from different backgrounds.

\begin{enumerate}
 \item \textbf{AdS/CFT correspondence:} A duality between gravity in asymptotically anti-de Sitter (AdS) spacetime and a conformal field theory (CFT) on the spacetime  boundary.
 \item \textbf{Black hole:} A  spacetime region bounded by a horizon (see below) that conceals a singularity where certain scalar functions of the  curvature diverge.
Mathematically, black holes are classical solutions to Einstein's equations or to another theory of gravity.  
 \item \textbf{Duality:} 
An equivalence between two seemingly distinct descriptions of the same physics. 
\item \textbf{Entanglement:} Two degrees of freedom in a pure quantum state are {\it entangled} if their wavefunction does not factorize. If they are in a mixed state, they are considered entangled if their density matrix is not separable. This relationship can lead to the measurement of correlations stronger than any achievable classically. Entangled states store information inaccessible via measurements on individual  parties.   The \emph{von Neumann} or \emph{entanglement} entropy quantifies entanglement between   systems.
\item \textbf{Event horizon:} A null hypersurface which bounds a region of trapped surfaces.  Matter and light cannot escape to infinity through the event horizon.
\item \label{item:fast_scrambler}
\textbf{Fast scrambler:} A quantum system that quickly distributes information across degrees of freedom.  Generic OTOCs (see  below) of fast scramblers decay rapidly at early times:
\begin{equation}
\text{OTOC}(t) \sim (\text{const.}) - \frac{ (\text{const.}) }{N} \,
e^{ \lambda_{\rm OTOC} \, t } + \mathcal{O} (N^{-2}) \, .
\label{eq:general_OTOC}
\end{equation}
where $N$ is the number of degrees of freedom and $t$ is time. $\lambda_{\rm OTOC}$ is defined via semiclassical expansion in a small parameter like $\hbar$ or $1/N$. At the \emph{scrambling time} $t_*\sim \lambda_{\rm OTOC}^{-1}\log (N)$, the first two terms in~\eqref{eq:general_OTOC} are equal.  Black holes and the large-$N$ SYK model are believed to be examples of {\it fast scramblers}, with the largest possible  $\lambda_{\rm OTOC}= 2\pi/\beta$~\cite{Maldacena:2015waa}.

\item \textbf{Holography:} A duality stating that gravitational physics in a bulk region is equivalent to a quantum theory on the region's boundary.  A {\it holographic map} between bulk and boundary quantities identifies bulk quantities with the dual boundary description.

\item \textbf{Jackiw-Teitelboim (JT) gravity:} 
A two-dimensional gravitational theory coupled to a dilaton (a real scalar field $\Phi$)~\cite{Mertens:2022irh}. 
Let $G_{\rm N}$ be Newton's constant; $\mathcal{M}$, a $2$-dimensional smooth manifold; $g$, the metric's determinant; $R$, the Ricci scalar; $L$, the AdS radius; $K$, the extrinsic curvature of the induced metric on the manifold's boundary, $\partial \mathcal{M}$; and $h$, the determinant of that induced metric.
The JT-gravity action, excluding topological contributions, is:
\begin{equation}
    S_{\rm JT} [g, \Phi] 
    \coloneqq  \frac{1}{16 \pi G_{\rm N} } \int_{\mathcal{M}} d^2 x \, \sqrt{-g} \: \Phi \left( R + 
    \frac{2}{L^2}  \right) 
    + \frac{1}{8 \pi G_{\rm N} } \int_{\partial \mathcal{M}} dx \, \sqrt{-h} \: \Phi \left( K - 
    \frac{1}{L}  \right) \, .
\end{equation}
AdS/CFT duality relates  JT gravity  and the low-energy SYK model.
\item \textbf{Kolmogorov complexity:}
The length of the shortest  program  producing a given output.
\item \textbf{Out-of-time-order correlator (OTOC):}
    A correlation function measuring the spread of many-body entanglement.
     Let $S$ be a quantum many-body system, and
     let $W$ and $V$ denote local unitary or Hermitian operators  that act on distant subsystems of $S$. Let $W(t)$ be the Heisenberg-picture time-evolved operator $W$ and let $\rho$ denote a state of $S$. The OTOC \cite{Maldacena:2015waa}\footnote{     The OTOC often diverges in QFT. One therefore \emph{regularizes} it by removing the $\rho$ from the trace's argument. Instead, most commonly, one places a $\rho^{1/4}$ after each remaining factor. Alternative regularization prescriptions differ in the power of $\rho$ and in  operator ordering.
     }
     \begin{align}
       F(t)=  {\rm Tr} \left(  W^\dag(t) \,  V^\dag \,  W(t) \,  V \, \rho \right) \, 
       \label{eq:OTOC_Maldacena}
     \end{align}
 quantifies the failure of $W(t)$ to commute with $V$, or how much a local perturbation $V$ later affects a far-away operator $W$. Texts often refer implicitly to the real part of the OTOC.
\item \textbf{Quantum chaos:} 
The definition of quantum chaos is a subject of debate. According to an early definition, a quantum-chaotic system has a semiclassical limit that is classically chaotic. Since then, quantum chaos has been defined in terms of nonintegrability, energy-level statistics, out-of-time-ordered correlators, spectral form factors and adiabatic gauge potential~\cite{Haake_10_Quantum,DAlessio_16_From,Gogolin_16_Equilibration,Pandey:2020tru}.
In the remainder of this review, we will argue that Nielsen (Sec.~\ref{ssec:Nielsen_complexity}) and Krylov/spread (Sec.~\ref{sec:Krylov}) complexities can also be used as diagnostic tools of chaos.
\item \textbf{Quantum circuit:} A sequence of quantum gates (see key notion~\ref{item:quantum_gate}), often represented with time, discretized into \emph{layers},
running from left to right. 
Multiple gates can act in each layer (if they commute  and if at most one gate acts on each qubit).  The number of layers is the circuit's \emph{depth}.
\item \textbf{Quantum complexity:} The least number of quantum gates (see key notion~\ref{item:quantum_gate} for the definition of quantum gate), selected from a chosen set required to (i) implement a target unitary transformation or (ii) prepare a target state from a given initial one. High-energy theorists sometimes call this quantity \textit{quantum computational complexity}, which can have a different meaning in computer science.
In sections providing \ analytic bounds, we will emphasize differences between the exact circuit complexity $\mathcal{C}_0(U)$ of a unitary $U$,
and the approximate circuit complexity $\mathcal{C}_{\delta}(U)$, the minimal number of gates to approximate $U$ up to  error $\delta$.
When this distinction is unimportant, we drop the subscript.
\item \label{item:quantum_gate} \textbf{Quantum gate:} An elementary unitary operation that acts on a quantum system, usually a set of qubits. A \emph{$k$-local quantum gate} acts on just $k$ qubits. Example 1-qubit gates include the Hadamard gate; and example 2-qubit gates include the controlled-NOT, or CNOT, gate~\cite{nielsen2010quantum}.
\item \textbf{Random state:} A state selected from a Hilbert space according to the \emph{Haar measure}. This measure is unitarily invariant; so (Haar-)random states are ``uniformly randomly''.
\item \textbf{Sachdev-Ye-Kitaev (SYK) model:}
An exactly solvable model of $N$ Majorana fermions $\psi_{i}$, where $i\in\{1,\ldots N\}$.
Let $J_{i_1 i_2 \dots i_q}$ denote the strength of a random coupling among an even number $q$ of fermions.
The SYK Hamiltonian has the form
\begin{equation}
H_{\rm SYK} \coloneqq i^{q/2} \sum_{1 \leq i_1 < i_2 \dots < i_q \leq N} 
J_{i_1 i_2 \dots i_q} \psi_{i_1} \psi_{i_2} \dots \psi_{i_q} \, .
\end{equation}
The SYK model admits simplifications if we take $N\rightarrow\infty$ and $q\rightarrow\infty$ with the ratio $\lambda_{\text{SYK}}=2q^2/N$ fixed.  This limit is often referred to as double scaled SYK (DSSYK).
\item \textbf{Tensor networks:} 
A graphical representation of quantum states~\cite{Cirac:2020obd} built from multilinear maps called \emph{tensors}. 
A tensor's inputs and outputs, labeled by indices, represent  degrees of freedom. Tensor networks are evaluated by \emph{contracting}  tensors by summing over indices.
Tensor networks assist variational studies of ground states, and can be used to bound entanglement entropies in a simple, graphical manner. One example, the \emph{multiscale entanglement renormalization ansatz} (MERA)~\cite{Vidal:2007hda},  efficiently approximates ground states of scale-invariant  systems. A continuous version (cMERA)  approximates  ground states of free CFTs~\cite{Haegeman:2011uy}.
\item \textbf{Thermofield-double state:}
Consider a quantum system with Hamiltonian $H$, and energy eigenstates $| n \rangle$ with energies $E_n$.
The thermofield-double state
\begin{equation}\label{eq:mytfd}
    | \mathrm{TFD} \rangle \coloneqq \frac{1}{\sqrt{Z_{\beta}}}  \sum_n e^{- \frac{\beta E_n}{2}} | n \rangle_1 | n^* \rangle_2 \, ,
\end{equation}
is defined on two copies of the system, where $Z_{\beta}$  is the partition function. In the last equation we defined $| n^* \rangle\equiv\Theta |n\rangle$ with $\Theta$ an anti-unitary operator, such as CPT, that commutes with the 
Hamiltonian, see \eg \cite{Cottrell:2018ash}. The conjugation is sometimes kept implicit in order to simplify the notation. The reduced theory on each copy is in a canonical, finite-temperature  state.

\item \textbf{Universal gate set:} A set of quantum gates  that  can be used to approximately effect every unitary operator arbitrarily precisely.
\end{enumerate}
\clearpage
\begin{figure}[ht]
\centering
\includegraphics
[scale=1]
{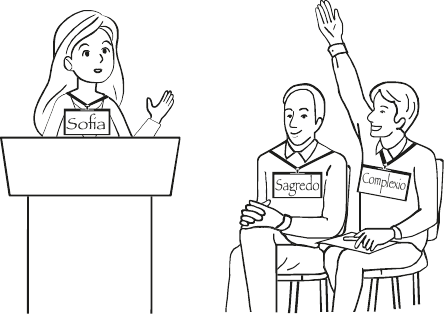}
\label{fig:CartoonDialogue}
\end{figure}

\section{Dialogue}
\label{sec:dialogue}

\noindent \textit{Sofia} is leading a discussion about complexity. \textit{Complexio} is an  audience member who seeks to clarify ideas in the literature,  sometimes with technical questions. \textit{Sagredo} mediates.

\begin{flushleft}
\textit{To the discerning reader}    
\end{flushleft}

\textit{Sagredo:}
Finally, we resolved to meet today to clarify the past decade's  developments concerning complexity.

\textit{Complexio:}
Sofia, can you clarify the definition of quantum complexity?

\textit{Sofia:}
Quantum complexity is defined in terms of a gate set and a reference state. You can think of the set as containing the gates implementable by some experimentalist. The reference state should be simple, or computationally easy to prepare. Common reference states include tensor products, especially a product of $\ket{0}$s. A state's quantum complexity is the least number of gates, drawn from that set, in any circuit that prepares the state from the reference.

\textit{Complexio:}
I have heard computer scientists invoke \emph{computational complexity} when discussing a classical or quantum algorithm. Is computer scientists' computational complexity the same as quantum complexity---which I believe is also called \emph{quantum computational complexity}, confusingly?

\textit{Sofia:}
The two concepts are easily conflated. Let us use just the term \emph{quantum complexity,} rather than \emph{quantum computational complexity,} to avoid confusion with computer scientists' terminology. Computational complexity is a concept applied in both classical and quantum computer science. It is the number of steps in an algorithm. In contrast, quantum complexity is the least number of steps in \emph{any} algorithm that prepares a target state or implements a target unitary.

\textit{Complexio:}
Why should I care about quantum complexity? 

\textit{Sofia:}
Complexity has diverse applications.
One can use it to characterize the dynamics of a quantum system and thermalization, solve optimization problems, capture chaotic properties, describe quantum information scrambling\ldots
Shall I continue?

\textit{Sagredo:}
These applications sound interesting, but complexity as a concept seems intangible. In most cases, I would think that it is difficult or impossible to compute.

\textit{Sofia:}
Fair objection.
Nevertheless, it is possible to make arguments that reveal how a state's quantum complexity changes under generic state evolution. Also, bounding complexity is easier than calculating it, and bounds suffice for many purposes.

\textit{Sagredo:}
Sofia, even finding bounds---other than the simple bounds known already---seems difficult. How can we make progress?

\textit{Sofia:}
At least two approaches can improve current bounds on complexity. One applies to random unitary circuits (see Sec.~\ref{sec_growth}-\ref{subsec_grwothofapproximate}), which are a toy model for chaotic dynamics. Invoking random unitary circuits, we can apply differential topology and algebraic geometry to manifolds formed by unitary transformations.  Also, when considering random unitary circuits, we can use unitary designs---probability distributions that cannot be distinguished from the Haar measure in polynomial time. 

The second approach (see Sec.~\ref{ssec:Nielsen_complexity}) centers on Nielsen's notion of complexity. Suppose we want to perform some unitary transformation on $n$ qubits. Many circuits implement that transformation. Now observe that the unitary group $\mathrm{SU}(2^n)$, equipped with a continuous norm, forms a differential manifold. Nielsen considers geodesic paths on this manifold from the identity to the transformation of interest. The geodesic is computed with respect to a ``complexity metric'' that specifies that distances are small in ``simple'' directions on the manifold and large in ``complex'' directions.  Here ``simple'' can be defined in whatever way is appropriate to the problem at hand -- for example, we might say that an operation is ``simple'' if it acts on less than $k$ qubits. The geodesic length in the complexity metric bounds the quantum complexity of the unitary transformation \cite{nielsen2010quantum,Brown:2022phc}. 
This bound can be useful, but may not be tight.

\textit{Sagredo:}
I have heard of other notions of complexity, such as Kolmogorov complexity. Can you define them?

\textit{Sofia:}
Certainly! Kolmogorov complexity is the length of the shortest computer program that  
produces a 
target output \cite{Kolmogorov}. Another notion, 
Krylov complexity, quantifies operator growth---an operator's spread across the space of operators during evolution in the Heisenberg picture~\cite{Parker:2018yvk}.
Similarly, spread complexity quantifies the expansion of states across the Hilbert space during time evolution \cite{Balasubramanian:2022tpr}. 

\textit{Complexio:}
Does Krylov or spread complexity offer any advantage over the other notions of complexity?

\textit{Sofia:}
Both have interesting physical applications and are relatively simple to calculate.  Krylov complexity was introduced to distinguish chaotic from integrable dynamics of finite-size quantum many-body systems that lack semiclassical limits~\cite{Parker:2018yvk}.
Spread complexity relates to spectral form factors, 
which characterize the energy spectra of many-body quantum-systems~\cite{Balasubramanian:2022tpr}, and provides tools for quantifying late time dynamics.  
To calculate the Krylov or spread complexity, we must tridiagonalize the Hamiltonian and there are  efficient numerical \cite{Parker:2018yvk,Balasubramanian:2022tpr} and analytical \cite{Balasubramanian:2022dnj,Balasubramanian:2023kwd} techniques for doing so.
Therefore, 
one can calculate these complexities relatively easily.

\textit{Complexio:}
Interesting story. However, quantum complexity still seems\ldots complex. Complicated. I wish there were a simple way to understand and apply it.

\textit{Sagredo:}
I have heard that the holographic duality is an equivalence between gravitational physics in a bulk geometry and a quantum theory defined on its boundary. Holographic duals often simplify difficult quantum problems.
Does quantum complexity have a holographic interpretation?

\textit{Sofia:}
We hope so! Quantum complexity has been proposed to have an Einstein-Rosen bridge's (ERB's) volume as a holographic dual. An ERB is a spacetime region that connects a maximally extended  Schwarzschild black hole's two boundaries. In the holographic context, the black hole is in an asymptotically Anti-de Sitter (AdS) spacetime and is dual to the thermofield-double state of a Conformal Field Theory (CFT) on the spacetime boundary.  The ERB's volume was proposed to be a holographic dual to the boundary state's quantum complexity \cite{Susskind:2014moa,Stanford:2014jda}.
More-complicated geometric quantities were conjectured to be dual to quantum complexity, too \cite{Brown:2015bva,Brown:2015lvg,Couch:2016exn,Belin:2021bga,Belin:2022xmt}.

\textit{Complexio:}
Why should I expect the volume to be related to the boundary state's complexity?
Does any evidence support the conjecture? 

\textit{Sofia:} 
Yes. The geometric quantities behave similarly to the complexity of a quantum state evolving under random dynamics.
One important behavior is linear growth until late times.  This is achieved by the proposed dual geometric quantities in a semiclassical gravitational setting. Another behavior is the switchback effect, in which a perturbation delays growth. 

More evidence emerges from a class of two-dimensional bulk models. There, a notion of wormhole length was formulated nonperturbatively. This length saturates at times exponentially long in the black hole's entropy~\cite{Iliesiu:2021ari} (unlike the classical theory where the length grows forever). This late time behavior mirrors the expected saturation of complexity at times exponentially long in the system size \cite{Aaronson:2016vto}.
Moreover, in~\cite{Berkooz:2018qkz,Lin:2022rbf,Rabinovici:2023yex,Balasubramanian:2024lqk}, the authors show that a state's spread complexity precisely matches the dual wormhole's classical volume at early times while also saturating at late times, at least in a two-dimensional gravitational setting.  What is more, in general dimensions and for any theory of gravity, the saturation of the wormhole's size at exponential values must be enforced by the finiteness of the black hole's entropy, as explained in \cite{Balasubramanian:2022gmo,Balasubramanian:2022lnw}.

\textit{Complexio:}
Are you sure that we are invoking the correct type of complexity?
How do you know if we should use Nielsen, Krylov, or another complexity?

\textit{Sofia:} 
The map I just mentioned---between a wormhole's volume and spread complexity---hints that spread complexity is appropriate. Yet Refs.~\cite{Caputa:2017urj,Caputa:2017yrh,Caputa:2018kdj,Chandra:2021kdv,Chagnet:2021uvi,Chandra:2022pgl,Erdmenger:2022lov,Arean:2024pzo,Caceres:2024edr} point to other notions of complexity. We are not yet certain.

\textit{Complexio:} The community sounds awfully uncertain about the correct notion of complexity in holographic contexts. Why bother working on the topic?

\textit{Sofia:} First, the gravitational and quantum settings would both merit analysis even in the absence of any duality.
We want to understand the structure of the interiors of black holes and to optimize quantum circuits. The notion of complexity has facilitated these goals. Second, we do hope to replace the many conjectures with a smaller number of theorems.

\textit{Complexio:}  
The community seems to me to be far from theorems. To support a bulk-boundary map for Nielsen's complexity, people constructed toy examples involving free QFTs. Yet free QFTs are not known to be dual to any gravitational system. Rather, certain strongly coupled QFTs are. So how can the toy examples support the map?

\textit{Sofia:}
We expect some qualitative properties of QFTs not to depend on the coupling strength.
An example is the structure of the ultraviolet divergences in the notion of complexity used in a QFT setting.
The UV divergence structure of the entanglement entropy offers another example.

\textit{Sagredo:}
Indeed, entanglement entropy provides a striking example of a holographic mapping that has passed important consistency checks. The dual to a boundary QFT state's entanglement entropy is the area of a minimal surface in 
the geometry dual to that state \cite{Ryu:2006bv}. The dual objects have been identified clearly in this example for various states.

\textit{Complexio:}
But I heard of a paper with the title ``Complexity=Anything'' which seemed to imply that different notions of complexity could match each of infinitely many gravitational quantities~\cite{Belin:2021bga,Belin:2022xmt,Jorstad:2023kmq}. Every such quantity exhibits linear growth and the switchback effect. But then, I can pick any conjecture I like, and all of them define a geometrical dual to quantum complexity.  Does this multiplicity make any sense? Is the cornucopia of such holographic conjectures useful in any way?

\textit{Sofia:}
Well, actually, as we discussed, quantum state complexity depends on a gate set and a reference state. We are free to choose each. This freedom mirrors the ambiguity in the complexity's gravitational dual.

Still, I understand the concern -- the catchphrase ``Complexity = Anything'' makes it sound like there is no concrete statement at all.  But as we discussed earlier, there is a precise relation between spread complexity and wormhole length at least in the case of two-dimensional gravity.  In terms of the ``Complexity = Anything'' idea, it would help to show that that the ambiguities in alternative definitions of complexity, at least within a single framework like circuit or spread complexity, precisely match the corresponding ambiguities on the gravitational side.  It could also be that some symmetries or other desirable properties could restrict the appropriate gravitational quantities to one geometric object, and in this case we would need a principle to select the corresponding notion of complexity.   One approach, tried in \cite{Belin:2018bpg}, was to use bulk symplectic forms to relate boundary quantities to circuit complexity.

\textit{Complexio:}
Most of the holographic complexity studies we just discussed were performed in the context of the AdS/CFT correspondence. Why have recent works focused on de Sitter spacetime? I thought that de Sitter spacetime's holographic properties were less understood than AdS spacetime's. 

\textit{Sofia:}
You are right; de Sitter holography is 
poorly understood. Nevertheless, 
it holds promise for understanding quantum gravity in expanding universes similar to ours. After all, we do not live in AdS spacetime. If a holographic complexity conjecture for de Sitter spacetime were correct, it could elucidate the quantum dial of such spacetimes.

\textit{Complexio:}
You mentioned earlier that the definition of quantum state complexity 
has ambiguities, and depends on a choice of gate set and a reference state. Does any guiding principle direct these choices?

\textit{Sofia:}
Yes: certain operations may be easier to implement, and certain states may be easier to prepare, than others in experiments.
I can illustrate with an example. 
We should penalize difficult-to-implement generators. 
Distributed computing offers an example of difficult operations: imagine performing a computation using multiple small quantum computers, or on widely separated qubits.  It is difficult to maintain quantum coherence across such separations.  It is easiest to keep the information processing coherent if one computer performs most of the computations on local qubits.  Any nonlocal operator, which acts on multiple nodes, deserves a large complexity penalty.  

In the alternative Krylov or spread complexity approaches, we can start with any operator or state.  But \cite{Balasubramanian:2022tpr} proved that, for a given initial state, these approaches yield a unique notion of complexity by minimizing the spread of the operator or state over all possible bases describing the evolving system.  In the case of spread complexity, different initial states (e.g., random states, thermal states, states localized on some eigenspace) can be used to probe different dynamical properties of the system.

\textit{Complexio:}
Interesting. I am excited to read whether we can simulate gravitational physics on a tabletop, using a quantum computer.

\textit{Sagredo:}
And let this be the final conclusion of our long discussion.
Above all, I shall impatiently read this review, to learn more about recent developments in complexity.

\clearpage
\section{Definition and time evolution of quantum complexity}
\label{sec:what_is_quantum_complexity}

A quantum circuit consists of gates. A gate is an elementary unitary operation that acts simultaneously on some (typically small) number of qubits (or other qudits---$d$-level quantum systems). Every $n$-qubit unitary operation can be implemented arbitrarily precisely, by selecting an appropriate combination of gates from a universal set~\cite{nielsen2010quantum}.
\textit{Quantum complexity} 
is heuristically defined as the minimum number of gates needed to perform a certain task. Examples include mapping an initial state to a final one. Experimentally implementing a gate costs a finite amount of time. Therefore, some quantum-complexity measures reflect the duration of the optimal algorithm for performing a task.

In this section, we present the characteristic features of the time evolution of quantum complexity under the action of a chaotic Hamiltonian. These features can be summarized as follows:
\begin{itemize}
    \item Complexity grows linearly 
    for a time exponential in the system's size.
    It saturates at an approximately constant value until a doubly exponential time.
    After that, Poincar\'{e} recurrences return the complexity to near-minimal values.
    
    \item The complexity exhibits a switchback effect: a small perturbation delays its linear growth by a scrambling time.
\end{itemize}

Sec.~\ref{subsec:counting} shows that one can predict the linear growth and switchback effect from simple counting arguments in a circuit model.
Section~\ref{sec_stages} discusses the saturation of complexity in the context of quantum many-body equilibration.
In Sec.~\ref{sec_growth}, we introduce \emph{random} circuits as simple models for chaotic dynamics. With increasing depth, such circuits mimic the properties of unitary transformations selected from the Haar measure.
This phenomenon can be exploited to make many of the heuristic arguments about the growth of complexity rigorous for random quantum circuits.
In Sec.~\ref{subsec:lineargrowth} and Sec.~\ref{subsec_grwothofapproximate}, we present two recent rigorous arguments that prove linear growth of quantum complexity for an exponentially long time.
In Sec.~\ref{subsec:barriersfortime-independent} we revisit the problem of proving linear growth of quantum circuit complexity for exponentially long time for the time-evolution of time-independent local Hamiltonians from the perspective of computational complexity theory. 
This perspective will show that significant progress in theoretical computer science is necessary to even prove superpolynomial (in the system's size) lower bounds on quantum complexity in this setting.

\subsection{Time evolution of complexity, according to counting arguments about quantum circuits}
\label{subsec:counting}

Consider a fast-scrambling quantum system evolving under a unitary transformation $U(t) \coloneqq e^{-iHt}$.
How does this unitary's complexity evolve in time? We answer this question with  a simple counting argument~\cite{Susskind:2014jwa,Brown:2016wib,Susskind:2018pmk,Chapman:2021jbh}.

\begin{figure}[ht]
\centering
\includegraphics[scale=0.62]{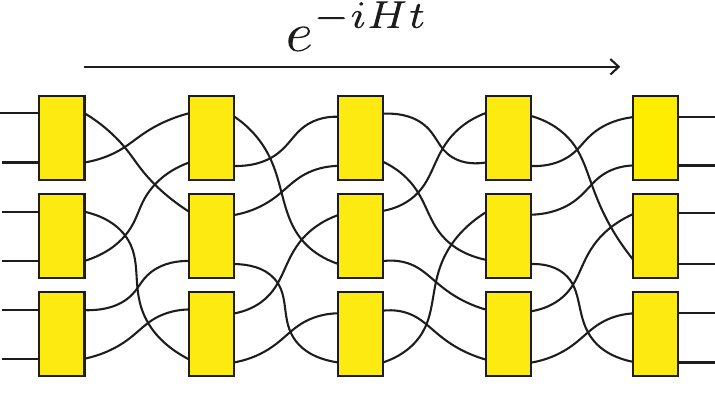}\hspace{50pt}
\caption{Illustration of the circuit that models unitary evolution following from a generic 2-local Hamiltonian.}
\label{fig:CircuitIllustration}
\end{figure}

To show how,  consider a system of $K\gg 1$ qubits.\footnote{We use qubits for concreteness; the argument extends to higher-dimensional qudits.}
Suppose that the Hamiltonian $H$ is fast-scrambling: information about initially localized perturbations spreads across the system in many-body entanglement quickly (in a time logarithmic in the system size~\cite{Susskind:2011ap,Lashkari_13_Towards}).
Also, suppose that $H$ is 2-local, effecting just 2-body interactions.\footnote{We just need the interaction to be $k$-local, for some $k \ll K$.  We are taking $k=2$ for concreteness.}
By discretizing time, we can regard the evolution as formed from two-qubit gates. At each time step, all the qubits are paired, and each pair is transformed by the action of a gate (Fig.~\ref{fig:CircuitIllustration}). 

What is the minimal number of gates needed to 
simulate evolution under $H$ for some time $t$?
If $t$ is, at most, exponentially large in $K$, the optimal circuit is believed to follow from discretizing the evolution under $H$. In other words, the discretized time evolution circuit likely does not contain gates that cancel each other, since the system is a fast scrambler.
Of course, such cancellations would prevent the circuit derived from the Hamiltonian from being the shortest.
Hence the complexity of a fast scrambler's $U(t)$ should evolve linearly in the number $n$ of circuit layers:
\begin{equation}\label{eq:linearcounting}
    \mathcal{C}  = K n/2 \, .
\end{equation}
This is easily turned into an equation for the complexity evolution as a function of time, since $n=t/\ell$ with $\ell$ the characteristic time scale for the application of a layer in the circuit.\footnote{In Sec.~\ref{subsec_grwothofapproximate} we introduce $T$ for a similar quantity: the number of layers in a 1D random quantum circuit in a brickwork layout.} A precise mathematical derivation of Eq.~\eqref{eq:linearcounting}
can be made in the context of random circuits (Sec.~\ref{subsec:lineargrowth}). 

\begin{figure}[ht]
\centering
\includegraphics[scale=1.1]{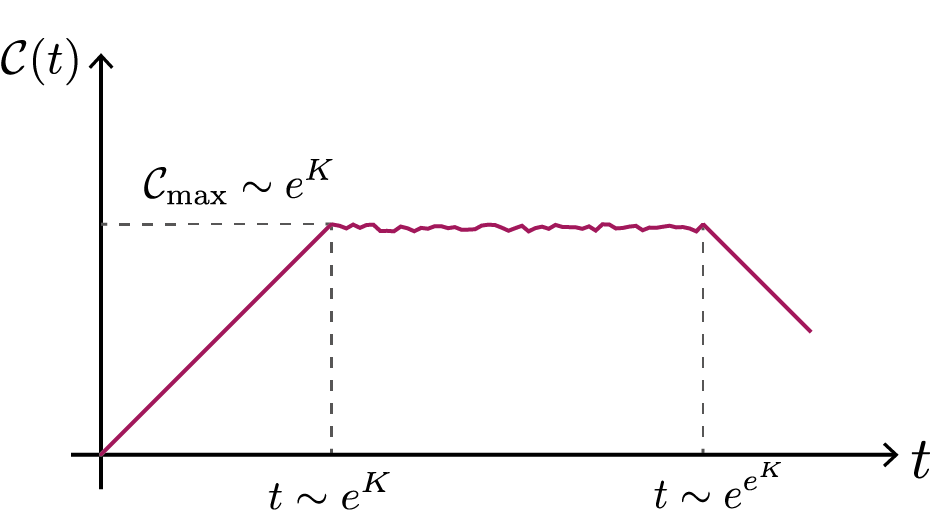}\hspace{50pt}
\caption{Illustration of the time evolution of complexity  for a generic 2-local fast-scrambling Hamiltonian.
}
\label{fig:TimeEvolGen}
\end{figure}

For how long does the unitary's complexity grow linearly?
Another counting argument answers this question, as shown in~\cite[Sec.~4.5.4]{nielsen2010quantum} and~\cite[Sec.~8]{susskind2018black}.
So far, we have reasoned about digitally simulating evolution under a 2-local Hamiltonian. Consider, instead, an arbitrary, finite, universal set of 2-qubit gates. Consider the set of all depth-$n$ circuits formed from those gates. Each circuit implements some unitary transformation.
The complexity of $U(t)$ stops growing linearly once the aforementioned circuits have explored most of the volume of the group $\mathrm{SU}(2^K)$.\footnote{This claim and the ones to follow are based on populating the group manifold with $\delta$ size balls, where $\delta$ is the tolerance. The following claims therefore hold when studying approximate circuit complexity $C_\delta$, see section \ref{sec:key_notions}. The exact derivation is reviewed in \cite{Chapman:2021jbh}.} 
This exploration is achieved by a time 
$t\sim \mathcal{O}(2^{2K})$. At this time, the complexity reaches a maximal value $\mathcal{C}_{\text{max}}\sim \mathcal{O}(2^{2K})$. The complexity is expected to then fluctuate around its maximal value.
After a time $\sim 2^{2^K}$, the complexity is expected to return to near-minimal value, in a Poincar\'{e} recurrence.
Figure~\ref{fig:TimeEvolGen} depicts these expected features of the evolution of complexity.

In the previous two paragraphs, we supposed that gates would not cancel each other's effects. If a circuit does contain gates that cancel, we call the cancellation a \emph{shortcut}: the circuit implements the same unitary as a shorter circuit that lacks the cancelled gates. The no-shortcut assumption also enables us to answer the question
\emph{how does complexity react to a perturbation?} 

Consider, again, the qubit system introduced above. Denote by $W$ a one-qubit unitary perturbation.
Its \emph{precursor operator} is
\begin{equation}
   W(t) \coloneqq U(t) W U(-t) \, .
\label{eq:precursor}
\end{equation}
Using this precursor, we can test how the system would behave if the perturbation $W$ acted an amount $t$ of time earlier.
What is the complexity of $W(t)$? If $W$ is the identity operator, then the precursor's complexity vanishes always: gates used to implement $U(t)$ always cancel gates used to implement $U(-t)$.

\begin{figure}[ht]
\centering
\includegraphics[scale=1.0]{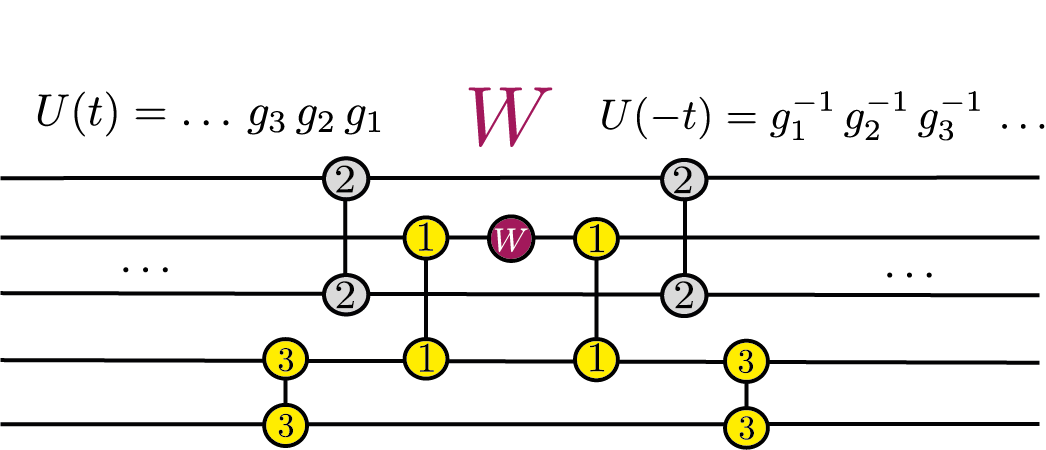}\hspace{50pt}
\caption{Circuit implementation of the precursor operator $W(t)$. Yellow gates are necessary for implementing $W(t)$. Gray gates are not necessary and cancel out, failing to appear in the optimal circuit.}
\label{fig:EpidemicCircuit}
\end{figure}

Gates cancel less if $W$ is nontrivial. Consider Fig.~\ref{fig:EpidemicCircuit}, a circuit that implements the precursor $W(t)$. The gates leftward of  $W$ represent the optimal construction of $U(t)$. That construction's first gate does not commute with the perturbation $W$. Therefore, gate 1 contributes to the complexity of $W(t)$. In contrast, gate 2 commutes with the perturbation. Therefore, the gate cancels between the circuit implementations of $U(t)$ and $U(-t)$. Hence gate 2 does not contribute to the complexity of $W(t)$. Like gate 2, gate 3 commutes with $W$. However, gate 3 does not (generally) commute with gate 1, since both act on the same qubits. Therefore, gate 3 likely contributes to the complexity of $W(t)$.

If a perturbation has influenced a given qubit (if the qubit is in the perturbation's light cone) by a given time, we call that qubit \emph{infected}. The number of infected qubits---the qubit \emph{epidemic}---grows with the circuit depth. Once 
all qubits are infected, we expect gates not to cancel any longer.
The epidemic crests at a time that we call the \emph{scrambling time}, $t_*$. After this time, as we will see, the complexity grows at twice the rate specified in Eq.~\eqref{eq:linearcounting}.

Again using counting arguments, we answer two questions:
\begin{enumerate}

   \item \label{item_s}
   How many qubits are infected at a given circuit layer? 
   
   \item \label{item_Ct}
   What is the precursor's complexity?    
\end{enumerate}

To answer these questions, we review and establish notation. As before, $\ell$ denotes a circuit layer's time duration. $s$ denotes the number of qubits infected after a time $t$, or after $n = t / \ell$ circuit layers.

Let us answer question~\ref{item_s}.
At any given layer, the $s$ infected qubits interact to some extent with the $K-s$ uninfected qubits. These interactions change the number of infected qubits by an amount $\Delta s$.
We expect on probabilistic grounds\footnote{At a given layer, all qubits are paired. The number of newly infected qubits is proportional to the number of previously uninfected qubits, times the probability that they are paired with infected qubits. A qubit can't be paired with itself; hence the $K-1$.}
\begin{equation}
    \label{eq_Delta_s}
    \Delta s = s \; \frac{K-s}{K-1} \, .
\end{equation}
We can turn this finite-difference equation into a differential equation for $ds/dt$: recall that the number of circuit layers is $n = t / \ell$. Therefore, $\Delta n = 1$, so $\Delta s = \frac{\Delta s}{\Delta n} = \ell \frac{\Delta s}{\Delta t}$. The infinitesimal version thereof is $\ell \frac{ds}{dt} = s \; \frac{K-s}{K-1}$, by the right-hand side of Eq.~\eqref{eq_Delta_s}.
We solve the differential equation, invoking the boundary condition $s(0)=1$ and neglecting small corrections of relative order $1/K$:
\begin{equation}
    \label{eq_st}
    s(t) = K \: \frac{e^{(t-t_*)/\ell}}{1+e^{(t-t_*)/\ell}} \, ,
\end{equation}
where we have introduced the scrambling time $t_* = \ell\log K$.

\begin{figure}[ht]
\centering
\includegraphics[scale=0.35]{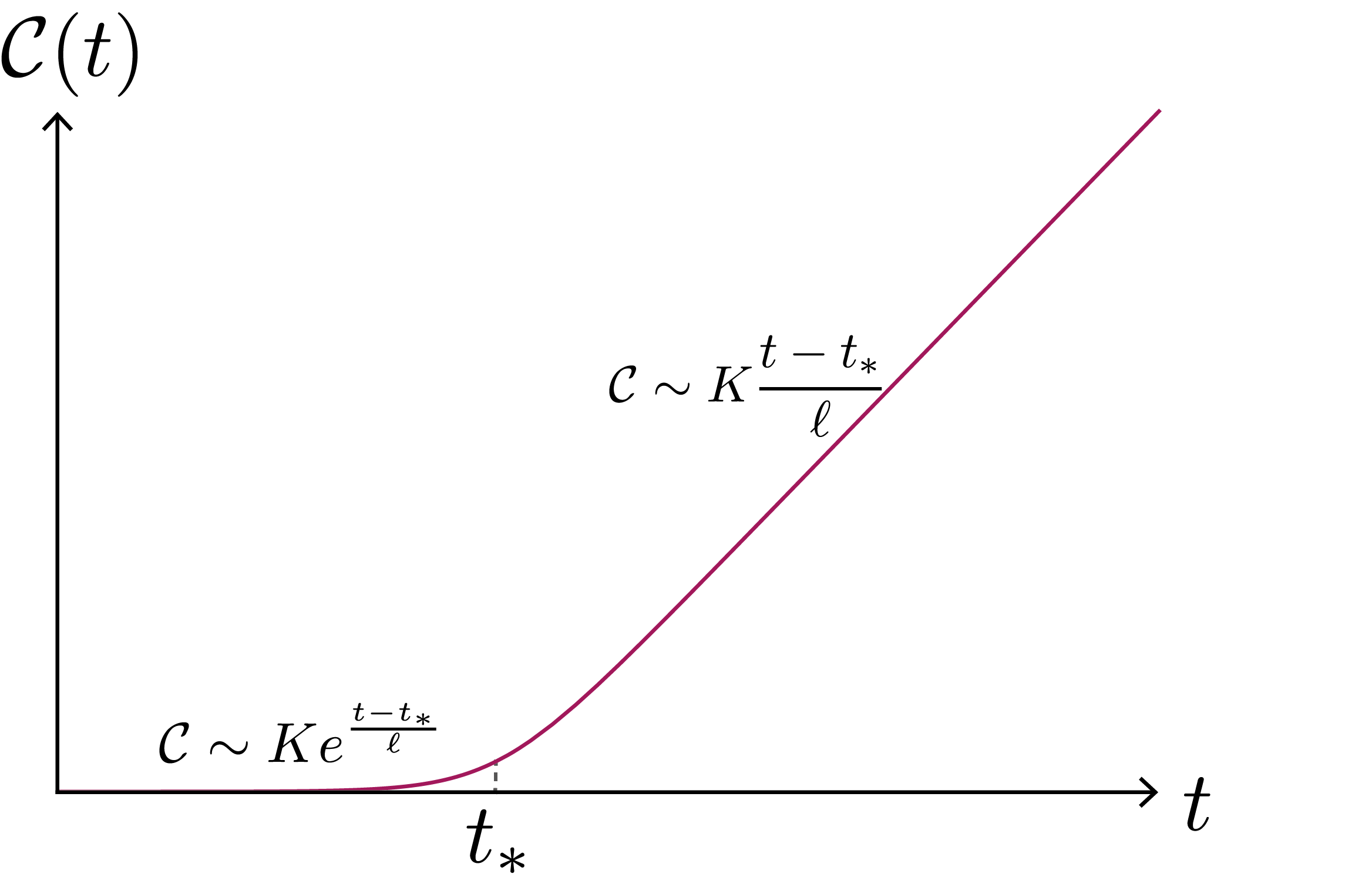}\hspace{50pt}
\caption{Precursor's expected complexity, as a function of time. The switchback effect is the delay, until $t \approx t_*$, in the complexity's linear growth.}

\label{fig:SwitchbackCt}
\end{figure}

Now, we can answer question~\ref{item_Ct}, calculating the precursor's complexity. The complexity---the number of gates---follows from how each infected qubit probably infects another qubit, via a gate, at each circuit layer $n'$. Layer $n'$ thereby appears to increase $W(t')$'s complexity by one-half the number $s(t')$ of qubits infected by (up until the end of) layer $n'$. However, if a gate appears on the left-hand side of $W(t')$ in Fig.~\ref{fig:EpidemicCircuit}, then the gate appears also on the right-hand side. Therefore, layer $n'$ actually increases $W(t)$'s complexity by $s(t')$.
To count the gates, therefore, we sum the number of qubits infected by the first time step, the number infected by the next time step, the number infected by the next time step, and so on, until the final circuit layer $n$. We convert this sum into an integral over the time $t$:
\begin{equation}
    \mathcal{C}(t) = \frac{1}{\ell} \int_0^t s(t') dt' \, .
\end{equation}
We substitute in from Eq.~\eqref{eq_st} and integrate:
\begin{equation}
    \mathcal{C}(t) = K \log(1+e^{(t-t_*)/\ell}) = \begin{cases}
K e^{(t-t_*)/\ell} & t\ll t_*\\
K(t-t_*)/\ell  & t\gg t_* 
\end{cases}  \, .
\label{eq:complexity_precursor}
\end{equation}
Initially, the complexity grows exponentially, as $e^{t / \ell}$. Yet the complexity is small throughout this stage, due to the 
$e^{-t_* / \ell}$ and the scrambling time's large size. Once $t$ is large enough, we say that the perturbation has \emph{scrambled}. Afterward, the complexity grows linearly, at twice the rate at which the unitary's complexity grows, because $W(t)$ contains two unitaries. Figure~\ref{fig:SwitchbackCt} illustrates the delay in the complexity's linear growth---the \emph{switchback effect}---encoded in the $-t_*$ in the second line of Eq.~\eqref{eq:complexity_precursor}.
Holographic complexity proposals reproduce the switchback effect and the initial exponential growth of complexity (see Sec.~\ref{ssec:evidence_holo_complexity}, particularly the discussion around Eq.~\eqref{eq:late_time_switchback}). This fact originally motivated the complexity conjectures discussed in Sec.~\ref{sec:holography}. 
An interesting question for future research is to investigate whether more formal arguments with random circuits can be used to reproduce the switchback effect.
The discussion above focused on a simple perturbation of one qubit at one instant, as well as on 2-qubit gates. These assumptions can be relaxed~\cite{Susskind:2014jwa,Brown:2016wib,Susskind:2018pmk,Chapman:2021jbh,Stanford:2014jda,Chapman:2018lsv}.

\subsection{Complexity saturation as a late-time stage of quantum many-body equilibration}
\label{sec_stages}

The second law of thermodynamics suggests that state complexity should grow under generic dynamics. Of course, the second law concerns entropy, not complexity, and originally described classical systems. Yet complexity seems like it ought to obey the spirit of the second law~\cite{Brown:2017jil}.  In fact, there is an analogy, with a late time saturation of complexity characterizing the final stage of quantum many-body equilibration.

To elaborate, we first review equilibration and thermalization. We say that a system has \emph{equilibrated} if it satisfies two necessary conditions: (i) Large-scale quantities, such as temperature and volume, remain constant. (ii) No net flow of anything, such as energy or particles, enters or leaves the system. We can also define the \emph{thermalization} of a system that exchanges only heat with a temperature $T$ environment as follows. Suppose that the system is classical, and consider measuring its energy. Let $p(E)$ denote the probability of obtaining the outcome $E$. The system thermalizes as 
$p(E) \to e^{-E / (k_{\rm B} T) } / Z_{\rm class}$. Here $k_{\rm B}$ is Boltzmann's constant, the partition function  $Z_{\rm class} \coloneqq  \int dE \, \mu(E) \, e^{-E / (k_{\rm B} T) }$ normalizes the distribution, and $\mu(E)$ denotes the density of states. Now, suppose that the system is quantum and evolves under a Hamiltonian $H$. The system thermalizes as its state (density operator) approaches 
$e^{-H / (k_{\rm B} T) } / Z_{\rm q}$, where
$Z_{\rm q} \coloneqq {\rm Tr} (e^{-H / (k_{\rm B} T) })$.
This canonical state is an example of a thermal state. Different thermal states arise if the system exchanges quantities $Q_\alpha$ other than heat with the environment; examples include particles. In the more general case, the thermal state becomes
$\propto e^{- \big( H - \sum_\alpha \mu_\alpha Q_\alpha \big) / (k_{\rm B} T) }$; the $\mu_\alpha$s denote effective chemical potentials.

To review the second law, we recall a classical monatomic 
gas in a box. Suppose that the gas particles are bunched together in a corner. The particles will spread across the box: the entropy of the gas, calculated with the Sackur-Tetrode equation, will end up greater than it was originally. 
The gas will equilibrate internally.
Furthermore, suppose that the particles interact. They will thermalize; a particle's phase-space distribution will approach the Maxwell--Boltzmann distribution.

Given the second law's ubiquity, one should expect generic quantum many-body systems to also equilibrate and thermalize internally. Quantum many-body systems undergo more stages of equilibration and thermalization than classical systems do. We will describe some stages, which end with complexity saturation. First, we discuss the kinds of systems that are expected to thermalize.

Which systems thermalize is difficult to predict and prove. Thermalization is often attributed to chaos, but there are many definitions of quantum chaos~\cite{Haake_10_Quantum,DAlessio_16_From,Gogolin_16_Equilibration}. 
According to early thinking, a system is quantum-chaotic if it has a chaotic semiclassical limit. Later studies defined chaos through energy-level statistics modeled by Wigner--Dyson distributions. Recent work has defined chaos through properties of out-of-time-ordered correlators~\cite{Swingle_18_Unscrambling} and spectral form factors~\cite{Dyson_62_Statistical,Leviandier_86_Fourier,Brezin_92_Spectral,Haake_10_Quantum}, reviewed below. 
Some authors acknowledge challenges of characterizing quantum chaos by writing ``chaos'' within quotation marks and taking the term to mean \emph{nonintegrability}~\cite{Murthy_22_Non}. Yet debate surrounds the definition of quantum nonintegrability, too. According to a common definition, a system is integrable if it has extensively many nontrivial conserved quantities~\cite{DAlessio_16_From}.\footnote{
Consider eigendecomposing an arbitrary quantum Hamiltonian: $H = \sum_j E_j \ketbra{j}{j}$. The Hamiltonian conserves every eigenprojector $\ketbra{j}{j}$. Furthermore, the number of eigenprojectors is exponentially large in the system's size. Yet these facts do not render $H$ integrable; if they did, every $H$ would be integrable. A conserved quantity must at least not be a simple sum of eigenprojectors in order to be considered nontrivial.}
To compound the ambiguity, many equilibration results follow from empirical observations, rather than from proofs, and govern specific systems, rather than general ones. We will therefore write ``chaotic'' within quotation marks in this subsection to communicate expectations about behaviors common to many systems called \emph{chaotic} or \emph{nonintegrable}.

In reviewing stages of quantum many-body equilibration, we refer to the following setup.
Consider a closed quantum system of $K \gg 1$ subsystems---for example, qubits. Let the Hamiltonian $H$ be ``chaotic''; and let the interactions be $k$-body, for some $k \in \{1, 2, \ldots, K\}$. Suppose that the system begins in a simple, pure nonequilibrium state, such as an $K$-qubit product state. 

We now define two quantities---the out-of-time-ordered correlator and spectral form factor---used to define stages of equilibration. Let $W$ and $V$ denote local, Hermitian or unitary operators localized far apart. Examples include Pauli operators that act on distant qubits. Define the Heisenberg-picture operator $W(t) \coloneqq e^{iHt} W e^{-iHt}$, and consider an arbitrary state $\rho$. The (four-point) out-of-time-ordered correlator is 
${\rm Tr} \LParen W^\dag(t) V^\dag W(t) V \rho \RParen$;
we discuss its significance below.
Define the partition function 
$Z(\beta) \coloneqq {\rm Tr} (e^{-\beta H} )$.
Taking $\beta, t \in \mathbb{R}$, then the spectral form factor is $| Z(\beta + i t) / Z(\beta) |^2$.
It measures correlations across a Hamiltonian's spectrum.

As time progresses, our paradigmatic system will likely pass the following mileposts:
\begin{enumerate}

   \item \textbf{Local equilibration}: Two-point correlators decay to $\sim 1 / e$ of their initial values. Time-ordered correlators relax to approximately their long-time values. Small subsystems thermalize. This equilibration occurs at the \emph{dissipation time} $t_{\rm D}$~\cite{Maldacena:2015waa}, also called the \emph{thermalization time}. The same timescale is called the \emph{collision time} if a system admits a quasiparticle description.
   Under certain conditions (e.g., if the interactions are highly nonlocal), $t_{\rm D}$ is constant in $K$, the number of degrees of freedom. Under other conditions, $t_{\rm D} \sim K$. For example, this latter scaling characterizes a one-dimensional system with nearest-neighbor interactions, by the Lieb--Robinson bound~\cite{Lieb_72_Finite}.

   \item \textbf{Quantum information scrambling}: Initially localized information spreads across the system through many-body entanglement. Out-of-time-ordered correlators decay from their $O(1)$ initial values~\cite{Swingle_18_Unscrambling}. This scrambling occurs around the \emph{scrambling time}, $t_*$. $t_*$ can be parametrically larger than $t_{\rm D}$, for example if there are all-to-all couplings~\cite{Lashkari_13_Towards}. A lower bound limits how quickly many-body entanglement can form:
   $t_* \geq \frac{\hbar}{2 \pi k_{\rm B} T} \, \log(K)$~\cite{Maldacena:2015waa}.
   This bound saturates only if $t_{\rm D}$ is constant.
   In the context of semiclassical, single-particle quantum chaos, an analogue of $t_*$ is called the \emph{Ehrenfest time}~\cite{Larkin_69_Quasiclassical}.

   \item \textbf{Breakdown of quantum--classical correspondence and rise in random-matrix theory's predictiveness}: This stage can begin around the same time as scrambling. Hence we have already seen one name for the time when this stage takes place: the \emph{Ehrenfest time} $t_{\rm E}$, also called the \emph{Thouless time} $t_{\rm Th}$. This time scale varies logarithmically with the system size:    $t_{\rm E} \sim \log(K)$.    How this stage manifests depends on the system and measures studied. An example  involves transport in electronic systems~\cite{Aleiner_96_Divergence,Schomerus_05_Quantum}: an initially narrow wave packet expands to cover a classically relevant length scale around $t_{\rm E}$. The quantum system's behavior diverges from its classical analogue; hence the name \emph{Ehrenfest time} alludes to Ehrenfest's theorem, which highlights a parallel between quantum and classical systems. In another example, at this stage, the spectral form factor stops declining and begins to ramp upward again: random-matrix theory begins to accurately predict the spectral form factor, which ceases to depend on microscopic details~\cite{Kos_18_Many,Winer_24_Spectral,81_Maslov_Semi}.\footnote{Certain semiclassical behaviors of some quantum systems may extend beyond the Ehrenfest time.~\cite{61_Maslov_Quasi,81_Chirikov_Dynamical,81_Shepelyanskii_Quasi}. This extension mirrors the ambiguities, described above, surrounding chaos. All these ambiguities motivated our formulation that the ``paradigmatic system will likely pass the following mileposts.''}
 
   \item \textbf{The spectral form factor changes from increasing linearly with time to remaining constant}: This change happens at the \emph{Heisenberg time} $t_{\rm H}$, or \emph{plateau time}~\cite{Kos_18_Many}. Denote by $\Delta$ the average inverse gap between consecutive eigenenergies. The Heisenberg time is $t_{\rm H} \coloneqq h / \Delta \sim e^K$~\cite{Schomerus_05_Quantum}.

   \item \textbf{Complexity saturation}: The state's complexity grows to $\sim e^K$. The saturation time has been conjectured to scale as $e^K$. This conjecture has been proven under certain conditions (Sec.~\ref{sec_growth}). 

\end{enumerate}

As a caveat: different authors use the same terms differently when referring to stages of quantum many-body equilibration.
Also, the above list is intended to be illustrative, not comprehensive; compiling all the stages~\cite{Gogolin_16_Equilibration,Hunter_18_Chaos,Bhattacharjee_24_Thermalization,Winer_24_Spectral,Nandy:2024htc,25_VallejoFabila_Timescales} would require another review. For example, subtleties in the spectral form factor define other stages; see Sec.~1.1 of~\cite{Hunter_18_Chaos}, and Sec.~1.4.3 of~\cite{Winer_24_Spectral}, for clear synopses and for references. Operator dynamics in Krylov space and the spread of wavefunctions require their own treatment~\cite{Nandy:2024htc,Parker:2018yvk,Balasubramanian:2022tpr}.
Additionally, if $K$ is finite, then revivals (Poincar\'e recurrences and complexity revivals) occur long after times $\sim e^K$. Such revivals temporarily undo the equilibration but are consistent with the second law for systems of finite size.

Nevertheless, the list above demonstrates two points: (i) Quantum many-body equilibration involves more stages than classical equilibration. (ii) Complexity saturation constitutes a late stage.

\subsection{Growth of complexity under random circuits}
\label{sec_growth}

Random circuits are a toy model for chaotic quantum dynamics.
We can prove random-circuit versions of conjectures originally formulated about the complexity of chaotic dynamics. When forming a random circuit, we choose the gates independently from a probability distribution on the 
unitary group, for example $\mathrm{SU}(4)$ if we have a four dimensional Hilbert space. Typically, we choose gates according to the Haar measure $\mu_{\rm H}$ on the unitary group, the unique distribution that is invariant under left- and right- multiplication by any unitary.
$\mu_{\rm H}$ generalizes the uniform distribution on an interval, 
as well as the Lebesgue measure, to (locally compact) Lie groups. $\mu_{\rm H}$ can be viewed as the uniform measure on these groups.

Known barriers such as the natural proof barrier discovered by Razborov and Rudich~\cite{razborov2004feasible,Gowersblog,arora2009computational} prevent  us from proving superpolynomial lower bounds on the circuit complexity of an explicit Boolean function.
The minimal number of elementary gates (such as AND, NAND, and XOR) in a Boolean function's implementation is called its circuit complexity.  A similar obstructions might inhibit us from proving superpolynomial lower bounds (in the system size) on the quantum complexity of quantum states. 
That is, the minimal number of $2$-local quantum gates required for construction of the target state from a similar reference. One can more easily prove lower bounds on the complexities of states prepared with random circuits as the latter quickly assume properties of the Haar measure,
counting arguments about which yield lower bounds on the circuit complexity.
In 1949 Claude Shannon~\cite{shannon1949synthesis} observed that the number of Boolean functions $f:\{0,1\}^n\to \{0,1\}$ outgrows the number of functions implementable with less than $\sim 2^n/n$ gates. 
A similar counting argument applies to Haar-random unitaries~\cite{preskill1998lecture}.

Sections~\ref{subsec:lineargrowth} and Section~\ref{subsec_grwothofapproximate} review two results about the growth of quantum circuit complexity.
Both arguments 
followed from adapting Haar-measure counting arguments to random circuits.  In a sense, these arguments offer to replace statements about vaguely defined  ``generic''  dynamics reviewed in Section \ref{subsec:counting} with well-defined statements about average dynamics.  These statements should apply to the dynamics of ``most'' systems and so should be ``generic'' in a well defined sense.  We will arrive at these results by thinking about the statistics/distribution of unitaries and  circuits, where  ``genericity'' acquires the sense of being ``very likely'' or ``with likelihood=1''.
Section~\ref{subsec:barriersfortime-independent} reviews 
why we still lack satisfying lower bounds on the quantum complexity of chaotic dynamics.

\subsection{Linear growth of exact circuit complexity}\label{subsec:lineargrowth}
Here, we review a proof of the linear growth of the exact circuit complexity~\cite{haferkamp2021linear} for random circuits.
More precisely, we show that the exact circuit complexity grows linearly with the number of layers, up to exponentially depth, with unit probability over the choice of circuits.
Moreover, the complexity saturates for exponentially deep circuits, again with unit probability.
The exact circuit complexity quantifies the minimal number of gates required to prepare a state or effect a unitary \textit{exactly}. The preparation and action of the unitary cannot include errors. Consequently, some unitaries arbitrarily close to the identity operator have exact circuit complexities of $e^{\Omega(K)}$,\footnote{In this section, we use big-Omega notation, $\Omega(X)$, to mean, roughly, ``a term that grows at least as quickly as $X$''.} where $K$ denotes the system size, or number of degrees of freedom (qubits).

We now sketch a proof of the conjecture~\cite{Brown:2017jil} that circuit complexity grows linearly for an exponentially long time, as applied to random circuits and the exact circuit complexity. We will compare the degrees of freedom in two sets of unitaries: the unitaries generated by the depth-$D$ circuits and the unitaries generated by smaller, depth-$D'$ circuits.
The basic idea echoes the fact that, in $\mathbb{R}^3$, a plane has no volume. In terms of a well-behaved probability measure on a manifold $M$, every lower-dimensional submanifold has a probability of $0$.
 Denote by $\mathcal{U}_R$ the set of all unitaries that can be generated by $R$ gates each.
Fortunately, $\mathcal{U}_R$, although not a manifold, turns out to be a semialgebraic set defined by constraints $\{f_i\geq 0, g_i >0, h_i=0\}$, where $f_i, g_i,h_i: {\rm SU}(D)\to \mathbb{R}$ are polynomials.
Every semialgebraic set 
has a well-defined dimension similar to a manifold's dimensionality.

We can apply the dimension-comparison logic above 
to random circuits: let $A\subset \mathcal{U}_R$ denote a semialgebraic set with $\dim (A)< \dim(\mathcal{U}_R)$.
If we draw elements of $\mathcal{U}_R$ from random circuits, 
 then elements of $A$ appear with probability $0$. 

One can lower-bound $\dim(\mathcal{U}_R)$ in multiple ways~\cite{haferkamp2021linear}.
Reference~\cite{li2022short} contains a particularly simple proof of this fact
for circuits in a brickwork architecture. The bound is $\sim R/K$, roughly the number of layers.
As a consequence, we find the following lower bound for the exact circuit complexity $\mathcal{C}_0(U)$ of a random quantum circuit. The minimal number of gates in an exact circuit implementation of $U$ is 
\begin{equation}
    \mathcal{C}_0(U)\geq \Omega(R/K) \, .
\end{equation}
with unit probability over the choice of random quantum circuit. Up to a factor $\sim 1/K$, this is exactly the behavior predicted by the informal counting argument in Sec.~\ref{subsec:counting}. Recall that heuristic arguments predict $\mathcal{C}_0(U)\geq nK/2$, where $n$ denotes the number of layers.
This result confirms the intuition that gates typically do not cancel in circuits.
In the next subsection, we discuss how to prove that even approximate cancellations have a low probability of occurring. The proof relies on unitary designs.

\subsection{Linear growth of approximate circuit complexity}\label{subsec_grwothofapproximate}

The simple argument of Sec.~\ref{subsec:lineargrowth} is fine-tuned to the exact circuit complexity. Using combinatorial methods, we can obtain similar results about more operational notions of circuit complexity such as the approximate circuit complexity: the minimal number of gates to approximate a state or unitary up to some fixed error. 
Below, we present such a combinatorial argument, using approximate unitary designs (defined below)~\cite{dankert2005efficient,gross2007evenly,low2010pseudo}. Unitary designs were first used to lower-bound quantum complexity in~\cite{brandao2021models,brandao2016local,Roberts:2016hpo}.

Unitary designs are probability distributions on the unitary group SU$(D)$. One cannot distinguish a unitary $t$-design from the Haar measure using only expectation values of degree-$t$ polynomials.
More formally, let $\nu$ denote any probability distribution on SU$(D)$. 
Denote by $\mathbb{E}_{U\sim\nu} f(U)$ the expectation value of a random variable $f:\mathrm{SU}(D)\to \mathbb{R}$, over unitaries drawn according to $\nu$.
We call $\nu$ a \emph{unitary $t$-design} if
\begin{equation}
    \underset{{U\sim\nu} }{\mathbb{E}} \LParen f(U,\overline{U}) \RParen =  \underset{{U\sim\mu_{\rm H}} }{\mathbb{E}} \LParen f(U,\overline{U}) \RParen \, ,
\end{equation}
for all balanced polynomials $f$ of degree $2t$.
\emph{Balanced} means that the $f$s are polynomials in the matrix elements of $U$ and $\overline{U}$ such that each monomial contains the same number of $U$ elements and $\overline{U}$ elements. 
This definition is equivalent to the operator equation
\begin{equation}\label{eq:definitiondesignsII}
    \underset{{U\sim\nu} }{\mathbb{E}}
    \left( U^{\otimes t}\otimes \overline{U}^{\otimes t} \right) 
    = \underset{{U\sim\mu_{\rm H}} }{\mathbb{E}}
    \left( U^{\otimes t}\otimes \overline{U}^{\otimes t} \right) .
\end{equation}
There are various definitions of \textit{approximate} unitary designs based on how to relax Eq.~\eqref{eq:definitiondesignsII}.

To see why unitary designs imply lower bounds on the approximate circuit complexity, we review why the circuit complexities of Haar-random unitaries are nearly maximal with the overwhelming probability $1-e^{-\Omega(4^K)}$. 
Consider a $K$-qubit system and an arbitrary $K$-qubit pure state $\ket{\psi}$. 
The $\varepsilon$-\emph{ball} $B_{\varepsilon}( \ket{\psi} )$ around $\ket{\psi}$ consists of the pure states $\ket{\phi}$ that differ from $\ket{\psi}$ in overlap by, at most, $\varepsilon$:  
$B_{\varepsilon}(\ket{\psi}) \coloneqq \{|\phi\rangle \in (\mathbb{C}^2)^{\otimes K}, \langle \phi|\phi\rangle =1, |\langle \psi|\phi\rangle|^2\geq 1-\varepsilon\}$. We must count the $\varepsilon$-balls that fit into SU$(2^K)$. The volume of a ball of radius $r$ in $\mathbb{R}^d$ grows as $\sim r^d$.
Similarly, the probability of drawing an element of the $\varepsilon$-ball $B_{\varepsilon}(\ket{\psi})$ from the Haar measure on $\{\ket{\psi}\in (\mathbb{C}^2)^{\otimes K}, \langle \psi|\psi\rangle=1\}$ is $\mathrm{Pr} \LParen B_{\varepsilon}(\ket{\psi}) \RParen \sim O(\varepsilon^{2^K})$.
Therefore, at least $2^{c\,2^K}$ (for some constant $c>0$ depending on $\varepsilon$) balls 
$B_{\varepsilon}(\ket{\psi})$ are needed to cover most of SU$(2^K)$. 
On the other hand, consider the circuits of $R$ gates chosen from a finite gate set $\mathcal{G}$. These circuits can implement $\leq |\mathcal{G}|^R$ unitaries. 
Therefore, the majority of the $2^{c\,2^K}$ $\varepsilon$-balls are outside the set of states that can be approximated with $o(2^K)$ gates. 
In other words, their approximate circuit complexity is exponential. 

The rough argument above applies to every ensemble of states: almost every state from any ensemble $\nu$ has a circuit complexity $\geq \max_{\psi} \log \big( \nu \LParen B_{\varepsilon}(\ket{\psi}) \RParen \big)$. 
Consider the ensemble of states generated by random circuits.
We can lower-bound 
almost every state's approximate circuit complexity
by upper-bounding $\max_{ \ket{\psi} } \{ \nu_K \LParen B_{\varepsilon}(\ket{\psi}) \RParen \}$.
The design property~\eqref{eq:definitiondesignsII} implies lower bounds on the approximate circuit complexity, via higher moments of the Haar measure~\cite{Roberts:2016hpo,brandao2021models,brandao2016local}.
By the above counting argument for Haar-random states, a unitary drawn from a design on SU$(2^K)$ has, with high probability, an approximate circuit complexity lower-bounded by $\Omega(Kt)$.

We can combine this bound with the quick convergence of random circuits to unitary designs.
Brown and Viola provided evidence for fast convergence by analyzing the spectra of moment operators via mean-field techniques~\cite{brown2010convergence}.
Reference~\cite{brandao2016local} then rigorously showed that brickwork random circuits of depth $T\geq C t^{9.5} \,[2Kt+\log_2(1/\varepsilon)]$\footnote{We use the parameter $T$ to denote  the number of layers in a brickwork layout. This parameter is similar to the parameter  $n$ presented in Sec.~\ref{subsec:counting}. This change in notation is to avoid confusion as $n$ is almost exclusively used for the system size in the context of quantum information theory.} are $\varepsilon$-\textit{approximate} unitary $t$-designs, for
a constant $C>0$ (whose form the authors calculated) and for all $t\leq O(2^{2K/5})$.
An approximate design is a probability distribution that relaxes the equality in Eq.~\eqref{eq:definitiondesignsII}.
In particular, the notion of approximation we get is surprisingly strong and implies relative errors (also called multiplicative errors) to Haar random unitaries for the outcome probabilities of any quantum experiment that queries the $t$ copies of $U$ in parallel.
See Refs.~\cite{brandao2016local,chen2024incompressibility} for details.
This relaxation of the design property is sufficient to imply a circuit lower bound of $\mathcal{C}_{\delta}(U)\geq \Omega(Kt)$ with probability $1-e^{-\Omega(Kt)}$.
In particular, the probability of significant short-cuts is small, not only in the system-size, but also in the number of copies $t$.
The exponent $10.5$ of $t$ in this bound 
was improved to $5+o(1)$ in~\cite{haferkamp2022random} and 
Ref.~\cite{chen2024incompressibility} reduced the $t$ dependence to an optimal, linear scaling.
More precisely, random circuits are $\varepsilon$-approximate unitary $t$-designs in depth
\begin{equation}\label{eq:designdepth}
    T\geq CK^3 \, [2Kt+\log_2(1/\varepsilon)] \, .
\end{equation}
We can now combine the design depth in Eq.~\eqref{eq:designdepth} with the lower bound of $\Omega(Kt)$ on the circuit complexity for unitaries drawn from a unitary $t$-design.
Random depth-$T$ circuits $U$ satisfy, with a probability $1-e^{-\Omega(T/K^3)} \, ,$
\begin{equation}\label{eq:complexitybound}
    \mathcal{C}_{\delta}(U)
    \geq \Omega(Kt)
    =\Omega \LParen K(T/K^4) \RParen
    =\Omega(T/K^3) \, .
\end{equation}
$\mathcal{C}_{\delta}(U)$ denotes the minimal number of $2$-local unitaries necessary to generate any $V$ that is $\delta$-close to $U$ in operator-norm: $||U-V||\leq \delta$.
Notice that the probability of significant short-cuts is negligible.
In particular, for very deep circuits, this becomes increasingly crucial as the number of random bits that are required to draw a circuit of depth $T$ scales like $\sim KT$.

Very recently, the $K$ dependence of the depth in the generation of approximate $t$-designs was improved to a near optimal scaling of $O(\log(K)\,t\,\mathrm{polylog}(t))$~\cite{schuster2024random}. 
While this would further improve the $K$ dependence in Eq.~\eqref{eq:complexitybound}, the circuit layout in Ref.~\cite{schuster2024random} is not a random brickwork circuit as discussed in this section and different $t$ require vastly different circuit layouts.
Improving the design depth of random quantum circuits in a standard brickwork layout to the same near optimal scaling is currently an open problem. 

Combinatorial arguments based on unitary designs are flexible:
the idea of using counting arguments via moment bounds was extended to more operational definitions of complexity, such as the minimal number of gates required to distinguish a state from the maximally mixed state~\cite{brandao2021models}.
The proof technique relies on unitary designs, as in ~\cite{brandao2016local,brandao2016efficient,brandao2021models,haferkamp2022random}. It offers a second advantage over the dimension-comparison technique of~\cite{haferkamp2021linear,li2022short}: the bound~\eqref{eq:designdepth} holds also if we draw the gates from a universal gate set---from a finite set $\mathcal{G}\subset {\rm SU}(4)$ 
whose gates can approximate every $K$-qubit unitary.
Moreover, the techniques of Ref.~\cite{chen2024incompressibility} yield similar results for other circuit layouts as well as randomly chosen layouts.

Similar techniques can be used to study the growth of quantum circuit complexity in models of time-dependent dynamics.
For example, a discretized model of Brownian motion converges to approximate designs as quickly as random circuits do~\cite{onorati2017mixing}.
Reference~\cite{jian2023linear,Guo:2024zmr} provides evidence for a similar convergence by continuous Brownian dynamics.  These continuous versions can be directly formulated in the context of holography \cite{Magan:2024aet,Magan:2025hce} as we  discuss in Sec.~\ref{sec:holography}.

This and the previous subsection mostly concern the growth of  quantum circuit complexity for random circuits. 
Using high moments (of the order $t\sim 4^K$), one can also bound the time at which quantum complexity saturates and even show that (with high probability) a recurrence to low circuit complexity happens after a doubly exponentially long time~\cite{oszmaniec2022saturation}.
More precisely, for random circuits, the approximate circuit complexity saturates near its maximal value after a depth $2^{5K}$ with high probability over the choice of circuit. The approximate circuit complexity exhibits a recurrence at a depth $d_{\mathrm{rec}}$ that satisfies $a_{\varepsilon}^{2^{2K}}\leq d_{\mathrm{rec}}\leq b_{\varepsilon}^{2^{2K}}$. The constants $a_{\varepsilon}$ and $b_{\varepsilon}$ depend only on the approximation error $\varepsilon>0$ in the approximate circuit complexity's definition.
To tightly bound the time at which saturation of circuit complexity occurs, one would have to extend the bound in~\cite{chen2024incompressibility} to extremely high moments ($t\sim 4^K$).
Such a tight bound is possible for a classical analogue of random quantum circuits called random reversible circuits~\cite{chen2024incompressibility}.
The latter generate random elements in the permutation group S$_{2^K}$ acting as permutations of $n$-bit string.

\subsection{Barriers to proving that quantum complexity grows superpolynomially at late times}\label{subsec:barriersfortime-independent}

Each of the methods discussed above exploits a counting argument. The argument concerns 
a placeholder quantity that roughly measures the ensemble's randomness. 
Finding such a quantity for an ensemble 
is far easier than finding a similarly growing quantity for individual circuits.
In the exact-circuit-complexity proof, the placeholder is the ``accessible dimension'' $\dim \mathcal{U}_R$. In the approximate-circuit-complexity proof, any quantity that signifies the convergence to designs will do, such as frame potentials~\cite{Roberts:2016hpo} or the expected overlaps $\mathbb{E}_{V_1,\ldots,V_d}|\langle \psi|V_d\cdots V_1|0^K\rangle|^2$ for random $2$-qubit gates $V_1,\ldots, V_d$. 
This strategy dominates all our proofs of lower bounds on circuit complexity beyond linear depth.
Indeed, this limitation might be fundamental as circuit complexity is known to be notoriously difficult to bound.

But these comments apply to individual states. What about time evolution under randomly drawn local Hamiltonians? This scenario is closer to the setting considered by Brown and Susskind~\cite{Brown:2017jil}, who conjectured that the quantum complexity grows linearly for an exponentially long time for a typical time-independent local Hamiltonian.
Still, the total number of distinct local time independent Hamiltonians is upper bounded by $2^{\mathrm{poly}(K)}$; so clearly the above counting argument fails to establish superpolynomial lower bounds on the approximate quantum circuit complexity in the system size.

The problem of proving lower bounds on the quantum complexity for the time-evolution of local Hamiltonians turns out to be related to the separation of computational-complexity classes:
Aaronson and Susskind~\cite{susskind2018black} point out that the circuit complexity of exponentially long-time evolution by local Hamiltonians is superpolynomial in the system size if and only if $\mathrm{PSPACE}\not\subset\mathrm{BQP}/\mathrm{poly}$.
$\mathrm{PSPACE}$ is the class of all computations that can be performed with polynomial memory. $\mathrm{BQP}/\mathrm{poly}$ is the class of all problems that can be solved with (nonuniform) families of polynomially sized quantum circuits.  Separations of complexity classes are notoriously difficult to prove, and few such separations are known.
Proving a separation of the kind $\mathrm{PSPACE}\not\subset \mathrm{BQP}/\mathrm{poly}$ appears to require a breakthrough in theoretical computer science. Therefore, even showing the existence of a local Hamiltonian for which the circuit complexity does not saturate after a polynomial time requires major technical advances.

\clearpage
\section{Paradigms for complexity I: Nielsen complexity}
\label{ssec:Nielsen_complexity}

As discussed in the previous section, one can define two notions of complexity for a unitary operator: exact and  approximate. Each notion has a downside.
For starters, exact complexity can behave counterintuitively. For example, a unitary close to the identity can have a large complexity as obtaining it exactly might require fine tuned combination of very many gates. On the other hand, approximate complexity involves a tolerance, which may seem arbitrary. One can resolve both issues by defining complexity in terms of a continuous quantity.

Nielsen \emph{et al.} identified the optimal implementation of a  unitary transformation $U$, using  \textit{complexity geometry} \cite{Nielsen1,Nielsen2,Nielsen3}, which we will now explain.  Consider the manifold of the unitary group equipped with an inner product on its tangent space that assigns a large norm to ``complex'' operations and a small norm to ``simple'' ones.  Equivalently, this inner product defines a ``complexity metric'' on the unitary group manifold. We have to choose what is complex and what is simple -- for example, a simple operation may be one that acts on a small number of qubits.  Now consider the time evolution operator of a physical system, $U(t) = \mathcal{\overset{\leftarrow}{P}}  \exp(-i \int_0^t dt' H(t'))$ where $H$ is the Hamiltonian and $\mathcal{\overset{\leftarrow}{P}} $ indicates path ordering.  Over the time interval $[0,T]$,  the time evolution operator follows a trajectory from the identity $U(0) = \id$ to $U(T)$. Nielsen \emph{et al.} defined the complexity of the unitary $U(T)$ as the length of the shortest geodesic from $\id$ to $U(T)$ on the unitary group manifold equipped with the complexity metric.  The length of this geodesic upper bounds the approximate circuit complexity and lower bounds the exact circuit complexity of implementing $U(T)$ as a composition of discrete gates \cite{nielsen2010quantum}.

Nielsen complexity has several appealing features:
\begin{enumerate}
   \item In Nielsen's framework, evolution along a continuous trajectory decomposes into a sequence of infinitesimal steps.  This decomposition is equivalent to \emph{Trotterization}, a technique for building a discrete circuit to simulate Hamiltonian time evolution~\cite{Lloyd96}.  Hence Nielsen's complexity is related to an important quantum-computational concept.
   \item    Using Nielsen's geometry, one can upper-bound the approximate circuit complexity, and lower-bound the exact circuit complexity, concepts 
   discussed in Sec.~\ref{sec:what_is_quantum_complexity}.
   \item Nielsen complexity is a geometric object and thus we can try to compute and analyze it with the methods of differential and algebraic geometry.
  \item  Nielsen complexity connects naturally to Hamiltonian control problems that are extensively studied in the quantum computation literature. 
   \item One has substantial freedom in equipping Nielsen's complexity geometry with a distance measure 
   on the unitary manifold.
   This freedom mimics the choice of a gate set in the definition of exact and approximate circuit complexity. 
   \emph{A priori}, it is difficult to compute Nielsen's complexity because the complexity geometry in interesting cases is generally high-dimensional and highly curved.  But different choices of metric can belong to the same
   equivalence class with the same long-distance behavior. Within an equivalence class, we can sometimes find modestly curved members  \cite{Nielsen3,Brown:2021uov,Brown:2022phc}, 
   facilitating the  calculation of Nielsen complexity.
   \item 
   The problem of calculating Nielsen complexity can become 
   exactly or approximately solvable  when we consider unitary representations of certain symmetry groups.
   Examples include unitary representations of symplectic and orthogonal groups (relevant respectively for free bosonic and fermionic quantum systems), the global conformal group, and circuits constructed from the Virasoro algebra of generators (relevant for spacetime transformations in CFTs). 
   \item Nielsen complexity, as a continuous quantity, naturally relates to continuous features of QFT and cMERA tensor networks~\cite{Haegeman:2011uy,Nozaki:2012zj}, as well as to physical time evolution. In fact, the parallel with cMERA motivated the earliest works on complexity in free QFT~\cite{Jefferson:2017sdb,Chapman:2017rqy}. 
\end{enumerate}

We start in Sec.~\ref{sec.Nielsen.intro} with the definition of Nielsen complexity. Section \ref{subsec:geometricNielsen} discusses possible choices of complexity metrics. In Sec.~\ref{ssec:constraints_cost_function}, we discuss the required  properties a complexity geometry should have in order to reproduce the linear time dependence and switchback effect for system evolution with a chaotic Hamiltonian. 
Section~\ref{sec:generalresults} presents general results on the time evolution of Nielsen's complexity on the unitary manifold. 
In Sec.~\ref{sec:NielsenBinding}, we discuss a choice of norm on the unitary manifold relevant for distributed computing, which highlights a connection between complexity and the entanglement entropy. We then turn to  studies of Nielsen complexity in  QFT in Sec.~\ref{sec:NielsenQFTALL}.

\subsection{Definition of Nielsen complexity \label{sec.Nielsen.intro}}
We will first define a complexity geometry for the  special unitary group $\mathrm{SU}(N)$ with $N=2^K$ describing transformations acting on states of the $K$-qubit system in Sec.~\ref{subsec:counting}.  As we will later see, we can also define complexity geometries for the global and local conformal groups relevant to conformal field theories, and for the symplectic and orthogonal groups relevant to free bosons and fermions.

We want to identify the optimal trajectory in $\mathrm{SU}(N)$ for generating a target unitary $U$ from the identity via the action of a Hamiltonian $H(t)$.  If $H$ and $t$ are the physical Hamiltonian and time, we can think of this as describing physical time evolution as  a trajectory on the unitary group.  But $H(t)$ could also be an alternate family of operations carried out over an auxiliary time, perhaps giving a quicker way to reach the target $U$ from the identity. Let $\{ T_I \}$ be a Hermitian basis for the Lie algebra of the group, $\mathfrak{su}(N)$.  We can expand the Hamiltonian in this basis as:
\begin{equation}
H(t) = \sum_I Y^I (t) \, T_I \, .
\label{eq:expansion_Ham}    
\end{equation}
The $Y^I(t)$ denote \emph{velocities} or \emph{control functions}. They are the components of the vector that is tangent at time $t$ to the trajectory generated by $H(t)$ through the group manifold.
Different trajectories represent different ways of generating $U$. We have parameterized each trajectory by $t \in[0,1]$, such that $U(t{=}0) = \id $ and $U(t{=}1)=U$.

At each point along the path, the generated unitary transformation has the form 
\begin{equation}
    U(t) = \mathcal{\overset{\leftarrow}{P}} \exp \left( -i \int_0^{t} dt' H(t') \right) \, .
\label{eq:generic_path_unitaries}
\end{equation}
The path ordering $\mathcal{\overset{\leftarrow}{P}}$ constructs the trajectory by multiplying the exponential terms from right to left.
Let  $F[Y^I(t)]$ be a \emph{cost} or \emph{distance} function defined on infinitesimal displacements on the group manifold via an inner product on the tangent space. The cost function encodes the difficulty of carrying out an operation that displaces the unitary operator in any given direction along the tangent space at a given point in the group manifold. The integrated cost/length of a trajectory is an integral of these infinitesimal costs.
\emph{Nielsen's  complexity} is defined as the minimal cost/length induced by the inner product, evaluated over all trajectories between $\id$ and $U$:
\begin{equation}
\mathcal{\mathcal{C}}_F [U] 
\coloneqq \min_{\lbrace Y^I(t) : \, U(0)= \id, \, U(1)=U \rbrace} \int_0^1 dt \;  F[Y^I(t)] \, .
\label{eq:unitary_complexity}
\end{equation}

The above definition of the complexity of unitary transformations induces a notion for the complexity of states in the Hilbert space on which the unitaries act.  To construct this quantity, assume that the $\mathrm{SU}(N)$ elements act on an $N$-dimensional Hilbert space $\mathcal{H}$.
We would like to define a quantum state complexity that
captures the difficulty of producing an arbitrary target state $\ket{\psi_\tar}$ from a fixed reference state $\ket{\psi_\refer}$.  Let us define \emph{Nielsen state complexity} as the lowest complexity of any special unitary transformation that evolves the reference to the target:
\begin{equation}
   \mathcal{C}^{\rm state}_F [\ket{\psi_T},\ket{\psi_R}] 
   \coloneqq \min_{ \lbrace U \in \mathrm{SU}(N) \, : \, \ket{\psi_\tar} = U \ket{\psi_\refer} \rbrace} \mathcal{C}_F [U] \, .
   \label{eq:def_compl_state}
\end{equation} 
It would be helpful if this quantity could be written directly in terms of a metric on the Hilbert space.  We will define such a metric below.

To define a metric on the Hilbert space, recall  first that quantum states should  be thought of as equivalence classes of elements that differ by a global phase. So we should first ask, what transformations in $\mathrm{SU}(N)$ leave a particular state invariant up to a phase, as such transformations should induce zero cost.   The set of such transformations is  the {\it stabilizer} of the state. Think of the given state as  a basis element of the Hilbert space.  Then transformations that only affect the remaining $N-1$ basis elements leave it invariant. Including  multiplication by a phase, the stabilizer of a state $\ket{\psi}$ is therefore the maximal proper subgroup $\mathrm{SU}(N-1) \times \mathrm{U}(1)$ of $\mathrm{SU}(N)$.

We can define equivalence classes in terms of the stabilizer as follows. Let $V$ denote an element in the $\ket{\psi}$ stabilizer. Consider right-multiplying any element $U \in \mathrm{SU}(N)$ by $V$. The resulting unitary, $U' \coloneqq UV$, lies in the same equivalence class as $U$: 
\begin{equation}
V \ket{\psi} = e^{i \phi} \ket{\psi} \quad \Rightarrow \quad
U'=U V \sim U \, .
\label{eq:def_stabilizer}
\end{equation}
Equation~\eqref{eq:def_stabilizer} defines a quotient  from the unitary group 
to the complex projective space $\mathbb{CP}^{N-1}$:
\begin{equation}
\pi \, : \, 
\mathrm{SU}(N) \mapsto \mathbb{CP}^{N-1} 
\coloneqq  \frac{\mathrm{SU}(N)}{\mathrm{SU}(N-1) \times \mathrm{U}(1)} \, .
\label{eq:quotient_map}
\end{equation}
We can now induce a norm on the tangent to the state space that accounts for this equivalence.

Suppose we have a state $|\psi(t)\rangle$, and we change it slightly with derivative $|\dot{\psi}(t)\rangle = -i H(t) |\psi(t)\rangle = -i (\sum_I Y^I(t) T_I ) |\psi(t)\rangle$. Certain linear combinations of $Y^I$ are fixed by fixing the state and its derivative, while others remain free; the latter correspond to the stabilizer of $\ket{\psi(t)}$. Minimizing the cost function over the stabilizer's degrees of freedom produces a norm on the tangent to the state space:
\begin{equation}
F^{\rm state} [|\dot{ \psi} (t)\rangle]_{\ket{\psi(t)}} 
\coloneqq \min_{\mathrm{ stab} \, \ket{\psi (t)} }  F[Y^I(t)] \, .
\label{eq:def_Fstate}
\end{equation}
In terms of this definition, we can express the state complexity~\eqref{eq:def_compl_state} as
\begin{equation}
\label{eq:statemin}
    \mathcal{C}^{\rm state}_F [\ket{\psi_\tar},\ket{\psi_\refer}] =\min_{{ \lbrace
    |\psi(t)\rangle \, : \, 
    |\psi(0)\rangle=\ket{\psi_\refer}, \,
    |\psi(1)\rangle=\ket{\psi_\tar} \rbrace }
    }  \int_0^1 dt \; F^{\rm state} [|\dot{ \psi} (t)\rangle]_{\ket{\psi(t)}}\, .
\end{equation}
We can use this formula to study geodesics in the state space without referring to unitaries.

\subsection{Geometric features of Nielsen complexity}\label{subsec:geometricNielsen}

The cost function $ F[Y^I]$ encodes the geometric properties of Nielsen's complexity, through the definition~\eqref{eq:unitary_complexity}.
Important cost functions 
have the form 
\begin{equation}
F_{p,\vec{q}} [Y^I] = \left( \sum_I q_I \, |Y^I|^p \right)^{\frac{1}{p}} \, .
\label{eq:cost_function_Finsler}
\end{equation}
The \emph{penalty factors} $q_I \geq 0$ are in one-to-one correspondence with the generators $T_I$ of the $\mathfrak{su}(N)$ algebra. They quantify the difficulty or cost of moving along each direction of the $\mathrm{SU}(N)$ tangent space.
If we think of the tangent directions as operations performed on states, we can choose penalty factors to reflect the difficulty of implementing the corresponding gates. 
For instance, few-qubit operators are naturally easier to implement than many-qubit operators. 
Therefore, it is natural to assign smaller penalty factors to the {\it easy} generators (that implement few-qubit operations) and larger penalties to the {\it hard} generators  that act on several qubits.  The terms \textit{easy} and \textit{hard} refer to the magnitude of the penalty factor associated with each generator.  A trajectory in the group manifold is then approximated by a circuit whose gates act for short time intervals.

Returning to the cost function~\eqref{eq:cost_function_Finsler}, consider taking $p = 1$. When $p=1$, $F_{p{=}1,\vec{q}} [Y^I]$ has a natural physical interpretation: it counts (with an appropriate measure) the number of gates used to build the continuous trajectory in the group manifold SU($N$). 
However, $F_{p{=}1,\vec{q}} [Y^I]$ has the disadvantage of not being smooth. We therefore cannot apply the calculus of variations to the geodesics.
When $p=2$, the norm~\eqref{eq:cost_function_Finsler} induces a Riemannian metric on the group manifold, 
simplifying the study of the geodesics.
In this case, the map~\eqref{eq:quotient_map} is a Riemannian submersion \cite{Auzzi:2020idm} (defined in~\cite{petersen2006riemannian}).
Heuristically, a Riemannian submersion is a smooth mapping from a higher-dimensional manifold to a lower-dimensional one. This mapping preserves distances in the directions perpendicular to the fibers.\footnote{The fibers are 
points in the higher-dimensional space that are mapped to the same point in the lower-dimensional manifold.}
In our setting, the fibers correspond to the maximal subgroup $\mathrm{SU}(N-1) \times \mathrm{U}(1)$. The submersion maps $\mathrm{SU}(N)$ to $\mathbb{CP}^{N-1}$.
These facts provide a systematic means of determining the metric induced by the cost-function minimization~\eqref{eq:def_Fstate} on the state space. 
The Nielsen complexity literature has mainly focused on $p=1,2$~\cite{Jefferson:2017sdb,Chapman:2017rqy,Khan:2018rzm,Hackl:2018ptj,Chapman:2018hou,Bhattacharyya:2018bbv,Guo:2018kzl,Brown:2019whu,Bernamonti:2019zyy,Caceres:2019pgf,Brown:2017jil,Caputa:2018kdj,Balasubramanian:2018hsu,Balasubramanian:2019wgd,Auzzi:2020idm,Chagnet:2021uvi,Basteiro:2021ene,Brown:2021uov,Balasubramanian:2021mxo,Brown:2022phc,Erdmenger:2022lov,Baiguera:2023bhm}.

One may wonder whether Nielsen's complexity is
related to discrete gate complexity. The answer is yes: Nielsen's complexity provides both upper and lower bounds on the gate complexity, provided the cost function satisfies certain conditions~\cite{Nielsen1,Nielsen2}. 
The authors of~\cite{Nielsen1,Nielsen2} further show that, by sufficiently penalizing the generators acting on several qubits, one can force the geodesic trajectory that reaches a unitary in the complexity geometry with arbitrary precision  to only use generators acting on one or two qubits.
The penalties required for this construction typically make the manifold highly curved. However, as we will discuss, Nielsen's complexity is universal at long distances within a certain equivalence class of metrics, and this equivalence class can be shown to also contain moderately-curved metrics.  A priori, this fact makes the computation of Nielsen's complexity easier by using tools from differential geometry. Finally, let us mention that when the motion in the unitary manifold is restricted to a small subgroup, exact or semi-analytic results can be obtained, which we discuss in later subsections.

\paragraph{Equivalence classes of metrics and the universality of penalty schedules}
There is substantial freedom in choosing a cost function. However, there is a large universality class of cost functions leading to  metrics on the group manifold that may vary significantly at short distances, but differ at large distances only polynomially in the path length and in the physical system's size~\cite{Brown:2021uov,Brown:2022phc}.  To simplify the rest of this subsection, we will take the penalty factors to  only depend on the number of qubits $k$ on which a gate acts, and denote them as $q_k$.  Sometimes $k$ is referred to as the weight of the gate -- $q_k$ should not be confused with the $q_I$ introduced in Eq.~\eqref{eq:cost_function_Finsler}. The notation $q_k$ means that generators $T_I$ that have the same weight $k$ share the same penalty factor.

Consider increasing any penalty factor while keeping the others constant. Above a critical value $\bar{q}_k$, the complexity ceases to depend on that penalty factor because we can implement this generator with weight $k$ as a composition of other, cheaper gates. 
The reason is as follows. Consider any fixed $m \geq 2$. Using the $m$-local gates, one can reconstruct all the $m'$-local gates for every $m'$ \cite{nielsen2010quantum}. 
Consequently, one can identify a certain \textit{critical schedule} $\lbrace \bar{q}_k \rbrace$ in terms of which we can define a universality class of cost functions that yield metrics with the same long distance properties~\cite{Brown:2021uov}.   Specifically, given the critical schedule, we consider penalty factors $\{ q_k \}$ satisfying $q_2 = \bar{q}_2$ and, for all $m > 2$, $q_m \geq \bar{q}_m$.   Metrics arising from this universality class of cost functions may differ greatly at short separations but  lead to approximately equal distances at long separations.
That is, short and long distance scales decouple. The critical schedule is conjectured to be the only universality class member for which
the conjugate points are pushed to infinity, and geodesics are straight lines for an exponential time in the number of qubits~\cite{Brown:2021uov}.\footnote{A conjugate point occurs along a geodesic if there is a local shortcut  from the start of the trajectory to some point along it (the notion of local shortcut will be more precisely defined below).  Such conjugate  points can be diagnosed by locally perturbing a geodesic.  If the perturbation does not change the length, the geodesic can be deformed to find a shorter geodesic between the same endpoints.  There can also be global obstructions to minimality of a geodesic, such as geodesics encircling a sphere in opposite directions along a diameter. We will discuss conjugate points in more detail in Sec.~\ref{sec:generalresults}.} Brown argued that the same universality class consists of metrics whose associated Nielsen complexity, measured by the geodesic length from the origin, is polynomially equivalent to the gate complexity. 
At the same time, it is possible to pick a metric inside the universality class such that the unitary manifold has modest curvature~\cite{Brown:2022phc}, thus making computations easier.

\subsection{Constraints on the cost function from expected time dynamics}
\label{ssec:constraints_cost_function}

In Sec.~\ref{subsec:counting}, we showed that quantum complexity for discrete time circuits exhibits two behaviors: linear growth which can last for an exponentially long time, and the switchback effect. These derivations relied on simple counting arguments. Afterwards, we quantitatively demonstrated this linear growth for random unitary circuits, by using algebraic geometry and unitary designs (Sections~\ref{sec_growth}--\ref{subsec:barriersfortime-independent}).  We would like to understand which complexity geometries reproduce the above key features of complexity when applied to the unitary time evolution generated by a chaotic Hamiltonian. 
To do so, we must select an appropriate metric on the $\mathrm{SU}(N)$ manifold. 
The requirements turn out to be~\cite{Nielsen3,Brown:2016wib,Brown:2017jil}:
\begin{enumerate}
\item The metric should be right-invariant, because the unitary manifold is homogeneous. 
\item The group manifold must have a negative average sectional curvature  \cite{Nielsen3,Brown:2016wib}, such that nearby geodesics diverge. This is necessary, but not sufficient, for establishing the above key features for chaotic and ergodic Hamiltonians.
\item  The geometry's sectional curvatures should scale as $1/K$,  where $K$ denotes the number of qubits, in order for the Nielsen complexity to display the switchback effect.
\end{enumerate}

Let us comment on the implications of these requirements. 
First, consider Eq.~\eqref{eq:def_Fstate} which yields a metric on the space of states rather than on unitaries. Do the above properties have a reflection in the induced geometry on the space of states? {In fact the induced metric is not homogeneous, complicating the study of the geodesics and curvature in the space of states~\cite{Brown:2019whu}. 
Nonetheless, as mentioned earlier, for the case of Riemannian cost functions, the map~\eqref{eq:quotient_map} is a Riemannian submersion \cite{Auzzi:2020idm} (defined in~\cite{petersen2006riemannian}).  Riemannian submersions relate   sectional curvatures in the space of unitaries and  sectional curvatures in the space of states, via the O'Neill formula~\cite{o1966fundamental}.  
The latter implies that the sectional curvature along a plane in the space of states is lower-bounded by the sectional curvature along an appropriate plane in the unitary manifold.

Second, the group manifold equipped with the Cartan-Killing metric---the standard metric on SU$(N)$---does not satisfy the above requirements because it has only constant positive sectional curvatures.  How can we construct a geometry that has negative sectional curvatures?   We can assign the penalty factors such that some of the $\mathfrak{su}(N)$ algebra's nonvanishing commutators have the structure
$[easy, \, easy] = \, hard$. The \textit{easy} and \textit{hard} refer to generators associated with small and large penalty factors in the cost function.
When the penalty factors are associated to generators such that the commutators take the above-mentioned form, then the Pythagorean theorem for curved space guarantees some negative sectional curvatures~\cite{Brown:2019whu}. The latter statement was proven in the contexts of the unitary manifolds of qubits and qutrits \cite{Brown:2019whu,Auzzi:2020idm}.\footnote{A qudit is a $d$-level quantum system, represented by a Hilbert space $\mathbb{C}^d$. A qubit has $d=2$; and a qutrit, $d=3$.}
Note that manifolds that are equipped with right-invariant metrics, like the ones we are interested in, must also have some positive sectional curvatures; otherwise, they are flat \cite{MILNOR1976293}. 

References~\cite{Balasubramanian:2019wgd,Balasubramanian:2021mxo} studied  time-dependent trajectories generated by the SYK Hamiltonian on the unitary manifold.  The authors showed that the geodesic distance from the origin to the time evolution operator is lower-bounded by a geodesic whose length grows linearly with time. This linear growth is truncated at conjugate points along the geodesic. There is a shorter path connecting the origin to a conjugate point than the one produced by extending the earlier time geodesic.  It is possible that the linear growth is truncated even earlier by global obstructions -- geodesics that wind in some other direction around the unitary group to the same point -- but these are much more difficult to study, as we will discuss in Sec.~\ref{sec:generalresults}.  Conjugate points in the critical complexity geometry described at the end of Sec.~\ref{subsec:geometricNielsen} are pushed to infinity \cite{Brown:2021uov}. Therefore, the linear growth of Nielsen's complexity in that geometry persists for a time exponential in the number of qubits.
In the context of spin chains, Brown showed that 
many complexity geometries have diameters that are exponentially large in the number of qubits \cite{Brown:2022phc}. This exponential relation is necessary for  complexity to saturate at an exponentially large value (Fig.~\ref{fig:TimeEvolGen}).

We have argued that Nielsen complexity exhibits one behavior characteristic of quantum complexity: the initial linear growth. Nielsen complexity also reproduces the other behavior---the switchback effect---in toy models~\cite{Brown:2017jil}.
Qualitative features similar to the switchback effect are also displayed by single-qubit and two-qubit systems~\cite{Caginalp:2020tzw}. However, a comprehensive picture of the emergence of the switchback effect from the complexity geometry is still lacking and is an open question for future research.

\subsection{Physical criteria for complexity growth: general results and applications}
\label{sec:generalresults}

How is Nielsen complexity related to standard concepts in physics, like energy spectra, correlation functions, and Schr\"odinger dynamics?  Several connections were uncovered by the authors of \cite{Balasubramanian:2019wgd,Balasubramanian:2021mxo}, who investigated  physical criteria that control the length of time over which Nielsen complexity will grow during Hamiltonian time evolution.   We will discuss their results below, and use this discussion as a way of illustrating how to perform practical calculations of Nielsen complexity.

For concreteness, let us consider a time-independent Hamiltonian acting on an $N$--dimensional Hilbert space.  Time evolution is then governed by the unitary operator $U(t) = e^{-iHt}$. We want to think of this operator as tracing a trajectory on the $\mathrm{SU}(N)$ group manifold from the  identity $\id$ at $t=0$ to $U(t)$.   Following Nielsen, we will regard the complexity of $U(t)$ as  measured by the length of the shortest geodesic from $\id$ to $U(t)$ in a certain complexity metric that we will define below.   

In fact, finding the shortest geodesic on $\mathrm{SU}(N)$ for general $N$ is a difficult problem in {\it any} metric  because the group manifold has a complicated fiber bundle topology. So, if we find one geodesic between two  group elements by solving the appropriate differential equation, there may well be others that start at the identity and go ``around'' the group in another direction to get to the same point. Some of the latter may even be shorter.  The simplest example of this occurs in $\mathrm{SU}(2)$, whose manifold is a 3-sphere.  Geodesics on the 3-sphere are great circles.  If we continue on a great circle past the point diametrically opposite to the initial one, there is a shorter geodesic that goes the other way around the sphere.  For general $\mathrm{SU}(N)$ there may be many more distinct paths between two group elements that exploit the complicated topology, and finding them all is an unsolved problem. 
The problem is even believed to be NP-complete from an algorithmic perspective.  While  finding all global shortcuts  on $\mathrm{SU}(N)$ equipped with a given complexity metric is intractable, the authors of \cite{Balasubramanian:2019wgd,Balasubramanian:2021mxo} pointed out that there is a class of shortcuts that we can analyze, associated to {\it conjugate points} along a geodesic.   Conjugate points imply that the geodesic we started with is a saddlepoint, not a minimum of the length, so that we can find a shorter path by descending from the saddle.   Below we will explain how to find conjugate points, and present physical criteria that delay their occurrence until exponential times under Hamiltonian evolution.

Our goal is to find geodesics on a Lie group manifold in a metric that penalizes ``hard'' generators that represent infinitesimal transformations that are difficult to implement.  First, consider the group generators that span the tangent space to the group manifold at the identity.  Consider the Cartan-Killing form $K_{IJ} \propto f_{IM}^L f_{JL}^M$ where $f_{IJ}^L$ are the group's structure constants defined using the commutation relation for the group generators $[T_I,T_J] = i f_{IJ}^L T_L$ with repeated indices summed.  Now define
\begin{equation}
G_{IJ} = \frac{c_I+c_J}{2} K_{IJ} \, ,
\label{eq:G_defined_K}
\end{equation}
with $c_I = \mathcal{O}(1)$ for ``easy'' directions in the group's tangent space at the identity, and $c_I = \mathcal{O}(e^{\epsilon S})$ for the ``hard'' directions. Here $\epsilon$ is any constant, and  $S = \ln N$ is the logarithm of the dimension of the Hilbert space on which the group acts.  We may take the hard directions to be generators that act on more than $k$ qubits for some $k$, or act non-locally, or are differentiated in some other way appropriate to the problem at hand.  For $\mathrm{SU}(N)$ we will split generators into the easy ones $\{ T_\alpha\}$ and the hard ones $\{T_{\dot\alpha}\}$, so that the diagonal metric is 
\begin{equation}
G_{IJ} = 
\begin{pmatrix} 
\delta_{\alpha\beta} & 0 \\ 
0 & (1 + \mu) \, \delta_{\dot\alpha \dot\beta}
\end{pmatrix}
~~~~~;~~~~~ \mu \sim e^{\epsilon S} \,.
\label{eq:SUNmetric}
\end{equation}
This is the metric, or inner product, on the tangent space at the identity.  If $X$ and $Y$ are tangents at any other point $U$ on the group manifold we define $g(X,Y) = G(XU^{-1},YU^{-1})$ where $G$ is a functional notation for the bilinear inner product defined in (\ref{eq:G_defined_K}).

With these definitions, the geodesic equation on $\mathrm{SU}(N)$ in Euler-Arnold form is \cite{Balasubramanian:2019wgd}\footnote{We relabeled the time $t$ along the circuit as $s$ to avoid confusion with the physical time of the system.}
\begin{equation}
G_{IJ} \frac{dV^J}{ds} = f_{IJ}^K \, V^J G_{KL} V^L \, ,
\label{eq:EulerArnold}
\end{equation}
with repeated indices summed.  Here, $V^I$ is tangent to the path,\footnote{The $V^I$ are analogous to the control functions $Y^I$ in equation \eqref{eq:expansion_Ham}; we use $V^I$ in this subsection and reserve $Y$ for the map described below Eq.~\eqref{eq:LNLJACOBI}.} $s$ is the parameter along the path, and $f_{IJ}^K$ are the structure constants of the group.
If we solve this for the tangents $V^I(s)$, then the geodesic path is
\begin{equation}
U(s) = \mathcal{\overset{\leftarrow}{P}} e^{\int_0^s ds' \, V^I(s') T_I} \, ,
\end{equation}
where $\mathcal{\overset{\leftarrow}{P}}$ represents path ordering. We consider the Nielsen complexity defined by 
\begin{equation}
\mathcal{C}(U(t)) = \min \int_0^1 ds \, \sqrt{G_{IJ} V^I(s) V^J(s)} \, ,
\end{equation}
where the minimum is taken over all geodesics from $I$ to $U(s)$. This is a generalization of  the complexity definition associated with the $p=2$ cost function in Eq.~\eqref{eq:cost_function_Finsler}.

This is a practical formalism for calculations.  For example, consider the simple case of $\mathrm{SU}(2)$ with generators $T_{1,2} = \gamma_{1,2}$ and $T_3 = i \gamma_1 \gamma_2$, where $\gamma_i$ are the standard Gamma matrices and $T_3$ is deemed ``hard'', while $T_{1,2}$ are easy \cite{Balasubramanian:2019wgd}. Geodesics take the explicit form
\begin{equation}
V^1(s) = v^1  \cos(v^3 \mu s) - v^2 \sin(v^3 \mu s)
~~;~~
V^2(s) = v^2  \cos(v^3 \mu s) + v^1 \sin(v^3 \mu s)
~~;~~
V^3(s) = \frac{v^3}{2} \, ,
\end{equation}
and the Nielsen complexity of going from $U(1) = \id$ to $U(1) = U_{{\rm target}}$ is
\begin{equation}
\mathcal{C} = \sqrt{ (v^1)^2 + (v^2)^2 + (1 + \mu) (v^3)^2 } \, ,
\end{equation}
where the initial velocities $v^I$ have to be chosen to get to the target at $s=1$. We can now pick geodesics for which $U(1) = e^{-i H t}$ with a Hamiltonian $H = J_1 T_1 + J_2 T_2$ built from the ``easy'' generators.  It is straightforward to then show that the complexity initially grows linearly as $\mathcal{C}(t) = t \sqrt{J_1^2 + J_2^2}$ as the Hamiltonian itself generates the shortest trajectory from the identity by avoiding the more expensive direction $T_3$.  The linear growth ends when the trajectory reaches the opposite pole of $\mathrm{SU}(2)$ from the identity because then there is a shorter geodesic to $e^{-iHt}$ that goes around the other side of the $S^3$ manifold of $\mathrm{SU}(2)$.  This causes the complexity to decrease back to  zero as $U(t) = e^{-iHt}$ returns to the origin. Continued time evolution leads to oscillations in the complexity.  We can view the return to zero complexity as the analog of a  recurrence in this simple integrable dynamics. Notice that the linear growth of complexity was truncated by a shortcut in the unitary group.  In this case the shortcut is a {\it global} obstruction to complexity growth -- it exists because of the non-trivial global topology of the group manifold.

Now consider $\mathrm{SU}(N)$ for general $N$. Following \cite{Balasubramanian:2019wgd,Balasubramanian:2021mxo} we will consider a representation of the group based on the algebra of Gamma matrices $\gamma_a$ with $a=0,\dots, N-1$ and $\{\gamma_a,\gamma_b\}= 2 \delta_{ab}$. We can construct  the $2^N$ generators of $\mathrm{SU}(N)$ as $T_{a_1\cdots a_m} = \gamma_{a_1} \cdots \gamma_{a_m}$ subject to the condition $a_p < a_q$ if $p<q$.  We will label the generators as $T_I$ with a multi-index $I \equiv (a_1 \cdots a_m)$.  We will later discuss a physical realization of the SYK model, constructed from $N$ Majorana fermions $\psi_a$ where $\gamma_a \sim \psi_a$.  In view of this identification, we will say that generators $T_\alpha$ with no more than $k$ $\gamma$ matrices in their definition are k-local, and that  generators $T_{\dot\alpha}$ with  more than $k$ $\gamma$ matrices in their definition are k-nonlocal.  We will also say that k-local generators are ``easy'' and that k-nonlocal generators are ``hard'' and are therefore penalized by $1+\mu$ in the metric in (\ref{eq:SUNmetric}).  Now suppose that the Hamiltonian is k-local: $H= \sum_\alpha J^\alpha T_\alpha$. Then the authors of \cite{Balasubramanian:2019wgd,Balasubramanian:2021mxo} showed from the Euler-Arnold equation (\ref{eq:EulerArnold}) that there is {\it always} a geodesic from the identity to $e^{-iHt}$ along the path taken by the physical time evolution.  In the notation of (\ref{eq:EulerArnold}), the components of the tangent vectors along this geodesic are $V^\alpha = J^\alpha t$ leading to the tangent vector $V = Ht$.   This means that the Nielsen complexity grows linearly in time,
\begin{equation}
\mathcal{C}(t) = t  \, \sqrt{\sum_\alpha (J^\alpha)^2}
\label{eq:GeneralLinearGrowth} 
\end{equation}
if the  time evolution trajectory, which is a geodesic, is also the {\it shortest} geodesic.

At least at the early stages of time evolution, we can be sure from the results discussed above that the trajectory  generated by the Hamiltonian is the shortest between $\id$ and $e^{-iHt}$.  How could this linear geodesic trajectory stop being the shortest?  One possibility is the existence of a geodesic loop that allows us to exploit the topology of $\mathrm{SU}(N)$ to arrive at the same destination through a different path.  This is what happened in the $\mathrm{SU}(2)$ example above.   Geodesic loops are difficult to work out in the general unitary group, so we will not attempt to do that.  However, note that in $\mathrm{SU}(2)$ the shorter path arising from a great cycle on the group manifold became relevant after the time evolution trajectory had traversed a diameter of the manifold. We expect something like that to be also generally true for geodesic loops of $\mathrm{SU}(N>2)$. Following Ref.~\cite{Brown:2022phc}, we also expect the diameter of $\mathrm{SU}(N)$ to be exponentially large in the number of degrees of freedom in complexity metrics satisfying reasonable consistency criteria.  This means that, all else being equal, we expect the geodesic loop obstruction to start to play a role in complexity growth after exponential time.  So we will focus on another kind of shortcut on the group manifold that could truncate the linear complexity growth in (\ref{eq:GeneralLinearGrowth}) far earlier. These shortcuts are associated to {\it conjugate points} on the linear Hamiltonian trajectory that we will now describe.

A conjugate point occurs on a geodesic $U(s): [0,1] \to \mathrm{SU}(N)$ with $U(0) = P$ and $U(1) = Q$ if a perturbation of the path produces a new curve with the same boundary conditions that satisfies the geodesic equation  to first order.  This implies that the geodesic we started with is a saddlepoint, not a minimum of the length, so that we can find a shorter path.  The actual minimizing geodesic signaled by the appearance of conjugate point can be a finite distance away.  Since conjugate points indicate converging geodesics, they cannot occur if all the sectional curvatures are negative.  There is a result of Milnor \cite{Milnor:1976pu} that says that most sectional curvatures of Lie groups are negative.  So some authors have considered high genus Riemann surfaces with a metric induced from the covering space, a hyperbolic disk, as a toy model of the unitary group manifold \cite{Brown:2016wib}.  In such a toy model there are no conjugate points, so they are irrelevant.  Others have constructed special metrics on the unitary group that are designed to push the conjugate points to infinity \cite{Brown:2021uov}.  But these approaches seem misleading about the structure of the unitary group because the main result of Milnor is that any right invariant metric on $\mathrm{SU}(N)$ with $N>2$ must have some positive sectional curvature, or else it is flat.  Indeed, as we will see, conjugate points are some of the earliest obstructions to complexity growth in free and integrable theories.

To find conjugate points we start with a linear path along the time evolution trajectory from $\id$ to $e^{-iHt}$.  This path has tangents $V(s) = H s$  which produce a line of unitaries $U(s) = e^{-istH}$ with $s=[0,1]$.  We want to perturb the trajectory $V(s) \to V(s) + \delta V(s)$ and still solve the Euler-Arnold equation to first order, with the same boundary conditions.   In first order perturbation theory, this gives the Jacobi equation.  In detail, let $\delta V = \delta V_L + \delta V_{NL}$ where the subscript $L$ indicates a projection to the k-local/easy subspace, and the subscript $NL$ denote a projection to the k-nonlocal/hard subspace.  Then the Jacobi equation implies: 
\begin{equation}\label{eq:LNLJACOBI}
i \frac{d \delta V_L(s)}{ds} = \mu t [H, \delta V_{NL}(s)]_L ~~~~~;~~~~~
i \frac{d \delta V_{NL}(s)}{ds} = \frac{\mu t}{1+\mu} + [H, \delta V_{NL}(s)]_{NL}
\end{equation}
where the subscripts $L$ and $NL$ again indicate projection to the k-local/easy and k-nonlocal/hard subspaces.

By solving this equation, we can work out how an initial perturbation of the trajectory at the origin changes the trajectory.  The authors of \cite{Balasubramanian:2019wgd} used the formal solution of the above equation to write down a superoperator $Y[\delta V(0)]$ mapping the initial perturbation of the trajectory, $\delta V(0)$, onto a perturbation of the endpoint.  Conjugate points occur when the endpoint is unchanged, namely when the superoperator $Y$ has a zero mode in the formalism of \cite{Balasubramanian:2019wgd}. We do not write the superoperator explicitly here because it has a complicated form that requires further definitions to explain, and instead refer the reader to \cite{Balasubramanian:2019wgd}. By working out criteria for the appearance of such zero modes, the authors of~\cite{Balasubramanian:2019wgd,Balasubramanian:2021mxo}  arrived at a number of general results for the appearance of conjugate points that are described below:

\begin{enumerate}
    [
    wide, 
    labelindent=0pt,
    label={\textbf{Result \arabic*: }},
    ref={Result \arabic*}]
\item 
For arbitrary local Hamiltonians with a finite complexity cost factor $\mu$, conjugate points must exist at finite distance along the linear geodesic. This is proved by starting with the result of \cite{Dowling:2006tnk} that conjugate points exist at finite distance when the complexity penalty $\mu$ vanishes.  The persistence of the conjugate points is then established by a continuity argument as $\mu$ increases, along with Morse theory on the space of paths on the group \cite{Balasubramanian:2021mxo}.  So the absence of conjugate points must be shown to establish linear complexity growth until times of order $t\sim e^S$.   

\item 
If a q-local Hamiltonian has an adjoint eigenoperator $O$ such that $[H,O] = \lambda O$ with $\lambda \in \mathbb{R}$ and $O \in \{ {\rm local/easy \ operators }\}$, then conjugate points occur at times $t_* = \frac{2\pi}{\lambda} \mathbb{Z}$ (where $\mathbb{R}$ are the reals and $\mathbb{Z}$ are the integers). This is proven in \cite{Balasubramanian:2021mxo} by explicit examination of the zero mode condition on the superoperator $Y$ mentioned above. \label{item:2}
\item 
If a q-local Hamiltonian has an adjoint eigenoperator $O'$ such that $[H,O'] = \lambda' O'$ with $\lambda' \in \mathbb{R}$ and $O' \in \{ {\rm nonlocal/hard \ operators } \}$, then conjugate points occur at times $t_* = \frac{2\pi (1 + \mu)}{\lambda'} \mathbb{Z}$. This is proven in the same way as \ref{item:2} above.
\item 
Let $M_{\alpha\beta}(t) = \int_0^1 ds \int_0^1 ds' \, {\rm Tr}[e^{i(s-s') t H} T_\alpha e^{-i(s-s') t H} T_\beta ]$, where $T_{\alpha}$ and $T_\beta$ are any two local/easy operators.  Suppose $M_{\alpha\beta}(t)$ has an eigenvector $X^\alpha$, with vanishing eigenvalue at some time $t^*$, then a conjugate point occurs at $t^*$.  We can think of $M_{\alpha\beta}$  as a temporally smeared, infinite temperature 2-point function of local operators.  To prove this, given the zero mode $X^\alpha$, the authors of~\cite{Balasubramanian:2021mxo} expanded an initial perturbation of the linear geodesic as $\delta V(0) = \sum_\alpha X^\alpha T_\alpha$ where $T_\alpha$ are the local/easy generators, and showed that the Frobenius norm of the superoperator $Y$ discussed above vanishes when it acts on this $\delta V(0)$.  This implies that the superoperator vanishes on this perturbation, which by construction implies a conjugate point.
\item   
Let $\ket{m}$ and $\ket{n}$ be any two energy eigenstates, and let $T_\alpha$ be the local/easy generators, and $T_{\dot\alpha}$ the nonlocal/hard generators.  Then linear growth of Nielsen complexity persists for a time of $\mathcal{O}(e^{\epsilon S})$, where $S$ is the logarithm of the Hilbert space dimension, if
\begin{equation}
R_{mn} = 
\frac{\sum_\alpha |\bra{m} T_\alpha \ket{n}|^2}{\sum_\alpha |\bra{m} T_\alpha \ket{n}|^2 + \sum_{\dot\alpha} |\bra{m} T_{\dot\alpha} \ket{n}|^2}
= e^{-2S} {\rm poly}(S) r_{mn}
\end{equation}
where $r_{mn} = \mathcal{O}(1)$ for $m\neq n$ and ${\rm poly}(S)$ means polynomial in $S$.  This criterion, which is called the {\bf Eigenstate Complexity Hypothesis} (ECH) (since it recalls the Eigenstate Thermalization Hypothesis), says that energy eigenstates cannot be mapped onto each other by easy operators.  This result was proven in \cite{Balasubramanian:2019wgd} by showing that if we assume ECH, the zero modes of the superoperator $Y$ occur at times $t^* = (2\pi/\Delta_{{\rm max}}) (1 + e^{\epsilon S})$,
where $\Delta_{\rm max}$ is the largest gap in the spectrum.\label{item:5}
\end{enumerate}

These theorems can be applied to the SYK model with $n$ Majorana fermions, which we will return to in  Sections~\ref{ssec:integrability_chaos} and \ref{ssec:holo_matching_2d}.   Here we briefly summarize the results of \cite{Balasubramanian:2019wgd,Balasubramanian:2021mxo} in this regard. The authors provide evidence that: ({\bf a}) The Nielsen complexity of the free SYK model (with only quadratic interactions) is bounded as $\mathcal{C}(t) \lesssim \sqrt{n}$; ({\bf b}) The Nielsen complexity of integrable deformations of the free SYK model is bounded as $\mathcal{C}(t) \lesssim {\rm poly}(n)$, meaning that the initial linear growth is guaranteed to truncate in polynomial time;  ({\bf c}) \ref{item:5} above, and numerical analysis of the spectrum and eigenstates of chaotic deformations of the free SYK model suggest that the Nielsen complexity of such systems is bounded as $\mathcal{C}(t) \lesssim e^{\epsilon n}$, meaning that linear complexity growth can persist for exponential time, if we neglect possible global obstructions from geodesic loops.  The finding ({\bf c}) applies to any theory in which the energy eigenstates are a Haar random unitary rotation of a ``standard'' or ``free'' basis of states.  The ETH hypothesis for chaotic theories suggests that their eigenstates may indeed be of this form.

\subsection{Relation between Nielsen complexity and entanglement: binding complexity}\label{sec:NielsenBinding}

The penalty factors in the definition of Nielsen's complexity can be attributed to the relative difficulty of applying certain generators. An interesting choice where analytic results can be obtained is the case in which our system is split into parts, and the penalty factors associated with generators acting within a given subsystem are much smaller than those of generators acting between multiple subsystems. 
This setting serves as a model to describe distributed quantum computing, where a large computer is divided into small quantum computers (called \textit{nodes}). 
The operations between nodes, which are possibly located far away from each other, require significantly more resources than operations within each subsystem. 

The limiting case, where only operations acting between subsystems have non-vanishing cost, was referred to as \textit{binding complexity} in  \cite{Balasubramanian:2018hsu}.  There, the authors suggested that binding complexity quantifies the difficulty of distributing entanglement among multiple parties.
The authors of Ref.~\cite{Balasubramanian:2018hsu} further suggested that the notion of binding complexity could serve as a measure for the robustness of entanglement. 
Here, one typically has in mind the $n$-party (with $n \geq 2$) generalizations of the W and GHZ states. The GHZ state is separable upon tracing out any subset of the parties, but the W state is not. In this sense, W states possess  more robust
entanglement.  The authors of \cite{Balasubramanian:2018hsu} argued that the binding complexity should scale as $\mathcal{O}(n)$ for GHZ states, and as $\mathcal{O}(n^2)$ for $W$ states. The paper also presented results for the binding complexity of Gaussian states of free QFT.  As we review next, we can get exact results for the binding complexity of spin chains in Nielsen's framework when the cost of generators acting within each subsystem is small enough~\cite{Baiguera:2023bhm}.
It turns out that cost functions satisfying this property relate Nielsen's complexity to the R\'enyi min-entropy in a precise way. Moreover, the above cost functions are bounded by the entanglement entropy of a state over a subregion~\cite{eisert2021entangling,Baiguera:2023bhm}.

To illustrate these results, let us begin with a formal definition of the binding complexity. Consider splitting a quantum system (for instance, a spin chain) into $m$ subsystems $A_{k = 1, 2, \ldots, m} \, .$
The \emph{binding complexity} is the minimal number of gates, each acting on a limited number of subsystems, needed to implement a target unitary with cost only incurred for gates acting between subsystems \cite{Balasubramanian:2018hsu}.
Below, we focus on 2-local gates which act on, at most, two subsystems each. Within Nielsen's framework, binding complexity has a close analogue, obtained by choosing a particular cost function on the unitary space.
To illustrate, consider a bipartite system whose Hamiltonian \eqref{eq:expansion_Ham} decomposes as\footnote{We denote the generators of the special unitary group in subsystem $A_1$ with index $a \in \lbrace 1, \dots, N_{A_1}^2 -1 \rbrace $  from the beginning of the alphabet, and the generators in $A_2$ with index $i \in \lbrace 1, \dots, N_{A_2}^2 -1 \rbrace$ from the middle part of the alphabet. 
We also denote with a sub(super)script $A_1, A_2$ the subsystem where the local generators act.\label{foot:ind}}
\begin{equation}
H(t) = Y^a_{A_1} (t) T^{A_1}_a \otimes \mathbf{1}^{A_2} +Y^i_{A_2} (t) \mathbf{1}^{A_1} \otimes T^{A_2}_i + Y^{ai} (t) T^{A_1}_a \otimes T^{A_2}_i \, .
\label{eq:decomposition_Hamiltonian}
\end{equation}
Each of the generators $T^{A_1}_a$ and $T^{A_2}_i$ acts on only one side of the system. In contrast, $T^{A_1}_a \otimes T^{A_2}_i$ can entangle the sides.
The velocities split into subsets 
$Y^I \coloneqq \lbrace Y^a_{A_1}, Y^i_{A_2}, Y^{ai} \rbrace$.
This division implies a decomposition of the penalty factors~\eqref{eq:cost_function_Finsler} into subsets
$q_I \coloneqq \lbrace q^{A_1}_a, q^{A_2}_i , q_{ai} \rbrace$. 

Having specified the setup, we define the \textit{Nielsen binding complexity} $\mathcal{BC}$. It is the unitary complexity~\eqref{eq:unitary_complexity} evaluated with the penalty factors
\begin{equation}
q^A_a = q^B_i = 0 
\quad \text{and} \quad
q_{ai} \geq 1 \, .
\label{eq:penalties_binding}
\end{equation}
These penalties impose greater costs on entangling (or \textit{non-local}) gates than on single-subsystem (or \textit{local}) operations.
Let us substitute the penalties~\eqref{eq:penalties_binding} into the cost function~\eqref{eq:cost_function_Finsler} and perform the minimization \eqref{eq:def_compl_state} to obtain a complexity norm on the space of quantum states. Two simplifications follow: first, the space of states contains null directions, defined as the loci of states interconnected via local unitaries  along which the cost function vanishes. 
Second, one can integrate out nondynamical degrees of freedom in the geodesic minimization. We obtain a norm $\mathcal{BF}[\lambda_k, \dot{\lambda}_k]$ on the space of states. $\mathcal{BF}$ depends only on a state's Schmidt coefficients $\lambda_k$  and their derivatives $\dot{\lambda}_k$. 

We can analytically compute the binding state complexity for certain choices of the penalty factors $q_{ai}$ in Eq.~\eqref{eq:penalties_binding}, and of the parameter $p$ in Eq.~\eqref{eq:cost_function_Finsler}.
A Riemannian cost function is our first example.
Let us write out the binding state complexity associated with the cost function \eqref{eq:cost_function_Finsler}, $p=2$, and the penalty factors \eqref{eq:penalties_binding} with $q_{ai}=1$ \cite{Baiguera:2023bhm}:
\begin{equation}
\mathcal{BC}^{\rm state}_{2, \rm hom} = \sqrt{\frac{2}{\mathcal{N}_{A_1} \mathcal{N}_{A_2} }} \, \mathrm{arccos} \left( e^{-\frac{1}{2}S_{\infty}(\ket{\psi_{\rm T}})} \right) \, .
\label{eq:bind_compl_GellMann}
\end{equation}
The \textit{hom} stands for the nonlocal generators' homogeneous penalties.  
The $\mathcal{N}_{A_k}$ are normalization factors defined such that $\tr ( T^{A_k}_a T^{A_k}_b ) = \mathcal{N}_{A_k} \delta_{ab}$. The $S_{\infty}$ denotes the R\'{e}nyi min-entropy of the reduced state of subsystem $A_1$ 
in the target state $\ket{\psi_{\rm T}}$.\footnote{The R\'{e}nyi min-entropy of the reduced state of a subsystem $A_1$ in the state $|\psi_{\rm T}\rangle$ is defined by $S_{\infty} (|\psi_{\rm T}\rangle)= -2 \log \bar{\lambda}_1$, where $\bar{\lambda}_1$ is the maximal Schmidt coefficient of the target state with respect to the same bipartite division.}

In our second example of an analytically computable binding complexity, we set $p=1$ in the cost function~\eqref{eq:cost_function_Finsler}.
This binding complexity does not depend on the penalty factors for nonlocal generators (up to an overall normalization), 
under certain assumptions
about the generators acting along the optimal trajectory.
Consequently, changing the penalty factors cannot yield arbitrary functions of the Schmidt coefficients. Rather, for certain choices of the generators acting along the trajectory, the binding state complexity is a fixed function of the Schmidt coefficients, independent of the penalty factors. 
This independence implies that, within a certain class of cost functions, there is a universal relation between the entanglement of a state in a subsystem and its binding complexity.

The relationship between binding complexity and Schmidt coefficients leads not only to exact results, but also bounds  other quantum-information quantities. 
For instance, one can lower-bound the binding state complexity, in terms of the entanglement entropy of a state over a subregion $A_1$ \cite{bravyi:2007,van:2013,Verstraete:2016,eisert2021entangling}.
Denote by $S_{A_1}$ the entanglement entropy of subregion $A_1$, by $d$ the dimensionality of the smaller subsystem's Hilbert space, and by $c$ a constant.
The $p{=}1$ binding state complexity upper bounds $S_{A_1}$ as follows:
\begin{equation} 
\label{Slwrbnds}
\frac{S_{A_1}}{c \log{d}} \leq \mathcal{BC}^{\rm state}_1[\ket{\psi_{\rm R}},\ket{\psi_{\rm T}}] \, .
\end{equation}
Binding complexity also provides lower bounds for other notions of complexity, such as the geometrically local complexity \cite{eisert2021entangling}.
Further details of the latter quantity will be discussed around Eq.~\eqref{eq:geom_local_circuits}.

Finally, let us mention that in the context of the holographic correspondence,  binding complexity has been conjectured to be dual to the interior volume of multi-boundary wormholes~\cite{Balasubramanian:2018hsu}.

\subsection{Complexity in quantum field theories}\label{sec:NielsenQFTALL}

As we will review in Sec.~\ref{sec:holography}, Nielsen's complexity experienced a revival following a conjectured relationship with  gravitational quantities in the context of the holographic AdS/CFT correspondence.  The holographic correspondence deals with Quantum Field Theories (QFTs) -- a description of many-body quantum systems in terms of fields. To allow a quantitative comparison with holographic conjectures, it was therefore necessary to develop a  method for calculating complexity in QFTs. Naturally, the first examples studied were those of free QFTs, first bosonic~\cite{Jefferson:2017sdb,Chapman:2017rqy} and then fermionic~\cite{Khan:2018rzm,Hackl:2018ptj}, as reviewed in subsection \ref{sec:timedepSYK}. While these calculations are too simple for an exact comparison with chaotic theories in general -- and with holographyic systems in particular -- this approach helped to bridge the gap between heuristic arguments and more rigorous calculations. This framework also inspired later studies of complexity in more complicated field theories, including weakly interacting theories \cite{Bhattacharyya:2018bbv, Cotler:2018ufx}, and conformal field theories, as described in subsections \ref{sec.2dcft} and \ref{sec.cft} below.

\subsubsection{Free quantum field theories \label{sec.free.qft}}\label{sec:timedepSYK}

The calculation of complexity in free field theories is based on Nielsen’s approach (see Sec.~\ref{sec.Nielsen.intro}). In this framework, the evolution of quantum states is treated continuously, with a cost assigned to each possible evolution. To simplify the problem, one often restricts  to unitary representations of finite-dimensional spaces. For free QFTs, this is typically achieved by restricting to transitions between Gaussian states.  Such transitions are underpinned by the symplectic group for bosonic Gaussian states and the orthogonal group for fermionic Gaussian states. 

Like many other quantities in QFT, complexity is expected to diverge because quantum correlations must be built at arbitrarily short distance scales. Therefore, a regulator should be introduced to handle short distances, or equivalently, large momenta. The exact result will depend on the choice of reference state, and in QFT, it is not immediately obvious which one to choose. Inspired by quantum computation, the reference state is often chosen to be completely unentangled, meaning that it lies outside the Hilbert space of the non-regulated theory, leading to the observed divergences.

The study of complexity in scalar bosonic QFT began with \cite{Jefferson:2017sdb, Chapman:2017rqy}. In these papers, the authors considered the following Hamiltonian for a scalar QFT in $d$-dimensional spacetime:
\begin{equation}\label{eq:freebosonH}
H = \frac{1}{2} \int d^{d-1} x \left[\pi(\vec x)^2 + (\vec\nabla \phi(\vec x))^2 + m^2 \phi(\vec x)^2 \right].
\end{equation}
The relevant symplectic transformations were constructed using the (typically discretized set of) phase space degrees of freedom 
$\hat \xi^a = (\phi(\vec x),\pi(\vec x))$.\footnote{These can alternatively be spanned using momentum-grid phase   space variables $\hat \xi^a = (\phi(\vec k),\pi(\vec k))$}
Symplectic transformations of the form 
\begin{equation}\label{eq:unitarysympl}
    \hat U(t) = \exp(-\frac{i}{2}\hat \xi^a k_{(ab)}(t)  \hat \xi^b),
\end{equation}
where $k_{(ab)}(t)$ is a symmetric matrix, move us between Gaussian states. A natural description of Gaussian states is in terms of their \emph{covariance matrix} $G^{(ab)}$ and \emph{displacement vector} $\omega^a$ defined using the first and second moments of the phase space variables
\begin{equation}
    \Tr \left(\hat \rho \hat \xi^a \hat \xi^b\right) = \frac{1}{2} \left(G^{(ab)}+i \Omega^{[ab]}\right)
    , \qquad \Tr \left(\hat \rho \hat \xi^a\right) = \omega^a,
\end{equation}
where $\hat \rho$ is the Gaussian density matrix, and $\Omega^{[ab]}$ encodes the anti-symmetric commutation relations of the phase space variables.
The covariance matrices were not the original formulation used to study complexity in bosonic QFT. Originally, the studies only focused on squeezing operations, and therefore expressing everything in terms of the wavefunction was sufficient. Nevertheless the use of covariance matrices permits for more general transformations between Gaussian states which was employed in later studies, see \eg \cite{Hackl:2018ptj,Chapman:2018hou}. We will therefore stick with this approach in our presentation.

The quantum circuits generated by the unitary transformation \eqref{eq:unitarysympl} can be recast in terms of their influence on the covariance matrix and displacement vector as follows
\begin{equation}
\begin{split}
  {\bf G}=S {\bf G} S^T, \quad 
  {\boldsymbol{\omega}}= S {\boldsymbol{\omega}},\quad \text{where} \quad  S \coloneqq e^{\mathbf{\Omega k}} \, .
  \quad 
\end{split}
\end{equation}
Here ${\bf G}, \boldsymbol{\omega}$ and ${\bf k}$ denote the covariance matrix, displacement vector and symmetric matrix in the circuit \eqref{eq:unitarysympl} and  explicit indices have been suppressed to shorten the notation. 
The control functions $Y^I$ in equation \eqref{eq:expansion_Ham} can be read off as follows
\begin{equation}
  Y^I = \frac{1}{2} \tr \left(\partial_\sigma S S^{-1} K_I\right) \, ,
\end{equation}
where $K_I$ is an orthonormal basis for the symplectic group. 
This setup is the input for the complexity calculation with the cost assigned according to equation
\eqref{eq:cost_function_Finsler}.

When evaluating complexity in QFTs, there are several  choices to be made. Initial studies \cite{Jefferson:2017sdb, Chapman:2017rqy} selected cost functions according to equation \eqref{eq:cost_function_Finsler}, with $q_I = 1$ and $p = 1,2$, and only squeezing operations were considered. The target state was taken to be the regularized ground state of the bosonic field theory \eqref{eq:freebosonH}.
Regularization was implemented  either via a lattice discretization or a sharp momentum cutoff. The reference state was chosen as the ground state of the Hamiltonian
\begin{equation}
H = \frac{1}{2} \int d^{d-1} x \left[\pi(\vec x)^2 +  \mu^2 \phi(\vec x)^2 \right],
\end{equation}
where $\mu$ is a characteristic energy scale. This Hamiltonian describes a set of independent, or unentangled, oscillators with a characteristic frequency $\mu$. When constructing correlations at all scales, it is natural to begin with a state whose frequency is of the order of the inverse lattice spacing, so we define $\tilde{\mu} \coloneqq \mu \delta$, where $\delta$ is the lattice spacing.

The original papers \cite{Jefferson:2017sdb, Chapman:2017rqy} focused primarily on the structure of divergences and found some degree of agreement with holographic results. In particular, they calculated the complexity of the vacuum state of a free bosonic theory, which reads:
\begin{equation}
\begin{split}
\mathcal{C}_{p=1}^{\text{upper bound}} &\ = \frac{1}{2} \int d^dk \left|\log (\omega_k/\mu)\right|
\simeq \frac{\mathcal{V}}{2\delta^{d-1}} |\log (\tilde \mu )| +\ldots
, 
    \\
    \mathcal{C}_{p=2} &\ = \frac{1}{2} \sqrt{ \int_k \left[\log (\omega_k/\mu)\right]^2} \simeq \frac{1}{2} \left(\frac{\mathcal{V}}{\delta^{d-1}}\right)^{\frac{1}{2}} |\log (\tilde \mu )| +\ldots
    ,
\end{split}
\end{equation}
where for $p=1$ only an upper bound was found. In these equations $\mathcal{V}$ stands for the spatial volume of the system, $\delta$ is an ultraviolet cutoff (for example a lattice spacing) and the dots stand for less divergent terms (see the review \cite{Chapman:2021jbh}).   These results should be contrasted with those obtained from holography \eqref{eq:hologvaccompres}. We will later see that the $\mathcal{C}_{p=1}$ expression matches well with the volume-law divergence found in holographic calculations, see Eq.~\eqref{eq:hologvaccompres}. 
Additionally, the choice of reference scale in the field theory parallels the choice of the counterterm scale $L_{ct}$ in holography, as we will see in, e.g., Equation~\eqref{eq:subregion_CA}.

Subsequent studies explored a variety of alternative cost functions, free theories with other fields such as charged bosons, fermions, and gauge fields, weakly interacting theories and other target states, including the thermofield-double state, and mixed states \cite{Balasubramanian:2018hsu, Hackl:2018ptj,Chapman:2018hou, Khan:2018rzm, Bhattacharyya:2018bbv, Guo:2018kzl, Cotler:2018ufx, Bernamonti:2019zyy, Caceres:2019pgf,Camargo:2020yfv,Hashimoto:2017fga, Chapman:2019clq, Ge:2019mjt,Camargo:2018eof}. 
These studies concluded that, in general, the structure of divergences observed in QFT complexity calculations qualitatively matches the results from holography. This is not particularly surprising, since at very short distances (or equivalently, near the cutoff scale), the precise details of the theory tend to become irrelevant. On the other hand, the time dynamics of the theory shows significant deviations from holographic results, which is to be expected, as free and chaotic theories typically exhibit very different dynamical behavior. In particular, a natural expectation is that in free systems of finite volume, after including penalties on Gaussian gates coupling distant parts of the system, complexity growth associated with Hamiltonian time evolution will terminate at times of the order of the volume (and even sooner without the  penalties~\cite{Chapman:2018hou}). In chaotic systems, we expect complexity growth to occur over significantly longer timescales, exponential in the volume.

While many basic insights were obtained from the above approach, it fails to reproduce many important properties of holographic complexity, including those related to temporal dynamics. This observation motivated a shift of focus towards the study of complexity in field theories more closely related to holography. The latter included attempts to study complexity in conformal field theories via a focus on their universal conformal symmetry, which we describe below.

\subsubsection{Two dimensions - complexity of Virasoro circuits \label{sec.2dcft}}

Nielsen's geometric approach is  appealing from a geometric point of view, and is tailor-made to mimic discrete quantum circuits built with gates available in  laboratories. However, it is not clear how to define gates in a generic QFT.  Indeed, we may need to consider all local operators, in view of the fact that an operator product can be represented as a sum over local operators via the operator product expansion. There is also the question of choosing a meaningful, yet tractable, cost function that gives rise to an optimization problem that is under control. 

Such challenges forced the community to approach the problem of Nielsen's complexity in QFT in a creative way.  Progress can be traced back to~\cite{Caputa:2018kdj}, whose authors proposed to study Nielsen's complexity for  conformal transformations of 2-dimensional CFTs, i.e., coordinate transformations that preserve the form of the metric up to an overall (coordinate-dependent) scalar factor.

The immediate advantage of this approach is that it focuses on universal properties shared by all CFTs, regardless of the value of the interaction strength.  As a result, it  holds promise for understanding the gravitational dual of complexity within the AdS/CFT correspondence.  Of course, this universality poses also a  restriction, as the resulting complexity measures will be insensitive to the detailed microscopic dynamics.

The connection between gates and local operators in a QFT occurs because conformal transformations are generated by a local operator that is universally present in any QFT, the energy momentum-tensor, smeared over a time-slice. A familiar example of a conformal transformation is time translation generated by the Hamiltonian, which in turn is an integral of energy density - the time-time component of the energy-momentum tensor.

In more detail, \cite{Caputa:2018kdj} and subsequent studies focused on  two-dimensional CFTs living on the Lorentzian cylinder. As illustrated in Fig.~\ref{fig:viraso_cases}, we parametrize the spatial circle of unit radius  by $\sigma$ and the physical time by $t$. In two dimensions, the energy-momentum tensor of a CFT has two independent components that commute with each other. In the following, we will predominantly focus on one of them only,  $T(\sigma) \equiv \frac{1}{2}T_{tt} + \frac{1}{2} T_{t\sigma}$. The algebraic structure underlying this subsection's construction is the Virasoro algebra
\be
[L_n,L_m]=(n-m)L_{n+m}+\frac{c}{12}n(n^2-1)\delta_{n+m,0} \, ,
\ee
whose generators $L_{j}$ are related to Fourier modes of $T(\sigma)$
\begin{eqnarray}
T(\sigma)&=&\sum_{n\in \mathbb{Z}} \left(L_n-\frac{c}{24}\delta_{n,0}\right)e^{ -i n \sigma} \, .
\end{eqnarray}
In the above expressions, $c$ is the central charge of the CFT, which can be thought of as measuring the number of degrees of freedom in the following sense. For a CFT containing $N$ free bosons, $c = N$, which is the number of bosonic degrees of freedom; whereas for a CFT containing $N$ free fermions, $c = \frac{N}{2}$. In holography, where the boundary CFTs are strongly coupled, $c \rightarrow \infty$. Since the algebra is the same for all CFTs with the same value of $c$, one explicitly sees here the point we introduced earlier: the properties of Virasoro algebra-based complexity will not differ between $N \rightarrow \infty$ free bosons and strongly-coupled holographic theories.

\begin{figure}
    \centering
    \includegraphics[scale=0.3]{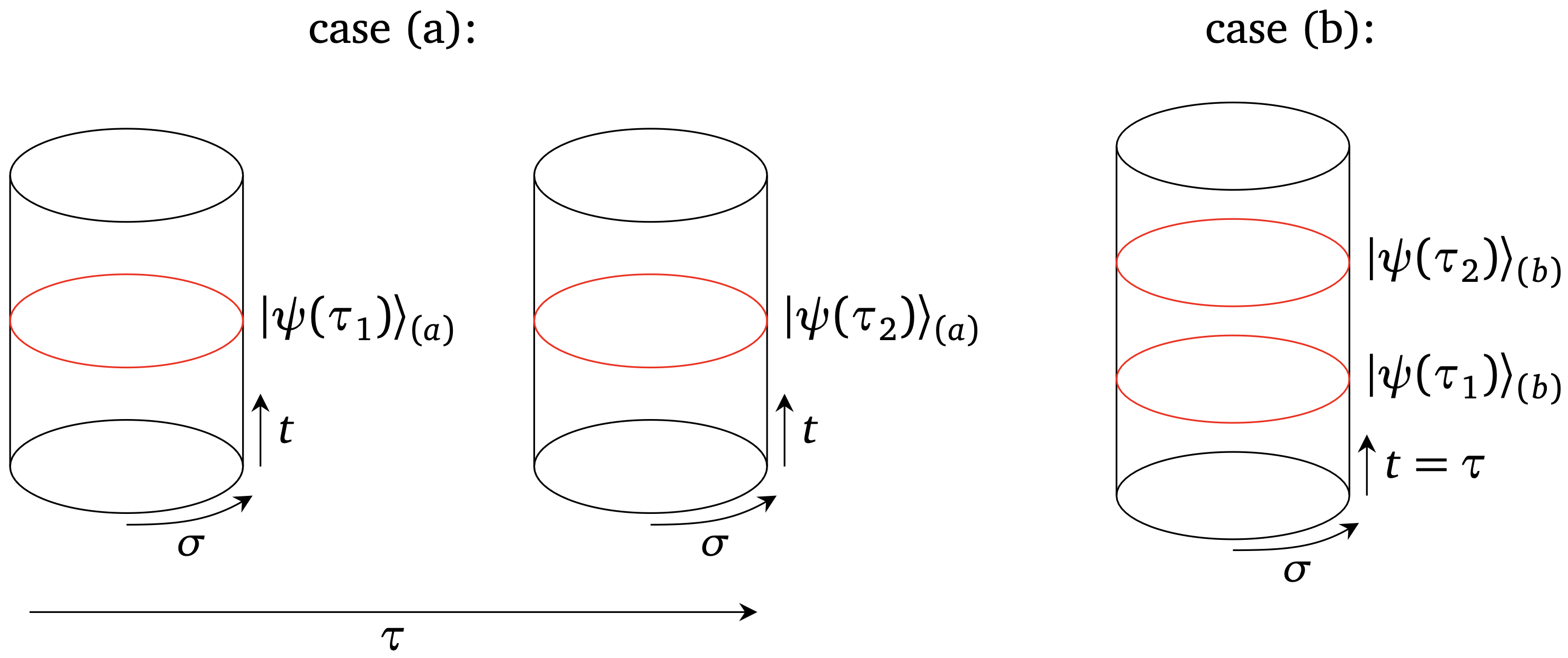}
    \caption{Two kinds of circuits that can be considered in the context of conformal transformations, or, more generally, QFT (in this latter scenario, one should think of $\sigma$ as a set of coordinates on spatial slices). Case (a): The circuit parameter~$\tau$ is an auxiliary parameter whose increase takes one CFT state to another CFT state. The transformation does not lead to a shift in the physical time~$t$. This is the setting addressed in~\cite{Caputa:2018kdj}. Case (b): The circuit parameter~$\tau$ is identified with the physical time~$t$, as considered for the first time in~\cite{Erdmenger:2021wzc}. Figure adapted from~\cite{Erdmenger:2021wzc}.}
    \label{fig:viraso_cases}
\end{figure} 

The setup of interest for~\cite{Caputa:2018kdj}, as well as subsequent works including~\cite{Erdmenger:2020sup,Flory:2020eot,Flory:2020eot,deBoer:2023lrd}, is given by continuous unitary circuits
\begin{equation}
U(\tau)=\cev{\mathcal{P}}\exp\left[\int^\tau_0Q(\tau')d\tau'\right] \, ,
\label{UtVir}
\end{equation}
acting on a CFT Hamiltonian eigenstate $|h\rangle$ with the label $h$ corresponding to a primary operator with conformal dimension~$h$ ($h = 0$ gives the vacuum state). The circuit generator in \eqref{UtVir} is a smeared component of the energy-momentum tensor operator over the full time slice 
\begin{equation}\label{gateV}
 Q(\tau)\coloneqq \int^{2\pi}_0\frac{d\sigma}{2\pi}\epsilon(\tau,\sigma)  T(\sigma)= \sum_{n\in \mathbb{Z}}\epsilon_n(\tau) \left(L_{-n}-\frac{c}{24}\delta_{n,0}\right) \, ,
\end{equation}
where
\begin{eqnarray}
\epsilon(\tau,\sigma)&\equiv&\sum_{n\in \mathbb{Z}} \epsilon_n(\tau)e^{ -i n \sigma}.
\end{eqnarray}
In comparison with Eq.~\eqref{eq:expansion_Ham} discussed in the context of qubit systems, the integral over $\sigma$ is the analogue of the summation over the $I$-index, the $\epsilon(\tau)$ factor is the tangent space velocity $Y^{I}(t)$, and the generators can be thought of as $T(\sigma)$. Alternatively, we can think of the Virasoro generators as defining circuit generators. In such case, the summation over them is the analogue of the sum over $I$ in Eq.~\eqref{eq:expansion_Ham}, and the tangent space velocity is related to $\epsilon_{n}(\tau)$. 

In considering the Nielsen's complexity, there is an important choice to be made regarding the interpretation of the~$\tau$ parameter~\cite{Erdmenger:2021wzc}. The first choice is to regard it as an auxiliary parameter in relation to the physical time~$t$, see case~(a) in~Fig.~\ref{fig:viraso_cases}. For this choice, relevant for~\cite{Caputa:2018kdj}, the function~$\epsilon(\tau)$ associated with the diffeomorphism $\sigma \rightarrow f(\tau,\sigma)$ is given by
\begin{equation}
\epsilon(\tau, \sigma) = \frac{\partial_{\tau} f(\tau, \sigma)}{\partial_{\sigma} f(\tau, \sigma)}.
\end{equation}
The complexity  minimization problem at hand now requires us to find a function of two parameters $f(\tau,\sigma)$ subject to the condition $f(\tau = 0,\sigma) = \sigma$ and $f(\tau = 1, \sigma) = f(\sigma)$, where $f(\sigma)$ is a  diffeomorphism that transforms the state $|h\rangle$ according to the CFT rules.

The papers~\cite{Caputa:2018kdj,Erdmenger:2020sup,Flory:2020eot,Flory:2020dja} uncovered and developed a very beautiful relation between optimizing cost functions of the form
\begin{equation}
\label{eq.CFTcost}
F[\epsilon] = \sqrt{\int_{0}^{2\pi} d\sigma \int_{0}^{2\pi} d\kappa \, \Pi(\sigma-\kappa)\, \epsilon(\tau,\sigma)\, \epsilon(\tau,\kappa)} \, ,
\end{equation}
and Euler-Arnold-type partial differential equations (PDEs) of  relevance for mathematical physics. In particular, for  penalty schedules of the form
\begin{equation}
\Pi(\sigma-\kappa) = a \, \delta(\sigma - \kappa) + b \,\delta''(\sigma-\kappa)
\end{equation}
with $a$ and $b$ constant and $\delta(\sigma-\kappa)$  the Dirac delta function, the solutions of the optimization problem are the solutions of the following paradigmatic PDEs~\cite{Flory:2020eot}
\begin{itemize}
\item Korteweg-de Vries equation predicting solitons: $a = 1$, $b = 0$,
\item Hunter-Saxton equation relevant for liquid crystal physics: $a = 0$, $b = 1$,
\item Camassa-Holm equation modelling wave breaking: $a = 1$, $b = 1$.
\end{itemize}
Furthermore, Refs.~\cite{Flory:2020eot,Flory:2020dja} considered the Fubini-Study cost function, which corresponds to Eq.~\eqref{eq.CFTcost} with
\begin{equation}
\Pi(\sigma-\kappa) = \frac{c}{32\sin^4\left((\sigma-\kappa)/2\right)} - \frac{h}{2\sin^2\left((\sigma-\kappa)/2\right)} \, ,
\end{equation}
and solved the associated optimization perturbatively for sample conformal transformations close to identity. This result was subsequently used for comparisons with holography, building on~\cite{Flory:2018akz}, see Sec.~\ref{sec:holography}.

There are further generalizations of the approach described above. Reference~\cite{Erdmenger:2020sup} connected with the rich literature of the Kac-Moody groups by incorporating an additional global symmetry. Furthermore, in Ref.~\cite{Erdmenger:2024xmj} a scalar primary operator was added as an additional generator on top of the energy-momentum tensor operator. The main limitation of the works prior to~\cite{Erdmenger:2024xmj} is that the previously described unitary circuits remain bounded inside a conformal family (the so-called \textit{Verma module}) of the Virasoro algebra. Results in~\cite{Erdmenger:2024xmj} concern trajectories that move between different Verma modules in a  CFT, revealing the dependence of the Fubini-Study cost of such circuits on the source function of the primary generator. This investigation constitutes an important  step towards arriving at a general picture of Nielsen's complexity with local operators viewed as generators, as advocated above.

Finally, another line of works devoted to the study of complexity in CFT by examining trajectories in which the physical time~$t$ is the circuit parameter~$\tau$~\cite{Erdmenger:2021wzc,Erdmenger:2022lov,deBoer:2023lrd}, see Fig.~\ref{fig:viraso_cases}~case~(b). In the context of the examples considered, this approach requires slightly different choices of the circuit velocity $\epsilon(t,\sigma)$. The main advantage of this approach is that it makes  direct contact with the spacetime (path integral) formulation of QFTs and with holography \cite{Erdmenger:2022lov}. 

\subsubsection{Circuits in the conformal group in $d\geq 2$ \label{sec.cft}}

While it is helpful to use the conformal algebra to generate circuits, as we have seen above, the calculations are still quite challenging for two dimensional CFTs as the conformal group in this case has an infinite number of generators. The problem simplifies for higher dimensional CFTs because the conformal group is finite dimensional.
The global conformal group $\mathrm{SO}(d,2)$ associated with symmetries of the vacuum of CFTs in $d$ spacetime dimensions also plays a key role in connection to holography, since it coincides with the local isometries of higher-dimensional AdS spacetime. Within the setup discussed in Sec.~\ref{sec.2dcft} this corresponds to the subgroup\footnote{The global conformal group for CFTs living in two spacetime dimensions, $\mathrm{SO}(d,2)$, factorizes into a product of two $\mathrm{SO}(1,2)$. The Virasoro generators considered in Sec.~\ref{sec.2dcft} contain (among other things)  one of these~$\mathrm{SO}(1,2)$. The other $\mathrm{SO}(1,2)$ originates from the remaining independent component of the energy-momentum tensor operator: $ \bar{T} \coloneqq \frac{1}{2}T_{tt} - \frac{1}{2}T_{t\sigma}$.} of the Virasoro algebra associated with the generators $L_{-1}$, $L_{0}$ and $L_{1}$. 

This framework is similar to the qubit setup  in Sec.~\ref{sec.Nielsen.intro}, with the exception that we now consider a unitary representation of a (possibly non-compact) Lie group $G$.\footnote{One can always build a unitary and finite-dimensional representation of a Lie group by using Euclidean generators.}
For concreteness, a generic trajectory on the conformal group can be constructed out of the generators $P_\mu$, $D$, $L_{\mu\nu}$ and $K_\mu$ (which generate translations, dilatations, rotations and special conformal transformations, respectively) as follows,
\begin{equation}
    U(s) = e^{i\alpha(s) \cdot P} e^{i \gamma_D(s) D} \left( \prod_{\mu<\nu} e^{i\lambda_{\mu\nu}(s)L_{\mu\nu}}  \right) e^{i\beta(s)\cdot K} \, ,
    \label{eq:unitary_CFT}
\end{equation}
where $\alpha_\mu$, $\gamma_D$, $\lambda_{\mu\nu}$ and $\beta_\mu$ are complex variables, and $s\in [0,1]$ is a path parameter along the circuit.
The reference state is chosen to be a primary state $|\Delta\rangle$, such that $D|\Delta\rangle = \Delta|\Delta\rangle$ and $K_\mu|\Delta\rangle=L_{\mu\nu}|\Delta\rangle =0$. 
The states reached by the unitaries \eqref{eq:unitary_CFT} live within the same conformal family; the extension to a larger family of CFT states is still an open question.

Let us discuss Nielsen's state complexity associated with a target state $|\psi_{\rm R} \rangle$, reached by acting on a primary state $| \Delta \rangle$ with a unitary operator of the form \eqref{eq:unitary_CFT}. The projection of a Riemannian right-invariant cost function from a Lie group $G$ to the coset space $G/H$ (where $H$ is the maximal proper subgroup of a state) is unique, and it is defined by a pseudo-Riemannian submersion~\cite{oneill1983semiriemannian,Besse:1987pua}.  This  provides a systematic procedure for inducing a metric from the conformal group $\mathrm{SO}(d,2)$ to the coset space $\mathcal{B}=\frac{\mathrm{SO}(d,2)}{\mathrm{SO}(2) \times \mathrm{SO}(d)}$.
Moreover, the projection procedure is equivalent to two other ways of generating a metric over the quotient: the minimization in Eq.~\eqref{eq:def_Fstate}, 
and the use of a geometric action associated with the geometry of coadjoint orbits \cite{CFTpen}. 
A coadjoint orbit is the space of all equivalent configurations that can be reached by acting with the symmetry group on a point in the dual space of the corresponding Lie algebra.
Physically, it represents a phase space corresponding to a fixed value of a conserved charge, where the symmetry group of the system allows movement within this space. 
Two cost function candidates were considered in the literature in this context~\cite{Chagnet:2021uvi} 
\begin{subequations}
\label{eq:cost_functions_coadj}
\begin{eqnarray}
F_{p=1, \vec{q}=\mathbf{1}} d\sigma &=& |\langle\psi_{\rm R}|U^\dagger dU |\psi_{\rm R}\rangle | \, , \\
 ds_{\rm FS} = F_{\rm FS} \, d\sigma &\equiv& \left(\langle \psi_{\rm R}|dU^\dagger dU |\psi_{\rm R}\rangle - |\langle\psi_{\rm R}|U^\dagger dU |\psi_{\rm R} \rangle |^2\right)^{1/2} \, . \label{eq:cost_FS}
\label{eq:cost_U1}
\end{eqnarray}
\end{subequations}
Both cost functions have some limitations because of the absence of penalty factors associated with different gates. Nevertheless, they provide a good starting point.
In particular, the Fubini-Study (FS) line element $ds_{\rm FS}$ given by Eq.~\eqref{eq:cost_FS} is the natural metric on the coset space $\mathcal{B}$ of the conformal group, whose geodesics are known.
The cost function  in Eq.~\eqref{eq:cost_U1} can be related to a K\"{a}hler metric associated with coadjoint orbits in the dual space~\cite{Chagnet:2021uvi}. 
It is still an open problem to generalize the results of~\cite{Caputa:2018kdj,Chagnet:2021uvi} to account for non-trivial penalty factors; this is the subject of ongoing research \cite{CFTpen}. Finally,  Ref.~\cite{Chagnet:2021uvi} found an explicit form of the complexity metric for higher-dimensional CFTs. In terms of the control functions in Eq.~\eqref{eq:unitary_CFT}, their result has a precise analogue in terms of distances between geodesics in AdS, as we will discuss in Sec.~\ref{subsec:Nielsenholo}.

\clearpage
\section{Paradigms for complexity II: Krylov and spread complexities}
\label{sec:Krylov}

In the previous sections, we explained notions of complexity that arise from thinking about physical time evolution as a computation. From this perspective, the complexity of a physical process is quantified by asking how many elementary or cheap operations are needed to effect it, given some fixed repertoire of operations or a  function quantifying their cost.  Alternatively, consider that   in classical physics, if we start with systems populating some small compact region of phase space, we might call the dynamics ``simple'' if these initial conditions are transported to some other small region of phase space at a later time.  We might call the dynamics ``complex'' if the initial conditions are spread out throughout the phase space by the dynamics.  A notion of quantum complexity quantifying this sort of spread would seem to have a natural relation to physical phenomena like thermalization and chaos in quantum many-body systems and QFT.

One approach in quantum mechanics is to quantify the growth of different operators over time via the notion of {\it Krylov complexity} developed in  \cite{Parker:2018yvk}.  Alternatively, we could ask how  different initial states dynamically explore the Hilbert space during unitary evolution via the notion of  {\it spread complexity} developed in \cite{Balasubramanian:2022tpr}. These approaches are related, but each of them is convenient for answering different questions.  In both approaches, the key insight is to recognize that, while physicists often analyze quantum systems by diagonalizing the Hamiltonian to extract the density of states, it is much more convenient to {\it tri-diagonalize} the Hamiltonian if we want to study dynamics.  The diagonal and immediately off-diagonal components of the triadiagonalized Hamiltonian are called the Lanczos coefficients, and {\it any} quantum dynamics can be written as a one-dimensional hopping chain in terms of the Lanczos coefficients and the associated Krylov basis for the Hilbert space.  Among other things, this representation of a quantum system makes it possible to explicitly quantify and compute how a quantum wavefunction spreads under Hamiltonian dynamics across the Hilbert space, or how an operator spreads in operator space.

Below, we review the background for Krylov and spread complexities, and explain how to compute these quantities. We  present  numerical methods,  analytical results in systems with a lot of symmetry, and a general analytical formula relating the density of states of a system to its Lanczos coefficients. We also present a formula for the correlations in the Lanczos coefficients of Random Matrix Theories.  Along the way, we relate these ideas to Nielsen complexity, and describe applications to the study of chaos and integrability. In the next Section, devoted to holography, we will describe a precise relationship between the length of wormholes in the two-dimensional JT gravity, and the spread complexity of the Thermofield Double state of the dual SYK model, see Sec.~\ref{ssec:holo_matching_2d}. Readers may also wish to consult the excellent recent review \cite{Nandy:2024htc}.

\subsection{Spread complexity of evolving states}
\label{ssec:Krylov_def}
 
Consider the unitary time evolution of some initial quantum state $\ket{\Psi_0}$ 
\be
\ket{\Psi(t)}=e^{-iHt}\ket{\Psi_0} = \sum^\infty_{k=0}\frac{(-it)^k}{k!}H^k\ket{\Psi_0} \, ,
\label{eq:ContTimeEv}
\ee
where $H$ is a time-independent Hamiltonian. Since the evolution is unitary, the state at any given time is just a fixed vector in the Hilbert space.  Standard methods, including the Out-Of-Time-Order-Correlation (OTOC, see definitions in Sec.~\ref{sec:key_notions}), quantify how time evolution spreads a collection of nearby states through the Hilbert space. At least at early times, this process can be characterized by computing quantities akin to the Lyapunov exponent. Instead, we will quantify how widely the trajectory of a {\it single} state explores the Hilbert space over time.  One way of doing this is to measure the spread of the support of a time-evolving state in some fixed basis.  Which basis should we pick?  The early work of Kolmogorov, defining the complexity of sequences as the size of the smallest Turing machine that produces them, suggested a paradigm: the complexity of a system should be measured in terms of the size of its minimal representation.  Taking this perspective, we could define the  complexity of an evolving state by minimizing the spread of the wavefunction over all possible bases. The authors of \cite{Balasubramanian:2022tpr} proved that there is a unique orthonormal minimizing basis throughout a finite time interval for continuous  evolution, and at
all times in the case of discrete evolution. This is the Krylov basis which can be constructed by a variety of methods described below.  

For discrete time evolution, a simple argument explains how the Krylov basis minimizes the spread of the wavefunction. Consider an initial state $\ket{\psi_0}$, and a basis supporting the state on a single basis element $\ket{K_0} = \ket{\psi_0}$.  Time evolution produces the state at time 1 as $\ket{\psi_1} = U_1 \ket{\psi_0}$, where $U_1$ is a unitary operator.  By picking the second basis element $\ket{K_1}$ to be proportional to the part of $\ket{\psi_1}$ orthogonal to $\ket{\psi_0}$, we can support $\ket{\psi_1}$ on two basis elements, $\{ \ket{K_0}, \ket{K_1} \}$.  Continuing in this way by recursively orthogonalizing the time-evolving state, we produce a basis that obviously minimizes the support of the state at any time.  The analogous proof for continuous time evolution is more subtle and is described in \cite{Balasubramanian:2022tpr}.

\subsubsection{Recursion method and tridiagonalization} 
\label{sec:recursion}
To represent the time evolving state (\ref{eq:ContTimeEv}) in the minimizing Krylov basis for a continuous time evolution, we write
\be
\ket{\Psi(t)}
=\sum^\infty_{k=0}\frac{(-it)^k}{k!}H^k\ket{\Psi_0}
=\sum_{n=0}^{\mathcal{K} - 1} \psi_n(t)\ket{K_n} \, ,
\label{State1}
\ee
where $\mathcal{K}$ is the dimension of the span of the time-evolving state,  $\ket{K_0} = \ket{\psi_0}$, and the remaining $\ket{K_n}$ are constructed by the iterative  Lanczos algorithm \cite{Lanczos:1950zz}:
\be
|A_{n+1}\rangle=(H-a_{n})|K_n\rangle-b_n|K_{n-1}\rangle,\qquad |K_n\rangle=b^{-1}_n|A_n\rangle\;.
\label{eq:Lrecursion}
\ee
The Lanczos coefficients $a_n$ and $b_n$ are defined as
\be
a_n=\langle K_n|H|K_n\rangle,\qquad b_n=\langle A_n|A_n\rangle^{1/2}\;,
\ee
with $b_0 = 0$.  This procedure is just a systematic way of applying Gram-Schmidt orthonormalization to the Krylov subspace $\{ H^k\ket{\Psi_0} \}$, and will terminate at some $n=\mathcal{K}$ for which $b_{\mathcal{K}}=0$. Thus $\mathcal{K}$, the dimension of the Krylov basis, measures the dimension of the subspace of the Hilbert space explored by time evolution of the initial state.

\paragraph{Tridiagonalization} The construction \eqref{eq:Lrecursion} implies that the Hamiltonian acts tridiagonally in the Krylov basis 
\be
H\ket{K_n}=a_n\ket{K_n}+b_n\ket{K_{n-1}}+b_{n+1}\ket{K_{n+1}} \, .\label{KBdef}
\ee
In other other words, the Hamiltonian in this basis can be written as a matrix with entries $a_n$ along the diagonal, and entries $b_n$ just above and below it. For finite dimensional systems, this is called the Hessenberg form of the Hamiltonian.  Numerically stable algorithms for computing the Hessenberg form of a matrix are implemented by standard libraries in Python and Mathematica (see \cite{Balasubramanian:2022tpr}). These methods improve upon the numerically unstable Lanczos procedure, but generally require a change of basis to represent the desired initial state in a canonical form required by the implemented algorithm.  We can use the Hessenberg form to extract the Lanczos coefficients without explicitly constructing the Krylov basis.

Finally, differentiating Eq.~\eqref{State1} with respect to $t$ and using Eq.~\eqref{KBdef}, we find the  Schr\"{o}dinger equation
\be\label{SchrodingerEq}
i\partial_t\psi_n(t)=a_n\psi_n(t)+ b_n\psi_{n-1}(t) +b_{n+1}\psi_{n+1}(t)  \;,
\ee
that we should solve with the initial condition $\psi_n(0)=\delta_{n,0}$. This equation describes the dynamics of an effective ``particle" hopping on a one-dimensional chain with sites labeled by the Krylov index $n\in\{0,1,2...\}$. The Lanczos coefficients $a_n$ determine the probability for staying at site $n$, whereas $b_n$ determine the probability of hopping to the left or right.  Thus, the procedure we have described reduces {\it any} quantum dynamics to an effective one-dimensional hopping chain -- i.e., a quantum Markov process -- where the states defining the chain and transition coefficients are determined from the initial state $\ket{\psi_0}$ and the Hamiltonian.

\subsubsection{Moment method and the survival amplitude}
\label{sec:momentmethod}

The Lanczos coefficients can be efficiently extracted from the moments of the Hamiltonian in the initial state 
\begin{equation}
\mu_n = \bra{K_0} (-iH)^n \ket{K_0} \, .
\label{eq:moments}
\end{equation}
To see why, suppose that a quantum system is localized, at $t=0$, in the initial state $\ket{K_0}$.  The Markov-like structure of Eq.~(\ref{KBdef}) means that the probability amplitude for remaining in $\ket{K_0}$ after one application of the Hamilton is $-ia_0$, while the amplitude for transitioning to $\ket{K_1}$ is $-ib_1$. 
Next, let us apply the Hamiltonian a second time. The amplitude to wind up in $\ket{K_0}$ again is $i^2(a_0^2 + b_1^2)$.  But we already know $a_0$ from the first step.  So, if we we also know the second moment of the Hamiltonian $\bra{K_0} H^2 \ket{K_0}$ through some means, then we can extract $b_0$ by subtracting $a_0^2$ and taking a square root. Proceeding recursively in this way, we can show that each successive moment allows the extraction of an additional Lanczos coefficient --  the $a_n$ are computed from the odd moments by dividing out products of already computed $b_n$s, and the $b_n$s are likewise computed from the even moments. Applied numerically, this procedure is potentially sensitive to rounding errors from repeated divisions, but there are methods for circumventing this potential obstacle \cite{Balasubramanian:2022tpr}.\footnote{One possible way to resolve this difficulty, \eg is to find squares of the Lanczos coefficients by using rational arithmetics~\cite{Uskov:2024xac}. }  We can use the moment method to compute the Lanczos coefficients without constructing the Krylov basis.

\subsubsection{Survival amplitude}
The moments of the Hamiltonian are all encoded in the so-called {\it survival} or {\it return} amplitude for the state to remain unchanged over time:
\be
S(t)\equiv\langle\psi(t)|\psi_0\rangle=\langle \psi_0|e^{-iHt}|\psi_0\rangle ~~~
\Longrightarrow ~~~
\mu_n = \bra{K_0} (-iH)^n \ket{K_0} = \frac{d^n}{dt^n} S(t) \Big|_{t=0}
\,
.
\ee
So, if we have access to the survival amplitude through some other means, we can use it as a moment generating function for the Hamiltonian, and then proceed as described above to extract the Lanczos coefficients without explicitly constructing the Krylov basis.  For example, consider a thermofield double state\footnote{The notation in Eq.~\eqref{eq:TFD_Krylov} is analogous to Eq.~\eqref{eq:mytfd} in the key notions. Here we stress the dependence of the state on the inverse temperature $\beta$.}   
\begin{equation}
\ket{\psi_\beta} = \frac{1}{\sqrt{Z_\beta}} \sum_n e^{-\beta E_n} \ket{n,n} \, ,
\label{eq:TFD_Krylov}
\end{equation}
an entangled state in two copies of the same system. Tracing over either copy yields the thermal density matrix with partition function $Z_\beta$ in one copy.  Suppose that the two copies have Hamiltonians $H_L$ and $H_R$. Time evolution by $H=(H_L + H_R)/2$ or separately by $H=H_{L,R}$ produces the state
\begin{equation}
\ket{\psi_\beta(t)} = e^{-iHt} \ket{\psi_\beta} = \ket{\psi_{\beta + 2it}}\,,
\end{equation}
and the associated analytically continued partition sum 
$Z_{\beta - it} = \sum_n e^{-(\beta - it) E_n}$.  A short calculation shows that the survival amplitude can be written as
\begin{equation}
S(t)= \braket{\psi_{\beta + 2 it}}{\psi_\beta} = \frac{Z(\beta-it)}{Z(\beta)} \, .
\label{eq:KrylovrelationSFF}
\end{equation}
Thus, given an analytic form for the partition sum of a system, we can analytically continue to compute the survival amplitude for the thermofield double, and hence its Lanczos coefficients following the procedure discussed above. It is also interesting to note that the square of this survival amplitude is precisely the spectral form factor
\begin{equation}
\mathrm{SFF}_{\beta -it} = \frac{|Z_{\beta - it}|^2}{|Z_\beta|^2}\,,
\end{equation}
which has been  used to study late time quantum chaos \cite{Guhr:1997ve,Cotler:2016fpe} in random matrix theory and quantum gravity, quantum speed limits   \cite{delCampo:2017bzr}, and other phenomena that depend on energy spectrum statistics.

\subsubsection{Coarse-grained Lanczos spectrum and  the density of states}
\label{sec:coarseLanczosSpectrum}
Suppose we tridiagonalize the Hamiltonian by the recursion method or the moment method, as described above.  The resulting Hamiltonian, written entirely in terms of the Lanczos coefficients, should still have the same spectrum as the original one.  So we may wonder whether we can directly relate the energy spectrum, or equivalently the density of states, to the tridiagonal Lanczos spectrum (i.e., the distribution of Lanczos coefficients).  Indeed, as we will discuss below, we can find such a relation in the limit of large system size, given some smoothness assumptions and after coarse-graining  the Lanczos spectrum \cite{Balasubramanian:2022dnj}.

Consider a quantum system with an $N$ dimensional Hilbert space.  Let $n$ be the Lanczos index and define $x=n/N$.  In the large $N$ limit, $x$ is a real number between $0$ and $1$ and it is convenient to write the Lanczos coefficients as functions $a(x)$ and $b(x)$ instead of as $a_{xN}$ and $b_{xN}$. Let us consider Hamiltonians and initial states for which $a(x)$ and $b(x)$ have a continuous large $N$ limit.  As we will explain, if the large $N$ limit is continuous in this way, there is an analytical formula relating  the density states of the theory  to the Lanczos spectrum, coarse-grained in a way that we describe below.

Recall from Sec.~\ref{sec:recursion}  that the Lanczos coefficients $a_n$ and $b_n$ describe hopping dynamics on a one dimensional chain defined by the Krylov basis.  Cut this chain into segments of length $L$ such that $L/N \to 0$ at large $N$.  For example, we can take $L\sim\sqrt{N}$. We are assuming that $a(x)$ and $b(x)$ with $x=n/N$ have a continuous large $N$ limit.  This requires that $a_n$ and $b_n$  vary slowly within the blocks of length $L$.  Coarse-graining over these blocks, which represent infinitesimal intervals of $x$, we can approximate the Hamiltonian as consisting of $N/L$ blocks within each of which the Lanczos coefficients are constant, i.e., it is Toeplitz.  A standard formula from linear algebra then tells us that the energy eigenvalues associated to the block of size $L$ are $E_k = 2b \cos(k\pi/(L+1)) + a$ for $k= 1, \dots, L$, where $a$ and $b$ are the constant values of the Lanczos coefficients in the block. Thus, keeping in mind from the Lanczos algorithm that we can take $a$ real and $b$ positive, the density of states within each segment is
\begin{equation}
\rho_{(a,b)}(E) = \frac{1/L}{|dE_k/dk|} = \frac{\Theta(4b^2 - (E-a)^2)}{\pi \sqrt{4b^2 - (E-a)^2}}\,,
\end{equation}
where $\Theta$ is the Heaviside step function.
The union of the block eigenvalues must reproduce the complete energy spectrum. In the large $N$ limit both $L \sim \sqrt{N}$ and $N/L$ are large, so we can write the full density of states as an integral over the blocks indexed by $x$:
\begin{equation}
\rho(E) = \int_0^1 dx \, \frac{\Theta(4b(x)^2 - (E-a(x))^2)}{\pi \sqrt{4b(x)^2 - (E-a(x))^2}}\,.
\label{eq:DensityLanczosAnalytic1}
\end{equation}
This analytical formula universally relates the density of states and the coarse-grained Lanczos coefficients for large systems. For the systems like the SYK model and Random Matrix Theories that are defined by ensembles of Hamiltonians, this formula will also relate the average density of states to the ensemble average of Lanczos coefficients for random initial states. The authors of \cite{Balasubramanian:2022dnj} give a more formal analysis for why this formula is true, and explain how to invert it to calculate $a(x)$ and $b(x)$ from the density of states.  The formula is valid for initial states for which the large $N$ limit of $a(x)$ and $b(x)$ is continuous, and so long as $x$ is not too close to the edge, i.e., $x\sim O(1/N)$. At the edge, we have to use the recursion or moment methods described above to find the Lanczos coefficients step by step.

\subsubsection{Defining spread complexity}
Suppose we have derived the Lanczos spectrum using one of the methods described above.  We can use it to solve the Schr\"{o}dinger's equation (\ref{SchrodingerEq}) for the wavefunction in Krylov basis \eqref{State1}. Squaring the $\psi_n(t)$ yields a probability distribution that quantifies the spread of the wavefunction over the Krylov basis
\be
p_n(t)\coloneqq |\psi_n(t)|^2,\qquad \sum^{\mathcal{K}-1}_{n=0} p_n(t)=1.
\label{eq:krylovdist}
\ee
From the 1d chain perspective, the wavefunction starts out localized on $\ket{K_0}$.  As time passes, the peak of the wavefunction moves to the right (higher Krylov index $n$) and spreads out.   The wavefunction has a sharp, tsunami-like edge as it moves outward \cite{Balasubramanian:2022tpr}. This recalls the sharp edge seen in the spread of entanglement after a quench \cite{Calabrese:2009qy,Abajo-Arrastia:2010ajo,Balasubramanian:2011ur,Balasubramanian:2010ce,Liu:2013qca}, at least in two-dimensional conformal systems and in theories with a holographic dual.  In chaotic systems like Random Matrix Models (RMTs), the spreading wavefunction retains this coherence until it ``bounces off'' the end of the Krylov chain ($n=\mathcal{K}$) and settles down to equilibrium with broad support on all the basis elements.  This bounce arises distinctively from the spectral correlations in chaotic theories.  This was illustrated in \cite{Balasubramanian:2022tpr} by comparing chaotic RMTs in the Gaussian Unitary, Orthogonal, and Symplectic Ensembles (GUE, GOE, GSE) with systems having the same density of states but Poisson, i.e., uncorrelated, spectral statistics.

Complete information about the spread of the wavefunction is contained in the distribution (\ref{eq:krylovdist}).  We could quantify the width of this distribution in various ways.  For example, we could consider moments of the distribution $\langle n^s\rangle=\sum_n n^s \,p_n(t)$.  The first moment, or average position in the Krylov chain, was called the {\it spread complexity} in \cite{Balasubramanian:2022tpr}
\be\label{eq:krylov_def}
\mathcal{C}_K(t)=\langle n\rangle=\sum_n n\,p_n(t),
\ee
and the analogous quantity for the spread of operators (which we will discuss below) is often referred to as the   Krylov complexity (its study was initiated in \cite{Parker:2018yvk}). These quantities have proved useful in characterizing many aspects of quantum dynamics and, as we will discuss in Sec.~\ref{sec:holography},  turn out to be geometrized  in the duality between the SYK model and Jackiw-Teitelboim gravity as  the wormhole's length.

While the spread complexity in (\ref{eq:krylov_def}) has proven useful in many applications, alternative measures of wavefunction spread surely have their uses as well. First of all there are the higher moments we mentioned above -- see, e.g., \cite{ Caputa:2021ori}.  Another  natural measure of spread would be the entropy of the distribution (\ref{eq:krylovdist}) or its exponential
\begin{equation}
H(t) = - \sum_{n=1}^{\mathcal{K} - 1} p_n(t) \ln p_n(t) ~~~~;~~~~ \mathcal{C}_H(t) \equiv e^{H(t)} \, ,\label{eq:KEntropy}
\end{equation}
This quantity was studied in the operator and state contexts in \cite{Barbon:2019wsy,Caputa:2021sib,Patramanis:2021lkx,Caputa:2025dep}, but has not yet been widely applied.

Of course, we could have defined a distribution like (\ref{eq:krylovdist}) in any basis.  The proofs in \cite{Balasubramanian:2022tpr} demonstrated that the Krylov basis minimizes both spread complexity $\mathcal{C}_K(t)$ and the entropic measure $\mathcal{C}_H(t)$ at least in some finite time interval starting at $t=0$.   The Krylov basis is also, in a sense, the most classical representation of the spreading wavefunction.   This is because, as shown in \cite{Basu:2024tgg}, the Wigner function associated to the one dimensional Krylov chain has the lowest possible negativity as compared to any other basis.  The Wigner function is the closest thing in quantum mechanics to a classical description of the state.  Indeed, a quantum system with a Wigner function that is everywhere positive for all times can be efficiently simulated on a classical computer through standard methods for describing statistical systems.  But if the Wigner distribution  has negative regions, quantum phenomena play an essential role, indicating a likely advantage for quantum computers in simulating these systems \cite{Gottesman:1998hu, Veitch:2012ttw, Aaronson:2004xuh, Mari:2012ypq}. 

Below we will apply these methods to many physical systems including particles moving on group manifolds, spin systems, the SYK model, random matrix theory (RMT), and Jackiw-Teitelboim gravity.  For lack of space we will not give a detailed discussion of applications to quantum billards \cite{Hashimoto:2023swv, Camargo:2023eev, Balasubramanian:2024ghv}, which are especially fascinating because such systems can have vanishing Lyapunov exponents, but can nevertheless manifest chaos.     For example, the triangular billiards studied in \cite{Balasubramanian:2024ghv} can be 
integrable, pseudointegrable or non-integrable, depending on the internal angles which control how bundles of classical trajectories do or do not diverge when bouncing off the walls and corners.  As such, billiard systems provide a tractable venue for studying how integrability and non-integrability affect the spread of states in a quantum system.  Billiards also give a way of studying the effects of symmetry on the spread of wavefunctions. For example,  pseudointegrable and non-integrable isosceles triangular billiards have independent sectors that are symmetric and antisymmetric under reflection across the diagonal, and these sectors separately reproduce characteristics of chaotic theories, although the complete dynamics approximates some aspects of an integrable system, e.g., a Poisson distributed spectrum \cite{Balasubramanian:2024ghv}. It is interesting to more broadly consider how symmetries and their breaking may affect the spread of the wavefunction.  We will give some examples of these effects below for particles on group manifolds and different random matrix symmetry classes.  The effects of discrete symmetries have been studied in \cite{Bhattacharya:2023yec} whose authors developed a framework for understanding the effects of parity and time-reversal symmetry, and their breaking, on the spread of wavefunctions.  Among other things they found interesting localization phase transitions on the Krylov chain.  In some of these systems such as the SYK model and RMTs, there are efficient techniques to calculate the average of the survival amplitude $\overline{S(t)}$ over an ensemble of theories. Note that in these cases $\overline{S(t)} \, \,\overline{S(t')}$ need not equal $\overline{S(t)S(t')}$. This similarly implies that correlations between Lanczos coefficients that affect the structure of the dynamics must be directly computed and will not be accessed by just computing the average of these coefficients over the pertinent ensemble.

Above we focused entirely on systems with time-independent Hamiltonians.  It would be interesting to extend the methods to systems with time-varying Hamiltonians.  We also only discussed closed systems and pure quantum states, but many of the techniques can be  extended to open systems and mixed states \cite{Alishahiha:2022anw,Caputa:2024vrn}, as reviewed in \cite{Nandy:2024htc}. Since the density matrix describing mixed states can be considered an operator on the Hilbert space, it is natural to think about this  in terms of the spread of operators, to which we turn now.

\subsection{Krylov complexity of evolving operators}
\label{sec:OperatorComplexity}

In Sec.~\ref{ssec:Krylov_def}, we followed \cite{Balasubramanian:2022tpr} to quantify how quantum states spread dynamically over the Hilbert space.  In this section,  we will discuss a similar approach to the spread of operators \cite{Parker:2018yvk}, which was introduced earlier than  Ref.~\cite{Balasubramanian:2022tpr}.   We started with a description of the spread of states because it is technically simpler to some extent, and the proofs of minimality of the Krylov basis were developed in that context.  But the relation between the operator-based methods of this subsection and the state-based methods of the previous section is analogous to the relation between the Heisenberg and Schr\"{o}dinger approaches to quantum mechanics --  either approach may be more convenient depending on the question we are studying.

The idea of defining operator size through an expansion in a natural basis was, to our knowledge, first explored in holographic systems such as the SYK model \cite{Roberts:2014isa,Magan:2016ehs,Roberts:2018mnp,Qi:2018bje}. This concept was later quantified in \cite{Parker:2018yvk}, whose authors studied operator spread in the Krylov basis, and proposed to quantify operator size in terms of the \emph{Krylov complexity} (the operator analog of (\ref{eq:krylov_def})). In this framework, we consider the Heisenberg evolution of an initial, ``simple'' operator $\mathcal{O}(0)$
\be
\partial_t\mathcal{O}(t)=i[H,\mathcal{O}(t)]\,.
\ee
This equation can be formally solved as
\be
\mathcal{O}(t)=e^{iHt}\mathcal{O}(0)e^{-iHt}\coloneqq e^{i\mathcal{L}t}\mathcal{O}(0)\,,
\ee
where in the second step we introduced the Liouvillian super-operator $\mathcal{L}=[H,\,\cdot\,]$. As before, we can expand the operator $\mathcal{O}(t)$ as a power series in $t$ that involves nested commutators of the Hamiltonian with $\mathcal{O}(0)$, or equivalently, various powers of the Liouvillian acting on $\mathcal{O}(0)$. As in Sec.~\ref{ssec:Krylov_def}, this series builds the Krylov subspace to which we want to apply the Lanczos procedure.  If the Hilbert space is $N$--dimensional, the associated operator space in which we carry out this procedure is $N^2$--dimensional because operators have matrix elements $O_{ab}$, with $a,b=1...,N$.

The procedure for extracting the Krylov basis for operators and the associated Lanczos coefficients is the same as described in Sec.~\ref{ssec:Krylov_def} for states, after vectorizing operators by mapping them into an auxiliary Hilbert space by the GNS (Gelfand-Naimark-Segal) construction (see, e.g., the discussion in \cite{Magan:2020iac}). In quantum information theory, such a vectorization also appears as the Choi-Jamiolkowski isomorphism, otherwise known as channel-state duality \cite{jamiolkowski1972linear, choi1975completely}. After vectorizing operators in this way, $\mathcal{O} \to |\mathcal{O})$, we also have to choose an inner product on the auxiliary Hilbert space.  The authors of \cite{Magan:2020iac} showed that the Krylov complexity defined with respect to the alternative choices behaves differently as a function of time.  However, the approach described in Sec.~\ref{ssec:Krylov_def} suggests that we should pick the inner product that produces a Krylov basis that minimizes the spread of operators.  As discussed in \cite{Balasubramanian:2022tpr},  the results of \cite{Magan:2020iac} imply that this minimization is achieved by the Wightman inner product.  Indeed, this was the choice advocated in \cite{Parker:2018yvk}, albeit for other reasons. Thus, we set
\be
(A|B)=\langle e^{\frac{\beta}{2}H}A^\dagger e^{-\frac{\beta}{2}H}B\rangle_\beta,\qquad \langle X\rangle_\beta\coloneqq \frac{1}{Z(\beta)}\Tr\left(e^{-\beta H}X\right),
\ee
where $Z(\beta)$ is the thermal partition function at inverse temperature $\beta=1/T$. For finite dimensional Hilbert spaces, we often use the $\beta\to 0$ version of this formula which is the Hilbert-Schmidt inner product. 

Finally, we map the operator evolution into a time-dependent state
\be
|\mathcal{O}(t))=e^{i\mathcal{L} t}|\mathcal{O}(0))=\sum^{\mathcal{K}-1}_{n=0}i^n\varphi_n(t)|K_n).
\ee
For Hermitian operators, the $a_n$'s (recall Eq.~\eqref{SchrodingerEq}), vanish and we derive the Schr\"{o}dinger equation
\be
\partial_t\varphi_n(t)=b_n\varphi_{n-1}(t)-b_{n+1}\varphi_{n+1}(t).
\ee
Solving it yields the probability $p_n(t)=|\varphi_n(t)|^2$ and the \emph{Krylov complexity}, which quantifies the operator size, is computed by~\eqref{eq:krylov_def}. Interestingly, \cite{Rabinovici:2020ryf} showed that the dimension $\mathcal{K}$ of the Krylov space for operators is bounded by 
\be
1\le\mathcal{K}\le N^2-N+1 \, ,
\ee
where $N$ is again the dimension of the Hilbert space.

As in  Sec.~\ref{ssec:Krylov_def}, the Lanczos coefficients can be computed by either the recursion method or the moment method. As explained above, the moments of the Hamiltonian can be extracted from the survival amplitude,\footnote{Sometimes the survival amplitude is referred to as the auto-correlation function in this context.} written in terms of the Wightman inner product as
\be\label{eq:returnOPKrylov}
S(t)=\langle e^{\frac{\beta}{2}H}\mathcal{O}(t)e^{-\frac{\beta}{2}H}\mathcal{O}(0)\rangle_\beta=\frac{1}{Z(\beta)}\sum_{n,m}|\langle n|\mathcal{O}|m\rangle|^2e^{-(\frac{\beta}{2}-it)E_n}e^{-(\frac{\beta}{2}+it)E_m} \, .
\ee
In the second equality,  we introduced resolutions of the identity in the energy basis and represented the answer in terms of the matrix elements of the operator.

The  spread of states and operators can be used as a probe for understanding the dynamics of quantum systems.  Below, we will show how to use this tool in a number of different systems.  General results can also be obtained.  For example, the authors of the pioneering paper  \cite{Parker:2018yvk}  performed numerical and analytical studies of  Krylov complexity in integrable and chaotic systems, including the SYK model, and found that the growth of Lanczos coefficients $b_n$ is at most linear in $n$ 
\be
b_n\le \alpha n+\kappa.
\ee
They conjectured that theories in which operator growth saturates this bound are chaotic, in the same sense as probed by the OTOC correlators (see Sec.~\ref{sec:key_notions} for a definition of these correlators). In particular, they related the coefficient $\alpha$ to the characteristic exponent $\lambda_K$ in an exponential growth of the Krylov complexity for chaotic theories, \ie $\lambda_K=2\alpha$. In the case of the SYK model, this exponent coincided with the Lyapunov exponent read from the OTOC, which we label with $\lambda_{\text{OTOC}}$. In other words, they observed that for the SYK model $\lambda_K=\lambda_{\text{OTOC}}$. This match will be further discussed in section \ref{ssec:integrability_chaos}. However, in many theories, including 2D CFTs, operator return amplitudes are universal and do not readily distinguish between integrable and chaotic models. 
Therefore, the hypothesis of \cite{Parker:2018yvk} requires refinement; indeed, as we will discuss below in Sec.~\ref{ssec:evo_spread_RMM}, the difference between chaos and integrability may be more apparent in the cross-correlations between the Lanczos coefficients and the effects of these correlations on the dynamics.

\subsection{Applications}
\label{ssec:Examples}

\subsubsection{Particles on group manifolds}\label{sec:SymmKryl}

Explicit solutions of the Lanczos algorithm can be obtained when there are dynamical symmetries governing  the time evolution on the 1D Krylov chain \cite{Parker:2018yvk,Balasubramanian:2022tpr,Dymarsky:2019elm,Caputa:2021sib,Muck:2022xfc}. Since the Krylov space representation depends on both the Hamiltonian and the initial state, this emergent symmetry need not arise from the fundamental symmetries of the model. As we have seen, the Lanczos coefficients can be constructed from the survival amplitude. Within the same model,  some scenarios may exhibit  dynamical symmetries, while others may not. Below, we present examples in which a dynamical symmetry  arises from the motion of a particle on a group manifold \cite{Caputa:2021sib,Balasubramanian:2022tpr}.

Consider a family of fictitious Hamiltonians based on the SL(2,$\mathbb{R}$) group,
\begin{equation}
H = \alpha (L_{-1} + L_1) + \gamma L_0 + \delta \id  ~~~;~~~ [L_0, L_{\pm 1}] = \mp L_{\pm 1} ~~~~;~~~~ [L_1,L_{-1}] = 2 L_0 \,.\label{HSL2R}
\end{equation}
Here $\gamma$, $\alpha$, $\delta$ are constants that depend on the details of the physical set-up such as the operator dimension, temperature or the initial state as well as the physical, evolving Hamiltonian.  Recall, that in the discrete series representation with scaling dimension $h$, $L_0 \ket{h,n} = (h+n) \ket{h,n}$, $L_{-1} \ket{h,n} = \sqrt{(n+1) (2h + n)} \ket{h,n+1}$, and $L_{+1} \ket{h,n} = \sqrt{n (2h + n -1)} \ket{h,n-1}$. Now suppose we start at $t=0$ with a highest weight state.  Then, applying the recursion method or the moment method, the Krylov basis elements coincide with the Lie algebra basis $\ket{K_n} = \ket{h,n}$ by construction. Using this we find that
\be
a_n=\gamma(n+h)+\delta,\qquad b_n=\alpha\sqrt{n(n+2h-1)}\,,
\ee
are the Lanczos coefficients. For coefficients of this form, one can solve the Schr\"{o}dinger equation \eqref{SchrodingerEq} analytically and a general form of the spread/Krylov complexity in this scenario is given by
\be
\mathcal{C}_K(t)=\frac{2h}{1-\frac{\gamma^2}{4\alpha^2}}\sinh^2\left(\alpha t\sqrt{1-\frac{4\gamma^2}{\alpha^2}}\right)\,.\label{CSL2R}
\ee
This form also appears in other contexts such as the SYK model at large $q$. Indeed, for the growth of operators with dimension $h\sim 1/q$, we have $\gamma=0$ and $\alpha=\pi/\beta$ \cite{Caputa:2021sib}. Quantum speed limits have also been used to show that this form leads to the fastest possible universal growth of operators  \cite{Hornedal:2022pkc}.
More generally, \eqref{CSL2R} can be divided into three different classes depending on the relative magnitude of the parameters. The class with $\gamma<2\alpha$ where complexity grows exponentially, the  class with $\gamma>2\alpha$ where complexity is periodic and the intermediate class with exactly $\gamma=2\alpha$ where the complexity grows quadratically. 

For compact semi-simple Lie groups, we generically end up with finite dimensional Krylov spaces. For example, we can consider the SU(2) algebra with Krylov basis representation of a tri-diagonal Hamiltonian
\be
H=\alpha(J_++J_-)+\gamma J_0+\delta \id \quad ;\quad [J_0,J_{\pm}]=\pm J_{\pm}\quad;\quad [J_+,J_-]=2J_0\,.\label{HSU2}
\ee
Evolution of the initial highest (lowest) weight state $\ket{j,-j}$ labeled by spin $j$ allows us to express the dynamics using coherent states of SU(2) and discrete series representations labeled by the spin $j$. Indeed, there are $2j+1$ Krylov basis vectors and Lanczos coefficients that read 
\be
a_n=\gamma(n-j)+\delta,\qquad b_n=\alpha\sqrt{n(2j+1-n)}\,.\label{anbnSL2R}
\ee
Again, constants $\gamma$, $\alpha$, $\delta$ and $j$,  depend on  details of the physical setup, i.e., the initial state, and physical Hamiltonian. After solving \eqref{SchrodingerEq} analytically, we find the corresponding  complexity is periodic in time
\be
\mathcal{C}_K(t)=\frac{2j}{1+\frac{\gamma^2}{4\alpha^2}}\sin^2\left(\alpha t\sqrt{1+\frac{\gamma^2}{4\alpha^2}}\right)\,.
\ee

Finally, we can work out a similar class of solutions controlled by the Heisenberg-Weyl algebra for which the Hamiltonian in the Krylov basis can be represented as
\be
H=\alpha (a^\dagger+a)+\gamma N+\delta \id \quad;\quad [a,a^\dagger]=1,\quad [N,a^\dagger]=a^\dagger,\quad [N,a]=-a\,, \label{HHW}
\ee
where $N = a^\dagger a$. The Hamiltonian is tri-diagonal in the  $\ket{n}$ basis of eigenstates of the number operator $N$, which is generated by the ladder operators $a$ and $a^\dagger$.  Thus, this basis is the infinite-dimensional Krylov basis for $H$. The Lanczos coefficients are
\be
a_n=\gamma n+\delta,\qquad b_n=\alpha\sqrt{n}\,,
\ee
and the complexity becomes
\be
\mathcal{C}_K(t)=\frac{4\alpha^2}{\gamma^2}\sin^2\left(\frac{\gamma t}{2}\right)\,.
\ee
For $\gamma\to 0$ it grows quadratically, but otherwise oscillates in time despite the Krylov basis being infinite-dimensional. Observe also that, in all the three examples, spread complexity is independent of $\delta$.

Let us stress again that the examples and Hamiltonians $H$ that we discussed above are effective descriptions of the state dynamics on the Krylov chain. Namely, there are physical setups where the evolution
\be
\ket{\psi(t)}=e^{-iH_{phys}t}\ket{\psi_0}\,,
\ee
yields a return amplitude $S(t)=\langle\psi(t)|\psi_0\rangle$, from which the moments and Lanczos coefficients can be fit into the forms described above and $\ket{\psi(t)}$ is effectively a coherent state for discrete series representations of some semi-simple Lie algebra. This way, the dynamics can be interpreted as motion on group manifolds governed by the Hamiltonians \eqref{HSL2R}, \eqref{HSU2} or \eqref{HHW}. Although such analytical scenarios are not generic and we have to rely on numerics, there are several interesting examples that fall into this class and we will discuss some of them below.

In \cite{Dymarsky:2019elm} some of these solutions were found by mapping the Schr\"{o}dinger equation into the Toda system that was then solved analytically. In \cite{Muck:2022xfc}, the symmetry approach was explained by recognizing that the Schrödinger equation can be effectively written as a three-term recursion relation for orthogonal polynomials in the Liouvillian (or the Hamiltonian). Consequently, when the Lanczos coefficients have a particular dependence on $n$, which makes this equation coincide with the defining relation of known polynomial families, analytical solutions for the wave functions and complexity can be obtained. 
For example, the SL(2,$\mathbb{R}$) family with $b_n$ given by \eqref{anbnSL2R} corresponds to Meixner orthogonal polynomials.
\begin{figure}[bth]
  \centering
  \includegraphics[width=5cm]{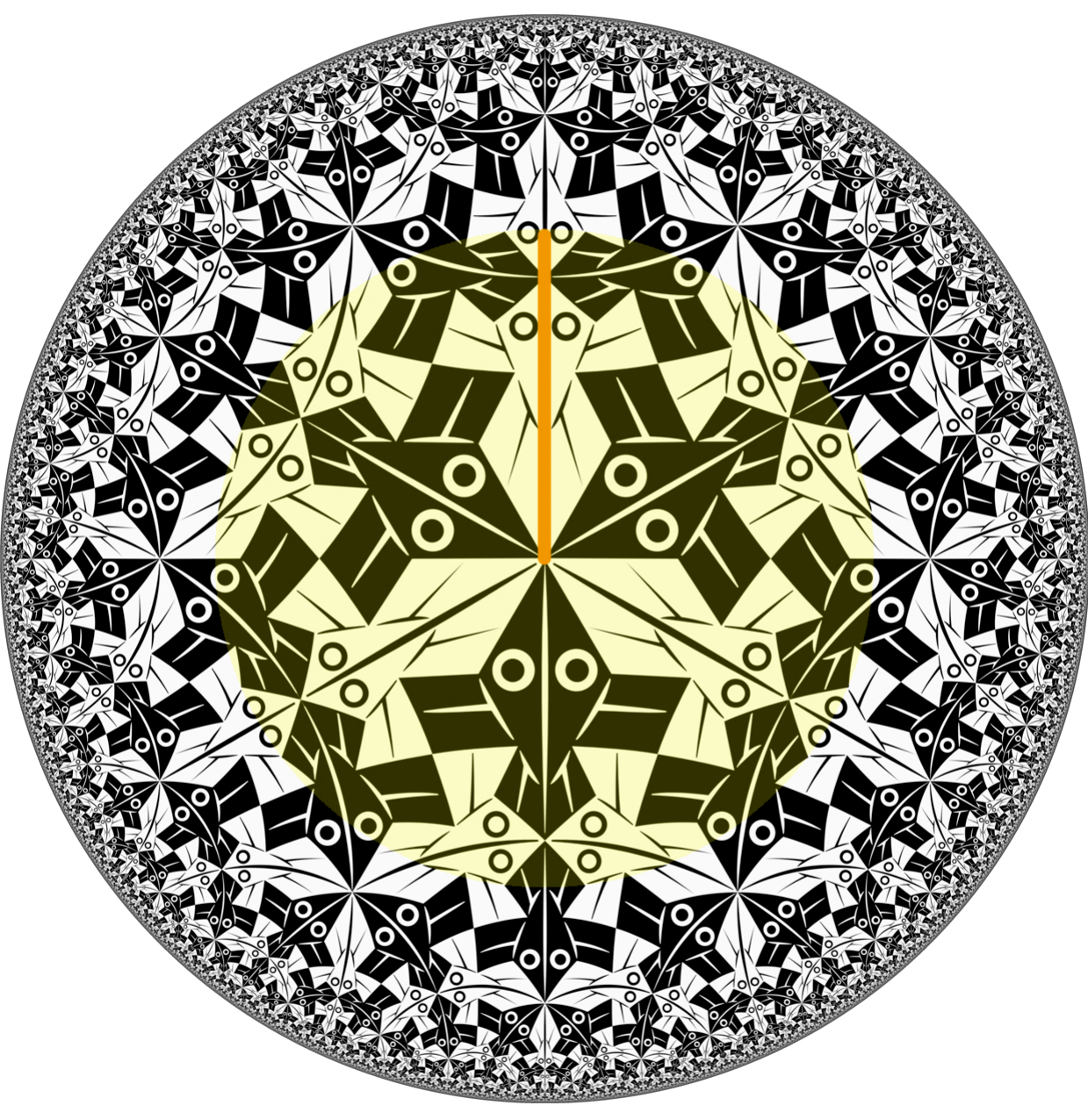}
  \caption{Operator growth and Krylov complexity for SL(2,$\mathbb{R}$) on the Fubini-Study geometry. Operator evolution becomes a geodesic (in orange) on the hyperbolic disc. Volume (in yellow) of the region enclosed by the particle's position is proportional to the Krylov complexity. }
\label{fig:Volume}
\end{figure}

For the symmetry classes discussed above, Krylov complexity admits a geometrical interpretation. Namely, the dynamics on the Krylov chain can be mapped onto motion in the phase space spanned by coherent states. Using the Fubini-Study (FS) metric on the corresponding phase spaces,  operator growth in the SL(2,$\mathbb{R}$), SU(2), and Heisenberg-Weyl cases corresponds to particle (geodesic) motion on the hyperbolic disk, the sphere, and the complex plane, respectively.
Moreover, Krylov complexity is proportional (with a factor of $\pi$) to the volume enclosed from the origin to the radius determined by the particle’s position at time $t$. An example of this for the SL(2,$\mathbb{R}$) symmetry is illustrated in Fig.~\ref{fig:Volume} (from \cite{Caputa:2021sib}). This result may seem counterintuitive given the conventional intuition from Nielsen’s definition of complexity, which associates complexity with geodesic length. Nevertheless, Krylov complexities in all of the symmetry examples above exhibit this feature. Interestingly, at late times, it is the Krylov entropy \eqref{eq:KEntropy}, that scales in a manner similar to the geodesic length in the Fubini-Study geometry \cite{Caputa:2021sib}.

\subsubsection{Operator growth in spin chains}
\label{ssec:op_growth_spin_chains}

Extensive numerical studies of operator growth and Krylov complexity have been conducted in \cite{Parker:2018yvk,Barbon:2019wsy,Rabinovici:2020ryf,Rabinovici:2021qqt,Rabinovici:2022beu}. In generic finite-dimensional, non-integrable systems, the growth of the Lanczos coefficients $b_n$ with $n$ exhibits three distinct regimes, which determine the scaling behavior of Krylov complexity. Denoting the number of degrees of freedom by $K$,\footnote{In some of the references, the number of degrees of freedom is denoted by $S$ and the size of the Hilbert space by $D$. We traded those symbols for $K$ and $N$ respectively, for consistency with other parts of the review. The only exception is the section about SYK where the number of degrees of freedom is denoted $N$ for consistency with the large literature on the topic.} the first regime is characterized by a linear growth of $b_n$'s, leading to the exponential growth of Krylov complexity. For infinite-dimensional systems, this exponential growth persists indefinitely. For finite dimensional systems, after times of order $t\ge\log(K)$ or  Krylov index $n\sim K$, the Lanczos coefficients saturate at a constant value $b_n\sim \Lambda K$, where $\Lambda$ is a parameter related to the system’s spectral bandwidth. In this regime, Krylov complexity grows linearly with time. 
The above-mentioned time evolution of the Lanczos coefficients resembles a switchback effect, see Fig.~\ref{fig:SwitchbackCt}. Indeed, a linear growth is achieved after an initial time, logarithmic in the number of degrees of freedom.
Finally, at exponentially large times $t\sim e^{K}$, the Lanczos coefficients gradually decrease toward zero, leading to the saturation of Krylov complexity.

Moreover, the authors of \cite{Rabinovici:2022beu} examined how operator growth differs between integrable and chaotic spin-chain models. The authors studied the XXZ spin chain with integrability-breaking deformations, allowing them to interpolate between the two regimes.   
Noting that the behavior of Krylov complexity depends on both the Hamiltonian and the choice of initial operator, they focused on examples of initial operators such as the sum of two spin operators located at some fixed distance from the ends of the chain.
In these examples, they found that the late-time plateau value is suppressed in the integrable phase, whereas in the chaotic phase, it is larger and eventually matches the prediction of random matrix theory for a system of the same dimension.  The authors argued that this suppression arises  
from a sort of Anderson localization in the one-dimensional Krylov chain arising from disorder in the Lanczos coefficients.  More details can be found in the review \cite{Nandy:2024htc}.

However, at least for generic initial operators/states, the plateau value cannot serve as a signature of chaos vs. integrability, for the reasons described below.  Recall from Sec.~\ref{sec:coarseLanczosSpectrum} that there is an analytical formula relating the density of states and the coarse-grained Lanczos spectrum.  This formula applies to cases where the Hamiltonian and initial state lead to  Lanczos coefficients that have a continuous large system limit, for example, if we are working with generic initial states with continuous support in the energy basis.  The authors of \cite{Balasubramanian:2023kwd} showed that for such states, the late time plateau of the spread complexity can be determined analytically in terms of the coarse-grained Lanczos spectrum. Since the relation between this spectrum and the density of states is universal, i.e., it applies to both integrable and chaotic theories which can have the same density of states, the plateau value cannot generally separate integrability and chaos, although it may do so for specially prepared initial conditions. That said, as we will see in Sec.~\ref{ssec:evo_spread_RMM}, the central phenomenon driving the results of \cite{Rabinovici:2022beu}, i.e., order vs disorder in the Lanczos coefficients, can be distilled into a sharp conjecture regarding the covariances of the Lanczos coefficients for chaotic theories \cite{Balasubramanian:2022dnj,Balasubramanian:2023kwd}.  These covariances affect the approach of the spread complexity to the late time plateau even in cases where the plateau value is the same between integrable and chaotic theories.   We will describe this conjecture and compare its predictions}explicitly with results for integrable and chaotic spin chains in Sec.~\ref{ssec:evo_spread_RMM}.

\subsubsection{SYK model}
\label{ssec:integrability_chaos}

The authors of \cite{Parker:2018yvk}, originally suggested Krylov complexity as a diagnostic of chaos. They proposed that this quantity would grow exponentially in chaotic systems because of the rapid spreading of operators expected there. In particular, according to \cite{Parker:2018yvk}, if the Lanczos coeffcients grow linearly at large $n$
\begin{equation}\label{eq:linearlanczos}
    b_n\sim \alpha n
\end{equation}
the Krylov complexity will grow exponentially at early times\footnote{By early, we mean up to times of the order of the logarithm of the number of the degrees of freedom in the system (or forever, if the system has infinitely many degrees of freedom).}
\begin{equation}
    \mathcal{C}_K(t)\sim e^{\lambda_K t}, \qquad \lambda_K \coloneqq 2\alpha \, .
\end{equation}
For chaotic systems where \eqref{eq:linearlanczos} holds, \cite{Parker:2018yvk} also proposed that the Krylov exponent   upper bounds  the Lyapunov exponent
\begin{equation}
    \lambda_{\text{OTOC}}\leq \lambda_K \leq \frac{2\pi}{\beta} \, .
\end{equation}
The first inequality was proven at infinite temperature in \cite{Parker:2018yvk} and, under the assumption of certain analytic and smoothness properties of the Lanczos coefficients, the right inequality was proven at finite temperature in \cite{Avdoshkin:2019trj,Gu:2021xaj}. The inequality $\lambda_{\text{OTOC}}\leq  \frac{2\pi}{\beta}$,  where $\beta=1/T$ is the inverse temperature, is the bound on chaos in quantum mechanical systems from \cite{Maldacena:2015waa} (the definition of $\lambda_{\text{OTOC}}$ is reviewed in  Sec.~\ref{sec:key_notions}).

These conjectures have been studied extensively in the SYK model, which has had important applications in the description of strange metals,  the study of quantum chaos and as a model for quantum gravity, see, e.g., \cite{Sachdev:1992fk,Kitaev:2015,Maldacena:2016hyu,Cotler:2016fpe,Roberts:2018mnp,Sarosi:2017ykf,Rosenhaus:2018dtp,Chowdhury:2021qpy}. The SYK Hamiltonian is constructed from $N$ Majorana fermions satisfying anti-commutation relations $\{\psi_i,\psi_j\}=\delta_{ij}$ with $i,j=1,\ldots,N$:
\begin{equation}
    H_q = (i)^{q/2}\sum_{1\leq i_1< i_2 < \ldots i_q\leq N} J_{i_1 i_2 \ldots i_q}\psi_{i_1} \psi_{i_2} \ldots\psi_{i_q} \, .\label{eq:SYKHamq}
\end{equation}
The couplings $J$ are real, and are drawn identically and independently from a Gaussian distribution  with zero mean and  fixed variance
\begin{equation}
    \langle J_{i_1 i_2 \ldots i_q}\rangle =0, \qquad 
    \langle J_{i_1 i_2 \ldots i_q}^2\rangle = \frac{2^{q-1}}{q} \frac{\mathcal{J}^2 (q-1)!}{N^{q-1}} \, .
\end{equation}
The SYK model approaches a nearly conformal fixed point at low temperatures, where the fermions acquire a conformal dimension of $1/q$ and the model becomes (nearly) maximally chaotic in the sense of the chaos bound \cite{Maldacena:2015waa}. The case $q=2$ is special -- it is integrable rather than chaotic.

The SYK model admits some simplifications in the limit of infinite $N$, and even more when we take $q$ to also be large.\footnote{
Here we mean that we first take the limit $N\rightarrow \infty$ and then we take $q\rightarrow \infty$. In the language of DSSYK, where $N$ is taken to infinity keeping the ratio $\lambda_{\text{SYK}} = 2q^2/N$ fixed, our limit has $\lambda_{\text{SYK}}\rightarrow0$.\label{footnote.DSSYK}}
In this case, the survival amplitude averaged over the SYK ensemble is known analytically and one can use it to extract an averaged Lanczos sequence.  At  large $N$, some quantities are self averaging including the thermal two-point function \cite{Chowdhury:2021qpy} and many texts, including the pioneering Ref.~\cite{Parker:2018yvk}, work under the assumption that the Krylov complexity is similarly self-averaging. These works extract the Lanczos sequence from  the ensemble averaged survival amplitude using the moment method described in Sec.~\ref{sec:momentmethod}.\footnote{Sometimes the survival amplitude is referred to as the auto-correlation function or thermal two-point function in this context.}  As we will see  below in the discussion of Random Matrix Theory in Sec.~\ref{ssec:evo_spread_RMM}, ensemble averages of this kind match the coarse-grained Lanczos coefficients described in Sec.~\ref{sec:coarseLanczosSpectrum}, but do not capture higher moments such as covariances of the Lanczos spectrum.
Applying the moment method to the ensemble averaged survival amplitude, the leading contribution to the Lanczos coefficients at large $q$ reads
\begin{equation}
    b_{n,\beta}^{\text{SYK}} = \begin{dcases*}
2\nu \beta^{-1} \sqrt{2/q} + O(1/q)
   & if  $n=1$\,, \\[1ex]
2\nu \beta^{-1}  \sqrt{n(n-1)} + O(1/q)
   & if $n>1$\,.
\end{dcases*}
\end{equation}
where $\nu$ is implicitly defined via the relation $\beta \mathcal{J} = 2\nu/\cos(\nu)$.

\begin{figure}[tbh]
    \centering
    \includegraphics[scale=0.6]{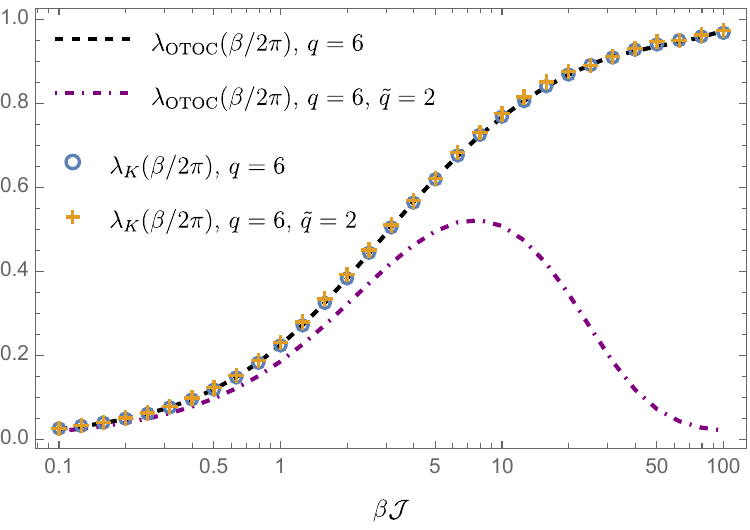}
    ~~
    \includegraphics[scale=0.59]{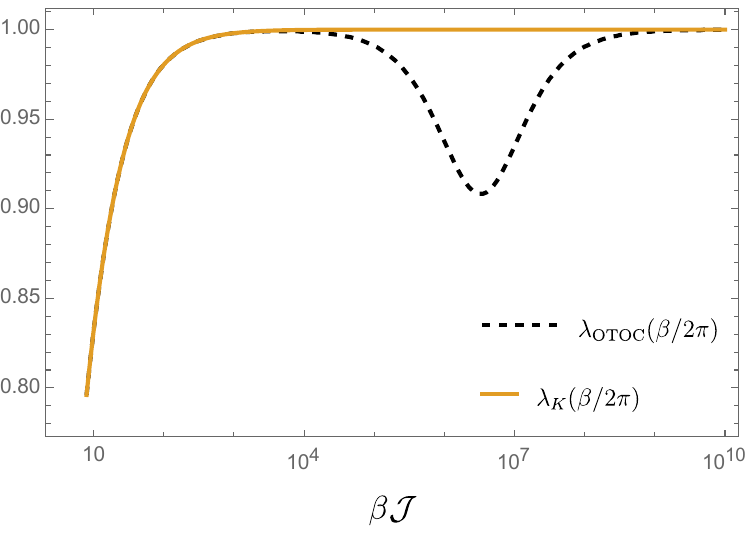}
    \caption{Lyapunov and Krylov exponents for flow SYK theories \eqref{eq:flow_SYK}. Left: a flow from $q=6$ to $q=2$ at large $N$ with $s=0.2$. The Lyapunov exponent indicates that chaos develops at intermediate temperature (in the intermediate IR) but integrable behavior is approached at much lower temperatures (in the deep infrared). Right: flow between two chaotic regimes at large $N$ and large $q$ where $\tilde q/q=2$ is kept fixed when taking the limit and $s=10^{-3}$. We observe two regimes of near maximal chaos with a clear transition indicated by the OTOC exponent's minima but not by the Krylov exponent. Plots adapted from \cite{Chapman:2024pdw}.}
    \label{fig:summary}
\end{figure}

In this limit, the leading large $q$ contribution to the Krylov exponent is exactly equal to the Lyapunov exponent for the OTOC correlators \cite{Parker:2018yvk}:
\begin{equation}
    \lambda_K=\lambda_{\text{OTOC}} = 4\nu/\beta.
\end{equation}
A particularly interesting case is the limit of low temperatures (large $\beta$). In this case both exponents approach the chaos bound $2\pi/\beta$. The  surprising equality between the Krylov and Lyaponov exponent for the OTOC suggested that the Krylov exponent, which is  simpler to compute than the OTOC, would provide a shortcut for computing the Lyapunov exponent of the latter. However, at large but finite $q$,  the equality fails at order $1/q$, that is \cite{Chapman:2024pdw}
\begin{equation}
    \frac{\beta}{2\pi}(\lambda_K-\lambda_{\text{OTOC}}) = \frac{4\pi^2}{3 q \beta J} .
\end{equation}
Larger deviations can also occur.
For example, consider deforming the SYK Hamiltonian by another SYK Hamiltonian with $\tilde q<q$
\begin{equation}\label{eq:flow_SYK}
    H_{\text{def}} = H_q +s H_{\tilde q}
\end{equation}
The perturbation is an infrared deformation of the original SYK since it has effective dimension  $\tilde q/q<1$ around the fixed point of the original Hamiltonian. This means that as we keep flowing to the infrared,  a new fixed point will appear that is dominated by the deformation. This type of deformation has been studied in, \eg  \cite{Garcia-Garcia:2017bkg,Kim:2020mho,Garcia-Garcia:2020dzm,Lunkin:2020tbq,Nandy:2022hcm,Menzler:2024ifs,
Jiang:2019pam, Anninos:2020cwo,Anninos:2022qgy,Louw:2023lpq}. In the context of chaos, we expect the infrared deformation to induce a transition between two chaotic regimes, or if $\tilde q=2$, between chaos and integrability along the RG flow controlled by the temperature parameter. We can try to diagnose these chaos-chaos or chaos-integrability transitions using the OTOC and Krylov exponents \cite{Chapman:2024pdw}. Fig.~\ref{fig:summary} show that while the Lyapunov exponent exhibits features indicative of such transitions, the Krylov exponent does not.

Perhaps this should not come as a surprise, as the starting point for extracting the Lanczos sequence in the above discussion was the ensemble averaged two point function and Lanczos coefficients.  As the next section suggests, to diagnose chaos we may need the ensemble fluctuations and correlations of the Lanczos sequence rather than their average.  

Additionally, the studies described above are investigating early time chaos.  As we will discuss below, the spread complexity of chaotic systems grows to a peak at late times that are exponential in the entropy,  and then slopes back down to a plateau.  Equivalently, the peak occurs at a time linear in the dimension of the Hilbert space, which scales as the exponential of the entropy.  The presence of this peak and slope depend on characteristic spectral correlations that are absent in integrable theories with Poisson spectra \cite{Balasubramanian:2022tpr}. We will discuss this below in the context of Random Matrix Theory, but all the characteristic behavior of spread complexity that we will describe applies equally to chaotic SYK models.

\subsubsection{Random Matrix Theory} 
\label{ssec:evo_spread_RMM}

A standard paradigm for quantum chaotic behavior is time evolution under a random matrix Hamiltonian drawn from the measure
\begin{equation}
\frac{1}{Z_{\beta_D,N}} e^{-\frac{\beta_D N}{4} \mathrm{Tr}[V(H)]}\,,
\label{eq:RMTensemble}
\end{equation}
where $H$ is a Hamiltonian matrix acting on an $N$--dimensional Hilbert space, $V(H) = \sum_{n=0}^\infty v_n \mathrm{Tr}(H^n)$ is a potential, and the partition sum $Z_{\beta_D,N}$ normalizes the measure.\footnote{In this section we return to using $N$ for the Hilbert space dimension rather than the number of degrees of freedom.} The Dyson index is $\beta_D=1,2,4$  depending on whether we choose $H$ to be Hermitian, real and symmetric, or quaternionic, corresponding to unitary, orthogonal, or symplectic ensembles.  If $V$ is quadratic, we say that the Random Matrix Theory (RMT) is described by the Gaussian Unitary, Orthogonal or Symplectic Ensembles (GUE,GOE, GSE).   We are going to study time evolution and spread complexity of a generic state in the basis in which the Hamiltonian is drawn.  Since the matrix is being drawn randomly, we can pick this to be the state $(1,0,0,\cdots)$ without loss of generality.

Suppose we diagonalize the Hamiltonian in the exponent of (\ref{eq:RMTensemble}) by a unitary transformation.  Famously, the change of variables introduces the Vandermode determinant $\Delta = \prod_{i<j} |\lambda_i - \lambda_j|^{\beta_D}$ in the measure, leading to a distribution over eigenvalues of the form
\begin{equation}
p(\lambda_1,\cdots\lambda_N) = 
Z_{\beta_D,N} e^{-\frac{\beta_D N}{4} \mathrm{Tr}[V(\Lambda)]}\prod_{i<j} |\lambda_i - \lambda_j|^{\beta_D} = 
Z_{\beta_D,N} e^{-\frac{\beta_D N}{4} \sum_n v_n \sum_k \lambda_k^n }\prod_{i<j} |\lambda_i - \lambda_j|^{\beta_D}\,,
\label{eq:EigvalDist}
\end{equation}
where $V(\Lambda)$ is the potential evaluated on the diagonal matrix of eigenvalues $\lambda_k$.  There is also an important relation between the potential $V$ and the ensemble averaged density of states:
\begin{equation}
\frac{1}{4} V'(\omega) = \mathrm{p.v.} \int dE \, \frac{\rho(E)}{\omega - E}\,,
\label{eq:PotDens}
\end{equation}
where $V'$ is the derivative of the potential, $\rho(E)$ is the density of stares, and $\mathrm{p.v.}$ indicates the principal value of the integral.  We will use these relations to work out expressions for the Lanczos spectrum.

As explained in \cite{Balasubramanian:2022dnj}, we can use (\ref{eq:PotDens}) to calculate the density of states from the potential of the RMT, and then we can use (\ref{eq:DensityLanczosAnalytic1}) from Sec.~\ref{sec:coarseLanczosSpectrum} to determine the coarse-grained Lanczos spectrum and the spread complexity.  This construction is averaged over the ensemble, but as shown in \cite{Balasubramanian:2022dnj}, at large $N$ this gives the same result as coarse-graining over nearby Lanczos indices in a single draw from the ensemble, i.e., using the coarse-graining procedure outlined in Sec.~\ref{sec:coarseLanczosSpectrum}.  But we can do better than this and determine the distribution of Lanczos coefficients.  For Gaussian RMTs, this distribution was worked out in \cite{Dumitriu_2002}.   The Jacobian arising from the basis transformation to get a tridiagonal Hamiltonian will be the same for any potential.  The authors of \cite{Balasubramanian:2022dnj} used this to show that for an RMT with an arbitrary potential like (\ref{eq:RMTensemble}) the distribution of Lanczos coefficients is 
\begin{equation}
p(a_0,\cdots a_{N-1}; b_0,\cdots b_{N-1}) \propto 
\left(\prod_{n=1}^{N-1} b_n^{(N-n)\beta_D - 1} \right)
e^{-\frac{\beta_D N}{4} Tr[V(H)]}\,,
\label{eq:LanczosDistribution}
\end{equation}
where within the trace,  $H$ is understood as tridiagonalized to the Lanczos form.

At large $N$ the distribution (\ref{eq:LanczosDistribution}) is sharply peaked. So we can use the method of saddle points to extract the ensemble average (the saddle point) and the covariances (from the quadratic expansion around the saddle point). The ensemble average precisely reproduces (\ref{eq:DensityLanczosAnalytic1}), which we already derived through other means, and which applies to any theory for which we know the density of states.  A formula for the covariances of the $a_n$ and $b_n$ is derived in \cite{Balasubramanian:2022dnj,Balasubramanian:2023kwd}.   We will not repeat the  derivation here for lack of space, but the result is 
\begin{equation}
\mathrm{cov}(a_i,a_{i+\delta}) =
4 \mathrm{cov}(b_i,b_{i+\delta}) =
\frac{1}{2\pi \beta_D N}
\int_0^{2\pi} dk \,  \frac{e^{2ik\delta}}{\lambda(k,a_i,b_i)}\,,
\label{eq:RMTcorrelations}
\end{equation}
with
\begin{equation}
\lambda(k,a_i,b_i) = \int dE \frac{V'(E) - V'(a_i)}{b_i (E - a_i)} \eta((E-a_i)/b_i,e^{ik})\,, ~~~~~~~~
\eta(x,t) = \frac{x}{\pi \sqrt{4-x^2}} \frac{1}{t + 1/t - x}\,.
\end{equation}
Among other things, these results show that the variance in $a(x)$ around the ensemble mean should be four times the variance in $b(x)$ for any RMT, where we are taking $x=n/N$ and coarse-graining as described in Sec.~\ref{sec:coarseLanczosSpectrum}. The authors of \cite{Balasubramanian:2022dnj} showed that these analytical formulae reproduce numerical results for RMTs with various potentials.

\begin{figure}[ht!]
\centering
\begin{tabular}{c}
  \includegraphics[width=0.60\linewidth]{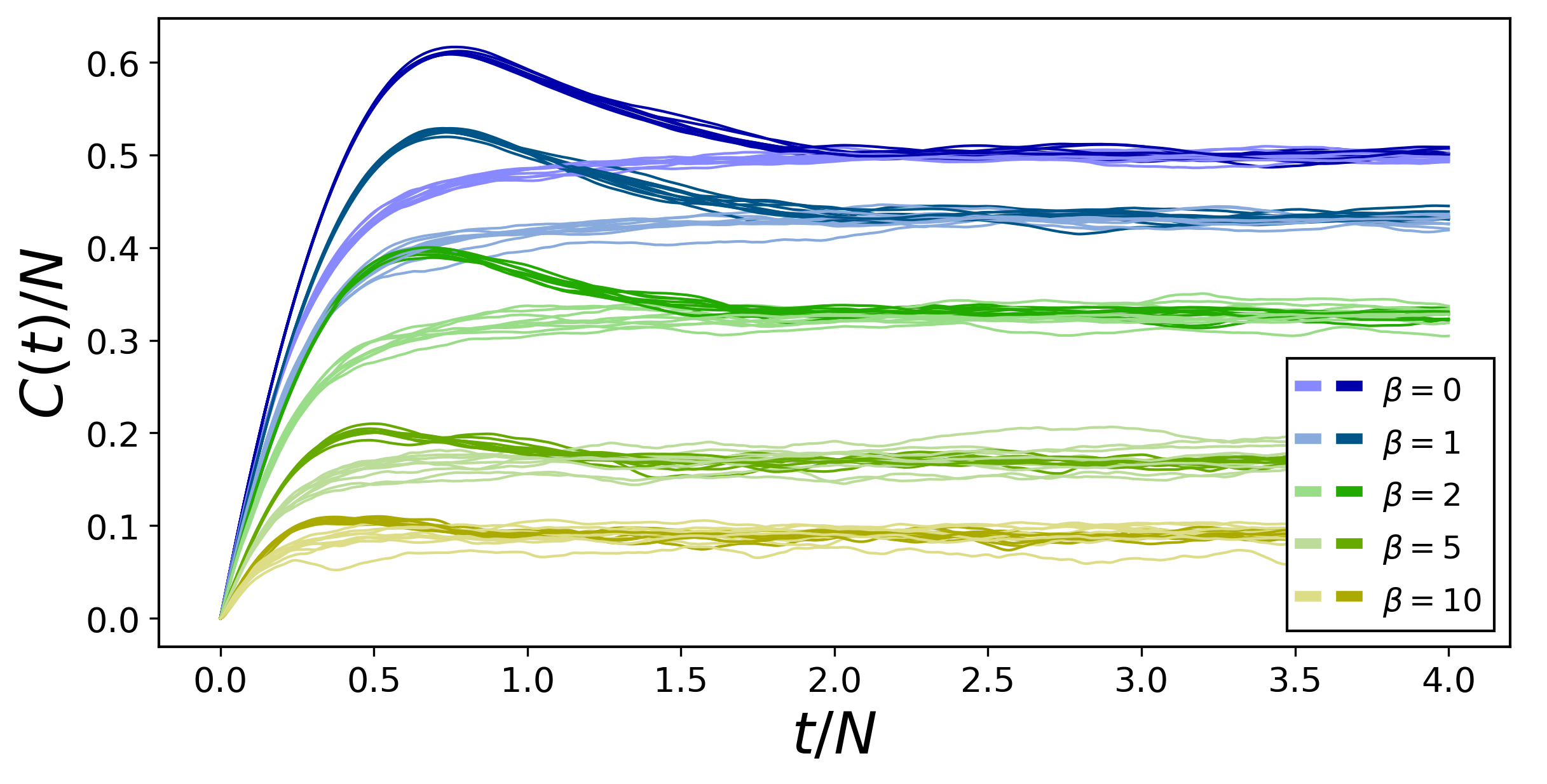}  
\end{tabular}
\caption{Quantum state complexity of the time-evolved thermofield-double state over an exponentially large period of time for different values of $N$ and $\beta$, as described in the main text. {\it Dark Hues:} GUE ensemble.  Going from highest (blue) to lowest (yellow)  curves  we have $\beta=\{0, 1, 2,5,10\}$.  In each case, we have plotted ensembles with $N=\{1024, 1280, 1536, 1792, 2048, 2560, 3072, 3584, 4096\}$.
Complexity grows linearly to a peak, followed by a downward slope to a plateau. 
{\it Light Hues:} Ensemble with the same density of states as GUE, but without correlations between eigenvalues. In this case, the curves  plateau without reaching a peak followed by a downward slope.
Figure taken from \cite{Balasubramanian:2022tpr}.}
\label{fig_KC_GUE_and_Spectral_plateau}
\end{figure}

Following \cite{Balasubramanian:2022tpr},   the spread complexity of the time-evolved thermofield-double state for a GUE Hamiltonian is plotted in Fig.~\ref{fig_KC_GUE_and_Spectral_plateau} for various matrix sizes $N$ and inverse temperatures $\beta$. The dynamics exhibit four characteristic regimes: a linear ramp up to a peak that is exponential in the entropy (and hence linear in $N$, the Hilbert space dimension), followed by a slope down to a plateau. The authors of \cite{Balasubramanian:2022tpr} showed that the peak and  slope arise from eigenvalue correlations known as spectral rigidity in chaotic theories. These correlations can be removed by drawing eigenvalues at random from the Wigner semicircle distribution -- this produces a spectrum without eigenvalue repulsion, like the Poisson spectra expected for integrable theories. The evolution with such a spectrum (Fig.~\ref{fig_KC_GUE_and_Spectral_plateau} in light hues) shows only the ramp and plateau.   Thus, the peak and slope are characteristic of late time chaos.   The same patterns appear in the late time spread complexity of the chaotic SYK model \cite{Balasubramanian:2022tpr}. 

Quantum chaotic systems are conjectured to have spectra that are well-described by RMT statistics \cite{Dyson_62_Statistical,PhysRevLett.52.1}.  In view of the results above, \cite{Balasubramanian:2023kwd} conjectured that: {\it Quantum chaotic systems display a Lanczos spectrum that is well described by RMT.}  As we discussed,  the coarse-grained Lanczos spectrum can be determined from just the density of states.  So the content of this conjecture lies in the correlations -- it is saying that the covariances of the Lanczos coefficients are described by the RMT formulae stated above. 

The conventional statements relating quantum chaos to Random Matrix Theory concern the statistics of nearby energy levels, and equivalently late time dynamics.   Correspondingly, the universal RMT results that we described above concern the ``bulk'' of the Lanczos spectrum, i.e., for indices of $O(d)$ where $d$ is the dimension of the Hilbert space.  In general  Lanczos  coefficients with larger indices control transitions to and between states that are further down in the Krylov chain that are populated later in the dynamics, although the precise connection between Lanczos index and time remains a topic of study.  As such, the universal behavior we described applies to the late time dynamics.  The Lanczos coefficients with indices that scale sub-extensively in $N$ must be computed separately by the recursion method and reflect non-universal and system-specific early time behavior even in systems that show late time chaos.
On a related note, a precise setup where RMT emerges in holography is the SYK model, and the Thouless time in this setup was derived using the Spectral Form Factor (SFF) in \cite{Nosaka:2018iat} (and is associated with the dip of SFF). Since the SFF plays a central role in  spread complexity computations \cite{Balasubramanian:2022tpr}, we also expect that the late time regime will correspond to the time after the peak of the spread complexity, and this was confirmed numerically in \cite{Balasubramanian:2022tpr,Balasubramanian:2023kwd}.


To test this conjecture, we can numerically compute the Lanczos spectrum of integrable and chaotic spin chains such as
\bea
    H_\text{int1}&=&A_\text{int1}\left(\sum_{i=1}^{K-1} X_iX_{i+1} + \sum_{i=1}^K C_1 Z_i\right)\;,\nonumber\\
    H_\text{cha1}&=&A_\text{cha1}\left(\sum_{i=1}^{K-1} X_iX_{i+1} + \sum_{i=1}^K C_2 Z_i +C_3 X_i + H_\text{site}\right)+B_\text{cha1}\;,
\eea
where $X_i$, $Y_i$, $Z_i$ are Pauli spin operators on site $i$; $K$ is the number of spins; $A_k$, $B_k$ and $C_k$ are numerical parameters; and  $H_\text{site}=0.5Z_3+0.3X_3$ is a one-site disorder operator that helps breaking parity and symmetry under the action of $\sum_i Z_i$. The model defined by $H_\text{int1}$ 
is integrable, while $H_\text{cha1}$
is chaotic as suggested by their level statistics, and have been explored in the literature \cite{Ba_uls_2011, Rabinovici:2022beu}. 
Indeed, as shown in \cite{Balasubramanian:2023kwd}, the analytical formulae developed above give an excellent description of the Lanczos coefficients in all cases, but the variances and covariances deviate significantly from the RMT formula for integrable spin chains, and in particular are much larger.

We can also write down an explicit formula for the height of the late time plateau in the spread complexity.   First, let $\hat{n} = \sum_n n \ket{K_n}\bra{K_n}$ be the position operator along the Krylov chain associated to an initial state $\ket{\phi}$.  Then, define the complexity of an energy eigenstate $\ket{E}$ relative to $\ket{\phi}$ as  $\mathcal{C}_E = \bra{E} \hat{n} \ket{E}$.  A short calculation  in \cite{Balasubramanian:2023kwd} shows that the height of the late time plateau in the spread complexity will be 
\begin{equation}
\mathcal{C}_{\rm plateau} = \sum_E p(E) \, \mathcal{C}_E\,,
\label{eq:plateauformula1}
\end{equation}
where $p(E) = |\langle E|\phi\rangle|^2$ is the initial probability of eigenstate $\ket{E}$.  If $p(E)$ is continuous, we can replace $\mathcal{C}_E$ by the local average in a small window over energies to get a coarse-grained eigenstate complexity $\overline{\mathcal{C}_E}$.  A calculation in \cite{Balasubramanian:2023kwd} shows that $\overline{\mathcal{C}_E}$ is entirely determined by Lanczos coefficients, which are in turn determined in terms of the density of states following Eq.~(\ref{eq:DensityLanczosAnalytic1}).  In terms of the coarse grained eigenstate complexities, and in the large system limit, we can take a continuum limit of (\ref{eq:plateauformula1}), which yields
\begin{equation}
\mathcal{C}_{\rm plateau} = \int dE \, \rho(E) \, p(E) \, \overline{\mathcal{C}_E}\,.
\label{eq:averageplateau}
\end{equation}
This means that for large systems and initial states with continuous support in the energy basis, the plateau of the spread complexity will not discriminate between chaotic and integrable theories with the same average density of states. 
However, if $p(E)$ is ``noisy'' rather than smooth, and the noise is negatively correlated with the noise in $\mathcal{C}_E$, the plateau can be significantly lower than the averaged value in (\ref{eq:averageplateau}). This lowering can be larger for integrable theories because, as we discussed above, the variance in the Lanczos is higher, leading to a form of Anderson localization of energy eigenstates along the Krylov chain (see also \cite{Rabinovici:2022beu}).  So if an eigenstate is ``centered'' further from the beginning of the chain it tends to have less overlap with the initial state for an integrable theory.   It would be interesting to investigate further how tuning the initial state leads to differences in the late time behavior for integrable versus chaotic theories because of variances and covariances in the Lanczos spectrum

These analyses have been extended to a broad class of integrable and chaotic models, including the SYK model and spin chains.  Key aspects such as the presence or absence of the peak and slope \cite{Erdmenger:2023wjg}, the peak height \cite{Baggioli:2024wbz}, and the system's approach to the plateau \cite{Alishahiha:2024vbf} have been advocated as  indicators of  chaos. However, it seems that  Krylov complexity \cite{Bhattacharjee:2022vlt} and spread complexity \cite{Huh:2023jxt} are not sensitive to saddle-dominated scrambling (exponential growth due to unstable saddle points in phase space), at least for the initial states/operators studied in these papers, a phenomenon that  OTOCs also fail to distinguish from genuine quantum chaos \cite{Xu:2019lhc}.

\subsection{Open problems and directions}
We conclude this section on the complexity of the spread of states and operators with a discussion of some open questions and progress towards answering them.

\paragraph{Krylov complexity in QFT}
As we discussed, Krylov complexity can be evaluated from the survival amplitude. If we  adopt the Wightman inner product for the reasons discussed in Sec.~\ref{sec:OperatorComplexity}, we can compute  the survival amplitude from the thermal two-point functions
\be
S(t)=\langle \mathcal{O}(t-i\beta/2)\mathcal{O}(0)\rangle_{\beta}.
\ee
These quantities are known exactly in some QFTs. For example, in 2D CFTs, thermal two-point correlators are universally determined by Ward identities and are the same for both integrable and chaotic CFTs.  In this case,  there is an infinite sequence of $b_n$ Lanczos coefficients that grow linearly for large $n$ \cite{dymarsky2021krylov}. At least at first sight, this serves as a counterexample to the universal operator growth hypothesis.  This issue was explored in \cite{Avdoshkin:2022xuw, Camargo:2022rnt}, where the authors argued that to meaningfully characterize operator growth in QFTs, one must first introduce appropriate cut-offs. They analyzed the effects of both UV and IR cut-offs, and concluded that the chaotic or scrambling properties of QFTs may be captured by $b_n$'s larger than the UV cut-off $\Lambda$. Meanwhile, the IR cut-off (e.g. modeled by introducing a mass) induced a separation (staggering) between even and odd Lanczos coefficients that grew linearly with different intercepts \cite{Camargo:2022rnt}.

Thus, distinguishing characteristic features of chaotic and integrable operator growth in continuum QFTs remains an open problem. 
A promising approach could be to employ von Neumann algebras and algebraic QFTs. Some progress in this direction has been made in \cite{Ouseph:2023juq, Gesteau:2023rrx}, but a direct connection with Krylov-basis methods has not been established. Alternatively, we could compute the covariances of the Lanczos coefficients in QFT, in view of the conjecture in \cite{Balasubramanian:2022dnj,Balasubramanian:2023kwd} that their structure separates chaos from integrability. In theories with a discrete spectrum, it may also help to project onto a microcanonical band of energies (see also \cite{Kar:2021nbm}) thereby rendering the accessible Hilbert space finite dimensional, thus enabling us to directly adopt the ideas and methods that have been developed for  the latter.

\paragraph{Connecting Nielsen and Krylov complexities}
The relation between Krylov and Nielsen's approaches has been debated in several works. In \cite{Lv:2023jbv}, the authors argued that Krylov complexity can be interpreted geometrically as a distance in a certain projection of the Fubini-Study geometry that is natural in quantum optics, making it amenable to a circuit complexity interpretation. However, as we already saw from the symmetry-based examples, Krylov complexity, as a function of time, is proportional to the volume in the Fubini-Study geometry.  As a volume, rather than a length, it does not respect some properties of geodesics, such as the triangle inequality. Such properties were suggested in \cite{Aguilar-Gutierrez:2023nyk} as an obstruction to  Krylov complexity serving as a distance measure.  However, it is not clear that the Fubini-Study setting is the one in which we should seek a relation with Nielsen's complexity. Indeed, \cite{Craps:2023ivc} argued that, in certain setups, there is a relation between the time average of spread complexity of state evolution and an upper bound on Nielsen's  complexity with specific fine-tuned penalty factors. Recalling that the Nielsen approach has ambiguities in the choice of basis of generators and cost function, it remains possible that some judicious choice could establish a direct relation between the Nielsen and Krylov/spread approaches to complexity, at least for some specific choices of states. Perhaps useful progress could be achieved by formulating Krylov/spread complexity approach to general, time-dependent Hamiltonians (see progress in \cite{Takahashi:2024hex}). This would be closer in spirit to the starting point in Nielsen's definitions and could allow us to use bounds on circuit complexity to constrain Krylov complexity.  Indeed, if we treat the evolution parameter as a  circuit time, we can mathematically think about the complexity of formation of states in the language of spread complexity \cite{Caputa:2022eye}. Thus it remains an interesting problem to systematically relate the complexity of state/operator spreading to circuit complexity.

\paragraph{Other systems} It would be interesting to extend the  spread/Krylov complexity approaches to new systems.  First, the generalization to time-dependent Hamiltonians is important to achieve for the reasons described above, and because we would like to understand systems that are driven by external sources.  Second, if we contemplate external sources, we should also consider open quantum systems, whose dynamics are controlled by Lindbladian evolution of the density matrix~\cite{Carolan:2024wov}.  There is a substantial literature on the Linbladian formalism which is partly reviewed in \cite{Nandy:2024htc}. The interaction  of an external bath with an open system certainly affects the structure of entanglement spreading (see, e.g.,  \cite{Agon:2014uxa}), and has generally been a subject of significant recent interest in condensed matter physics through the study of measurement-driven phase transitions (see, e.g., \cite{block2022measurement}).  The spread complexity of measurement-induced non-unitary dynamics has been explored in \cite{Bhattacharya:2023yec}.   Meanwhile, \cite{Das:2024zuu} have uncovered various bounds and relations pertaining to the state/operator complexity of purifying a mixed state. Another interesting avenue is to study the spread complexity of Carrollian theories of relevance for flat space holography.  Likewise it would be interesting to study the kind of chiral theories that appear in the AdS/CFT dual to highly rotating extremal black holes \cite{Guica:2008mu,Balasubramanian:2003kq,Balasubramanian:2009bg} and the matrix models that appear as the dual to the 11-dimensional M-theory \cite{Banks:1996vh,Balasubramanian:1997kd,Polchinski:1999br}. In this direction, spread complexity has been studied in the planar limit of holographic theories in \cite{Das:2024tnw}.

\paragraph{Applications to the physics of black holes}  In Sec.~\ref{ssec:quantitative_matching_holo} we will discuss how spread complexity of the double scaled SYK model is directly related to the length of the dual wormholes in  two-dimensional Jackiw-Teitelboim  gravity.   But we could ask a broader question.  Many lines of argument suggest that the microstates of black holes are immensely complex and may appear to be hidden behind a horizon simply because simple probes cannot discriminate between them -- see, e.g., \cite{Balasubramanian:2005mg,Balasubramanian:2005kk,Balasubramanian:2006jt,Balasubramanian:2006iw}.  Similarly, a number of authors have argued that in the AdS/CFT correspondence  excitations behind the horizon are encoded in the CFT by exponentially complex operators \cite{Engelhardt:2021mue,engelhardt2022finding} in a non-isometric manner \cite{Akers:2022qdl}.   The information stored in Hawking radiation is also supposed to be encoded in an extremely complex manner, with the complexity expressed  in the dual gravitational theory in terms of geometric obstructions preventing asymptotic observers from accessing the information \cite{Brown:2019rox}.  At least in toy models of black hole evaporation like \cite{Penington:2019kki}, such considerations can be explicitly realized and characterized in information-theoretic and complexity-theoretic terms \cite{Balasubramanian:2023xdp}.  But black holes can form from collapse of simple states like an ordered cloud of particles.  So, the immense complexity of black hole microstates, and of the subsequent encoding of the initial state in the Hawking radiation, must be generated dynamically by evolution under the Hamiltonian of the theory.  An important question is to understand how this happens.  Perhaps the techniques of Krylov/spread complexity, or of Nielsen complexity from the previous section, can help.  One  recent result shows that if a holographic CFT time evolution operator is approximately pseudorandom or Haar random on a low energy subspace of states, there must be an event horizon in the dual theory of gravity \cite{Engelhardt:2024hpe}. Likewise, it seems that any region of AdS with sufficiently high holographic encoding complexity will be robustly protect from low-rank measurements, and hence be inaccessible to low-complexity observers \cite{Balasubramanian:2023xdp}.  It would be interesting to understand how the dynamical evolution of simple states can produce complex states with such properties.

\clearpage

\section{Quantum complexity and space-time: a more concerted approach}\label{sec:holography}

Connections between geometry and entanglement were first proposed by Ryu and Takayangi \cite{Ryu:2006bv} and were made covariant by Hubeny, Rangamani and Takayanagi \cite{Hubeny:2007xt}.  These authors showed that the entanglement entropies of subregions of the theory dual to an AdS spacetime are geometrized by the areas of certain extremal surfaces of the dual geometry. These developments led to major insights into the holographic duality, for example elucidating constraints on the entanglement patterns in quantum states that admit gravitational dual descriptions \cite{Hayden:2011ag}.  An analogy between the geometry of AdS spacetimes dual to conformal field theories and the graphical structure of tensor networks for constructing quantum states also provided an impetus, offering a schematic understanding of the hierarchical pattern of short and long range entanglement in holographic states \cite{Swingle:2009bg}.  Such developments motivated an important proposal by van Raamsdonk that spacetime is actually knitted together by the underlying patterns of quantum entanglement \cite{VanRaamsdonk:2010pw}.  Subsequent developments have illustrated many constraints on quantum states with holographic duals including new inequalities \cite{Hayden:2011ag} and  properties such a
 quantum error correction \cite{Almheiri:2014lwa} (see \cite{Rangamani:2016dms} for a review) that have a bearing on the complexity of constructing such states from a simple reference.

Others proceeded to study the dynamics of thermalization via the AdS/CFT correspondence, especially considering the evolution of entanglement and the corresponding dual geometric quantities after quantum quenches \cite{Abajo-Arrastia:2010ajo,Balasubramanian:2011ur,Balasubramanian:2010ce,Balasubramanian:2011at,Liu:2013qca,Liu:2013iza}. In all these cases, an initial rise of entanglement entropy ends in saturation after a thermalization time. These studies also led to an examination of the time evolution of entanglement entropy in thermo-field double (TFD) states dual to eternal black holes, i.e., to ERBs between two asymptotically AdS universes \cite{Hartman:2013aa,Maldacena:2013xja}. The result, computed in terms of certain extremal surfaces passing through the ERB, shows that entanglement entropy grows with time in a TFD state until it saturates at  equilibrium~\cite{Hartman:2013aa}.  However,
the black hole interior keeps growing long after the entanglement saturates. Therefore, Susskind suggested that the growth of the black hole's interior 
should be encoded in the AdS/CFT correspondence by a dual quantum mechanical quantity that increases for a time exponential in the system's size.  What quantity could conceivably keep increasing after a system reaches thermal equilibrium?  It is  known in computer science that some computations carried out by the action of local operators can require exponentially long time or exponentially many actions of the operators, quantified as the ``complexity'' of the resulting state.   If we think of time evolution as a computation performed by the action of a local Hamiltonian, one might guess that the complexity of the evolving quantum state, quantified in some way, may be related to the size of the black hole's  interior~\cite{Susskind:2014rva}.

As we discussed in Secs. \ref{sec:what_is_quantum_complexity}, \ref{ssec:Nielsen_complexity}, and \ref{sec:Krylov},  circuit complexity, Nielsen's complexity, and Krylov/spread complexity all show a long period of linear growth for chaotic systems and also display the switchback effect (see discussion in Sec.~\ref{ssec:op_growth_spin_chains}). 
Therefore, any dual geometric object must also show these dynamics. Consistently with this, the initial conjecture was that the volume of the ERB wormhole is dual to some notion of time-evolving complexity of the dual Thermo-field Double state \cite{Susskind:2014rva}.  However, as we discuss in Sections \ref{ssec:holo_compl_proposals} and \ref{ssec:evidence_holo_complexity},  a number of different geometric quantities all  reproduce linear growth and the switchback effect \cite{Stanford:2014jda,Brown:2015bva,Brown:2015lvg,Couch:2016exn,Belin:2021bga,Belin:2022xmt}.  We also discuss a proposal in the case of two-dimensional gravity for a non-perturbative definition of the size of the ERB wormhole \cite{Iliesiu:2021ari} that saturates  at times exponential in the black hole entropy, reproducing the expected late time saturation of complexity  (Secs.~\ref{sec:what_is_quantum_complexity},\ref{ssec:Nielsen_complexity} and \ref{sec:Krylov}). Strictly speaking, the various candidate holographic duals of  complexity described in Sec.~\ref{ssec:holo_compl_proposals} all have divergences that require regulation.  We understand these divergences as arising in the dual CFT from the continuum nature of the theory, just like the divergences of entanglement entropy.  We explain these considerations in  Sec.~\ref{ssec:time_indep_properties}.  Proposed extensions of the ideas to subregions, geometries with defects, and black holes in de Sitter space are summarized Sec.~\ref{ssec:generalization_holo_complexity}.

All the geometric quantities proposed as duals to some notion of complexity in Sec.~\ref{ssec:holo_compl_proposals} are diffeomorphism-invariant, and hence we expect them to map  onto  precise gauge-invariant quantities in the dual theory.  Is there a precise map to some particular notion of complexity? In other words, can we go beyond a qualitative agreement with the linear growth and switchback effect seen in any reasonable definition of complexity, to write an equation between gravitational and CFT quantities?  Very recently,  precise relations have been found between the size of 
an Einstein-Rosen Bridge in two-dimensional Jackiw-Teitelboim gravity and the spread complexity of a dual double-scaled SYK model, matching both the linear early time classical growth and the expected late time quantum saturation \cite{Berkooz:2018qkz,Lin:2019kpf,Rabinovici:2023yex,Nandy:2024zcd,Balasubramanian:2024lqk}. There are also  equalities relating  the rate of change of Krylov/spread complexity and the momentum of particles in AdS spacetime.  These results are reviewed in Sec.~\ref{ssec:quantitative_matching_holo}.  Sec.~\ref{sec:NielsenAndGeometry} discusses some ideas for relating Nielsen's complexity to dual geometries.  Despite these findings, it is difficult to match geometrical observables in the gravitational setting with quantum complexity in field theory.  This difficulty led to questions about the foundations of quantum computation in the context of the AdS/CFT correspondence. The existence of pseudorandom states, discussed in Sec.~\ref{ssec:pseudorandomness}, can be used to argue that some entries in the holographic dictionary will be hard to compute, or alternatively, that quantum gravity efficiently solves problems that will require exponentially long times for a quantum computer. This unexpected line of research was revealed through the study of quantum complexity.

\subsection{Holographic complexity proposals}
\label{ssec:holo_compl_proposals}

\begin{figure}[ht]
    \centering
    \subfigure[CV]{\label{subfig:CV} \includegraphics[width=0.3\textwidth]{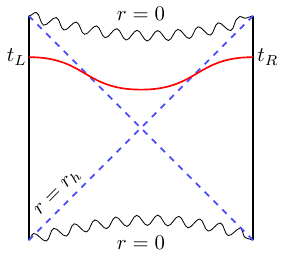}} \,
    \subfigure[CA, CV2.0]{ \label{subfig:CA} \includegraphics[width=0.3\textwidth]{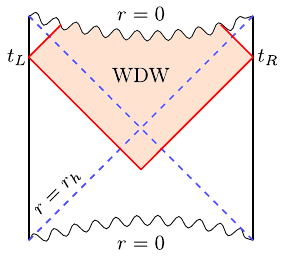}}  \,
 \subfigure[CAny]{ \label{subfig:CAny} \includegraphics[width=0.3\textwidth]{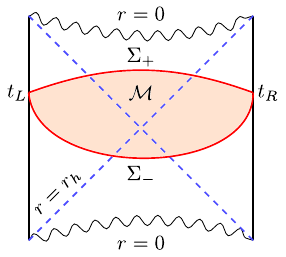}}
    \caption{Geometric objects defining holographic complexity proposals drawn inside the Penrose diagrams of the Schwarzschild-AdS black hole. 
    Pictures adapted from \cite{Brown:2015lvg,Belin:2022xmt}. 
    (a) Maximal-volume codimension-one surface used in the CV proposal [Eq.~\eqref{eq:CV_proposal}]. (b) WDW patch used to compute CA and CV2.0 in Eqs.~\eqref{eq:CA_conjecture} and~\eqref{eq:CV20}. (c) Bulk region $\mathcal{M}$ with codimension-one boundaries $\Sigma_{\pm}$ defined by extremizing Eq.~\eqref{eq:functional_geom_CAny} in the CAny proposals. 
    }
    \label{fig:holo_complexity}
\end{figure}

We begin by defining the holographic complexity proposals (see Fig.~\ref{fig:holo_complexity}).

The \textit{complexity equals volume} (CV) proposal  
\cite{Stanford:2014jda,Susskind:2014moa} declares that holographic complexity is proportional to the maximal volume $\mathcal{V}$ of a codimension-one spacelike surface $\mathcal{B}$ anchored at the 
$(d-1)$-dimensional boundary slice $\Sigma_{\rm CFT}$ where a CFT state is defined (see fig.~\ref{subfig:CV}), \ie\footnote{In the context of CV conjecture, Ref.~\cite{Couch:2018phr} proposed to measure this geometric observable in terms of the maximal time from the horizon to the final slice (approached in the late stages of evolution). This notion also shares many similarities with the standard CV proposal, and at the same time removes certain ambiguities between small and large black holes in AdS spacetime. }
\begin{equation}
\mathcal{C}_V (\Sigma_{\rm CFT}) = \max_{\partial \mathcal{B} = \Sigma_{\rm CFT}} \left[ \frac{\mathcal{V}(\mathcal{B})}{G_N \ell_{\rm bulk}}  \right] \, ,    
\label{eq:CV_proposal}
\end{equation}
where 
$\ell_{\rm bulk}$ is a length scale introduced to maintain dimensional consistency. In the case of AdS spacetime, $\ell_{\rm bulk}$ is often identified with the AdS radius $L$. 
The maximal surface anchored on both boundaries of an eternal black hole penetrates the black hole interior, but avoids the singularity and regions with high curvature. This makes the maximal volume  reliable in the semiclassical approximation, without the need to make further assumptions.

Other holographic proposals involve gravitational observables evaluated on the Wheeler-De Witt (WDW) patch, \ie the bulk domain of dependence of a spacelike surface anchored at the boundary slice $\Sigma_{\rm CFT}$ (see fig.~\ref{subfig:CA}).
The first conjecture, \textit{complexity equals action} (CA), associates 
complexity with the on-shell gravitational action $I_{\rm WDW}$ \cite{Brown:2015bva,Brown:2015lvg}
\begin{equation}
    \mathcal{C}_A (\Sigma_{\rm CFT}) = \frac{I_{\rm WDW}}{\pi \hbar} \, .
    \label{eq:CA_conjecture}
\end{equation}
The gravitational action receives contributions from the null boundaries of the WDW patch, as required to make the variational principle in general relativity well-defined (see \cite{Lehner2016NullAct} or appendix A of \cite{Carmi:2016wjl}). 
The second proposal, \textit{complexity=volume 2.0} (CV2.0), identifies complexity as the spacetime volume of the WDW patch \cite{Couch:2016exn}
\begin{equation}
\mathcal{C}_{2.0V}(\Sigma_{\rm CFT}) = \frac{\mathcal{V}_{\rm WDW}}{G_N \ell^2_{\rm bulk}} \, .
\label{eq:CV20}
\end{equation}
Each of these proposals presents ambiguities.
CV and CV2.0 require a length scale $\ell_{\rm bulk}$ in the holographic dictionary.
The CA proposal contains contributions from the null boundaries of the WDW patch, whose normal vectors depend on an arbitrary normalization constant. The last dependence can be removed by including certain counterterms in the gravitational action, but those in turn come equipped with their own (arbitrary) length scale, which we denote $L_{\rm ct}$~\cite{Lehner2016NullAct}. When studying black holes, the CA and CV2.0 proposals probe regions near the black hole singularity. A delicate cancellation between a diverging metric factor and the vanishing volume of the sphere in the angular directions near the singularity leads to a result which does not diverge. Nevertheless, the validity of the formula in the semiclassical approximation is uncertain since the region near the singularity is not well understood in this limit.

The above ambiguities were taken as features of the holographic complexity proposal, rather than bugs, since computational complexity is also subject to several ambiguities: for instance, the choice of allowed gates in a circuit, the tolerance to reach a target state, the cost of gates \textit{etc.}, see Sec.~\ref{sec:what_is_quantum_complexity}. 
This perspective was taken a step further with a recent proposal that went under the name \textit{complexity equals anything} (CAny), \ie complexity may be dual to any one out of an infinite class of geometric observables.
CAny requires a codimension-zero bulk region $\mathcal{M}$ with future (past) boundary $\Sigma_{+} (\Sigma_-)$ anchored at the CFT slice $\partial \Sigma_{\pm} = \Sigma_{\rm CFT}$.
One defines the following functional (see fig.~\ref{subfig:CAny}) \cite{Belin:2021bga,Belin:2022xmt,Jorstad:2023kmq} 
\begin{equation}
    W_{G_2,F_{2,\pm}} (\mathcal{M}) = 
   \frac{1}{\ell_{\rm bulk}} \int_{\mathcal{M}} d^{d+1} x \, \sqrt{-g} \, G_2(g_{\mu\nu}) + \int_{\Sigma_+} d^d x \, \sqrt{h} \, F_{2,+}(g_{\mu\nu},X^{\mu}_+)  + \int_{\Sigma_-} d^d x \, \sqrt{h} \, F_{2,-}(g_{\mu\nu},X^{\mu}_-)  \, ,
   \label{eq:functional_geom_CAny}
\end{equation}
where $G_2, F_{2,\pm}$ are scalar functions, $g_{\mu\nu}$ is the bulk metric and $X^{\mu}_{\pm}$ are embedding coordinates for the inclusion of the codimension-one boundary surfaces $\Sigma_{\pm}$ inside the background geometry.
The functional \eqref{eq:functional_geom_CAny} selects a specific bulk region by imposing the extremization condition
\begin{equation}
    \delta_{X_\pm} \left[    W_{G_2,F_{2,\pm}} (\mathcal{M})  \right] = 0 \, ,
    \label{eq:extremization_CAny}
\end{equation}
where the shape of the boundaries $\Sigma_{\pm}$ is varied. When more than one solution exists, the one with maximal $W_{G_2,F_{2,\pm}}$ is chosen.
Once the unique codimension-zero region $(\bar{\mathcal{M}}, \bar{\Sigma}_{\pm})$ solving Eq.~\eqref{eq:extremization_CAny}  is identified, holographic complexity is computed by the observable 
\begin{equation}
\begin{aligned}
    & \mathcal{C}_{\rm Any} [G_1 ,F_{1,\pm},\bar{\mathcal{M}}](\Sigma_{\rm CFT}) = \frac{1}{G_N \ell_{\rm bulk}^2} \int_{\bar{\mathcal{M}}} d^{d+1} x \, \sqrt{-g} \, G_1 (g_{\mu\nu}) \\
    & + \frac{1}{G_N \ell_{\rm bulk}}  \int_{\bar{\Sigma}_+} d^d x \, \sqrt{h} \, F_{1,+}(g_{\mu\nu},X^{\mu}_+)  
    + \frac{1}{G_N \ell_{\rm bulk}}  \int_{\bar{\Sigma}_-} d^d x \, \sqrt{h} \, F_{1,-}(g_{\mu\nu},X^{\mu}_-)  \, ,
    \end{aligned}
    \label{eq:CAny_observable}
\end{equation}
where $G_1, F_{1, \pm}$ are scalar functions.
In summary, the CAny conjecture consists of two steps: an extremization governed by the function \eqref{eq:functional_geom_CAny} to select the relevant bulk region in the geometry, followed by the computation of the gravitational observable in Eq.~\eqref{eq:CAny_observable}.
CAny defines a large class of physical observables, compatible with qualitative features of complexity such as a linear growth and the switchback effect, that can be built out of the building blocks available in the gravitational system.

One can recover CV, CA and CV2.0 from CAny.
To recover CV and CV2.0, let us take the functionals \eqref{eq:functional_geom_CAny} and \eqref{eq:CAny_observable} to be the same (\ie $G_1=G_2$ and $F_{1, \pm}=F_{2,\pm}$), together with constant scalar functions:
\begin{equation}
    \mathcal{C}_{\rm gen} = \frac{1}{G_N \ell_{\rm bulk}} 
    \left[ \frac{\alpha_{\rm B}}{\ell_{\rm bulk}} \int_{\mathcal{M}} d^{d+1} x \, \sqrt{-g} + \alpha_+ \int_{\Sigma_+} d^d x \, \sqrt{h} 
     + \alpha_- \int_{\Sigma_-} d^d x \, \sqrt{h} \right] \, .
     \label{eq:Cgen_CAny}
\end{equation}
CV and CV2.0 are recovered by making different terms in Eq.~\eqref{eq:Cgen_CAny} dominant.
The CV prescription is recovered by keeping $\alpha_-, \alpha_{\rm B}$ constant while performing the limit $\mathcal{C}_{V} = \lim_{\alpha_+ \rightarrow \infty} \left( \alpha_+^{-1} \mathcal{C}_{\rm gen}  \right) \, .$  
In other words, only the future codimension-one surface $\Sigma_+$ survives, the extremization problem imposes that the slice has maximal volume, and the physical observable is the volume itself.
To recover CV2.0, one fixes the bulk coefficient $\alpha_{\rm B}$ and computes $ \mathcal{C}_{2.0V} = \lim_{\alpha_{\pm} \rightarrow 0} \left( \alpha_B^{-1} \mathcal{C}_{\rm gen}  \right) \, .$
In this limit, the boundary surfaces $\Sigma_{\pm}$ are pushed towards the null surfaces originating from the CFT slice $\Sigma_{\rm CFT}$, thus composing the WDW patch.
One can also recover CA from CAny, but now the functionals \eqref{eq:functional_geom_CAny} and \eqref{eq:CAny_observable} should differ from each other.
Specifically, one needs to extremize $\mathcal{C}_{\rm gen}$ in Eq.~\eqref{eq:Cgen_CAny} in the limit $\alpha_{\pm} \rightarrow 0$ to identify the WDW patch, but use the gravitational action \eqref{eq:CA_conjecture} as the observable which computes the  complexity.

\subsection{Time-dependent properties of the holographic proposals}
\label{ssec:evidence_holo_complexity}

In this subsection, we show that all the holographic complexity proposals defined in Sec.~\ref{ssec:holo_compl_proposals} universally exhibit the behaviors required for a correspondence with complexity,  \ie linear growth at late times and the switchback effect in a black-hole background.

We display these features for the CAny proposal, since it encodes all the other proposals. 
Consider the planar Schwarzschild-AdS black hole solution 
\begin{equation}
    ds^2 = -f(r) dt^2 + \frac{dr^2}{f(r)} + r^2 d \vec{x}^2 \, , 
    \qquad
    f(r) = \frac{r^2}{L^2} \left( 1 - \frac{r_h^d}{r^d} \right) \, ,
    \label{eq:BH_geometry}
\end{equation}
where $r_h$ is the horizon radius.
This two-sided geometry is dual to the thermofield-double state $|\mathrm{TFD} \rangle$ of two decoupled boundary CFTs (see Eq.~\eqref{eq:mytfd}).
The isometries of the background \eqref{eq:BH_geometry} allow us to express the complexity observable $\mathcal{C}_{\rm Any}$ in Eq.~\eqref{eq:CAny_observable} as follows,
\begin{equation}
    \mathcal{C}_{\rm Any}(t) = \frac{V_x}{G_N L} \sum_{\varepsilon=+,-} \int_{\Sigma_{\varepsilon}} d\sigma \, \mathcal{L}_{\varepsilon} (r, dr/d\sigma, dv/d\sigma) \, 
    \label{eq:CAny_thermofield_case}
\end{equation}
where $\sigma$ is an intrinsic radial parameter, $V_x = \int d^{d-1} x$ is the transverse volume, and $v=t + r^*(r)$ is an infalling null coordinate defined in terms of $dr^*=dr/f(r)$.

Since $v$ is cyclic for $\mathcal{L}_{\varepsilon}$ in Eq.~\eqref{eq:CAny_thermofield_case},\footnote{This means that $\mathcal{L}_{\varepsilon}$ depends on $\dot v \coloneqq dv/d\sigma$, but not on $v$.} the conjugate momenta $P_v^{\varepsilon} = \partial\mathcal{L}_{\varepsilon}/\partial\dot{v}$ are conserved along the profiles of $\Sigma_{\varepsilon}$. 
The late time regime $t \rightarrow \infty$ is achieved when each of the codimension-one surfaces $\Sigma_{\pm}$ hug a final slice at constant radial coordinate, such that the momenta approach constant values, up to exponentially-suppressed terms. This leads to the linear growth \cite{Belin:2021bga,Belin:2022xmt}
\begin{equation}
    \lim_{t \rightarrow \infty} P^{\pm}_v = P^{\pm}_{\infty} - \mathcal{O}(e^{-t}) \, , \quad  \Rightarrow \quad
   \lim_{t \rightarrow \infty} \mathcal{C}_{\rm Any} \approx \frac{V_x}{G_N L} \left( P^+_{\infty} + P^-_{\infty}  \right) t \, .
    \label{eq:linear_growth_CAny}
\end{equation}
One can further show that $P^{\pm}_{\infty} \propto TS$, where $T$ is the Hawking temperature and $S$ the entropy.

Let us consider the switchback effect, see \eg \cite{Stanford:2014jda,Chapman:2018dem,Chapman:2018lsv}. We modify the thermofield-double state as follows \cite{Shenker:2013pqa,Shenker:2013yza}
\begin{equation}
    | \Psi (t_L, t_R) \rangle = e^{-i H_L t_L -i H_R t_R} \, 
    W_L (t_n) \dots W_L (t_1) | \mathrm{TFD} (t=0) \rangle \, ,
    \label{eq:perturbed_TFD}
\end{equation}
where $W_L$ are low-energy perturbations acting on the left boundary.
These perturbations carry energies of the order of the black hole's temperature (thermal-scale perturbations), which is assumed to be much smaller than the black hole's mass. 
Let us assume that $|t_{k+1}-t_k| > t_*$, where $t_*$ is the scrambling time, and the label $k = 0, 1, \dots, n+1 $ in the inequalities includes the boundary times via the definitions $t_0=-t_R$ and $t_{n+1}=t_L$. 
We also assume that, in the sequence of times $t_0,\ldots t_{n+1}$ there are $n_{\rm sb}$ switchbacks (also called \textit{time folds}): in other words, the arguments of the absolute values $|t_{k+1} - t_k|$ change sign $n_{\rm sb} \leq n$ times. 
To keep the bulk solutions as simple as possible, we assume that
the perturbations are  approximately spherically symmetric. The bulk geometry dual to the perturbed state \eqref{eq:perturbed_TFD} can be modeled by a black hole with $n \geq 1$ shock waves produced by null matter.
In this setting, any CAny observable in the late time limit reads \cite{Belin:2022xmt}
\begin{equation}
  \lim_{t_L,t_R \rightarrow \infty}  \mathcal{C}_{\rm Any}  \propto \frac{V_x}{G_N L} \left(P^+_{\infty} + P^-_{\infty}  \right) \left( |t_R+t_1| + |t_2-t_1| + \dots 
   + |t_L-t_n| - 2 n_{\rm sb} t_* \right) \, .
   \label{eq:late_time_switchback}
\end{equation}
The contribution $-2 n_{\rm sb} t_*$ reflects the spreading of the perturbations into the system during a time scale measured by the scrambling time $t_*=\frac{1}{2\pi T}\log \left( M/E \right) \approx \frac{1}{2\pi T}\log S$, where $E\ll M$ is the energy of the shock, and $S$ the black hole's entropy. 
This characterizes the switchback effect for the CAny observables. At late boundary times $(t_L, t_R)$, holographic complexity will grow linearly with the sum of the boundary times $t_L+t_R$, but with a delay induced by the scrambling time of the perturbation. This phenomenon was already observed in earlier studies of the CV, CV2.0 and CA proposals \cite{Stanford:2014jda,Brown:2015lvg,Couch:2016exn}. 
This is not a surprise. As we said, those proposals can be obtained from CAny by appropriate limiting procedures.

\paragraph{Behavior at early and intermediate times}
Let us discuss other refined properties of the holographic conjectures, associated with early and intermediate times compared to the scrambling time and the inverse temperature scale.
It was originally speculated that the black hole mass $M$
should provide an upper bound on the complexity rate $d\mathcal{C}/dt \leq 2M/\pi$, based on Lloyd's bound on the computational speed \cite{Lloyd_2000}.
This statement was supported by the the full time evolution of CV in 
an eternal AdS black hole background, since the volume grows at a monotonically increasing rate that approaches a constant  from below at late times (\eg see Fig.~7 in \cite{Carmi:2017jqz}). This constant exactly saturates Lloyd's bound for large black holes.
However, the rate of CA reaches a maximum at a finite time before approaching from above a constant final value saturating  Lloyd's bound (\eg see Fig.~22 in \cite{Carmi:2017jqz}). 
Therefore, the intermediate regime in the evolution of CA violates the bound.   Other counterexamples have also been found~\cite{Swingle:2017zcd,Yang:2017czx,Couch:2017yil,Auzzi:2018zdu,Auzzi:2018pbc,An:2018xhv,Alishahiha:2018tep,Bernamonti:2021jyu,Wang:2023ipy,Aguilar-Gutierrez:2023ccv}.
It was argued in \cite{Cottrell:2017ayj} that a black hole should be modeled by the composition of simple gates (\ie close to the identity), and is thus incompatible with the hypotheses used to derive the Lloyd's bound. 
Overall, the adaptation of Lloyd's bound to holographic complexity is an open problem.

\begin{figure}[ht]
    \centering
 \includegraphics[width=0.5\linewidth]{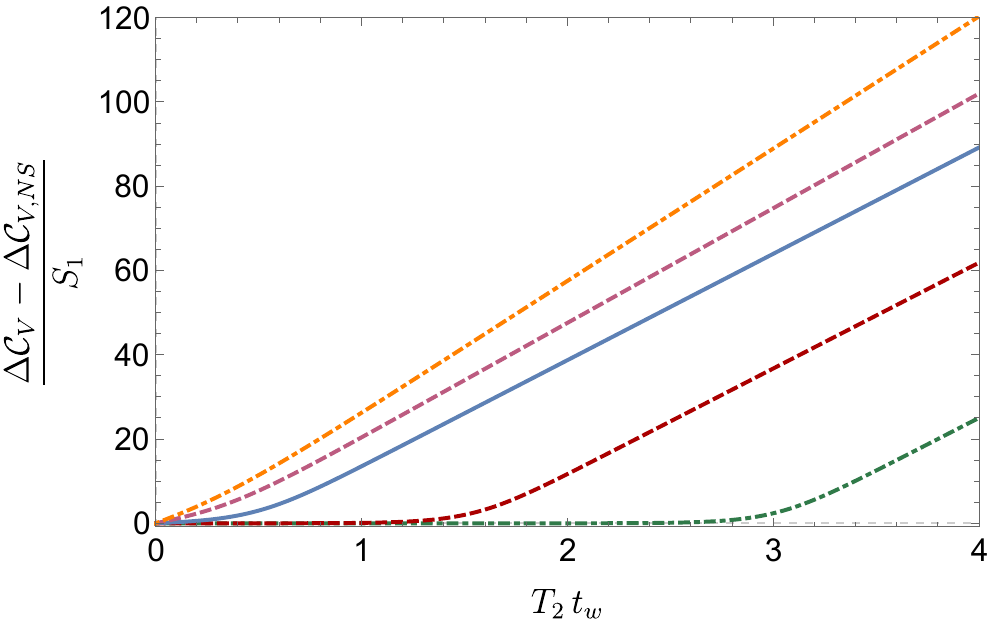}
    \caption{Volume complexity $\Delta \mathcal{C}_V$ in three bulk (two boundary) dimensions, compared to the value $\Delta \mathcal{C}_{V, NS}$ of the complexity in the Neveu-Schwarz vacuum, as a function of the boundary insertion time $-t_w$ of the shock wave. $S_1$ is the thermal entropy of the black hole solution before the shock wave insertion; $T_2$ the Hawking temperature after the insertion of the shock.
    Picture taken from Fig.~19 of reference~\cite{Chapman:2018lsv}. 
    From bottom (green) to top (orange), the colors correspond to increasing value of the energy of the shockwave as follows: $E_{\text{shock}}/(2 M_1)=1.5 \text{ (orange)},0.625 \text{ (pink)},0.105 \text{ (blue)},10^{-4} \text{ (red)},10^{-8}$ \text{ (green)}.}
    \label{fig:Vaidya}
\end{figure}

Next, one can study the reaction of the holographic observables to shock waves corresponding to the insertion of null matter at arbitrary boundary times \cite{Chapman:2018dem,Chapman:2018lsv}. Earlier, we assumed that the insertion times $t_k$ of the perturbations (which included the boundary times $t_L, t_R$) were separated by an interval larger than $t_*$ (see the discussion below equation \eqref{eq:perturbed_TFD}).  
Here we will not make this assumption. 
For simplicity, let us fix $t_L=t_R=0$.
We consider the case of a single \textit{time fold} at the (negative) time $-t_w$ when the shock is inserted, and study the dependence of physical quantities on this time.
In these cases, CV and CA show a plateau, where complexity is nearly constant, at early times $t_w\ll t_*$.  But these quantities approach linear growth at later times $t_w\gg t_*$.
The plateau becomes longer   when the perturbation has smaller energy, showing signatures of scrambling in chaotic systems.\footnote{Alternatively, one can study the full dependence of CV and CA on the boundary times, \eg see Fig.~11 and Fig.~17 in Ref.~\cite{Chapman:2018lsv}. In this case, holographic complexity admits a plateau region around $t_L=t_R=0$, which becomes longer when the shock is inserted at earlier times. This is another manifestation of the switchback effect.}
The scrambling time can be extracted by studying the plateau size, and turns out to be $t_{*} \approx \frac{1}{2\pi T}\log(M/T)$ (where $T$ is the Hawking temperature), as expected.
Applying the CV conjecture, the results are summarized in Fig.~\ref{fig:Vaidya}.
Furthermore, one can study the precise early 
$t_w$ dependence of holographic complexity.
A careful analysis (summarized in Fig.~27 of \cite{Chapman:2018lsv}) shows that the complexity grows exponentially
$ \mathcal{C}(t_w)\approx\exp(2\pi(t_w-t_*)/\beta )$, with a characteristic exponent which is exactly equal to the Lyapunov exponent of maximally chaotic systems (see $\lambda_{\rm OTOC}$ in the key notion \ref{item:fast_scrambler}) \cite{Maldacena:2015waa}. 
Remarkably, all these behaviors of the quantities proposed as duals to holographic complexity agree with Eq.~\eqref{eq:complexity_precursor}, obtained from the simple circuit models presented in Sec.~\ref{subsec:counting}. 
The matching is evident by comparing Figs.~\ref{fig:Vaidya} and~\ref{fig:SwitchbackCt}.
Finally, \cite{Chapman:2018lsv} extends previous studies to cases where the shocks have energies above the thermal scale.

\paragraph{Beyond semiclassical gravity: late-time behavior} 
As we have shown earlier in this subsection, all the holographic complexity proposals by construction display a linear growth until late times.
This result was derived in a semiclassical regime: we considered the black-hole geometry \eqref{eq:BH_geometry} as a solution of general relativity in asymptotically AdS spacetime, without introducing quantum gravity corrections.
As a result, holographic complexity could only be explored until times exponential in the black hole's entropy.
Quantum complexity, on the other hand, is expected to saturate and reach a plateau after the linear growth, until double-exponential times in the black hole's entropy (see the discussion around Fig.~\ref{fig:TimeEvolGen}).
This expectation from quantum circuits naturally leads to the following questions: can one compute quantum corrections to any of the geometric observables discussed so far? 
Does the inclusion of quantum corrections cause the holographic complexity proposals to saturate at late times?

In the case of the CV conjecture, the answer to both questions is affirmative for two-dimensional models of dilaton gravity, including JT gravity.
Reference~\cite{Iliesiu:2021ari} proposed a nonperturbative definition of the ERB's length that includes quantum corrections from surfaces with higher topologies.
We denote by $\gamma$ any non self-intersecting geodesic with length $\ell_{\gamma}$, $\Delta>0$ a regulator, and $\langle \dots \rangle$ the 
sum over geodesics defined over surfaces of any topology within the gravitational path integral. 
The nonperturbative length $\ell$ of the ERB reads
\begin{equation}
    \langle \ell \rangle \coloneqq  \lim_{\Delta \rightarrow 0} 
    \Big\langle \sum_{\gamma} \ell_{\gamma} e^{- \Delta \ell_{\gamma}}  \Big\rangle \, .
    \label{eq:length_exp_Iliesiu}
\end{equation}
One may worry that this quantity is divergent, but it turns out that its time dependent part $\langle\ell\rangle-\langle\ell\rangle_{t=0}$, which is all we need, is finite and independent of the regulator. 
Denoting by $S_0$ the leading-order entropy of a black hole in JT gravity, the ERB's length grows linearly until it saturates at a time and value both of order $e^{S_0}$ (see Fig.~1 and Fig.~5 of \cite{Iliesiu:2021ari}).
This behavior matches the time evolution of quantum complexity depicted in Fig.~\ref{fig:TimeEvolGen}, but with a difference.
The variance of the ERB's length is negligeble at times $t \sim \mathcal{O}(e^{S_0})$, but it monotonically grows until becoming of the same order as $\langle \ell \rangle$ at times  $t \sim \mathcal{O}(e^{2S_0})$.

Recently, the authors of~\cite{Miyaji:2025yvm} made further progress in defining a nonperturbative version of holographic complexity.  Rather than focusing on the length expectation value \eqref{eq:length_exp_Iliesiu}, they studied the spectral decomposition of $\langle e^{- \Delta \ell} \rangle$, and identified the latter quantity as a generating function to calculate quantum complexity. The generating function shows a slope-ramp-plateau structure similar to the case of the SFF, see Figs.~6-7 of \cite{Miyaji:2025yvm}.
In the limit $\Delta \rightarrow 0$, the ramp disappears from the time evolution  (Fig.~12 of \cite{Miyaji:2025yvm}), leading to the characteristic linear growth and late time saturation of complexity. Additional definitions leading to a late time saturation are based on spread complexity \cite{Balasubramanian:2024lqk}, and will be discussed in Sec.~\ref{ssec:holo_matching_2d}.

A general mechanism for  saturation of Einstein-Rosen  Bridges (ERBs) in {\it any} dimension was  proposed in \cite{Balasubramanian:2022gmo,Balasubramanian:2022lnw}, the authors of which constructed a basis of  microstates for eternal black holes in any dimension, and for any theory described at low energies by general relativity.  These states,  consisting of ERBs of different lengths supported internally by shells of matter, are perturbatively orthogonal.  Hence, they naively describe an infinite space of ERBs of arbitrary lengths.  However, topology-changing wormholes in the gravitational path integral lead to small quantum overlaps which 
make these states linearly dependent, so that they span a Hilbert space of dimension precisely equal to the exponential of the Bekenstein-Hawking entropy.  This means that we can pick a complete basis of wormholes of bounded length; so, long wormholes of the kind that appear at late times in the classical geometry can be regarded as superpositions of short bridges.  This suggests a quantum mechanical saturation of wormhole length in any dimension, similarly to the saturation mechanism for the thermal two-point function in 2d JT gravity via tunnelling to baby universes with shorter ERBs \cite{Saad:2019pqd}. If long ERBs should be regarded as superpositions of small bridges in this way, we should also ask if there is any definite notion of wormhole volume  that can be identified directly with complexity.  We will answer this question in Sec.~\ref{ssec:holo_matching_2d}.

\subsection{Time-independent properties of the holographic proposals}
\label{ssec:time_indep_properties}

To understand the time-independent predictions of the complexity conjectures, we need to deal with the UV divergences that these observables display as a consequence of approaching the AdS boundary.
On the field theory side, complexity is expected to diverge due to the short distance correlations. 
The structure of these divergences is a robust feature, which cares very little about whether the theory is weakly or strongly coupled. It is therefore ideally suited for comparing the holographic conjectures to simple set-ups in free field theory, which we reviewed in section \ref{sec.free.qft}. In holography, these divergences are regularized by introducing a short distance cutoff near the boundary of the asymptotically AdS spacetime, while in QFT they can be regularized, for instance, by placing the theory on a lattice with spacing $\delta$.

A similar divergent behavior appears when studying the entanglement entropy. In that case, the leading divergence behaves as an area law\footnote{Note that area law here refers to the term dominant in a short distance UV cutoff expansion (small $\delta$), not the term dominant in a large entangling region volume limit, as is sometimes the case in the condensed matter literature. The two expansions can differ when an additional scale is present, e.g., a temperature.} $A/\delta^{d-2}$ with $A$ the area of the entangling surface, $\delta$ a short distance cutoff and $d$ the spacetime dimension of the boundary field theory. This is true both in holographic settings and in free field theories \cite{Nishioka2009RTrev,Casini2009RevFreeEnt}. 
Furthermore, in the case of a smooth entangling surface, the divergences jump in powers of two, capturing different geometric features of the entangling surface. 
Universal contributions, depending on the anomalies of the theory, appear as coefficients of either the logarithmic or the constant parts of the expansion in the short-distance cutoff.

\begin{figure}[ht]
\centering
\subfigure[]{\includegraphics[scale=0.055]{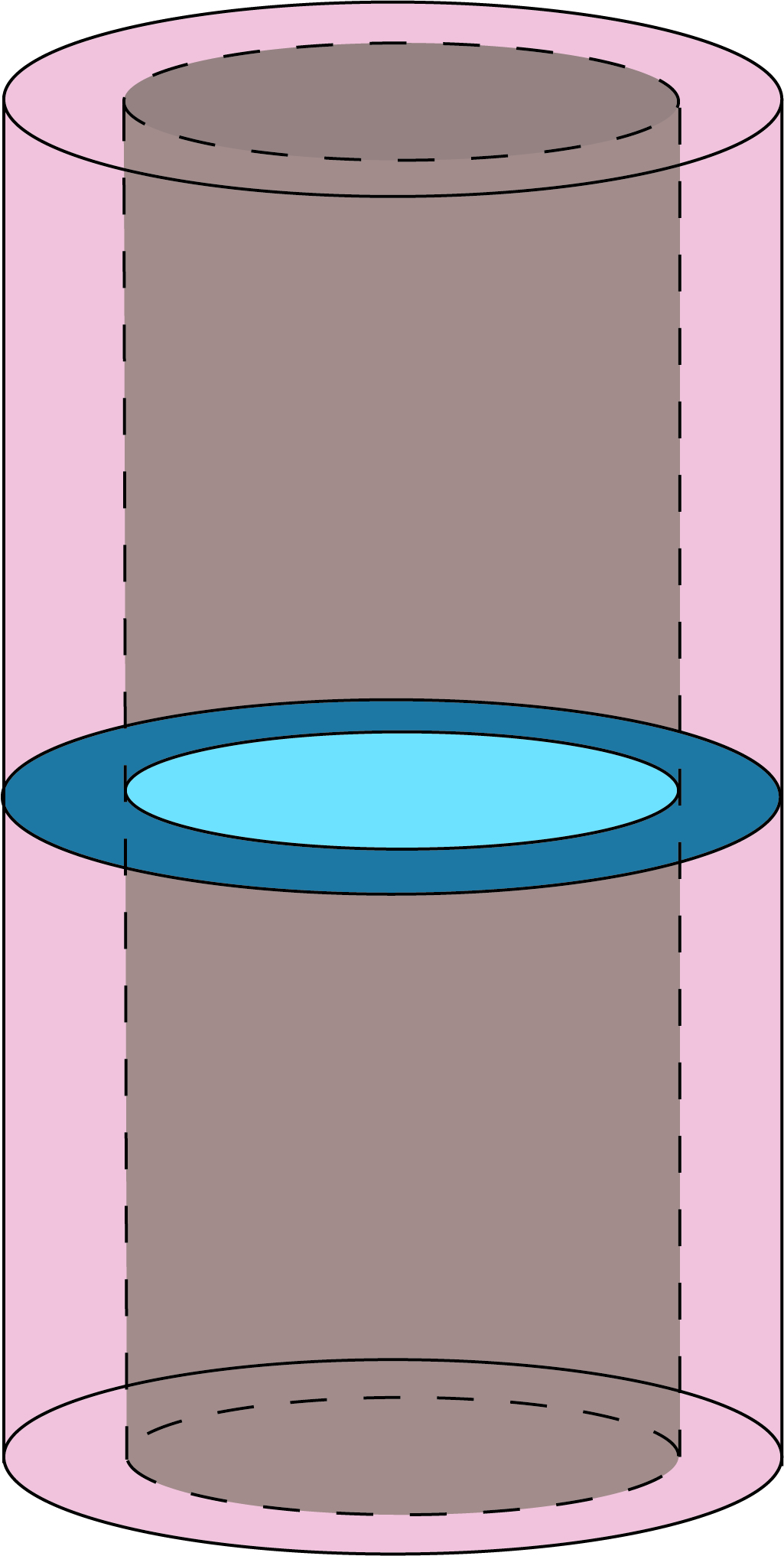}} \qquad \qquad
\subfigure[]{\includegraphics[scale=0.055]{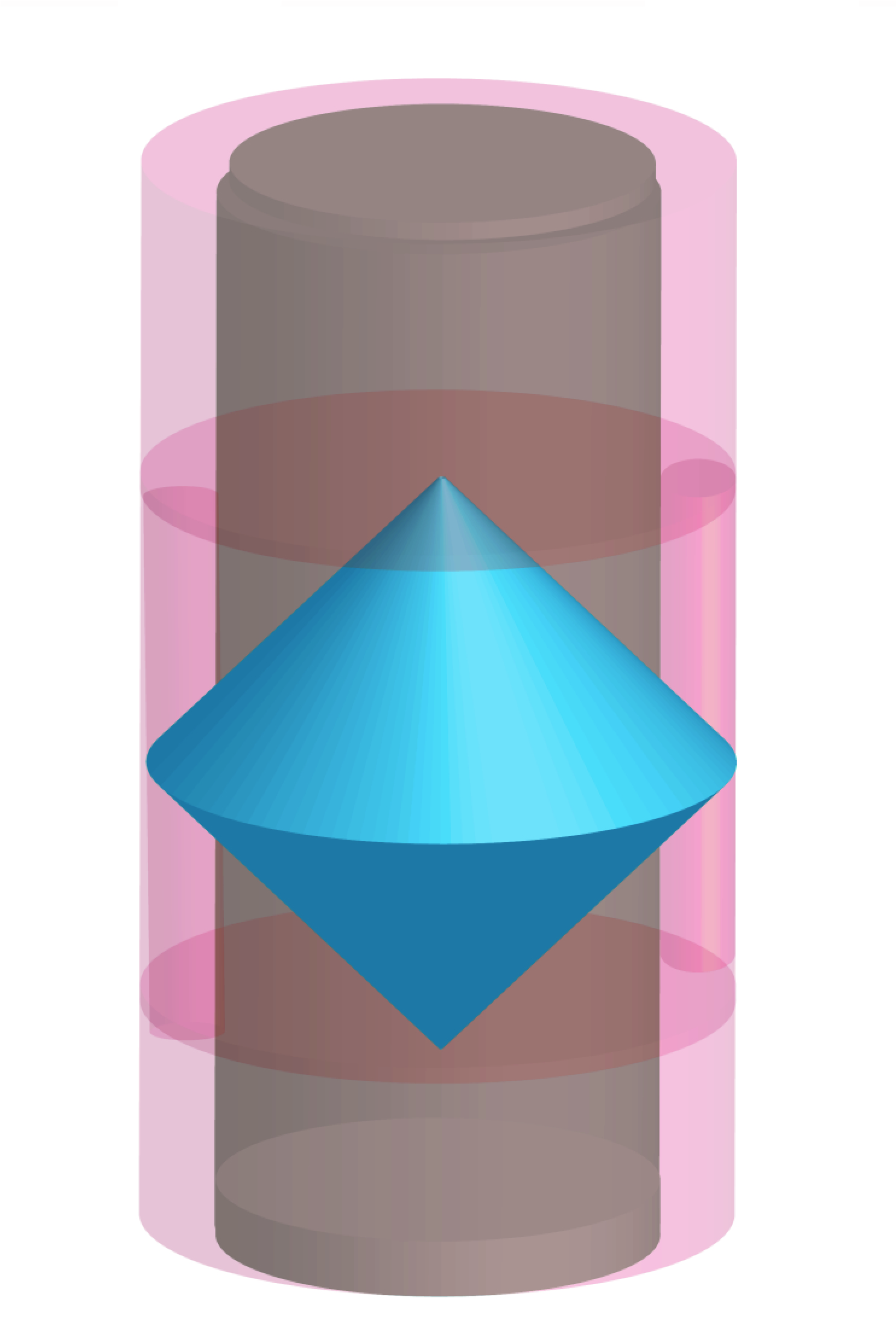}} 
\caption{Illustration of the CV (left) and CA (right) proposals in empty AdS spacetime for a state living on a constant-time boundary slice. The geometric quantities are evaluated inside the brown region with ends fixed by the short distance cutoff, illustrated in the figure as the pink region. Figure adapted from~\cite{Chapman:2016hwi}.
}
\label{fig:emptyCVCA}
\end{figure}

The above results motivated the study of UV divergences for the complexity of a vacuum state.
In \cite{Carmi2017DivC}, the authors explored the CV and CA proposals in empty AdS spacetime, see Fig.~\ref{fig:emptyCVCA}. They found that these quantities always diverge proportionally to the volume $\mathcal{V}$ of the system as 
\begin{equation}\label{eq:hologvaccompres}
    \mathcal{C}\propto k_d C_T \mathcal{V}/\delta^{d-1} +\ldots
\end{equation}
with $C_T$ the central charge, $k_d$ a coefficient which depends on the dimension and on certain definitional ambiguities, and the dots stand for less divergent terms.
Subleading divergences in the complexity appear in jumps of two powers of the cutoff and accompany various geometric features of the boundary slice of interest. 
These predictions are consistent with the results for free QFTs reported in Sec.~\ref{sec.free.qft}.

Next, one can consider complexity for more general states, such as the thermofield-double state dual to the two-sided eternal black hole \eqref{eq:BH_geometry}. 
In this case, one defines the \textit{complexity of formation} as the excess complexity required to construct this state compared to the complexity of two copies of the vacuum, \ie  $\mathcal{C}_{\text{form}} \coloneqq \mathcal{C}(|\mathrm{TFD}(t=0)\rangle)-2 \mathcal{C}(|\text{vac}\rangle)$. This quantity was evaluated in \cite{Chapman:2016hwi} using the CV and CA proposals and it was found that in the high temperature regime and in dimensions $d>2$, it grows linearly with the entropy of the system, \ie  $\delta \mathcal{C}=k_d S$.
Conversely, when $d=2$ the complexity was found to be a fixed constant, independent of the temperature.

\subsection{Generalizations of the holographic proposals}
\label{ssec:generalization_holo_complexity}

We focus on three generalizations: \textbf{(1)} the extension to the case of subregions, \textbf{(2)} settings with defects or boundaries,  and \textbf{(3)}  applications to de Sitter spacetime.  It is harder to define and test ideas about holographic complexity in these settings than in the standard ones discussed in the previous subsections. Nevertheless, they add information about which notion of quantum complexity should be expected on the field theory side of the holographic duality.\footnote{Field theory studies of complexity related to these three extensions appear in, \eg  \cite{Agon:2018zso,Caceres:2019pgf} for subregions, and \cite{Chapman:2018bqj} for  defects.  See also \cite{Czech:2016nxc} for a relevant tensor network construction and \cite{Lin:2022nss} for quantum duals to de Sitter spacetime. These constructions usually involve simplified models such as counting arguments for random circuits or even Gaussian free field theories and so can only be compared to the holographic results in a qualitative way.}

\paragraph{Subregions}
Consider a mixed state associated with a subregion $\mathcal{A}$ of the boundary. 
The CV proposal can be extended to this case by computing the maximal volume of a codimension-one surface delimited by the subregion $\mathcal{A}$ and its HRT surface \cite{Alishahiha:2015rta}.
According to the bulk reconstruction paradigm (see \cite{Harlow:2018fse} for a review), the information carried by the reduced density matrix $\rho_{\mathcal{A}}$ of a mixed state is holographically encoded by the entanglement wedge (EW), which is the bulk domain of dependence of a region that is enclosed within $\mathcal{A}$ and its HRT surface \cite{Headrick:2014cta}.
Therefore, in the case of CA (CV2.0) conjecture, subregion complexity was defined as the gravitational action (spacetime volume) in the bulk region given by the intersection of the WDW patch with the Entanglement Wedge \cite{Carmi:2016wjl}.

The general structure of the CV and CA proposals has been classified for a ball-shaped subregion $\mathcal{B}$ in arbitrary dimensions, giving a leading UV divergence similar to the full system \cite{Carmi:2016wjl}. That is, $\mathcal{C}_{\rm sub} \propto  k_d C_T \mathcal{V}(\mathcal{B})/\delta^{d-1},$  
where $k_d$, the same constant as in eq.~\eqref{eq:hologvaccompres}, carries information on various ambiguities and the dimensionality of the system, $C_T$ is the central charge, and $\mathcal{V}(\mathcal{B})$ is the volume of the ball. 
In general, the subregion complexity proposals have been shown to be superadditive in a pure state;\footnote{Consider any holographic complexity conjecture $\mathcal{C}(\Sigma)$ defined on a spacetime region anchored to a boundary Cauchy slice $\Sigma$. Let us split the Cauchy slice $\Sigma$ into a subregion $\mathcal{A}$ and its complement $\bar{\mathcal{A}}$. Denote with $\mathcal{C}(\mathcal{A})$ and $\mathcal{C}(\bar{\mathcal{A}})$ the corresponding subregion complexities, we say that holographic complexity is superadditive if
\begin{equation}
    \mathcal{C} (\Sigma) > \mathcal{C}(\mathcal{A}) + \mathcal{C} (\bar{\mathcal{A}}) \, ,
\end{equation}
and subadditive if the inequality is inverted.
}
CA is an exception, since one can tune the  counter term scale $L_{\rm{ct}}$ for the null-boundaries (defined below equation \eqref{eq:CV20}) to either get a subadditive or superadditive behaviour \cite{Caceres:2019pgf,Agon:2018zso,Alishahiha:2018lfv,Caceres:2018blh,Auzzi:2019fnp}. However, restricting the counter term scale $L_{\rm{ct}}$ such that the leading divergence in CA is positive, results in CA being superadditive too. 
Subsequent studies of holographic complexity associated with subregions were carried out, \eg in \cite{Alishahiha:2015rta,Carmi:2016wjl,Ben-Ami:2016qex,Abt:2017pmf,Roy:2017kha,Roy:2017uar,Bakhshaei:2017qud,Agon:2018zso,Alishahiha:2018lfv,Caceres:2018luq,Abt:2018ywl,Chen:2018mcc,Bhattacharya:2019zkb,Caceres:2019pgf,Auzzi:2019fnp,Auzzi:2019vyh,Auzzi:2019mah,Sato:2021ftf}. Relatedly, proposals for a microscopic dual interpretation were investigated, \eg in Refs.~\cite{Agon:2018zso,Chapman:2018hou,Balasubramanian:2018hsu,Caceres:2019pgf,Chapman:2019clq,Camargo:2018eof}.

The proposed formulae for subregion complexity have an interesting behavior in the BTZ black hole geometry.
Given a boundary subregion composed by $q$ segments of total length $\ell_{\rm tot}$, the CV proposal reads \cite{Abt:2017pmf}
\begin{equation}
    \mathcal{C}_V = \frac{2c}{3} \left[ \frac{\ell_{\rm tot}}{\delta} + \pi \left( q - 2 \chi  \right)  \right] \, ,
    \label{eq:subregion_CV_BTZ}
\end{equation}
where $c$ is the central charge, $\chi$ the Euler characteristic of the maximal surface, and $\delta$ a UV cutoff regulator.
Interestingly, this formula is topological, \ie it does not depend on the temperature. This feature is not shared by subregion CV in higher dimensions \cite{Alishahiha:2015rta,Ben-Ami:2016qex}.

The subregion CA and CV2.0 conjectures for a \emph{single} segment of length $\ell$ in the BTZ background were studied in 
\cite{Auzzi:2019vyh}
\begin{subequations}
\begin{equation}
    \mathcal{C}_A = \frac{c}{6 \pi^2} \log \left( \frac{L_{\rm ct}}{L} \right) \frac{\ell}{\delta} - \log \left( \frac{2 L_{\rm ct}}{L} \right) \frac{S_{\rm EE}}{\pi^2} + \frac{c}{24} \, ,
    \label{eq:subregion_CA}
\end{equation}
\begin{equation}
    \mathcal{C}_{2.0V} = \frac{2c}{3} \frac{\ell}{\delta} - 4 S_{\rm EE} - \frac{\pi^2}{6}  c \, ,  \qquad
     S_{\rm EE} = \frac{c}{3} \log \left[ \frac{1}{\pi T \delta}  \sinh \left( \pi T \ell \right) \right] \, ,
     \label{eq:subregion_CV20}
\end{equation}
\end{subequations}
where $L_{\rm ct}$ is a length scale introduced by a counterterm on null surfaces in the gravitational action, see the discussion below Eq.~\eqref{eq:CV20}, and $S_{\rm EE}$ is the entanglement entropy of the segment.
These results present an appealing feature in that the complexity contains a piece proportional to the entanglement entropy plus an additional constant which could naively depend on the topology as in the CV conjecture results \eqref{eq:subregion_CV_BTZ}. 
However, an analysis of CA and CV2.0 conjectures in the case of multiple segments in the BTZ background reveals an intricate dependence on the subregion sizes and temperature (see Eqs.~(4.23)--(4.24) of \cite{Auzzi:2019vyh}), which does not follow the same pattern for the dependence of complexity on the entanglement entropies as the one in Eqs.~\eqref{eq:subregion_CA}--\eqref{eq:subregion_CV20} \cite{Auzzi:2019vyh,Camargo:2020yfv}. 
Moreover, we do not expect a simple linear relation between complexity and entanglement entropy in dynamical situations, since, as we argued at the beginning of Sec.~\ref{sec:holography} the two quantities evolve very differently with time.

\paragraph{Defects and boundaries}
Defects and boundaries are extended objects that break the translational invariance of a system. They provide important probes in QFT, and are ubiquitous in condensed matter systems. 
Moreover, they can be engineered holographically, thus providing a playground to test the complexity conjectures.
We analyze the UV divergences of CV and CA in three settings.

The first is the two-sided Randall-Sundrum (RS) model, a gravity solution where two patches of AdS spacetime are glued together along a thin codimension-one brane \cite{Chapman:2018bqj}.
The second is the AdS/BCFT model, where an end-of-the-world brane, corresponding to a boundary in the dual CFT, delimits vacuum AdS \cite{Braccia:2019xxi,Sato:2019kik}. 
The last is Janus AdS, a dilatonic deformation of AdS spacetime dual to an interface CFT and arising from the dimensional reduction of a solution in type IIB supergravity \cite{Auzzi:2021nrj,Auzzi:2021ozb}.

We summarize the main features of the CV and CA conjectures in these geometries 
in the case of a subregion of length $\ell$ in 2+1 dimensions, see Table~\ref{tab:results_defects}.
The results refer to the \textit{complexity of formation} $\Delta\mathcal{C}$, defined as the complexity in the geometry with a defect (or boundary), minus the complexity evaluated in empty AdS spacetime. 
This quantity identifies the additional contributions due to the presence of the boundary or defect.
Scanning the table, we note that the UV divergence is always logarithmic in the length regulator $\delta$, independently of the kind of defect (or boundary) under consideration.
Therefore, the CV conjecture is, to some degree, more universal than the CA proposal; the latter exhibits a different degree of UV divergences depending on the specific defect or boundary under consideration.
 We further observe that Janus geometry is the only case where the 
 degree of divergence  is universal across the different holographic proposals. 
Janus defects describe an interface CFT with a smooth holographic dual which can be obtained as a reduction of type IIB supergravity. This fact may suggest that they provide a better setup for investigating holographic complexity (or any other holographic quantity).

\begin{table}[ht]   
\begin{center}    
\begin{tabular}  {|c|c|c|} \hline  &  $\Delta \mathcal{C}_V$ & $\Delta \mathcal{C}_A$  \\ \hline
\rule{0pt}{4.9ex} $2$-sided Randall-Sundrum  & $ \frac{2}{3} c \,  \eta_{\rm RS}  \,
\log \left(    \frac{\ell}{\,  \delta} \right)
 + \text{finite} $  & $ 0 $  \\
\rule{0pt}{4.9ex} AdS$_3$/BCFT$_2$ & $\frac{2}{3} c \,  \eta_{\rm BCFT}  \,
\log \left(    \frac{\ell}{\,  \delta} \right)
 + \text{finite} $  &  $ \mathrm{finite}   $     \\ 
 \rule{0pt}{4.9ex}  Janus AdS$_3$  & $ \frac{2}{3} c \,  \eta_{\rm JAdS}  \,
\log \left(    \frac{\ell}{\,  \delta} \right)
 + \text{finite}  $ & $ \frac{2}{3}  c \, \tilde{\eta}_{\rm JAdS} \log \left( \frac{\ell}{ \delta} \right) + \mathrm{finite} $   \\[0.2cm]
\hline
\end{tabular}   
\caption{Subregion complexity of formation $\Delta \mathcal{C}$ for an interval of length $\ell$ in the cases of CV and CA, compared to empty AdS spacetime.
The coefficients of the log divergences $\eta, \tilde{\eta}$ depend on the details of the defect or boundary, see \cite{Chapman:2018bqj,Braccia:2019xxi,Sato:2019kik,Auzzi:2021nrj,Auzzi:2021ozb}. } 
\label{tab:results_defects}
\end{center}
\end{table}

Holographic complexity was studied in Janus AdS geometry under different regularization schemes \cite{Baiguera:2021cba}.
In dimensions $d=2,4$, the authors found a scheme-independent logarithmic divergence, while in odd-dimensions a scheme-independent finite term was observed instead.
This result parallels the behavior of the entanglement entropy of smooth subregions, see \eg  
 \cite{Casini2009RevFreeEnt}.

\paragraph{Holographic complexity in de Sitter space}

De Sitter spacetime is a maximally symmetric solution of Einstein's equations with a positive cosmological constant. It describes the early stages of evolution of our universe, and possibly the late stages if there is a positive cosmological constant (\eg see the reviews \cite{Spradlin:2001pw,Anninos:2012qw,Galante:2023uyf}). 
Since there is no timelike boundary in de Sitter spacetime, it is challenging to find a holographic interpretation of this geometry in terms of a microscopic quantum mechanical system.
Nevertheless, the physical relevance of this spacetime stimulated several attempts \cite{Gibbons:1977mu,Bousso:1999dw,Banks:2000fe,Bousso:2000nf,Banks:2001yp,Strominger:2001pn, Balasubramanian:2001nb, Balasubramanian:2002zh,
Banks:2002wr,Leblond:2002ns,Leblond:2002tf,Parikh:2002py,Dyson:2002nt,Dyson:2002pf,Parikh:2004wh,Banks:2005bm,Freivogel:2005qh,Banks:2006rx,Lowe:2010np,Anninos:2011af,Fischetti:2014uxa,Anninos:2017hhn,Anninos:2018svg,Banks:2018ypk,Dong:2018cuv,Lewkowycz:2019xse,Banks:2020zcr,Mirbabayi:2020grb,Anninos:2020cwo,Anninos:2020hfj,Susskind:2021omt,Susskind:2021dfc,Coleman:2021nor,Susskind:2021esx,Shaghoulian:2021cef,Shaghoulian:2022fop,Lin:2022nss,Banihashemi:2022htw,Silverstein:2022dfj,Chandrasekaran:2022cip,Witten:2023xze,Witten:2023qsv,Rahman:2023pgt,Narovlansky:2023lfz,Batra:2024kjl,Verlinde:2024znh,Verlinde:2024zrh,Tietto:2025oxn}.
Recent developments suggest that the inclusion of an artificial timelike boundary in de Sitter spacetime is needed for the consistency of thermodynamics \cite{Banihashemi:2022htw}, microstate counting \cite{Lewkowycz:2019xse,Coleman:2021nor,Batra:2024kjl}, and to take into account the role played by an observer \cite{Anninos:2011af,Chandrasekaran:2022cip,Witten:2023xze,Witten:2023qsv,Balasubramanian:2023xyd,Balasubramanian:2021wgd, Tietto:2025oxn}.
These observations led to a revival of \textit{static patch holography}, according to which a dual quantum theory lives on a codimension-one timelike surface (the \textit{stretched horizon}) located just inside the cosmological horizon \cite{Bousso:2000nf,Parikh:2004wh,Banks:2006rx,Dong:2018cuv,Shaghoulian:2021cef,Susskind:2021esx,Susskind:2021dfc,Shaghoulian:2022fop,Rahman:2023pgt,Narovlansky:2023lfz,Verlinde:2024znh,Verlinde:2024zrh}. 
The holographic complexity proposals introduced in subsection \ref{ssec:holo_compl_proposals} can be applied to de Sitter spacetime by requiring that the geometric objects of interest are anchored at the stretched horizons on the two sides of the geometry \cite{Susskind:2021esx}.
The study of these generalizations is speculative, but may hint towards the expected properties of the quantum mechanical dual to de Sitter space.

A main novelty, compared to the AdS case, is that CV, CV2.0 and CA conjectures in dS space include contributions from spacetime regions close to timelike infinity, and therefore  diverge at a finite critical time, \eg see Eq.~(3.17) of~\cite{Jorstad:2022mls}.\footnote{One can regularize both holographic complexity and its rate by introducing a cutoff surface at future infinity of dS spacetime. After introducing the cutoff, holographic complexity grows linearly, \eg see Eq.~(3.30) of \cite{Jorstad:2022mls}.} 
It was argued in \cite{Susskind:2021esx} that this behavior should be interpreted as a \textit{hyperfast} growth of complexity corresponding to circuits where the gates involve a large number of qubits at each step in the time evolution (\eg see \cite{Lin:2022nss}).
The hyperfast growth also occurs in two dimensions \cite{Anegawa:2023wrk} and  in models of inflation where a bubble of de Sitter spacetime is inside an AdS geometry \cite{Auzzi:2023qbm}.\footnote{Note however, that in two dimensions, the CV proposal yields a diverging rate of change in complexity as the critical time is approached, but the complexity itself remains finite in this limit.}\footnote{Holographic complexity has also been studied in other cosmological models, for instance the case of Kasner-like singularities~\cite{Narayan:2024fcp}. }
However, there are two significant exceptions to this trend. First, there is a class of codimension-one CAny observables that exhibits linear or exponential growth persisting forever, without a divergent rate at finite time \cite{Aguilar-Gutierrez:2023zqm}.
Furthermore, in any dimension of empty de Sitter space, the volume observable relevant to the CV conjecture does not strictly speaking ``exist'' if we anchor the relevant surfaces on the observer's world line. To address this, we can instead anchor the surface to a ``stretched horizon'' at some radius $r_{\text{st}}$.  But in this case the volume stops ``existing'' at a time $t_{\rm crit}$ proportional to $L \, \mathrm{arctanh} (r_{\text{st}}/L)$, where $L$ is the de Sitter scale. So, in addition to the dependence on $L$ and therefore on temperature, the critical time also depends on the radius of the ``stretched horizon'' that regularizes the problem. If the stretched horizon is very close to the de Sitter horizon, it can be made arbitrarily large \cite{Jorstad:2022mls}.  Similarly, the CV conjecture is strictly speaking only defined for a short time of the order of the inverse temperature in the case of two-dimensional centaur geometries, \ie gravitational models where de Sitter spacetime is glued to an asymptotic AdS region with a standard timelike boundary \cite{Chapman:2021eyy}, because the  maximal volume surfaces no longer exist after the critical time unless they are regularized in some way. It would be interesting to understand whether these behaviors of the CV observables have an interpretable origin in a dual description.

Next, let us consider the switchback effect.
In the case of a finite-energy shock wave perturbing a black hole solution in asymptotically de Sitter spacetime, the standard complexity conjectures (CV, CV2.0 and CA) admit a plateau around $t=0$ when complexity is approximately constant \cite{Baiguera:2023tpt,Baiguera:2024xju}, similar to the AdS case.\footnote{The de Sitter shock waves considered in \cite{Baiguera:2023tpt,Baiguera:2024xju} correspond to the (spherically symmetric) ejection of mass into the cosmological horizon. As a consequence, the cosmological horizon is pushed further away from an observer sitting at some fixed radius. While this setting may seem somewhat peculiar, such geometries are consistent with the null energy condition and provide a simple setup in which complexity can be studied.}
Notably, the duration of this regime increases when the shock crosses the stretched horizon at earlier times (see figs.~23, 25 in \cite{Baiguera:2023tpt}), and corresponds to special geometrical configurations that arise because the Penrose diagram of de Sitter spacetime grows taller when a null pulse carrying positive energy is inserted in the bulk \cite{Gao:2000ga}.
The critical time at which complexity starts growing significantly in CV, CV2.0 and CA conjectures is always delayed by the presence of a shock wave moving along the cosmological horizon \cite{Anegawa:2023dad} (see also \cite{Faruk:2025bed}). More precisely, the duration of the above-mentioned plateau region is $t_{pl} = 4(t_w - t_*)$ where $-t_w$ is the time at which the shockwave is inserted and $t_*\sim \frac{1}{T_C}\log S_C$  is the analogue of the scrambling time (with $T_C,S_C$ the temperature and entropy of the cosmic horizon for the Schwarzschild-de Sitter geometry to which the shock is inserted)  \cite{Baiguera:2023tpt,Baiguera:2024xju}. This result is similar to the one found in the AdS case.
Reference \cite{Aguilar-Gutierrez:2023pnn} further showed that the codimension-one CAny observables displaying linear growth at late times present a delay in this evolution satisfying the same structure obtained in Eq.~\eqref{eq:late_time_switchback} in the AdS case.
All these behaviors support the existence of a switchback effect for the holographic conjectures applied to de Sitter spacetime. 
It would be useful to match the above behaviors with those of appropriate circuit models, similarly to the arguments of section \ref{subsec:counting}. 
In this regard, it is important to note that the nature of scrambling in de Sitter spacetime has been a subject of debate, see \eg Refs.~\cite{Aalsma:2020aib,Geng:2020kxh,Chapman:2021eyy,Milekhin:2024vbb}.

\subsection{Quantitative matches relating spread complexity and gravitational observables}
\label{ssec:quantitative_matching_holo}

In this section, we will review proposals for quantitative matches between notions of complexity in the boundary theory and geometric observables in the dual bulk.  The basic challenge is to frame the definition of complexity in terms of the holographic dictionary~\cite{Gubser:1998bc,Witten:1998qj} in order to get a corresponding quantity in the bulk gravity.  A notable recent success is the matching between the length of wormholes in Jackiw-Teitelboim gravity and  spread complexity in a dual Double Scaled SYK model.

\subsubsection{Spread complexity in double-scaled SYK model equals wormhole length}
\label{ssec:holo_matching_2d}

The SYK model was defined above in Sec.~\ref{ssec:integrability_chaos} and we repeat the formulae here for convenience in a slightly different notation that will be useful for our purposes. The model contains $N$ Majorana fermions $\psi_i$, $i=1,\cdots,N$,\footnote{We normalize them as
$
\left\lbrace \psi_i,\psi_j  \right\rbrace = \psi_i\psi_j+\psi_j\psi_i=2\delta_{ij}
$.}
interacting via the Hamiltonian \eqref{eq:SYKHamq}. The couplings $J$ are real, independently distributed, Gaussian random variables with first and second moments 
\be 
\langle J_{i_1 i_2 \hdots i_p} \rangle =0\,,\,\,\,\,\,\,\,\,\,\,\,\,\,
\langle J^2_{i_1 i_2 \hdots i_p} \rangle=(1-e^{-\lambda}){\begin{pmatrix}
N\\
p
\end{pmatrix} }^{-1} \mathcal{J}^2\;,
\label{eq:SYKcouplings}
\ee
where $\lambda=\frac{2p^2}{N}$. The scaling of the couplings is tuned so that the density of states is bounded between $-2\mathcal{J}$ and $2\mathcal{J}$ in the large-$N$ limit with $p$ fixed (see \cite{Sarosi:2017ykf} for a review). Here we focus on the so-called Double Scaled SYK (DSSYK) model in which $N,p\to\infty$ with $\lambda = 2p^2/N$ fixed (see the review \cite{Berkooz:2024lgq}). Below, we will also refer to a triple scaling limit that is defined by $\lambda\ll 1$ and energies $E/\mathcal{J}\ll 1$. In this limit the theory is governed by the Schwarzian action \cite{Maldacena:2016hyu,Lin:2022rbf}.

To compute the spread complexity of the DSSYK model we need the Lanczos coefficients.   As we discussed, these can be computed either by tridiagonalizing the Hamiltonian by the moment method, or through an integral formula applied to the density of states.  To apply the moment method we have to calculate the expectation values of powers of the Hamiltonian in the initial state which we take to be the thermo-field double state.  So we have to calculate
\begin{equation}
M_{2k} =
\langle \mathrm{Tr}(H^{2k}) \rangle
= i^{kp} \sum_{I_1\cdots I_{2k}} \langle J_{I_1}\cdots J_{I_{2k}} \rangle \mathrm{Tr}(\psi_{I_1} \cdots \psi_{I_{2k}})\,,
\end{equation}
where we introduced the notation $\psi_I = \psi_{i_1} \psi_{i_2}\cdots \psi_{i_p}$.
We have only written the even moments because the odd moments vanish. To finish the computation we have to Wick contract the multi-indexed fermion terms in pairs.   As discussed in \cite{Erdos:2014zgc,Berkooz:2018qkz},  the Feynman diagrams describing these contractions are called {\it chord diagrams} and can be visualized by placing $2k$ points on the circumference of a circle, and then contracting the dots in pairs. Carrying out these contractions, we arrive at the moments  
\be
M_{2k}=(1-q)^k\mathcal{J}^{2k} \sum_{\text{diag. 
 with k chords}} q^{\text{number of intersections}}=\langle 0|T^{2k}|0\rangle\, ,
\ee
where $q=e^{-\lambda}$.
The second equality expresses the moments in terms of a tridiagonal transfer matrix $T$ acting on $\ket{0}$, an auxiliary state with zero chords, that represents the thermofield double state at infinite temperature \cite{Berkooz:2018qkz}.  Essentially, by construction $T$ is the tridiagonalized Hamiltonian; the apparently auxiliary chord basis is precisely the Krylov basis of Sec.~\ref{sec:Krylov}.  Now, following the moment method, we can immediately read off the Lanczos coefficients. They are 
\be
b_n=\mathcal{J} 
\sqrt{1-q^n} \, .
\label{eq:doubscalLanncz}
\ee
We can then solve the Schrodinger equation, and compute spread complexity.  In the double scaling limit $N\to \infty$, and hence the Hilbert space is infinite-dimensional. Correspondingly, using the Lanczos coefficients in (\ref{eq:doubscalLanncz}), we find a spread complexity that grows linearly forever (see Sec.~\ref{sec:Krylov}).  Meanwhile, we can apply an additional low-energy scaling limit \cite{Berkooz:2018qkz}. This so-called triple scaling limit is effectively a continuum limit for the chord basis in which $T$, the tridiagonalized Hamiltonian, reduces to a boundary description of a bulk Jackiw-Teitelboim (JT) gravity. In this limit, the authors of~\cite{Lin:2022rbf,Rabinovici:2023yex} showed that the quantity calculating the length of the wormhole in the TFD state of JT gravity is precisely the spread complexity discussed above.  This construction establishes a precise correspondence between wormhole length in gravity and spread complexity in this limit of JT gravity and the dual DSSYK model.

However, this construction leads to a puzzle.  At any finite $N$, the Hilbert space should be finite-dimensional, and hence the spread complexity should saturate at late times, together with the quantum analog of wormhole length. How can we see this?  Fortunately, the results of \cite{Erdos:2014zgc} demonstrate that the moments of the SYK Hamiltonian described above arise from the distribution
\be \label{densityDSSYK}
\rho(E\vert q)=\frac{2}{\pi \sqrt{4\mathcal{J}^2-E^2}}\prod\limits_{k=0}^{\infty}\left[\frac{1-q^{2k+2}}{1-q^{2k+1}}\left(1-\frac{q^k E^2}{\mathcal{J}^2(1+q^k)^2}\right)  \right] \;.
\ee
In other words, this is the density of states at a given $q$.  Following the discussion in Sec.~\ref{sec:coarseLanczosSpectrum},  we can directly compute the Lanczos spectrum from this density of states at large but finite $N$ via an integral formula.  This procedure was carried out in \cite{Nandy:2024zcd,Balasubramanian:2024lqk}, where the authors demonstrated that the Lanczos coefficients for the SYK model match the computations of \cite{Berkooz:2018qkz} (described above) for an initial range of Krylov indices, but then decline systematically to zero.  The decline occurs over the whole range of Krylov indices up to the dimension of the Hilbert space.  This descent to zero for the Lanczos coefficients causes the spread complexity to saturate at late times.  If we maintain the identification between wormhole length and spread complexity that applies at early times, we would say that the wormhole length is saturating at late times.
From the bulk JT gravity point of view, this decline thus corresponds to a non-perturbative quantum effect in gravity that corrects the steady increase in the Lanczos coefficients and wormhole size expected from the early-time classical description of the system.  Of course, at late times we could imagine alternative non-perturbative definitions of the wormhole length in JT gravity that also agree with the early-time identification with spread complexity.  Interesting recent developments along this line are also presented in \cite{Iliesiu:2021ari,Iliesiu:2024cnh,Miyaji:2025yvm} which were discussed in Sec.~\ref{ssec:evidence_holo_complexity}. 

In a related vein, Ref.~\cite{Heller:2024ldz} utilized new developments identifying the gravity dual of DSSYK~\cite{Blommaert:2024ydx} to show that the relation between spread complexity and the bulk volume extends past the limits used in the studies of~\cite{Rabinovici:2023yex}. In particular, this relation extends to the full DSSYK regime at arbitrary temperatures. The key finding of~\cite{Heller:2024ldz} was that the match with the volume on the gravity side required computing the spread complexity for the preparation of a finite-temperature state using the Euclidean path integral. This is opposed to including the Euclidean state preparation in the reference state, and then assigning non-vanishing complexity only to the purely Lorentzian evolution. Furthermore, since the calculations in~\cite{Heller:2024ldz} also applied to the quantum regime, the authors were able to extract the form of the leading correction to the CV proposal originating from the contributions of quantum fields in a two-dimensional bulk geometry. This correction was crucially needed for the quantitative match with the boundary spread complexity. Interestingly, the correction exhibits a behavior compatible with the expectations for the spread complexity of bulk quantum fields, but a precise relation remains unknown.

These findings suggest that Krylov-basis definitions of complexity are natural from a bulk perspective, and could provide a non-perturbative definition of bulk volumes. A crucial next step would be to extend these studies to higher dimensions, beginning with AdS$_3$. While the powerful exact results available in DSSYK may not extend directly, recent progress with Virasoro TQFTs~\cite{Collier:2023fwi} could offer insights into the bulk Hilbert space and its relation to the Krylov basis.   Another approach could be to exploit the observation in \cite{Gautason:2025ryg} that at late times the volume of an extremal  slice through the ERB of any spherically-symmetric black hole reduces to a geodesic length in an effective two-dimensional JT gravity theory.  This suggests that we may be able to directly export the results relating spread complexity and wormholes in JT gravity to higher dimensions.

\subsubsection{Rate of spread-complexity growth and proper momentum}

The above comparisons between the Krylov basis definitions of complexity and holography were made in the context of low-dimensional toy models. This progress was possible due to the explicit definition of the thermofield-double state in the double-scaled SYK model using a transfer matrix and its $q$--deformed algebra. However, extending this approach to higher-dimensional holographic setups may be  challenging. Nevertheless, several universal insights about operator growth and spread complexity can already be tested in higher dimensions. One of the most concrete examples is the relation between the growth rate of operator complexity (or size) and the radial momentum of its dual massive particle in AdS spacetime. Intuitively, we expect that the time derivative of complexity (the rate of complexity) of the Heisenberg operator $\mathcal{O}(t)$ should be proportional to the radial momentum of the particle
\be
\partial_t \mathcal{C}(t)\sim P_{radial}.\label{CPrOp}
\ee
Since $\mathcal{C}$ is the mean position in the Krylov chain, this relation is telling us that the rate of progress along this chain is dual to the radial momentum.
This sort of relation has been argued for in various ways in~\cite{Susskind:2014jwa,Susskind:2018tei,Susskind:2019ddc,Magan:2018nmu,Lin:2019kpf}, including qualitative models of operator growth using epidemic models \cite{Susskind:2020gnl}. Furthermore, Refs.~\cite{Barbon:2020uux,Barbon:2020olv,Barbon:2019tuq} essentially proved this relation under the assumption of the CV proposal. However, at that time, a precise definition of complexity satisfying Eq.~\eqref{CPrOp} was missing.

By now, there are  setups where this formula can be precisely evaluated on both sides of the holographic correspondence. Firstly, in the SYK model at large $q$, one can directly compute the expectation value of the size operator \cite{Roberts:2018mnp,Lin:2019qwu} that can be defined in terms of two copies of fermions (left and right in the thermofield-double construction, denoted with subscripts $l$ and $r$ respectively) as
\be
\hat{S}=i\sum_j\psi^j_l\psi^j_r+\frac{N}{2}\,,
\ee
and in \cite{Lin:2019qwu}, authors showed that its rate of change in time (i.e., time derivative) is proportional to the radial momentum of a particle in AdS$_2$ (see also \cite{Brown:2018kvn}).

Recently, this correspondence was also confirmed 
between the spread complexity in 2D CFTs and the proper radial momentum of particles in AdS$_3$ spacetime. 
Reference~\cite{Caputa:2024sux} studied a local operator quench where a CFT state is locally excited by a primary operator with conformal dimension $\Delta$, and evolved unitarily with the Hamiltonian. Holographically, this setting corresponds to a particle with mass $m=\Delta$ propagating from the boundary towards the bulk, with the particle's initial position corresponding to the operator's energy regulator $\varepsilon$. Interestingly, the rate of spread complexity for this dynamical state is related to the particle's momentum via
\be
\partial_t \mathcal{C}_K(t)=-\frac{1}{\varepsilon}P_\rho(t).
\ee
However, the key observation is that the momentum must be computed in the proper radial distance coordinate $\rho$ (for example, this relation would not hold in Poincaré coordinates). This relation between Krylov complexity and momentum has also been analyzed from a slightly different perspective in \cite{Fan:2024iop,He:2024pox}, and the connection between the position on the Krylov chain $n$ and proper radial distance in the bulk was further substantiated in \cite{Dodelson:2025rng}.

Finally, the evolution of spread complexity following local quench protocols in 2D CFTs was studied in \cite{Caputa:2025dep}. The initial state is given by two thermal states on semi-infinite lines joined together at $x=0$ and evolved with the CFT Hamiltonian. The crucial object, as discussed in Sec.~\ref{sec:Krylov}, is the return amplitude that reads
\be
S(t)=\left(\frac{\sinh\left(\frac{\pi (t+2{\rm i}\varepsilon)}{\beta}\right)}{\sinh\left(\frac{2\pi {\rm i}\varepsilon}{\beta}\right)}\right)^{-c/8}\,,
\ee
where $\beta$ is the inverse temperature, $\varepsilon$ is the UV regulator and $c$ is the central charge of the 2D CFT. Interestingly, the return amplitude leads to Lanczos coefficients that depend on the central charge
\be
a_n=\frac{2\pi}{\beta \tan\left(\frac{2\pi\varepsilon}{\beta}\right)}\left(n+\frac{c}{16}\right),\qquad b_n=\frac{\pi}{\beta\sin\left(\frac{2\pi\varepsilon}{\beta}\right)}\sqrt{n\left(n+\frac{c}{8}-1\right)}\,.
\ee
These are examples of the SL(2,$\mathbb{R}$) analytical solutions discussed in Sec.\,\ref{sec:SymmKryl}. We can see that, similarly to the SYK example, the large--$c$ limit effectively constrains the range of $n$. Indeed, if we first extract the naive large--$c$ limit, Lanczos coefficients resemble the Heisenberg-Weyl example with $b_n\sim\sqrt{n}$ and quadratic growth of complexity. This is correct as long as $n\ll c/8$. If $n$ is comparable or larger than $c/8$, we have to take the full answer first.

This can be seen from the full answer for the spread complexity which is proportional to the central charge of the CFT as follows
\be
\mathcal{C}_{K}(t)=
    \frac{c\,\beta^2}{32\pi^2\varepsilon^2}\sinh^2\left(\frac{\pi t}{\beta}\right).\label{SCLC}
\ee
It grow quadratically only for early times, but at late times it grows exponentially.

This local quench setup is described holographically using the AdS/BCFT correspondence \cite{Takayanagi:2011zk}, where the end-of-the-world (EOW) brane is time-dependent. Interestingly, the tip of the EOW brane, which probes the deepest part of the AdS bulk, follows a geodesic of a massive particle (with heavy mass $m=c/16$). The radial momentum of this particle, computed in the proper distance coordinate, precisely satisfies
\be
\partial_t \mathcal{C}_{K}(t)=-\frac{1}{2\varepsilon}P_\rho(t).
\ee
These explicit checks suggest that, depending on the initial state, it may be possible to engineer precise holographic setups where the spread complexity rate of change can be directly matched with a quantity in gravity and then integrated. The overall dependence on the central charge in \eqref{SCLC} is particularly encouraging and enters similarly as in the single-interval entanglement entropy in 2D CFTs that can be reproduced from the Ryu-Takayanagi formula \cite{Ryu:2006bv}.

To finish, let us point out that an intriguing perspective on the Krylov basis was also presented in \cite{Basu:2024tgg}, where, using Wigner functions, the authors argued that the Krylov basis is ''the most classical'' basis in Quantum Field Theory, potentially making it a natural candidate for describing the evolution of semi-classical states in the bulk code subspace.  

\subsection{Nielsen's complexity and holography}\label{subsec:Nielsenholo}
\label{sec:NielsenAndGeometry}

There are also attempts to understand holographic complexity by exploiting equivalence between bulk and boundary symplectic forms \cite{Belin:2018fxe,Belin:2018bpg,Belin:2022xmt},
inspired by behavior under conformal transformations in Ba\~nados geometries \cite{Flory:2018akz,Flory:2019kah}.\footnote{Alternative attempts, where gravity arises as a consequence of spacetime optimizing the computational cost of its own dynamics, are discussed in \cite{Pedraza:2022dqi,Chandra:2021kdv,Carrasco:2023fcj}. }

The relation to symplectic forms also motivated the connection between quantum complexity in CFTs and gravity for the Fubini-Study cost function obtained in section \ref{sec.cft}. To understand this relation, we observe that one can map trajectories built from the unitary operators \eqref{eq:unitary_CFT} of a CFT$_d$ to the geodesics of a massive particle of mass $m$ in AdS$_{d+1}$ spacetime.
The Fubini-Study metric \eqref{eq:cost_FS} receives a natural geometric interpretation in terms of the minimal $\delta X_{\rm min}$ and maximal $\delta X_{\rm max}$ perpendicular distances between infinitesimally nearby geodesics (as illustrated in Fig.~1 of Ref.~\cite{Chagnet:2021uvi})
\begin{equation}
\label{eq:interpCFTFS}
d s^2_{\rm FS} = \dfrac{m^2}{2} \left(\delta X^2_{\rm min} + \delta X^2_{\rm max} \right) \, .
\end{equation}
It would be interesting to generalize the relation~\eqref{eq:interpCFTFS} to complexity associated with states that are related by a finite transformation. Furthermore, the connection to cost functions other than the Fubini-Study one is a subject of an ongoing investigation~\cite{CFTpen}.

The previous approach described the entire circuit (starting from a scalar primary state and employing CFT generators as gates) in terms of a massive particle trajectory in AdS. This approach does not immediately lend itself to an intuitive AdS/CFT interpretation in which quantum states are viewed as living on the boundary of AdS space. For this reason, it is desirable to seek a formulation of the state–complexity problem for general excited states in holography, where the states along the circuit can be more directly mapped to the boundary of AdS. Naively, the holographic complexity picture would require two distinct bulk geometries for the reference and target states (and, in fact, for any state along the circuit), since gravity would dictate the state’s time evolution in a way not necessarily aligned with its evolution along the circuit. However, it turns out that in certain cases it is possible to embed both the reference and target states in the same holographic geometry and connect them through unitary time evolution on the boundary, which can be viewed as a circuit in physical time. Although finding such a geometry is not straightforward in general, the problem simplifies when considering two states that are very close to each other. This is precisely the setting of~\cite{Erdmenger:2022lov}, which proposed the following idea.

For circuits in which the circuit parameter is the physical time in QFT, one layer of the circuit is simply given by the instantaneous QFT Hamiltonian. This Hamiltonian is defined in terms of the energy-momentum tensor components smeared along a time slice, e.g. $H = \int d^{d-1} x \, T_{tt}(\vec{x})$. The infinitesimal increment of the Fubini-Study distance~\eqref{eq:cost_FS} is proportional to the variance of the Hamiltonian in an instantaneous state. However, given that the Hamiltonian is a sum of local operators, its variance is given by a sum of two-point functions of local operators. In AdS/CFT, the holographic dictionary~\cite{Gubser:1998bc,Witten:1998qj,Skenderis:2008dh,Skenderis:2008dg} provides a systematic geometric way of computing correlation functions of the boundary QFTs. As a result, the Fubini-Study cost~\eqref{eq:cost_FS} associated with one layer of time evolution on the boundary is always geometric by the virtue that all its ingredients are geometric. This statement therefore expresses an exact mapping between two geometries: the auxiliary complexity geometry of the Fubini-Study cost on one hand, and the gravitational geometry of the bulk on the other. While the above statement is geometric, the  bulk manifestation of the Fubini-Study cost that leads to it is rather complicated and implicit. However, in the context of holographic two-dimensional CFTs and states represented holographically as solutions of the Einstein's equations with negative cosmological constant in three bulk dimensions, one can relate
 two-point functions of the energy-momentum tensor to geodesic lengths in the bulk~\cite{Erdmenger:2022lov}. This leads to a much more explicit expression for the Fubini-Study cost and a direct connection to the kinematic space program~\cite{Czech:2015qta,Czech:2016xec,deBoer:2016pqk}.


\subsection{Computational pseudorandomness and the holographic dictionary}
\label{ssec:pseudorandomness}

In all the above discussions, we mainly considered the volume of wormholes as a dual description of some notion of the complexity of time-evolving thermofield double state. It was observed by Susskind~\cite{Susskind:2014rva} that the steady growth of a wormhole's volume poses a challenge in the context of the AdS/CFT correspondence:
the dual quantity to the wormhole's volume cannot be a local observable as those are expected to equilibrate and hence saturate quickly in a quantum theory. 
Instead, \cite{Susskind:2014rva} proposed that the dual of the wormhole's volume is a global property of the dual state such as quantum complexity. 
However, from a complexity-theoretic point of view, there appears to be a mismatch between these two quantities. 
The volume of a wormhole is a simple quantity that can be easily calculated from the gravitational description. 
In comparison, quantum complexity is notoriously difficult to compute or even bound.
Making this mismatch concrete, Ref.~\cite{Bouland:2019pvu} argues that either quantum gravity cannot be simulated efficiently on a quantum computer, or some entries in the holographic dictionary require exponentially long computations.  The former would violate the extended Church-Turing thesis -- the widely held belief that all physical processes can be efficiently simulated on a quantum computer.
More precisely, Ref.~\cite{Bouland:2019pvu} models the dual CFT as a quantum system initialized in a thermofield-double state $|\mathrm{TFD}\rangle$  dual to a zero-volume wormhole. 
This state is undergoing time evolution under a fixed Hamiltonian $H_{\mathrm{CFT}}$ such that $|\mathrm{TFD}\rangle$ is a low-energy state for $H_{\mathrm{CFT}}$.
However, in this time evolution, the system undergoes a number of random `shocks' acting locally on single qubits:
\begin{equation}\label{eq:evolutionwithshocks}
    e^{-i t_1H_{\mathrm{CFT}}} S_1  e^{-i t_2 H_{\mathrm{CFT}}}S_2\cdots e^{-i t_{l} H_{\mathrm{CFT}}}S_le^{-i t_{l+1} H_{\mathrm{CFT}}} |\mathrm{TFD}\rangle.
\end{equation}

In Ref.~\cite{Bouland:2019pvu}, the authors conjecture that the time-evolution with shocks in Eq.~\eqref{eq:evolutionwithshocks}, while possible to efficiently simulate on a quantum computer, is computationally indistinguishable from a Haar random state.
Such an asymmetry between the easiness of preparing a state and the hardness of distinguishing it from a random state is called quantum pseudorandomness~\cite{ji2018pseudorandom,ma2024construct}.
More precisely, an ensemble of states, parameterized by a random key $k$ is called pseudorandom if 1) it can be efficiently prepared by polynomial-sized quantum circuits and 2) any quantum computer requires superpolynomial time to distinguish the ensemble from Haar random states.
This implies that even if the CFT evolution under shocks is computationally simple, its output states may appear indistinguishable from truly random states, making certain bulk properties computationally inaccessible from the CFT side. In particular, quantum circuit complexity cannot be computable using only polynomial-time quantum computations and measurements, as such computations would be able to rapidly distinguish pseudorandom and random states, which should be impossible. The existence of pseudorandom states was proven in Ref.~\cite{ji2018pseudorandom} under mild cryptographic assumptions.

We can now consider two time evolutions as in Eq.~\eqref{eq:evolutionwithshocks} for vastly different times, both undergoing shocks. Pseudorandomness implies that the two resulting states cannot be efficiently distinguished from Haar random states and, therefore, from each other.
However, the two states will correspond to wormholes of different volumes in any holographic dictionary, and the volume is efficiently computable from the AdS state using only coarse-grained properties of the metric.
Ref.~\cite{Bouland:2019pvu} argues that this tension can only be resolved if either 1) quantum gravity cannot be  efficiently simulated on a quantum computer or 2) the holographic dictionary cannot be computed in polynomial time. 
Notice that this argument actually does not require the Volume$=$Complexity conjecture.
In fact, the existence of pseudorandomness implies that any quantity that distinguishes between long and short time evolutions under shocks must be computationally hard to estimate. 

A central concept in the AdS/CFT correspondence is the relation between entanglement entropy of subregions in the CFT and the area of a corresponding extremal surface in AdS space, as specified by the Ryu-Takayanagi formula~\cite{Ryu:2006bv}.  The existence of this formula suggests that there may be a way of reconstructing AdS geometry from patterns of entanglement in the CFT -- a task that we may call {\it geometric reconstruction} (see, \eg \cite{Hamilton:2005ju,Hamilton:2006az,Hamilton:2006fh,Kabat:2011rz} for earlier approach to reconstructing local operators). With this in mind, suppose that we consider CFT states that satisfy the restrictive holographic entropy cone inequalities~\cite{bao2015holographic}.  These inequalities define a cone in the space spanned by the entanglement entropies of subregions and are a necessary requirement for geometrization of entanglement by the Ryu-Takayanagi formula. 
It was shown in Ref.~\cite{akers2024holographic} that ``geometric reconstruction'' is hard even for states whose entanglement entropy satisfies the holographic entropy cone.
Similar to the argument made in Ref.~\cite{Bouland:2019pvu} this hardness result uses pseudorandomness and, more precisely, the concept of pseudoentanglement.
Pseudoentanglement~\cite{aaronson2022quantum,gheorghiu2020estimating} refers to the phenomenon that even exponentially large gaps in entanglement can be computationally hard to detect~\cite{aaronson2022quantum}, which indicates that entanglement is an ``unfeelable'' quantity, similar to quantum circuit complexity.
In particular, Ref.~\cite{akers2024holographic} uses recently constructed low-entangling pseudorandom unitaries~\cite{schuster2024random} to show that two states with arbitrarily different bulk geometries can correspond to indistinguishable states in the CFT. 
As the construction in Ref.~\cite{akers2024holographic} does not require event horizons, the result suggests that reconstruction of the bulk geometry might even be hard outside the event horizon.
Previously, the presence of an event horizon was identified \cite{susskind2020horizons} as a way to reconcile the Church-Turing thesis with the duality of volume and complexity: an observer that needs to cross the event horizon to estimate the volume will then be unable to communicate the findings to an outside observer. This would be consistent with an exponentially hard-to-estimate dual quantity, such as complexity. Likewise, the authors of \cite{Engelhardt:2024hpe} argued that when the time evolution operator of a holographic CFT is approximately pseudorandom (or Haar random) on some code subspace, then there must be an event horizon in the corresponding bulk dual.  
The apparent tension with \cite{akers2024holographic} may be resolved by the fact that  the construction in the latter does not take the CFT Hamiltonian into account and produces high-energy states.

These results raise questions about the broader landscape of computational complexity in holography, particularly in relation to the Python’s Lunch conjecture~\cite{brown2020python}. 
This conjecture posits that if a geometry contains a local minimal surface that is not globally minimal, then finding the CFT operator corresponding to a local bulk operator inserted in the region surrounded by the local minimal surface is exponentially hard.
The conjecture suggests that if the bulk contains locally minimal but not globally minimal surfaces, then efficient application of operators in the CFT is exponentially complex.
The \textit{strong Python's lunch conjecture} then further posits that locally minimal surfaces are the only obstruction to efficient ``operator reconstruction''~\cite{akers2024holographic}.
A follow-up~\cite{engelhardt2022finding} shows that the strong Python's lunch conjecture is compatible with the findings of Ref.~\cite{Bouland:2019pvu} on the reconstruction of a wormhole's volume.   
Similarly, \cite{Gyongyosi:2022vaf,Balasubramanian:2022fiy} show that it is exponentially hard to decode the  state of a field in the interior of a black hole from the distant Hawking radiation entangled with it, as expected from the Hayden-Preskill scenario~\cite{Yoshida:2017aa}.
The results of Ref.~\cite{akers2024holographic} then suggest that the complexity of operator reconstruction in the CFT does not always align with the complexity of bulk geometry reconstruction.
These questions about the complexity of the holographic dictionary, inspired by the volume=complexity, require and deserve future investigations.

Finally, recent developments relate the growth of the volume of Einstein-Rosen bridges to the randomness of the underlying microscopic quantum gravity state \cite{Magan:2024aet,Magan:2025hce}. Concretely, the authors of \cite{Magan:2024aet,Magan:2025hce} found gravitational duals to continuous random circuits (see Sec.~\ref{subsec_grwothofapproximate} for a survey of random circuits in the context of circuit complexity). The resulting geometries are long wormholes with matter inhomogeneities, dubbed ER-caterpillars. The construction allowed the authors to demonstrate a relation between the length of the ER bridge and the degree of microscopic randomness of the state as a function of time.  More precisely, \cite{Magan:2024aet,Magan:2025hce} showed that an ER-caterpillar of length $l$ has an underlying quantum state well-approximated by a quantum state k-design with $k \propto l$ when $l$ is large.  It would be interesting to understand the implications of this construction in the context of the pseudorandomness mentioned above.

\clearpage
\section{Paradigms for complexity III: tensor-network-inspired definitions of complexity}
\label{sec:complexity_QFT}

Path-integral complexity \cite{Caputa:2017urj,Caputa:2017yrh}  was  the first approach to measuring complexity in quantum field theories that was motivated by holography.  This approach arose 
in the context of tensor networks, which provide variational ansatze for many-body quantum wavefunctions. In particular, optimized tensor networks such as, e.g., MERA \cite{Vidal:2007hda,Vidal:2008zz} or its continuous counterpart cMERA \cite{Haegeman:2011uy,Nozaki:2012zj}, give rise to a geometric structure that reflects the entanglement patterns of a quantum state.  An analogy between the tensor network approach to constructing quantum states and the structure of the holographic duality had already been noted by Swingle \cite{Swingle:2009bg}.  This proposal seeded a new line of research 
that led  Susskind  \cite{Susskind:2014moa} to argue that the growth of the volume of Einstein-Rosen bridges should be related to complexity, estimated by the number of tensors (the effective volume of the network) in the tensor-network representation of the time-evolved thermofield-double state \cite{Hartman:2013aa}. 
In Sec.~\ref{sec.pathintegralopt}, we review  a notion of complexity in CFT called \textit{path-integral complexity} inspired by the parallelism with tensor networks.  In Sec.~\ref{ssec:holo_path_integral} we discuss the holographic interpretation of this quantity in AdS spacetime.

\subsection{Complexity from path-integral optimization}
\label{sec.pathintegralopt}

Motivated by these developments, Ref.~\cite{Caputa:2017urj} aimed at extending intuitions from the study of tensor networks to general, strongly-interacting QFTs and to use these ideas to define a natural measure of complexity using path integrals. A brief review of their construction follows.

In QFTs, wavefunctions can be prepared by performing path integrals over physical space and Euclidean time. They are computed by integrating over all field configurations, subject to specified boundary conditions. As an example, consider two-dimensional CFTs on a flat, Euclidean plane $\mathbb{R}^2$ with coordinates $(\tau,x)$, and collectively denote all fields in the CFT by $\Phi(\tau,x)$. The ground state wave function $\Psi_{\rm CFT}[\Phi(x)]$, which in QFTs is a functional of the configuration at the time slice $\tau=0$, $\Phi(x)\coloneqq \Phi(x,\tau=0)$,  is defined as 
\begin{equation}
\Psi_{\rm{CFT}}[\Phi(x)] =\int \prod_{-\infty<\tau\leq 0,x}[D\tilde{\Phi}(\tau,x)] e^{-S_{\text{CFT}}[\tilde{\Phi}]}\delta(\tilde{\Phi}(0,x)-\Phi(x)),\label{PI2DCFT}
\end{equation}
where $S_{\rm CFT}$ is the Euclidean action of the 2D CFT and $D\tilde{\Phi}$ is the path integral's measure.\footnote{Formally one may need to deform the theory by putting it in a finite volume or introducing a small mass so that the spectrum is gapped and the path integral~\eqref{PI2DCFT} acts as a proper vacuum projector.}

The main new idea in \cite{Caputa:2017urj} was to introduce a non-trivial background metric on the space where the path integral \eqref{PI2DCFT} is performed (keeping the boundary conditions for the fields fixed), and to interpret this metric as an (unoptimized) density of gates of a \textit{continuous tensor network}. For our 2D CFT example, all the metrics can be put 
into the Weyl-flat form parametrized by a scalar function $\phi(\tau,x)$ as follows:
\be
ds^2=e^{2\phi(\tau,x)}(d\tau^2+dx^2),\qquad e^{2\phi(0,x)}=\frac{1}{\epsilon^2}\coloneqq e^{2\phi_0} \, .
\label{Curm}
\ee
The second equation is the boundary condition for the Weyl-factor to reduce to the original flat space metric (with lattice spacing $\epsilon$) as $\tau\to0$. 

\begin{figure}[ht]
\begin{center}
    \includegraphics[scale=0.3]{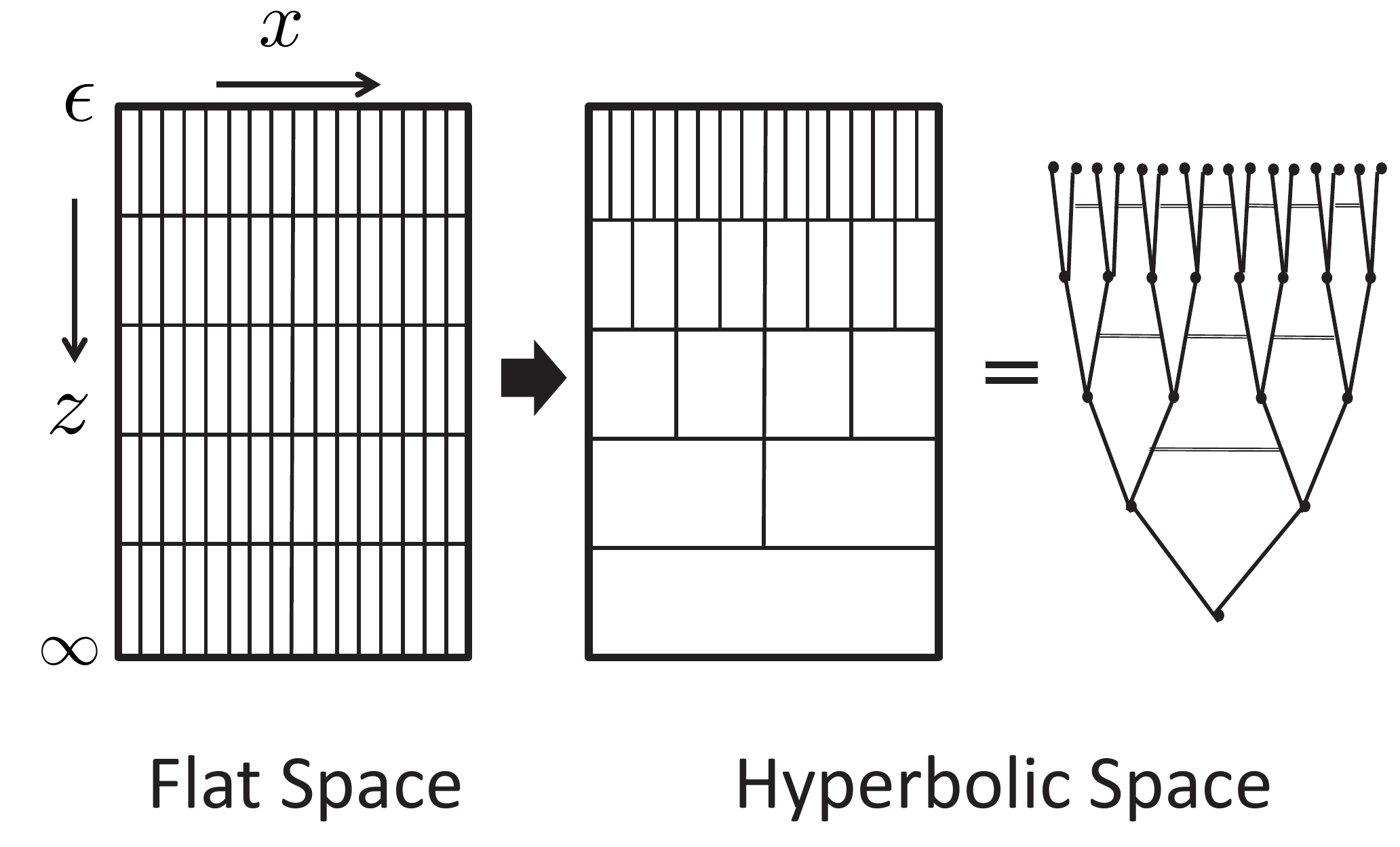}
    \end{center}
    \caption{A cartoon explaining path-integral optimization in CFTs as an analog of tensor-network optimization. Figure from \cite{Caputa:2017urj} with Euclidean time denoted as $z=-\tau$. }
    \label{figure:PIO}
\end{figure}

Next, we introduce the actual optimization procedure. Suppose we compare the wave functions computed with a background flat metric (denoted with $ \Psi_{\text{CFT}}^{\phi_0}[\Phi(x)]$) and with a curved metric (denoted with $ \Psi_{\text{CFT}}^{\phi}[\Phi(x)]$), they are proportional to each other 
\be
\Psi_{\text{CFT}}^{\phi}[\Phi(x)]=e^{S_{\mathcal{L}}[\phi]-S_{\mathcal{L}}[\phi_0]} \,  \Psi_{\text{CFT}}^{\phi_0}[\Phi(x)] \, . \label{PathINT}
\ee
The proportionality factor between the two wave functions is given by an exponent of the Liouville action 
\bea
S_{\mathcal{L}}[\phi]=\frac{c}{24\pi}\int^\infty_{-\infty} \!dx\! \int^{0}_{-\infty}\! d\tau\! \left[(\partial_x\phi)^2+(\partial_\tau\phi)^2
+ \mu e^{2\phi}\right], ~~ \label{LVac}
\eea
where $c$ is the central charge of the 2D CFT.\footnote{The coefficient $\mu$ of the potential term can be set to $1$ by shifting $\phi$.}
Intuitively, the proportionality factor in \eqref{PathINT} can be interpreted as a measure of how many repetitions of path-integration per site we perform to construct the state. Hence, the \textit{optimal path integral} is selected by minimizing the Liouville action with respect to $\phi$, i.e., picking the metric that solves the Liouville equation
\be
\label{eq.LiouvilleEOMS}
4\partial\bar{\partial}\phi(w,\bar{w})=\mu e^{2\phi(w,\bar{w})} \, ,\qquad w \coloneqq \tau+ix \, , 
\ee
with the boundary condition \eqref{Curm}. This procedure is called the \textit{path-integral optimization}. Recall that we are interpreting the metric as encoding the density of gates in the tensor network. One attempt to make this precise showed a correspondence between unitary and isometric tensors of the (c)MERA network and the kinetic and potential terms in the Liouville action \cite{Caputa:2017yrh,Czech:2017ryf}. Likewise, \cite{Camargo:2019isp,Caputa:2018kdj} argued that the Liouville action can be derived by counting the stress tensor gates with appropriate cost functions within Nielsen's geometric framework.

For example, for the CFT vacuum that is obtained by performing a Euclidean path integral on the half-plane, the optimal geometry is given by the hyperbolic plane.
Similar results can be obtained for path integrals on the circle, that prepare the vacuum state of a CFT on the circle; as well as path integrals on a strip of size $\beta$ used to prepare the thermofield-double state with inverse temperature $\beta$. 
In the former case, the background obtained via the path-integral optimization procedure is the hyperbolic geometry of the Poincar\'{e} disc. In the latter case, the background geometry is the hyperbolic strip, also called the Euclidean trumpet geometry. 

Similar optimization was also defined for the vacuum and thermal states of higher-dimensional CFTs \cite{Caputa:2017yrh}, but with a prescription to optimize only over the Weyl factor of the background metric on which the CFT path integral is evaluated (unlike in 2D, this is only a subset of possible geometries in higher-dimensions). Again, it was found that optimal metrics have a constant negative Ricci scalar curvature i.e., are hyperbolic.

It is natural to think about these metrics as  continuous counterparts of effective tensor-networks geometry that emerges after the optimization procedure in many-body systems. In fact, in the spirit of tensor network discussions in holographic CFTs, we can interpret the metrics from the optimization of path-integrals in $CFT_d$ as particular slices of their higher-dimensional $AdS_{d+1}$ dual. We will return to this point in Sec.~\ref{ssec:holo_path_integral}.

Another idea from~\cite{Caputa:2017urj,Caputa:2017yrh},  dubbed \textit{path-integral complexity}, was to interpret the Liouville action as a novel measure of complexity for wave functions prepared using Euclidean path integrals. The main reason to advocate this interpretation was that the on-shell Liouville action is proportional to the volume of the space on which the path integral is computed. Indeed, after fixing the normalization such that the integral of the Weyl factor $e^{2\phi}$ over a unit area (a square region of size $\epsilon$ in the original metric) 
has a value 1 (see detailed discussion in \cite{Bhattacharyya:2018wym}), we can interpret the volume of the Euclidean space as the number of tensors in the optimal network. The on-shell action associated with the CFT vacuum state reads
\be
\mathcal{C}_{\Psi_0}=\text{Min}[S_{\mathcal{L}}[\phi]]=\frac{cL}{12\pi\epsilon},\qquad L=\int dx \, .
\label{eq:action2d_path}
\ee
This result is also consistent with qualitative expectations and holographic conjecture \cite{Stanford:2014jda}, which estimates the leading divergence in complexity of the CFT state by the spatial volume. 

Later on, researchers extended the path-integral optimization method to other settings.
As we already mentioned, \cite{Caputa:2017urj,Caputa:2017yrh} proposed a higher-dimensional generalization of the action \eqref{eq:action2d_path}. The
the optimization of 2D  CFTs deformed by relevant operators was studied in \cite{Bhattacharyya:2018wym}, where a modified Liouville action was minimized over both metrics and local deformation couplings. 
The authors of \cite{Caputa:2020mgb} investigated path integrals for inhomogeneous CFTs, that are usually interpreted as 2D CFTs on non-trivial background metrics. From the perspective of the optimization, the starting point is the CFT on a non-trivial background metric ($\hat{g}_{ab}$ instead of $\delta_{ab}$) and then consider $g_{ab}=e^{2\phi}\hat{g}_{ab}$ as the metric for optimization. This leads to additional background Ricci curvature term $\hat{R}$ in the Liouville equation which can nevertheless be solved using standard uniformization techniques.\\
Soon after \cite{Caputa:2017urj}, reference~\cite{Czech:2017ryf} argued that a variation of path integral complexity in holographic 2D CFTs leads to Einstein's equations in the bulk of AdS$_3$. Moreover, the Liouville action (or its covariant version, called the Polyakov action) motivated
the treatment of Nielsen's complexity by introducing trajectories on the unitary manifold generated by sequences of 2D conformal transformations \cite{Caputa:2018kdj} (see Sec.~\ref{sec.2dcft}). 

The formulation in terms of a Liouville action also inspired a series of developments in the tensor-network community. 
They used the path integral approach to cMERA to define tensor networks in terms of the background geometry on which the state is prepared \cite{Milsted:2018yur,Milsted:2018san}. In particular, they showed how the background metric parameters translate into details of the tensor network. This idea was later used in~\cite{Camargo:2019isp} to argue for a Nielsen-type interpretation of the Liouville action in terms of an appropriate choice of the cost function on the complexity geometry. As opposed to the developments on Nielsen's complexity discussed in Sec.~\ref{ssec:Nielsen_complexity}, the work~\cite{Camargo:2019isp} necessarily utilized  circuits that contained both unitary and Hermitian gates associated with components of the energy-momentum tensor operator. This is similar in spirit to the results discussed in Sec.~\ref{ssec:holo_matching_2d} of \cite{Heller:2024ldz} and has the same origin: Euclidean time evolution is an efficient and natural method for preparing useful states. This points towards a mathematical generalization of the Nielsen's complexity that allows for exponentiation of anti-Hermitian (giving rise to unitary transformations), as well as Hermitian matrices (giving rise in particular to Euclidean time evolution).

Finally, the results in Ref.~\cite{Camargo:2019isp} pointed out that a general cost function for path integral optimization will also contain positive powers of the cut-off~$\epsilon$ associated with higher derivatives of $\phi$ in the spirit of an effective field theory. However, the solutions of Eq.~\eqref{eq.LiouvilleEOMS} acquire derivatives that are themselves of the order of $1/\epsilon$, breaking the hierarchy within these more general cost functions that left  the contribution~\eqref{LVac} as the dominant term.  Thus we need  a more systematic inclusion of the UV cut-off~$\epsilon$, bringing us to the subject of the next section.

\subsection{Holographic path-integral optimization }
\label{ssec:holo_path_integral}

\begin{figure}[ht]
\begin{center}
    \includegraphics[scale=0.28]{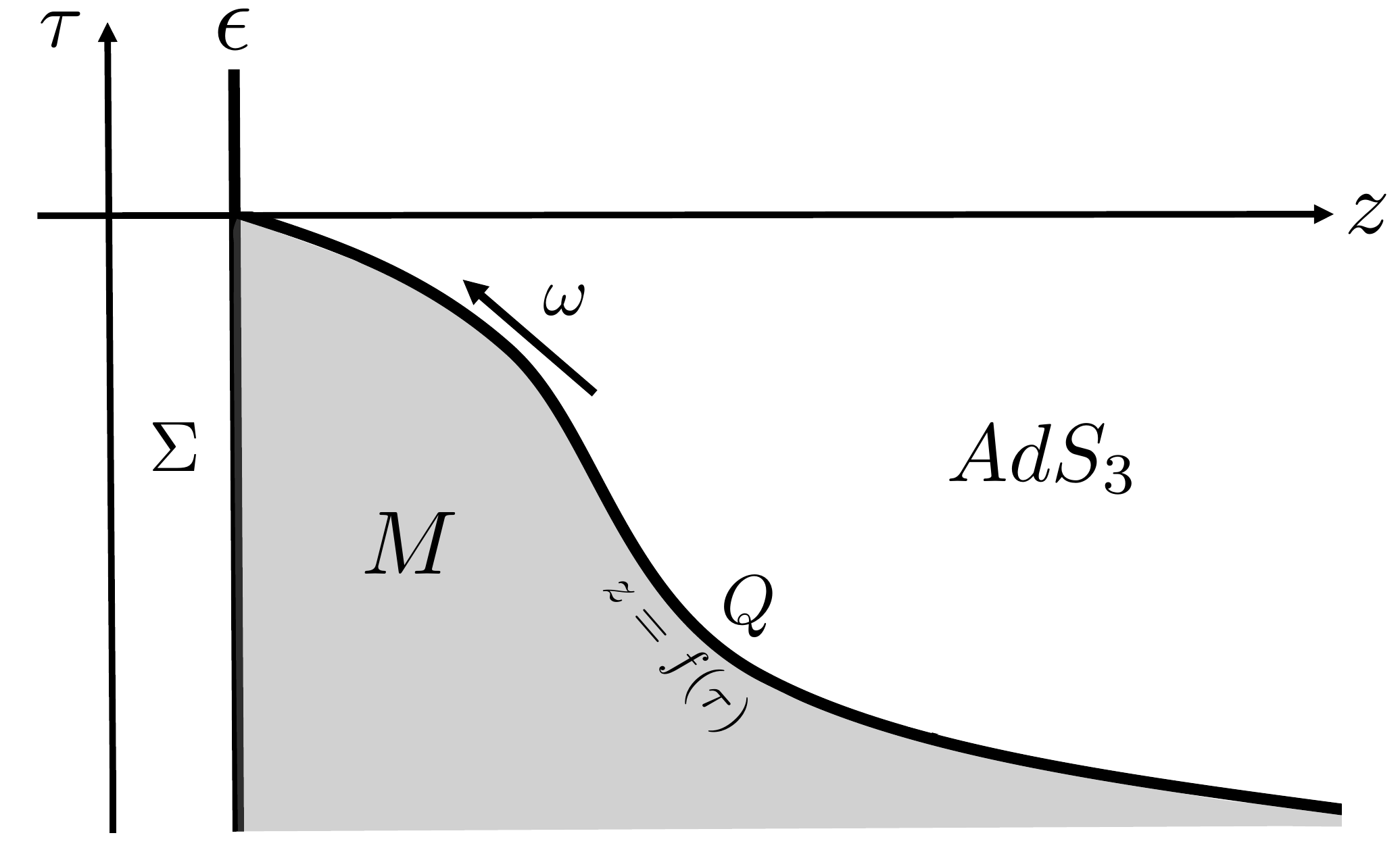}
    \end{center}
    \caption{Holographic proposal for the gravity dual of the path-integral optimization, adapted from \cite{Boruch:2020wax}.}
    \label{figure:HPIO}
\end{figure}

The optimization process described above starts with a CFT path integral in a flat geometry, and effectively places it on a particular hyperbolic metric. 
After an appropriate identification of the Euclidean time with a radial coordinate, this metric can be interpreted as a slice of AdS spacetime. It is then natural to ask whether this procedure, when applied to holographic CFTs, could have a dual geometric interpretation within the AdS/CFT correspondence. If successful, the embedding in holography could shed new light on the interpretation of the gravity dual of the Liouville action in terms of path-integral complexity of a CFT state.

With this motivation, references~\cite{Boruch:2020wax,Boruch:2021hqs}  proposed a holographic dual of the path-integral optimization procedure. 
Their construction was closely related to the framework of Anti-de Sitter/Boundary Conformal Field Theory (AdS/BCFT) \cite{Takayanagi:2011zk} in which one considers only a portion of the holographic geometry bounded by an end-of-the-world surface. 

More precisely, consider a computation of the Euclidean Hartle-Hawking wavefunction in a portion $M$ of AdS spacetime on Fig.~\ref{figure:HPIO}. The bulk region is bounded by the boundary cut-off surface $\Sigma$ located at $z=\epsilon$ (where $z$ is the radial coordinate of the AdS), and a generic spacelike surface $Q$ parametrized by $z=f(\tau)$ (homogeneous in the $x$ direction). The induced metric on $Q$ can be put into a Weyl-flat form by introducing an appropriate coordinate $\omega$ (i.e. a choice of gauge)
\be
ds^2=e^{2\phi(\omega)}(d\omega^2+dx^2)\,.\label{Met2DHH}
\ee
The HH wave function is then a functional of the induced metric defined by path integral over all bulk metrics with this boundary condition on $Q$ (as well as the implicit boundary condition on $\Sigma$)
\be
\Psi_{HH}[\phi]=\int [D g_{\mu\nu}]e^{-I_G[g]}\delta(g_{ab}|_Q-e^{2\phi}\delta_{ab})\,,
\ee
where $I_G[g]$ stands for the Einstein-Hilbert action with the boundary Gibbons-Hawking term. Computing this path integral is in general a formidable task but, to the leading order in $1/G_N$, the semiclassical HH wave function becomes the exponent of (minus) the on-shell gravitational action itself.
Mimicking the AdS/BCFT construction, we can also supplement the Gibbons-Hawking term by a constant tension parameter $T$ on $Q$ modeling certain types of bulk matter (and boundary entropy in AdS/BCFT). 

Now, the main observation of~\cite{Boruch:2020wax,Boruch:2021hqs} was that a maximization of the semiclassical HH wave function with respect to the choice of $\phi(\omega)$ is equivalent to imposing the Neumann boundary conditions (in analogy with AdS/BCFT) on $Q$ 
\be
K_{ab}-Kh_{ab}=-Th_{ab}\,,
\ee
where $K$ is the trace of extrinsic curvature $K_{ab}$ on $Q$. Moreover, performing these steps for HH wave functions in Poincare, global and (static) black hole spacetimes, yields ``maximizing" metrics \eqref{Met2DHH} that match those in the CFT optimization \eqref{Curm}, after the following identifications:
\be
\tau=\omega,\qquad \frac{c}{3}=\frac{2l}{G_N}\qquad \mu=1-\frac{T^2}{(d-1)^2}\,.
\ee
This agreement was conjectured (interpreted) as the correspondence between the maximization of HH wave functions in AdS and the path integral optimization in holographic CFTs. It is also interesting to comment that parameter $\mu$ in the Liouville action effectively quantifies the amount of optimization, ranging from the unoptimized original metric for $\mu=0$ and the optimal for $\mu=1$. This way, tension $T$ can be seen as its holographic counterpart as it parametrizes a family of surfaces $Q$ starting from the boundary surface $\Sigma$ for $T=-(d-1)$ up to surface $\tau=0$ for $T=0$ (see \cite{Boruch:2020wax} for detailed discussion).

The second key insight from this construction was a UV completion of the Liouville action, which could be interpreted as specifiying the leading $1/\epsilon$ contribution to path-integral complexity (see the discussion in \cite{Camargo:2019isp}). Specifically, \cite{Boruch:2020wax,Boruch:2021hqs} argued that the gravity action in region $M$ (see Fig.~\ref{figure:HPIO}), supplemented with appropriate boundary and Hayward corner terms, provides a finite-cutoff completion of the Liouville action. As a consistency check, taking the "boundary" limit of the gravity action reproduced both the Liouville action \eqref{LVac}, and its higher-dimensional generalization \cite{Caputa:2017yrh}.  

A closely related setup was later explored in \cite{Caputa:2020fbc} as an approach to  tensor networks in holographic CFTs. The authors demonstrated that the CFT on the slices $Q$ above could be explicitly constructed using $T\bar{T}$ deformation with a coupling that depends on Euclidean time,\footnote{This deformation later became the starting point of the ``Cauchy-slice holography" \cite{Araujo-Regado:2022gvw}.} and could be described within finite cut-off holography \cite{McGough:2016lol}. Parallel works~\cite{Chandra:2021kdv,Chandra:2022pgl} also argued for $T\bar{T}$-like interpretation of path integral complexity in holographic CFTs.  The key distinction between \cite{Boruch:2020wax,Boruch:2021hqs} and  \cite{Caputa:2020fbc} was the way in which the partition functions of holographic CFTs deformed by $T\bar{T}$ were computed.  Namely, in the $T\bar{T}$ context, the on-shell action was evaluated in the bulk region from the interior up to $Q$; whereas in the holographic path-integral optimization context, the action was evaluated in the region $M$ attached to the boundary. Nevertheless, as functionals of the induced metric on $Q$ (i.e. the Weyl factor $\phi(\omega)$), gravitational actions in both computations are the same; therefore, their maximization yields the same metrics $\phi(\omega)$. In other words, their maximization selects a particular slice $z=f(\tau)$ of asymptotically, locally $\mathrm{AdS}$ spacetimes that has constant extrinsic curvature $K$, the so-called constant-mean-curvature (CMC) slices.

Interestingly, slices of AdS spacetime with constant extrinsic curvature also appear naturally in the context of bulk and boundary symplectic forms~\cite{Belin:2018fxe}. Reference~\cite{Belin:2018bpg} proposed the same kind of slices to match the volume of the Einstein-Rosen Bridge in three-dimensional black hole geometries with a boundary notion of Nielsen complexity.

In summary, path integral optimization defines complexity via a partition function, a formulation which lends itself naturally to a description via the  holographic map between CFT and gravitational observables. It would be interesting to make this map more precise.

\clearpage
\section{Quantum complexity in quantum information theory and many-body physics}
\label{sec:qi}

Quantum information (QI) theory and complexity inform each other in many ways. We focus on three. First, QI theory shows that complexity governs a resource consumed in operational tasks (Sec.~\ref{sec_qi_operational}).
Second, complexity distinguishes topological phases (Sec.~\ref{sec_qi_topo}).
Third, a QI-theoretic quantity---entanglement entropy---bounds complexity under certain conditions (Sec.~\ref{sec_qi_entnglmt}).\footnote{We already discussed another connection between complexity and entanglement in the context of Nielsen's complexity in section \ref{sec:NielsenBinding}. There, we saw that very special definitions of complexity are completely fixed by the entanglement and R\'enyi entropies.}
Finally, the probability distribution over bitstrings $|x\rangle, x\in\{0,1\}^n$ induced by measuring a state also serves as a tool for bounding complexity (Sec.~\ref{ssec:concentration}).
Other QI-complexity intersections that we do not discuss extensively below include the bounding of complexity in terms of the quantum Wasserstein distance~\cite{Li_22_Wasserstein}, the relationship between complexity and the expressivity of quantum circuits~\cite{Haug_21_Capacity}, and applications of complexity to quantum machine learning~\cite{Haug_24_Generalization}.

\subsection{Operationalism and resource theory}
\label{sec_qi_operational}

The authors of~\cite{Brown:2017jil} suggested that complexity controls a resource useful in operational tasks. Let $\rho$ denote an arbitrary quantum state of a system $S$ and let $\mathcal{C}_0(\rho)$  denote the exact circuit complexity of $\rho$: the least number of basic operations required to prepare $\rho$.  Let $\mathcal{C}_{\rm max}$ denote the greatest possible complexity of any state on $S$. For an $N$-qubit system, $\mathcal{C}_{\rm max}\sim2^N \, $. Define the \emph{uncomplexity} of $\rho$ as the gap $\mathcal{C}_{\rm max} - \mathcal{C}_0(\rho)$. Uncomplexity serves as a resource in quantum computation. For example, consider running a quantum computation on an $n$-qubit register. The natural starting point is the zero-complexity state $| 0\rangle^{\otimes n}$. It serves analogously to clean scrap paper on which we write scratchwork when performing classical computations. 

Brown and Susskind conjectured that one can define a resource theory for quantum uncomplexity~\cite{Brown:2017jil}. A \emph{resource theory} models scenarios 
in which agents are restricted to perform only certain operations. For example, in the resource theory of entanglement, agents can perform only local operations and classical communications (LOCC). The allowed operations are called \emph{free}, and are modeled as costing nothing. Anything not free is a \emph{resource}; it may be consumed to enable an operational task. In the entanglement theory, for instance, entangled states are resources. Agents may consume a maximally entangled pair of qubits to teleport quantum information, using LOCC.

A resource theory of quantum uncomplexity was defined in~\cite{YungerHalpern:22Resource}. For simplicity, suppose that an agent in the resource theory has $n$ qubits (the setting can be generalized). 
The agent can attempt to perform any two-qubit gate $U$ (the setting can be generalized to $k$-qubit gates, for any $k \leq n$, and to gates in a particular set). However, noise corrupts every gate implementation. Hence the gate $\tilde{U}$ that is realized is chosen randomly from an $\epsilon$-ball (a small open set) about $U$. $\tilde{U}$ is called a \emph{fuzzy gate}. The compositions of all fuzzy gates are the fuzzy operations. Every agent will have a  tolerance for how fuzzy, or unknown, the state can become. This tolerance limits how many fuzzy gates the agent is willing to perform---and so curtails the complexities of the operations that will be performed on the system.

Resource theories can be used to formulate operational tasks and to quantify their optimal efficiencies. An example is \emph{uncomplexity extraction}~\cite{YungerHalpern:22Resource}: let $\delta \geq 0$ denote an error tolerance; and $\rho$, an arbitrary $n$-qubit state. One extracts uncomplexity from $\rho$ by distilling a state $\delta$-close to $| 0 \rangle^{\otimes k}$ in trace distance. How large a $k$ is achievable? The answer depends on the \emph{complexity entropy}. 

The complexity entropy quantifies how random $\rho$ looks to an agent who can perform only computationally restricted measurements~\cite{YungerHalpern:22Resource,Munson:2024Complexity}. In QI theory, a measurement is modeled with a \emph{positive-operator-valued measure} $\{ Q_\alpha \}$~\cite{nielsen2010quantum}. The \emph{measurement operators} $Q_\alpha$ are positive-semidefinite, $Q_\alpha \geq 0$, and normalized as $\sum_\alpha Q_\alpha = \mathbbm{1}$. Define as a \emph{zero-complexity measurement operator} any $Q_\alpha$ that acts on each qubit either (i) as the identity or (ii) by projecting the qubit onto $\ketbra{0}{0}$: 
$\bigotimes_{j = 1}^n  ( \ketbra{0}{0} )^{\alpha_j}$,
if $\alpha_j \in \{0, 1 \}$, and
$( \ketbra{0}{0} )^0 \coloneqq \mathbbm{1}$.
Let $r \in \mathbbm{Z}_{\geq 0}$ denote a fixed integer.
Consider acting with $\leq r$ gates and then a zero-complexity measurement operator. This process effects a \emph{complexity-$r$ measurement operator}, the set of which we denote by $M_r$.
We apply these definitions as follows. Consider performing a binary hypothesis test between $\rho$ and $\mathbbm{1} / 2^n$ (the maximally mixed state, and hence the maximally random, state). We receive one of the two states and must guess which we received. We perform a 
measurement $\{Q, \mathbbm{1} - Q\}$. (In the notation above, $Q = Q_1\in M_r$, and $\mathbbm{1} - Q = Q_2$.) If the $Q$ outcome obtains, we guess that we received $\rho$; and, if the $\mathbbm{1} - Q$ outcome, then $\mathbbm{1} / 2^n$. A type I error occurs if, upon receiving $\rho$, we wrongfully guess $\mathbbm{1} / 2^n$. Let $1 - \eta$ upper-bound the maximum allowable type I--error probability: ${\rm Tr}(Q \rho) \geq \eta$. A type II error occurs if, upon receiving $\mathbbm{1} / 2^n$, we wrongfully guess $\rho$. The minimal type II--error probability ${\rm Tr}(Q \mathbbm{1}) / 2^n$, subject to the type I constraint, motivates the \emph{complexity entropy}:
\begin{align}
   \label{eq_Comp_ent}
   H_{\rm c}^{r,\eta} (\rho)
   \coloneqq  \min_{\substack{Q \in M_r, \\
                   {\rm Tr}(Q \rho) \geq \eta} }
   \left\{ \log_2 \LParen  {\rm Tr}(Q) \RParen \right\} .
\end{align}

The complexity entropy controls the number of uncomplex qubits extractable from $\rho$ via fuzzy operations~\cite{YungerHalpern:22Resource}. Let $\epsilon$ quantify the gates' fuzziness, and let the tolerance $\delta \geq r \epsilon$. For every parameter value 
$\eta \in [1 - (\delta - r \epsilon)^2, \, 1]$, 
one can extract 
$k = n - H_{\rm c}^{r, \eta}(\rho)$ qubits $\delta$-close to $|0\rangle^{\otimes n}$, using fuzzy operations.
Conversely, one can extract at most $n - H_{\rm c}^{r, 1-\delta}$ such qubits. These results follow a pattern common in QI theory: entropies quantify the optimal efficiencies with which operational tasks can be performed.

Entropies are the workhorses of QI theory. That an entropy can quantify complexity, however, was far from obvious, before~\cite{YungerHalpern:22Resource}. For example, consider an $n$-qubit system prepared in a pure tensor-product state $\rho(0) = \ketbra{ \psi(0)}{\psi(0) }$. Under a nonintegrable Hamiltonian, the state evolves to $\rho(t) = \ketbra{ \psi(t) }{ \psi(t) }$ in a time $t$. The state's complexity grows, yet the von Neumann entropy $S_{\rm vN} \LParen \rho(t) \RParen 
= - {\rm Tr} \LParen \rho(t) \log_2 \rho(t) \RParen$
remains zero. Every subsystem's reduced state has an $S_{\rm vN}$ that likely grows, but such entropies saturate long before $\rho(t)$'s complexity does. Hence commonly used entropies do not track complexity. Avoiding this pitfall, the complexity entropy~\eqref{eq_Comp_ent} quantifies how random $\rho(t)$ \emph{appears} to  a computationally restricted observer. The observer-centric approach was pioneered in~\cite{Brandao:2019sgy}, through a metric called the strong complexity. Other precursors to the complexity entropy include~\cite{Hrastad:1999Pseudorandom,Chen:2017Computational}, motivated by pseudorandomness and cryptography. Reference~\cite{Munson:2024Complexity} introduces variations on $H_{\rm c}^{r,\eta} (\rho)$. These variations quantify the optimal efficiencies of more information-processing and thermodynamic tasks, including erasure and decoupling. 

These results offer hope that complexity entropies might quantify the operational efficiencies of holographic tasks. For example, consider an agent Alice falling into a black hole. Debate surrounds the possibility that she will encounter a firewall. According to one proposal, a firewall will consume her if the black hole state is dual to a CFT state whose complexity is not increasing~\cite{Brown:2017jil}.\footnote{A subtlety is that, when a black hole's complexity stops increasing, classical general relativity is expected to stop describing the black hole accurately.}
Tossing a qubit into the black hole doubles the CFT state's uncomplexity, enabling the complexity to return to increasing---and saving Alice.  This story might be generalized beyond qubits with help from $H_{\rm c}^{r, \eta}$.

\subsection{Circuit complexity and topological order}
\label{sec_qi_topo}

Phases of matter characterize how large numbers of particles can organize.
This includes the solid, liquid and gas phases.
In this section, we will briefly explain how topological order and quantum circuit complexity are related.

Until the early 80s it was thought that all phases of matter fit into Landau's framework of symmetry breaking~\cite{landau1937phys,landau1950symmetry}.
Roughly, Landau's theory asserts that a phase transition  occurs when a symmetry is broken, like the spontaneous change from a continuous symmetry in a chaotic/liquid phase to a discrete symmetry in a solid phase. 
Here, every phase is described by a pair of symmetry groups $G_{H}$ and $G_{\psi}$, where $G_{H}$ are all symmetry operations of the Hamiltonian and $G_{\psi}\subseteq G_{H}$ is the symmetry group of the ground state $\psi$. 
A phase transition  occurs if the symmetry group $G_{\psi}$ changes under an order parameter (such as the strength of a magnetic field). 

Starting with a plethora of inequivalent chiral spin states all satisfying the same symmetries,
it was discovered that quantum Hamiltonians exhibit an even richer theory of phase transitions. 
For an extensive review about topological phases, see Wen~\cite{wen2013topological}.
Since then an entire ``zoo of topological phases'' has been  discovered~\cite{wen2017colloquium}.
Topological phases refer to gapped ground states that show vast differences in their entanglement structure but are otherwise seemingly featureless (and therefore indistinguishable by symmetry groups).

In the following we will show how quantum circuit complexity is related to the concept of topological order. 
States exhibiting topological order are gapped ground states of local Hamiltonians that cannot be reduced to a product state continuously without closing the gap.
In particular, they are in a topological phase distinct from all product states, also called the trivial phase.
More precisely, consider a gapped local Hamiltonian $H_0$ with a unique ground state $|\psi\rangle$.
$|\psi\rangle$ can always be transformed into a product state via a continuous path consisting of local Hamiltonians $H(s)$ for $s\in [0,1]$ with $H(0)=H_0$. For $H_{0}$ acting on $n$ qubits, we simply define $H(s)= (1-s)H_0+ s \sum_{i=1}^n Z_i$, where $Z_i$ denotes the Pauli $Z$ matrix acting on the $i$th qubit/spin.
Clearly, the unique ground state of $H(1)$ is $|0^n\rangle$ and $H(1)$ is gapped.
However, it is very likely that at least one of the $H(s)$ has a degenerate ground state, and thus closes the gap.
A state is in the trivial phase if such a continuous path exists such that the gap above the unique ground state $|\psi(s)\rangle$ is uniformly lower bounded by a positive constant along the path.
A state is topologically ordered if no such path exists.
A continuous path of Hamiltonians with a uniformly lower bounded spectral gap is called \textit{adiabatic}.

How does this relate to quantum circuit complexity? It was observed in Ref.~\cite{chen2010local} that the above definition of topological order of a ground state $|\psi\rangle$ in terms of the absence of adiabatic paths is equivalent to a lower bound that grows in $n$ on the circuit depth for all circuits preparing $|\psi\rangle$ from a product state.

It turns out that that the existence of an adiabatic path is equivalent to the existence of a \textit{local unitary transformation}~\cite{chen2010local}:
\begin{equation}
    |\psi(s)\rangle = \mathcal{T}\left(e^{-i\int_0^s \tilde{H}(r)\mathrm{d}r}\right)|\psi(0)\rangle,
    \label{eq:timeord_state}
\end{equation}
where $\tilde{H}(r)$ is a local Hamiltonian and $\mathcal{T}$ denotes the path-ordering operator.
On the other hand, the existence of such a local unitary transformation implies the existence of an adiabatic path via\footnote{The notation in Eq.~\eqref{eq:timord_unitary} is analogous to Eq.~\eqref{eq:generic_path_unitaries}, which defined a trajectory implementing a unitary operator in the complexity geometry.}
\begin{equation}
    H(s) \coloneqq U(s) H U^{\dagger}(s), \qquad U(s)=\mathcal{T}\left(e^{-i\int_0^s \tilde{H}(r)\mathrm{d}r}\right).
    \label{eq:timord_unitary}
\end{equation}
It is easy to see that the Hamiltonians $H(s)$ are all gapped and the unitaries $U(s)$ preserve the spectrum of $H$ and therefore the spectral gap as well. This is because, if $\lambda$ is an eigenvalue of $H$ with eigenvector $|\psi\rangle$, then $U(s)|\psi\rangle$ is an eigenvector of $U(s)HU(s)^{\dagger}$ with the same eigenvalue $\lambda$ because $U(s)HU(s)^{\dagger}U(s)|\psi\rangle=U(s)H|\psi\rangle=\lambda U(s)|\psi\rangle$.
The relation to circuit complexity now follows from the fact that the operator $U(1)$ can be approximated by a constant depth quantum circuit via Trotterization.
Trotterization proceeds by splitting the time evolution (here the integral $\int_{0}^1 ds \, \tilde{H}(s)$) into small pieces, which are approximately local. 
The error this approximation introduces can be made arbitrarily small by increasing the circuit depth. 
In addition to this general result, different notions of quantum complexity presented in this review, including Nielsen and Krylov complexities, have also been explored in examples of specific systems exhibiting topological phase transitions, see, e.g., \cite{Liu:2019aji,Xiong:2019xoh,Ali:2018aon,Caputa:2022eye}.

\subsection{Bounding complexity growth with entanglement}
\label{sec_qi_entnglmt}
Quantum circuit complexity of a state quantifies the elementary resources necessary to prepare it from a product state.
Therefore, complexity also quantifies the resources necessary to transform the state back into a product state.
Clearly, therefore, complexity can be viewed as a measure of entanglement, which is ultimately the failure of a state to be of product form.
As such, we can compare it to the entanglement entropy, arguably the most prominent measure of entanglement.
A high entanglement entropy along a bipartite cut does not imply a high circuit complexity: the maximally entangled state between two systems each consisting of $n$ qubits can be prepared on a quantum computer with a constant depth quantum circuit. 
Even if we consider the maximum entanglement entropy over all bipartite cuts of the qubits, we find that it saturates trivially at a maximum value of $n$, whereas the circuit complexity can be exponentially large, of order 
$\Omega(2^n/n)$ (compare Section~\ref{sec_growth}). On the other hand, we can easily show the existence of quantum circuits with superpolynomial circuit complexity in the system size but no entanglement between some parts of the system. 
For example, simply apply a Haar random unitary to $\log^{1+\varepsilon}(n)$ many qubits and leave all the other qubits untouched.
So the circuit complexity and the entanglement entropy are not equivalent; but what can we say about their relationship?

Entanglement growth implies early-time bounds on complexity during chaotic evolution. More precisely, it turns out that entanglement \textit{does} imply non-trivial circuit lower bounds, but entanglement needs to be present along any bipartite cut~\cite{eisert2021entangling}.
The reason is that the Schmidt rank along any bipartite cut can only be increased by a constant factor. In particular, if a state $|\psi\rangle$ has a Schmidt rank of $\mathrm{SR}_{i}$ for a bipartite cut of all qubits into two sets $A=\{1,\ldots,i\}$ and $B=\{i+1,\ldots,n\}$, then we know that at least $c\log(\mathrm{SR}_{i})$ gates (each acting on a qubit in $A$ and a qubit in $B$) are required to prepare $|\psi\rangle$.   Suppose that we only want to lower bound the size of any geometrically local circuit that approximates the state $|\psi\rangle$. A geometrically local circuit can only apply gates to neighbouring qubits $(i,i+1)$. 
Then, any gate that acts on $A$ and $B$ necessarily acts on $(i,i+1)$.
We can thus apply the above argument for all $n$ Schmidt ranks $\mathrm{SR}_i$ without overcounting and find:
\begin{equation}
    \min_{\text{geom. loc. circuits}}\#\mathrm{gates}\geq \sum_{i=1}^n c\log(\mathrm{SR}_i).
    \label{eq:geom_local_circuits}
\end{equation}
At best, the left hand side is $\Omega(n^2)$. This is because each Schmidt rank $\mathrm{SR}_i$ is trivially bounded by the state space dimension $2^n$. Therefore, it is impossible to prove exponential circuit complexity with this technique.
Using similar tools, complexity was lower-bounded in one-dimensional models composed by noninteracting fermions~\cite{Aravinda:2023pcp}.  
Reference~\cite{Milekhin:2024mce} argued that the area of Hartman-Maldacena surfaces (\ie codimension-two minimal surfaces homologous to a boundary subregion \cite{Hartman:2013aa}) lower-bounds the Schmidt rank of the unitary time evolution operator.
In turn, the Schmidt rank lower-bounds the depth of any boundary circuit effecting such boundary unitary.

Next, let us mention a relation between circuit complexity and the so-called \textit{embezzlement of entanglement}. 
The latter terminology refers to a process in which entanglement is  extracted from a resource system by using local unitary operators, with approximately no detectable change to the resource state \cite{Hayden_embezz}. 
The embezzlement error estimates the amount by which this process changed the resource state.
References~\cite{vanLuijk:2024fdw,vanLuijk:2024nnx} argued that any state of a relativistic QFT could serve as a resource for the process, where the embezzlement error can be made arbitrarily small.
Recently, the author of \cite{Schwartzman:2024zim} showed that circuit complexity is lower-bounded by a quantity proportional to the inverse of the embezzlement error. 
In other words, circuit complexity acts as an obstruction to perfect embezzlement.

Entanglement, the subject of this subsection until now, does not suffice to enable quantum-computational speedups: classical computers can efficiently simulate even some highly entangled states' evolutions. The power behind certain entangled states, which enables quantum-computational speedups, is called \emph{magic} and \emph{nonstabilizerness}. A possible relationship between magic and circuit complexity has been suggested recently.

To introduce the relationship, we first overview relevant background~\cite{97_Gottesman_PhD,nielsen2010quantum}. Consider an $n$-qubit system. The \emph{Pauli group} consists of $4^n$ \emph{Pauli strings}, each a tensor product of $n$ Pauli operators and/or identity operators. The \emph{Clifford group} is the Pauli group's normalizer; each Clifford gate transforms every Pauli string into a Pauli string.
The Hadamard, phase, and controlled-NOT gates span the Clifford group.
The \emph{stabilizer states} result from acting on $\ket{0}^{\otimes n}$ with Clifford gates.
According to the Gottesman-Knill theorem, classical computers can efficiently simulate every quantum computation formed only from Clifford gates (such that the quantum computer always remains in the space of stabilizer states)~\cite{Gottesman:1998hu}.
Consequently, quantum-computational speedups stem from nonstabilizer states. The extra spice added by these states is called magic or nonstabilizerness~\cite{Veitch_2014}. This topic has recently advanced rapidly (e.g.,~\cite{Liu:2020yso,White:2020zoz,Leone:2021rzd,Bu:2022ozl,Wang:2023uog,Tirrito:2023fnw,Turkeshi:2023lqu,Frau:2024qmf,Cao:2024nrx,Hoshino:2025jko}).
One can quantify magic in various ways, including with the \textit{stabilizer R\'{e}nyi entropy}, an entropic measure defined in terms of the Bell basis~\cite{Leone:2021rzd,Hoshino:2025jko}. 
The authors of \cite{Bu:2022ozl} define a \emph{magic power} that quantifies a unitary's ability to behave unlike a Clifford gate---to evolve a Pauli string to a nontrivial linear combination of Pauli strings.
The magic power lower-bounds the unitary's circuit complexity~\cite{Bu:2022ozl}. The latter is quantified with Nielsen's circuit complexity, particularly the cost function~\eqref{eq:cost_function_Finsler} evaluated at $p = q_I = 1$.

\subsection{Bounding complexity via concentration}
\label{ssec:concentration}
Another way to bound the circuit complexity of a state and, in particular, the \textit{depth} necessary to prepare it is via correlations in the output probability distributions.
Let $|\psi\rangle=U|0^n\rangle$ be a state on $n$ qubits prepared by a constant depth quantum circuit $U$. 
The state $|\psi\rangle$ can be used to define a probability distribution over the bitstrings  $x\in\{0,1\}^n$ via the Born rule $p_x \coloneqq |\langle x|\psi\rangle|^2$.
We can show that the probability distribution is concentrated in the Boolean cube.
To make this more precise, we can define the orthogonal projector
\begin{equation}
    \Pi_{< k} \coloneqq \sum_{\substack{x\in\{0,1\}^n\\ |x|< k}}|x\rangle\langle x|,
\end{equation}
and $\Pi_{>k}$ defined similarly.
Here $|x|$ denotes the Hamming weight of $x$, i.e. the number of 1s in the bitstring.
Then, we can show that 
\begin{equation}\label{eq:concentrationinshallowcircuits}
    \langle \psi|\Pi_{> (1-c)n}|\psi\rangle\langle \psi|\Pi_{<c n}|\psi\rangle \approx 0
\end{equation}
for a constant $c>0$.
In other words, the mass of the output probability distribution cannot be in two far away regions.

We briefly sketch a proof of the concentration explained above, where we follow the argument presented in Ref.~\cite{anshu2023concentration}.
Readers that are only interested in how to apply such a statement to circuit complexity can skip this paragraph.
$|\psi\rangle$ is the unique ground state of the commuting, gapped and $l$-local Hamiltonian $H=\sum_{i} \mathbbm{1}-U\mathbbm{1}_{i-1}\otimes |0_i\rangle\langle 0_i|\otimes \mathbbm{1}_{n-i}U^{\dagger}$, where $l\leq 2^{\mathrm{depth of}\, U}$.
We can use that Chebyshev polynomials $C_m$ (of the first kind) satisfy $C_m(x)\leq e^{-m\sqrt{1-h}}$ for $x\in [0,h]$ and $C_r(1)=1$.
Notice that the spectrum of $C_m\left(\mathbbm{1}-\frac{H}{n}\right )$ consists of the values $C_m(1-\lambda_i/n)$, where the $\lambda_i$ are the eigenvalues of $H$. 
Therefore, plugging in that the spectral gap of $H$ is $1$, we find for the second highest eigenvalue of $C_m(1-H/n)$ the following bound:
\begin{equation}\label{eq:polyapproximation}
\left|\left||\psi\rangle\langle\psi|- C_m\left(\mathbbm{1}-\frac{H}{n}\right )\right|\right|_{\infty}\leq e^{-m/\sqrt{n}}.
\end{equation}
Choosing $m=c'n$ for another constant $c'>0$ makes this small.
However, we  notice that the operator $C_m\left(\mathbbm{1}-\frac{H}{n}\right )$ is a sum over $lc'n$-local terms. 
In particular, for a string of Hamming weight $<cn$, we have that each bitstring that has overlap with $C_m\left(\mathbbm{1}-\frac{H}{n}\right )|x\rangle $ has Hamming weight at most $(c+lc')n$.
Combining this limitation on the increase in Hamming weight with the approximation~\eqref{eq:polyapproximation} yields
\begin{equation}
    \langle \psi|\Pi_{> (1-c)n}|\psi\rangle\langle \psi|\Pi_{<c n}|\psi\rangle\approx   \langle \psi|C_m\left(\mathbbm{1}-\frac{H}{n}\right )\Pi_{>cn}|\psi\rangle \approx 0.
\end{equation}

Why does this property allow us to prove lower bounds on the depth necessary to prepare a state? 
Consider for an extreme example the GHZ state $|\mathrm{GHZ}\rangle = \frac{1}{\sqrt{2}}(|0^n\rangle +|1^n\rangle)$. 
Clearly, the output probability distribution $p_x$ is a fair coin toss between $0^n$ and $1^n$ and 
\begin{equation}
\langle \mathrm{GHZ}|\Pi_{> (1-c)n}|\mathrm{GHZ}\rangle\langle \mathrm{GHZ}|\Pi_{<c n}|\mathrm{GHZ}\rangle=\frac14,
\end{equation}
for any $c>0$, which is in direct contradiction to the concentration property in Eq.~\eqref{eq:concentrationinshallowcircuits}.
But Eq.~\eqref{eq:concentrationinshallowcircuits} is a direct consequence of the assumption that $|\mathrm{GHZ}\rangle$ was prepared by a constant depth circuit.
In fact, we can show that circuits of depth $\Omega(\log(n))$ are required to prepare $|\mathrm{GHZ}\rangle$. 

This strategy based on concentration in output probability distributions,was used in the proof of Freedman and Hasting's NLTS (``No Low-Energy Trivial States'') conjecture~\cite{freedman2013quantum}  due to Anshu, Breuckmann and Nirkhe~\cite{anshu2023nlts}.
The NLTS theorem asserts that Hamiltonians exist such that any state with an energy density below some constant threshold is topologically ordered in the sense of Section~\ref{sec_qi_topo}.
More precisely, there is a family of $k$-local Hamiltonians $H_n=\sum_{i=1}^m h_i$ acting on $n$ qubits with $m=O(n)$ and an  $\varepsilon>0$ such that all states $|\psi\rangle$ with $\langle\psi|H|\psi\rangle/n\leq \varepsilon$ are topologically ordered.
In particular, this result shows that topological order can persist at constant temperatures.

The techniques discussed in this subsection can only produce polynomial lower bounds on the circuit complexity of a state.
More precisely, the parent Hamiltonian $H=\sum_{i} \mathbbm{1}-U\mathbbm{1}_{i-1}\otimes |0_i\rangle\langle 0_i|\otimes \mathbbm{1}_{n-i}U^{\dagger}$ will be of locality $\sim n$, which does not result in a contradiction with concentration as $\Pi_{>(1-c)n}H\Pi_{<cn}\neq 0$.
\clearpage
\section{Epilogue}
\label{sec:conclusions}

Sofia, Sagredo, and Complexio have now read the  the review and are discussing their takeaways at a journal club.

\vskip 2mm

\textit{Sagredo:}
I have been impatiently awaiting your arrival, as 
I need guidance to identify the major opportunities in quantum-complexity research.
Also, it would be helpful if Sofia could summarize 
the main achievements  so far.

\textit{Sofia:}
I am happy to overview the main findings.
The review  explored several notions of quantum complexity: exact and approximate circuit complexities, Nielsen's geometric approach, notions based on operator and state spreading, and definitions inspired by tensor networks. 
These notions have deepened our understanding of the evolution of quantum systems, including  thermalization and differences between chaos and integrability. Within the AdS/CFT correspondence, complexity appears to be related to certain gravitational observables, which can perhaps elucidate the physics of the region behind black hole horizons.

\textit{Complexio:}
The gravitational quantities seem related to complexity only qualitatively. Is it worth spending so much research effort on hand waving?

\textit{Sofia:} As a matter of fact, Refs.~\cite{Lin:2022rbf,Rabinovici:2023yex,Balasubramanian:2024lqk,Heller:2024ldz} built on \cite{Berkooz:2018qkz} to identify a quantitative relationship recently. 
Section~\ref{ssec:holo_matching_2d} described this breakthrough:
a double-scaled SYK model's spread complexity is dual to the size of an ERB in two-dimensional theories of gravity. The duality, being exact, is a major achievement.

\textit{Complexio:}
Still, the result has a narrow range of validity.
In two bulk dimensions, the graviton is not even dynamical. Should we truly regard this result as a success?

\textit{Sofia:} The duality is simplest in two bulk dimensions, and this simplicity facilitated the proof.
Yet \cite{Balasubramanian:2024lqk} reinterprets the matching in a language that should enable a generalization to higher dimensions.  Additionally, insight into two-dimensional theories has advanced long-standing problems, famously including the black-hole-information paradox~\cite{Penington:2019npb,Almheiri:2019psf,Almheiri:2019hni,Almheiri:2020cfm}. The similar insight here will hopefully enable similar advancements.

\textit{Complexio:} Those opportunities sound appealing. Still, what if nobody proves a duality between spread complexity and ERB volume in higher-dimensional theories? 

\sloppy

\textit{Sofia:} 
Regardless of the holographic duality, studying complexity benefits the quantum-information and quantum-gravity communities.
As the review showed, quantum complexity can help us simulate time evolutions, identify quantum chaos, and more.
In gravitational systems, geometric observables can illuminate the black-hole interior, black-hole evaporation, and causal properties of a spacetime.

\fussy 

\textit{Sagredo:}
You mentioned chaos. Do I recall correctly from Sec.~\ref{ssec:integrability_chaos} that Krylov and spread complexities diagnose chaos?

\textit{Sofia:}
They do to an extent. 
Consider an ensemble of theories, such as the SYK model, distinguished by random values of one or more parameters. Variances associated to Krylov and spread complexities  diagnose chaos~\cite{Balasubramanian:2022dnj,Balasubramanian:2023kwd,Chapman:2024pdw}.  Furthermore, individual chaotic theories, even if we do not consider an ensemble, have more regular tridiagonal spectra (also called Lanczos spectra) than integrable ones. In fact, there is a precise conjecture for the functional form of the tridiagonal spectral covariances in chaotic theories \cite{Balasubramanian:2023kwd}.
Whether and how precisely Krylov complexity distinguishes chaotic from integrable QFTs remains an open question~\cite{Avdoshkin:2022xuw,Camargo:2022rnt}, although again the key distinction should lie in the fluctuations of the tridiagonal spectrum of the Hamiltonian.
Recent advances appear also in Refs.~\cite{Ouseph:2023juq,Gesteau:2023rrx}, which leverage von Neumann algebras and algebraic QFTs. 

\textit{Complexio:} Do the other notions of complexity that we have heard about---apart from Krylov complexity---diagnose chaos?

\textit{Sofia:}
Yes, Sec.~\ref{ssec:constraints_cost_function} discussed constraints on the complexity metric for finite-dimensional quantum system so that Nielsen complexity shows distinguishing features for chaotic dynamics. Similarly, Sec.~\ref{sec:generalresults} established physical criteria that lead to long-term Nielsen complexity growth in metrics that penalize complex versus simple generators. Systems that satisfy these criteria, which include some that show spectral features of chaos like level repulsion, show linear growth of Nielsen complexity for exponential time, at least up to possible obstructions arising from the global structure of the unitary group that are hard to study.  This long-term complexity growth has its roots in the negative average sectional curvature of the complexity geometry on the unitary group manifold~\cite{Brown:2016wib,Brown:2019whu,Balasubramanian:2019wgd,Auzzi:2020idm}.
References~\cite{Brown:2021uov,Brown:2022phc} discussed various classes of complexity geometries that penalize complex operations as desired.  These geometries have a negative average sectional curvature, and are constructed to facilitate analytical computation of geodesics as desired for the study of Nielsen complexity.

\textit{Complexio:} So we can learn about chaos from quantum complexity. Can we also learn from complexity about the holographic dictionary?

\textit{Sofia:}
Certainly---we have known for years that quantum information theory can reveal which part of the bulk encodes which feature of the boundary theory. This research program relates  to the claim that slices of AdS geometry naturally arise in tensor networks \cite{Swingle:2009bg}. The claim inspired notions of complexity discussed in Sec.~\ref{sec.pathintegralopt}. These notions may help us identify holographically dual quantities.

\textit{Sagredo:}
As your response highlights, researchers have devoted substantial work to AdS spacetimes. However, we live in an expanding universe. Can complexity help us understand quantum-gravity theories for expanding universes, such as our universe and de Sitter space?

\textit{Sofia:}
Perhaps. Whether de Sitter spacetime has a quantum dual remains an open question. If the answer is yes, then the methods of the AdS/CFT duality may port over to de Sitter spacetime, perhaps with some modifications. If so, we can
translate quantum-gravity questions into conceptually simpler questions about non-gravitating quantum systems. For now, people are conjecturing such a translation, as well as conjecturing about duals of complexity~\cite{Gibbons:1977mu,Bousso:1999dw,Banks:2000fe,Bousso:2000nf,Banks:2001yp,Strominger:2001pn,Banks:2002wr,Leblond:2002ns,Leblond:2002tf,Parikh:2002py,Dyson:2002nt,Dyson:2002pf,Parikh:2004wh,Banks:2005bm,Freivogel:2005qh,Banks:2006rx,Lowe:2010np,Anninos:2011af,Fischetti:2014uxa,Anninos:2017hhn,Anninos:2018svg,Banks:2018ypk,Dong:2018cuv,Lewkowycz:2019xse,Banks:2020zcr,Mirbabayi:2020grb,Anninos:2020cwo,Anninos:2020hfj,Susskind:2021omt,Susskind:2021dfc,Coleman:2021nor,Susskind:2021esx,Shaghoulian:2021cef,Shaghoulian:2022fop,Lin:2022nss,Banihashemi:2022htw,Silverstein:2022dfj,Chandrasekaran:2022cip,Witten:2023xze,Witten:2023qsv,Rahman:2023pgt,Narovlansky:2023lfz,Batra:2024kjl,Verlinde:2024znh,Verlinde:2024zrh,Tietto:2025oxn,Chapman:2021eyy,Jorstad:2022mls,Auzzi:2023qbm,Anegawa:2023wrk,Anegawa:2023dad,Baiguera:2023tpt,Aguilar-Gutierrez:2023zqm,Baiguera:2024xju,Aguilar-Gutierrez:2023pnn,Aguilar-Gutierrez:2024rka,Aguilar-Gutierrez:2024nau}. 
Hopefully, exploring holographic complexity in de Sitter spacetime will indicate which properties the spacetime's quantum dual should have, if it exists.

Such a development would unlock further opportunities for future work: some researchers expect the quantum dual of de Sitter space to be an open quantum system~\cite{Anninos:2020cwo,Anninos:2022qgy}.  If this is the case, we may learn about  complexity of dynamics in open quantum systems through de Sitter holography. Several works have already begun unpacking complexity in open quantum systems~\cite{Bhattacharyya:2021fii,Bhattacharyya:2022rhm,Bhattacharya:2022gbz,YungerHalpern:22Resource,Liu:2022god,Schuster:2022bot,Bhattacharjee:2022lzy,NSSrivatsa:2023pby,Bhattacharyya:2024duw,Ermakov:2024xtn,Schuster:2024jds,Guerra:2025dam}.
 
\textit{Complexio:}
Also on the topic of holography, according to the review, holographers called for the development of quantum-information-theoretic tools, including a resource theory for quantum complexity (see Sec.~\ref{sec_qi_operational}). 
Quantum information theorists defined the resource theory in~\cite{YungerHalpern:22Resource}. In turn, the resource theory led to new entropic measures of complexity~\cite{YungerHalpern:22Resource,Munson:2024Complexity}. Can these made-to-order tools now benefit holography?

\textit{Sofia:}
I hope so. For example, holographers have imagined an observer Alice falling into a black hole and fearing that she will hit a firewall. If Bob throws a maximally mixed qubit into the black hole, holographers have noted, the qubit will alleviate Alice's danger to an extent. We can now progress beyond that simple qubit. What if Bob throws a quantum system, of arbitrary dimensionality, in an arbitrary state? How much will he alleviate Alice's danger? Our new quantum-information-theoretic tools should enable general, quantitative answers.

\textit{Complexio:} The partnership between quantum information theorists and high-energy physicists is promising.
However, I heard that computer scientists 
have been suspicious of the duality proposed between volume and quantum complexity in the AdS/CFT correspondence. 

\textit{Sofia:}
You heard correctly. This skepticism led to the research topic that debuted in~\cite{Bouland:2019pvu}: the holographic dictionary's complexity. According to Sec.~\ref{ssec:pseudorandomness}, translating between complexity and volume takes a lot of time unless quantum gravity allows for more computational power than a quantum computer.
Computer scientists have proposed that the holographic dictionary might be highly complex. Their argument extends beyond the complexity=volume conjecture to other holographic complexity conjectures.
Followup work focused on the duality between area and entanglement entropy~\cite{akers2024holographic}. Still, we are not certain for which entries the holographic dictionary is complex. Ramifications will include our ability to study quantum gravity through quantum computing.  

\textit{Sagredo:} These holographic concerns bring to mind your earlier claim that, even in the absence of any duality, quantum complexity merits studying within quantum information theory. Sofia, you have already pointed out some open quantum-information-theoretic problems centered on quantum complexity. Can you list more?

\textit{Sofia:}
Certainly; I have three in mind. According to Sec.~\ref{sec:what_is_quantum_complexity}, physicists do not know how random quantum circuits can realize the switchback effect. Such a realization would strengthen the relationship between quantum chaos and random quantum circuits.

Second, we should lower-bound the circuit complexities of states in families that we can describe efficiently. We already have a lower bound $\sim n\,\log(n)$ as discussed in Sec.~\ref{ssec:concentration}.
Such an endeavor may encounter obstacles that classical computer scientists face. 
More precisely, pseudorandom functions could hinder the establishment of such circuit-complexity lower bounds.
The reason, roughly, is the following: imagine a quantity that distinguishes low-complexity-functions from high-complexity-functions. This quantity must, itself, be hard to compute.
Otherwise, one could easily use the quantity to distinguish pseudorandom functions (low complexity) from random functions (high complexity).
The resulting obstacles for lower-bounding classical circuit complexity are called \emph{natural proof barriers}~\cite{razborov2004feasible}.
We might develop the quantum analog of such natural proof barriers by strengthening a recent construction of pseudorandom unitaries~\cite{ma2024construct}.

Third, researchers expect that  the low-energy states of some local Hamiltonians have superpolynomial circuit complexities.
However, proving an unconditional superpolynomial lower bound 
would come close to separating the complexity classes $\mathrm{QMA}$ (Quantum Merlin-Arthur) and $\mathrm{QCMA}$ (Quantum-Classical Merlin-Arthur)~\cite{aaronson2007quantum}.
QMA contains decision problems that one can verify with a quantum proof; and QCMA, decision problems that a quantum computer can verify with a classical proof.
Unconditionally separating QMA and QCMA, one would achieve a monumental breakthrough in theoretical computer science by also separating the classes P and PSPACE (see \cite{aaronson2016p} for details). 
A more modest goal is to separate QMA from QCMA ``relative to an oracle''.
An oracle is a black box that solves a powerful problem in one step.
Often, unconditional statements about complexity classes can be made if all Turing machines are given access to such an oracle. 
The problem of separating QMA from QCMA relative to increasingly weaker oracles was the subject of significant efforts in the past years~\cite{aharonov2002quantum,aaronson2007quantum,fefferman2015quantum,natarajan2024distribution,li2023classical,liu2023non,liu2024qma,zhandry2024toward}.

Let me mention, as an aside, that the mathematics community has studied problems related to circuit optimization.
They have focused on the minimum number of steps required to reach all the points on certain group manifolds. We can recognize this number as a type of complexity. Finding gates that minimize this complexity amounts to solving the \emph{golden-gates problem}~\cite{PARZANCHEVSKI2018869,sarnak2637letter,evra2022ramanujan}.
That problem, in turn, relates to the famous mathematical problem of sphere packing. So complexity is really everywhere!

\textit{Sagredo:}
And let this conclude today's discussion. Clearly, many opportunities await us. I look forward to reading a future review that satisfies our curiosity about them!
\clearpage

\pagebreak

\section*{Acknowledgments}
\noindent We would like to thank Brian Swingle and Beni Yoshida for their involvement in the early stages of this review. We are grateful to our collaborators on the topic of quantum complexity who significantly contributed to both the knowledge development described here, as well as to our viewpoints on open questions in this area. Furthermore, we thank Igal Arav, Andrea Legramandi, Javier Magan, Tim Schuhmann and Paolo Stornati for valuable discussions during the writing process. We especially thank Saskia Demulder, Damian Galante, Tim Schuhmann, Tal Schwartzman and Zixi Wei  for insightful discussions and comments on the draft. 
The authors gratefully acknowledge the organizers and participants of the workshop ``EuroStrings 2024'' in Southampton, United Kingdom, where part of this work was carried out. The work of S.~B. was supported by the INFN grant ``Gauge Theories and Strings'' (GAST), via a research grant on ``Holographic dualities, quantum information and gravity.'' 
V.~B. is supported in part by the DOE
through DE-SC0013528 and QuantISED grant DE-SC0020360, and in part by the Eastman
Professorship at Balliol College, University of Oxford.
The work of S.~B. and S.~C. was supported by the Israel Science Foundation (grant No. 1417/21), by the German Research Foundation through the German-Israeli Project Cooperation (DIP) grant ``Holography and the Swampland,'' by Carole and Marcus Weinstein through the BGU Presidential Faculty Recruitment Fund, and by the ISF Center of Excellence for theoretical high-energy physics. S.~C. is also supported by the ERC starting Grant dSHologQI (project number 101117338). P.~C. is supported by NCN Sonata Bis 9 2019/34/E/ST2/00123 grant and by an ERC Consolidator grant (number: 101125449/acronym: QComplexity). Views and opinions expressed are, however, those of the authors only and do not necessarily reflect those of the European Union or the European Research Council. Neither the European Union nor the granting authority can be held responsible for them. N.~Y.~H. received support from the National Science Foundation 
(QLCI grant OMA-2120757).
J.~H. is supported by the Harvard Quantum Initiative (HQI).
M.~P.~H. has received support from the European Research Council (ERC) under the European Union’s Horizon 2020 research and innovation programme (grant number: 101089093 / project acronym: High-TheQ).

\appendix

\section{Guide to acronyms}
\label{app_Acronyms}

\begin{longtable}{r|l}
\textbf{Acronym} &  \textbf{Full form} \\[1.2mm] \hline
\rule{0pt}{4ex}    
AdS &   anti-de Sitter \\[1.2mm] 
$\mathcal{BC}$ &   binding complexity \\[1.2mm] 
BCFT &    boundary conformal field theory \\[1.2mm]
CA &  complexity=action \\[1.2mm]
CAny &   complexity=anything \\[1.2mm]
CFT    &      conformal field theory  \\[1.2mm]
cMERA &  continuous multiscale entanglement renormalization ansatz \\[1.2mm]
CV &   complexity=volume \\[1.2mm]
CV2.0 &   complexity=volume 2.0 \\[1.2mm]
DSSYK & double-scaled Sachdev-Ye-Kitaev \\[1.2mm]
ECH & eigenstate complexity hypothesis \\[1.2mm]
EOW & end-of-the-world \\[1.2mm]
ERB &  Einstein-Rosen bridge \\[1.2mm]
EW & entanglement wedge \\[1.2mm]
FS &  Fubini-Study \\[1.2mm]
GHZ &  Greenberger-Horne-Zeilinger \\[1.2mm]
GUE &  Gaussian unitary ensemble \\[1.2mm]
HRT & Hubeny-Rangamani-Takayanagi  \\[1.2mm]
IR &     infrared \\[1.2mm]
JT & Jackiw-Teitelboim \\[1.2mm] 
LOCC &   local operations and classical communication \\[1.2mm]
MERA &  multiscale entanglement renormalization ansatz \\[1.2mm] 
NLTS & no low-energy trivial state \\[1.2mm]
OTOC &  out-of-time-ordered correlator \\[1.2mm]
QCMA & quantum classical Merlin-Arthur \\[1.2mm]
QFT                   & quantum field theory  \\[1.2mm]
QM                & quantum mechanics  \\[1.2mm]
QMA & quantum Merlin-Arthur \\[1.2mm]
RMT & random matrix theory \\[1.2mm]
SFF &  spectral form factor \\[1.2mm]
SYK & Sachdev-Ye-Kitaev \\[1.2mm]
TFD &  thermofield-double \\[1.2mm]
TN & tensor network \\[1.2mm]
UV &     ultraviolet \\[1.2mm]
WDW &   Wheeler-De Witt \\[1.2mm]
\end{longtable}

\bibliography{references.bib}

\end{document}